\documentclass[11pt,a4paper]{article}
\pdfoutput=1

\usepackage[colorlinks=true, linkcolor=black!50!blue, urlcolor=blue, citecolor=blue, anchorcolor=blue]{hyperref}
\usepackage[font=small,labelfont=bf,margin=0mm,labelsep=period,tableposition=top]{caption}
\usepackage[a4paper,top=1.5cm,bottom=2cm,left=1.5cm,right=1.5cm,bindingoffset=0mm]{geometry}

\usepackage{placeins,setup}
\usepackage{graphicx}
\usepackage{float}
\usepackage{afterpage}
\usepackage{epsfig}
\usepackage{amssymb}
\usepackage{amsmath}
\usepackage{bm}
\usepackage{multirow}
\usepackage{url}
\usepackage{xcolor}
\usepackage{float}
\usepackage{afterpage}
\usepackage{ulem}
\usepackage{url}
\usepackage{booktabs,multirow,longtable}
\usepackage{cite}

\makeatletter
\def\thickhline{%
             \noalign{\ifnum0 =`}\fi\hrule \@height \thickarrayrulewidth \futurelet
             \reserved@a\@xthickhline}
\def\@xthickhline{\ifx\reserved@a\thickhline
                \vskip\doublerulesep
                \vskip -\thickarrayrulewidth
                \fi
                \ifnum0 =`{\fi}}
\makeatother
\newlength{\thickarrayrulewidth}
\usepackage{xcolor}
\definecolor{mtplotlib1}{HTML}{1F77b4}
\definecolor{mtplotlib2}{HTML}{FF7F0E}
\definecolor{mtplotlib3}{HTML}{2CA02C}
\definecolor{mtplotlib4}{HTML}{D62728}
\usepackage{tikz}
\usetikzlibrary{shapes,arrows,positioning}
\tikzstyle{block}=[rectangle, draw, fill=mtplotlib1!30, text width=\textwidth, 
                   text centered, rounded corners, minimum height=2em, 
                   minimum width=10em]
\tikzstyle{block1}=[rectangle, draw, fill=mtplotlib2!30, text width=0.4\textwidth, 
                   text centered, rounded corners, minimum height=2em, 
                   minimum width=10em]
\tikzstyle{block2}=[rectangle, draw, fill=mtplotlib3!30, text width=0.4\textwidth, 
                   text centered, rounded corners, minimum height=2em, 
                   minimum width=10em]
\tikzstyle{block3}=[rectangle, draw, fill=mtplotlib4!30, text width=0.4\textwidth, 
                   text centered, rounded corners, minimum height=2em, 
                   minimum width=10em]

\tikzstyle{line} =[-latex']
\tikzstyle{decision} = [diamond, draw, fill=red!20, text width=7.5em, text centered,  inner sep=0pt, minimum height=2em, aspect=4]

\newcommand{\be}{\begin{equation}}
\newcommand{\ee}{\end{equation}}
\newcommand{\bea}{\begin{eqnarray}}
\newcommand{\eea}{\end{eqnarray}}
\newcommand{\bi}{\begin{itemize}}
\newcommand{\ei}{\end{itemize}}
\newcommand{\ben}{\begin{enumerate}}
\newcommand{\een}{\end{enumerate}}

\newcommand{\lc}{\left[}
\newcommand{\rc}{\right]}
\newcommand{\lp}{\left(}
\newcommand{\rp}{\right)}

\def\frac#1#2{{{#1}\over {#2}}}
\def\gsim{\mathrel{\rlap{\lower4pt\hbox{\hskip1pt$\sim$}}
    \raise1pt\hbox{$>$}}}       
\def\lsim{\mathrel{\rlap{\lower4pt\hbox{\hskip1pt$\sim$}}
    \raise1pt\hbox{$<$}}}

\newcommand{\draft}[1]{}

\def\beq{\begin{equation}}
\def\eeq{\end{equation}}

\numberwithin{equation}{section}
\numberwithin{figure}{section}
\numberwithin{table}{section}

\usepackage{tabularx}
\newcolumntype{C}[1]{>{\centering\arraybackslash}p{#1}}

\begin{document}
\vspace{-2.0cm}
\begin{flushright}
Nikhef-2021-028\\
BONN-TH-2021-14
\end{flushright}
\vspace{0.3cm}

\begin{center}
  {\Large \bf nNNPDF3.0: Evidence for a modified partonic  structure in heavy nuclei}\\
  \vspace{1.1cm}
  {\small
    Rabah Abdul Khalek,$^{1,2}$
    Rhorry Gauld,$^{3}$
    Tommaso Giani,$^{1,2}$
    Emanuele R. Nocera,$^{4}$\\
    Tanjona R. Rabemananjara,$^{1,2,5}$
    and Juan Rojo$^{1,2}$
  }\\
  
\vspace{0.7cm}

{\it \small
  ~$^1$Department of Physics and Astronomy, VU Amsterdam, 1081HV Amsterdam,
  The Netherlands\\[0.1cm]
  ~$^2$Nikhef Theory Group, Science Park 105, 1098 XG Amsterdam,
  The Netherlands\\[0.1cm]
  ~$^3$Physikalisches Institut, University of Bonn, D-53115 Bonn,
  Germany\\[0.1cm]
  ~$^4$The Higgs Centre for Theoretical Physics, University of Edinburgh,\\
  JCMB, KB, Mayfield Rd, Edinburgh EH9 3JZ, Scotland\\[0.1cm]
  ~$^5$Tif Lab, Dipartimento di Fisica, Universit\`a di Milano\\
  and INFN, Sezione di Milano, Via Celoria 16, I-20133 Milano, Italy\\[0.1cm]
}
\vspace{1.0cm}

{\bf \large Abstract}

\end{center}  

We present an updated determination of nuclear parton distributions (nPDFs) from a global 
NLO QCD analysis of hard processes in fixed-target lepton-nucleus
and proton-nucleus together with collider proton-nucleus experiments.
In addition to neutral- and charged-current deep-inelastic 
and Drell-Yan measurements on nuclear targets, we consider the information provided by the 
production of electroweak gauge bosons, isolated photons, jet pairs, and charmed mesons in 
proton-lead collisions at the LHC across centre-of-mass energies of 5.02 TeV (Run I) and 8.16 TeV (Run II).
For the first time in a global nPDF analysis, the constraints from these various processes are 
accounted for both in the nuclear PDFs and in the free-proton PDF baseline.
The extensive dataset 
underlying the nNNPDF3.0 determination, combined with its model-independent parametrisation, 
reveals strong evidence for nuclear-induced modifications of the partonic structure of heavy nuclei,
specifically for the small-$x$ shadowing of gluons and sea quarks, as well as the large-$x$ anti-shadowing 
of gluons.
As a representative phenomenological application, we provide predictions for 
ultra-high-energy neutrino-nucleon cross-sections, relevant for data interpretation at neutrino 
observatories.
Our results provide key input for ongoing and future experimental programs, from that of heavy-ion 
collisions in controlled collider environments to the study of high-energy
astrophysical processes.

\clearpage

\tableofcontents

\clearpage

\section{Introduction}
\label{sec:introduction}

The parton distribution functions (PDFs) of nuclei, known as nuclear PDFs
(nPDFs)~\cite{Rojo:2019uip,Kovarik:2019xvh,Ethier:2020way}, are essential to a
variety of experimental programs that collide nuclei (or nuclei with protons)
at high energies~\cite{Fischer:2014wfa}.
At the Large Hadron Collider (LHC),
nPDFs are required as a theoretical input to the heavy ion program that aims to
disentangle cold from hot nuclear matter effects, making the detailed
characterisation of the latter possible. Ongoing and upcoming heavy
ion runs include high-luminosity proton--lead (pPb) and lead--lead (PbPb)
collisions~\cite{Jowett:2018yqk}, dedicated proton--oxygen (pO) and
oxygen--oxygen (OO) runs~\cite{Brewer:2021kiv}, and a fixed-target
mode~\cite{Hadjidakis:2018ifr}, where energetic proton beams collide with a
nuclear gas target filling the detectors.
A reliable determination
of nPDFs is also critical in various astrophysical processes, for instance to
make theoretical predictions of signal~\cite{Cooper-Sarkar:2011jtt,
  Bertone:2018dse,Garcia:2020jwr,Connolly:2011vc} and background~\cite{Gauld:2015yia,Garzelli:2016xmx,
  Zenaiev:2019ktw} events in neutrino--nucleus scattering as
measured at high-energy neutrino observatories such as
IceCube~\cite{IceCube:2006tjp} and
KM3NeT~\cite{KM3Net:2016zxf}. In the longer term, nPDFs will be probed
at the Electron Ion Collider (EIC)~\cite{AbdulKhalek:2021gbh,Anderle:2021wcy,
  Khalek:2021ulf}, by means of GeV-scale lepton scattering on light and heavy
nuclei, and further tested at the proposed Forward Physics Facility
(FPF)~\cite{Anchordoqui:2021ghd}, by means of TeV-scale neutrino scattering on
heavy nuclear targets.

In addition to their phenomenological role in the modelling of high-energy collisions involving
nuclei, nPDFs are also a means
towards an improved understanding of Quantum Chromodynamics (QCD) in the
low-energy non-perturbative regime.
Indeed,
a global analysis of nPDFs provides a
determination of the behaviour of nuclear modifications which are flavour,
atomic mass number $A$, and $x$ dependent.
This information is critical to
isolate the source of the different physical mechanisms that are responsible
for the modification of nPDFs in comparison to their free-nucleon counterparts,
such as (anti-)shadowing, the EMC effect, and Fermi motion.
A detailed
understanding of these effects is also necessary to provide
a robust baseline in the search for more exotic
forms of QCD matter, such as the gluon-dominated Color Glass
Condensate~\cite{Gelis:2010nm}, which is predicted to have an enhanced
formation in heavy nuclei.

Knowledge of nPDFs also enters, albeit indirectly, the precise determination of
free-proton PDFs. These are typically determined from global analyses that
include, among others, measurements of neutrino--nucleus structure functions.
These measurements uniquely constrain quark flavour separation at intermediate
and large $x$.
Uncertainties due to nPDFs may inflate the overall free-proton
PDF uncertainty by up to a factor of two at large $x$ when properly
taken into account~\cite{Ball:2018twp,Ball:2020xqw,Ball:2021leu}. More precise
nPDFs can therefore lead to more precise free-proton PDFs.

Taking into account these considerations, and building upon previous
studies by some of us~\cite{AbdulKhalek:2019mzd,AbdulKhalek:2020yuc}, here we
present an updated determination of nuclear PDFs from an extensive global
dataset: nNNPDF3.0. This determination benefits from the inclusion of new data
and improvements in the overall fitting methodology.
Concerning experimental data, a broad range of measurements from pPb
collisions at the LHC  are considered. Notably, CMS dijet~\cite{CMS:2018jpl}
and LHCb $D^0$-hadron~\cite{LHCb:2017yua} production data is included in both the
proton baseline and the nPDF determination, revealing gluon PDF shadowing at
small $x$.
We also explicitly remove all deuterium data ($A= 2$) from the 
proton baseline fit and include it in the nPDF fit. Concerning the methodology,
we follow the NNPDF3.1 approach~\cite{Ball:2017nwa} (but with an extended
dataset) for the proton baseline fit, while we supplement the nNNPDF2.0
methodology~\cite{AbdulKhalek:2020yuc} for the nuclear fit with a number 
of technical improvements in our machine learning framework, in particular
with automated hyperparameter optimisation~\cite{Carrazza:2019mzf}.

The combination of these various developments leads to substantially
improved nPDFs for all partons in a wide range of $x$ values. In particular,
we obtain strong evidence of nuclear shadowing for
both the gluon and quarks in the region of small-$x$ and large-$A$ values, as
well as of gluon anti-shadowing at large $x$ and large $A$.
As customary,
the nNNPDF3.0 parton sets are made publicly available for all
phenomenologically relevant values of $A$ via the
{\sc\small LHAPDF}~\cite{Buckley:2014ana} interface.

The outline of this paper is the following. In Sect.~\ref{sec:expdata} we
provide details of the experimental measurements and theoretical computations
used as input for the nNNPDF3.0 determination. The updates in fitting
methodology are described in Sect.~\ref{sec:fitting}.
Sect.~\ref{sec:results} presents the main results of this work, namely the
nNNPDF3.0 parton sets, their comparison with other determinations
(EPPS16~\cite{Eskola:2016oht} and nCTEQ15WZSIH~\cite{Duwentaster:2021ioo}),
and an assessment of the dependence of the extracted nuclear modifications
with respect to flavour, $A$, and $x$. The stability of the nNNPDF3.0 fit is
studied in Sect.~\ref{sec:stability}, where variants based on different
kinematic cuts, proton PDF baseline sets, and theoretical settings are
presented.
As a representative phenomenological application of nNNPDF3.0, in Sect.~\ref{sec:uheneut} we provide
predictions for the ultra-high-energy neutrino-nucleon cross-section for
several $A$ values as relevant for (next-generation) large-volume neutrino
detectors.
We indicate which nPDF sets are made available via
{\sc\small LHAPDF}, summarise our findings, provide usage prescriptions 
for the nNNPDF3.0 sets, and discuss some avenues for
possible future investigations in Sect.~\ref{sec:summary}.

Four appendices complete the paper. App.~\ref{app:conventions} summarises the
notation used through the paper; App.~\ref{app:asymmetry} clarifies the
kinematic transformation required to analyse pPb measurements in different
reference frames; App.~\ref{app:datacomp} collects several data/theory
comparisons for measurements included and not included in nNNPDF3.0;
and App.~\ref{app:RWex} presents a validation of the reweighting method used to
assess the impact of LHCb $D$-meson measurements.

\section{Experimental data and theoretical calculations}
\label{sec:expdata}

In this section we discuss the experimental data and theoretical calculations
that form the basis of the nNNPDF3.0 analysis.
We present an overview of the
nNNPDF3.0 dataset and general theoretical settings, and then focus on
the new measurements that are added in comparison to nNNPDF2.0.
For each
of these, we describe their implementation and the calculation of the
corresponding theoretical predictions. 

\subsection{Dataset overview}
\label{sec:dataset_overview}

The nNNPDF3.0 dataset includes all the measurements that were part of
nNNPDF2.0. These are the neutral-current (NC) isoscalar nuclear fixed-target
deep-inelastic scattering (DIS) structure function ratios measured, for a range
of nuclei, by the NMC~\cite{NewMuon:1995cua,NewMuon:1995tgs,NewMuon:1996yuf,
  NewMuon:1996gam}, EMC~\cite{EuropeanMuon:1988lbf,EuropeanMuon:1989mef,
  EuropeanMuon:1987obv,EuropeanMuon:1992pyr},
SLAC~\cite{Gomez:1993ri}, BCDMS~\cite{Alde:1990im,BCDMS:1987upi} and
FNAL~\cite{E665:1995xur,FermilabE665:1992rbv} experiments; the charged-current
(CC) neutrino-nucleus DIS cross-sections measured, respectively on Pb and Fe
targets, by the CHORUS~\cite{CHORUS:2005cpn} and NuTeV~\cite{NuTeV:2001dfo}
experiments; and cross-sections, differential in rapidity, for inclusive gauge
boson production in pPb collisions measured by the ATLAS~\cite{ATLAS:2015mwq}
and CMS~\cite{CMS:2015zlj,CMS:2015ehw,CMS:2019leu} experiments at the
LHC. These datasets are discussed in~\cite{AbdulKhalek:2020yuc}, to which
we refer the reader for further details.

In addition, two different groups of new measurements are incorporated into
the nNNPDF3.0
dataset. The first group consists of DIS and fixed-target Drell--Yan (DY)
measurements that involve deuterium targets, specifically the
NMC~\cite{NewMuon:1996uwk} deuteron to proton DIS structure functions; 
the SLAC~\cite{Whitlow:1991uw} and BCDMS~\cite{BCDMS:1989qop} deuteron
structure functions;  and the E866~\cite{NuSea:2001idv} fixed-target DY
deuteron to proton cross-section ratio.
This group also includes the
fixed-target DY measurement performed on Cu by the E605
experiment~\cite{Nagarajan:1991zz}.
In nNNPDF2.0 these datasets entered the
determination of the free proton PDF used as baseline in the nuclear PDF fit.
As will be discussed in Sect.~\ref{sec:fitting}, in nNNPDF3.0 we no longer
include them in the proton PDF baseline but rather choose to include them
directly in the nuclear fit.
This approach is conceptually more consistent, as
it completely removes any residual nuclear effects from the proton baseline.
Information loss on quark flavour separation for proton PDFs due to the
removal of these datasets is partially compensated by the availability of
additional measurements from proton--proton (pp)
collisions~\cite{Ball:2021leu}, specifically concerning new
weak gauge boson production data.

The second group of new measurements entering nNNPDF3.0 consists of LHC pPb
data.
Specifically, we consider forward and backward rapidity fiducial
cross-sections for the production of $W^\pm$ bosons measured by
ALICE~\cite{ALICE:2016rzo} at $\sqrt{s}=5.02$~TeV in the centre-of-mass (CoM)
frame; forward and backward rapidity fiducial cross-sections for the production
of $Z$ bosons measured by ALICE and LHCb at
5.02~TeV~\cite{ALICE:2016rzo,LHCb:2014jgh} and
8.16~TeV~\cite{ALICE:2020jff}; the differential cross-section for the
production of $Z$ bosons measured by 
CMS at 8.16~TeV~\cite{CMS:2021ynu}; the ratio of pPb to pp
differential cross-sections for dijet production measured by CMS 
at 5.02~TeV~\cite{CMS:2018jpl}; the ratio of pPb to pp differential
cross-sections for prompt photon production measured by ATLAS at
8.16~TeV~\cite{Aaboud:2019tab}; and the ratio of pPb to pp differential
cross-sections for prompt $D^0$ production measured at forward rapidities
by LHCb at 5.02~TeV~\cite{LHCb:2017yua}. The impact of the last measurement
on nNNPDF3.0 is evaluated by means of Bayesian
reweighting~\cite{Ball:2010gb,Ball:2011gg}, while that of all of the others
by means of a fit, see Sect.~\ref{sec:fitting}.
All these measurements are discussed in detail in
Sect.~\ref{sec:new_LHC_data} below. 

\begin{table}[!t]
  \centering
  \footnotesize
  \renewcommand{\arraystretch}{1.35}
\begin{tabularx}{\textwidth}{Xlcccc}
  \toprule
  Process
  & Dataset
  & Ref.
  & $n_{\rm dat}$
  & Nucl. spec.
  & Theory
  \\
  \midrule
  \multirow{3}{*}{NC DIS}
  & NMC\,96
  & \cite{NewMuon:1996uwk}
  & 123/260
  & $^2$D/p
  & {\tt APFEL}
  \\
  & SLAC\,91
  & \cite{Whitlow:1991uw}  
  & 38/211
  & $^2$D
  & {\tt APFEL}
  \\
  & BCDMS\,89
  & \cite{BCDMS:1989qop}  
  & 250/254
  & $^2$D
  & {\tt APFEL}
  \\
  \midrule
  \multirow{2}{*}{Fixed-target DY}
  & FNAL\,E866
  & \cite{NuSea:2001idv}
  & 15/15
  & $^2$D/p
  & {\tt APFEL}
  \\
  & FNAL\,E605
  & \cite{Nagarajan:1991zz}
  & 85/119
  & $^{64}$Cu
  & {\tt APFEL}
  \\  
  \midrule
  \midrule
  \multirow{4}{*}{Collider DY}
  & ALICE $W^\pm$, $Z$ (5.02~TeV)
  & \cite{ALICE:2016rzo}
  & 6/6
  & $^{208}$Pb
  & {\tt MCFM}
  \\
  & LHCb $Z$ (5.02~TeV) 
  & \cite{LHCb:2017yua}
  & 2/2
  & $^{208}$Pb
  & {\tt MCFM}
  \\
  & ALICE $Z$ (8.16~TeV)
  & \cite{ALICE:2020jff}
  & 2/2
  & $^{208}$Pb
  & {\tt MCFM}
  \\
  & CMS $Z$ (8.16~TeV)
  & \cite{CMS:2021ynu}
  & 36/36
  & $^{208}$Pb
  & {\tt MCFM}
  \\
  \midrule
  Dijet production
  & CMS p--Pb/pp (5.02~TeV)
  & \cite{CMS:2018jpl}
  & 84/84
  & $^{208}$Pb
  & {\tt NLOjet++}
  \\
  \midrule
  Prompt photon production
  & ATLAS p--Pb/pp (8.16~TeV)
  & \cite{Aaboud:2019tab}
  & 43/43
  & $^{208}$Pb
  & {\tt MCFM}
  \\
   \midrule
  Prompt $D^0$ production
  & LHCb p--Pb/pp (5.02~TeV)
  & \cite{LHCb:2017yua}
    & 37/37
  & $^{208}$Pb
  & {\tt POWHEG}
  \\
  \bottomrule
\end{tabularx}

  \vspace{0.3cm}
  \caption{The new measurements included in nNNPDF3.0 with respect to
     nNNPDF2.0.
    For each dataset, we indicate the name used throughout the
    paper, the reference, the number of data points $n_{\rm dat}$ after/before
    kinematic cuts, the nuclear species involved, and the codes used to
    compute the corresponding theoretical predictions.
    The datasets in the upper (lower) part of the table correspond to the
    first (second) group described in the text.}
  \label{tab:nNNPDF30_dataset}
\end{table}

The main features of the new datasets entering nNNPDF3.0 (in comparison to
nNNPDF2.0) are summarised in Table~\ref{tab:nNNPDF30_dataset}.
For each process we indicate the name of the datasets used throughout the paper,
the corresponding reference, the number of data points after/before kinematic
cuts (described below), the nuclear species involved, and the code used to compute
theoretical predictions. The upper (lower) part of the table lists the
datasets in the first (second) group discussed above.

\begin{figure}[!t]
  \centering
  \includegraphics[width=0.9\textwidth]{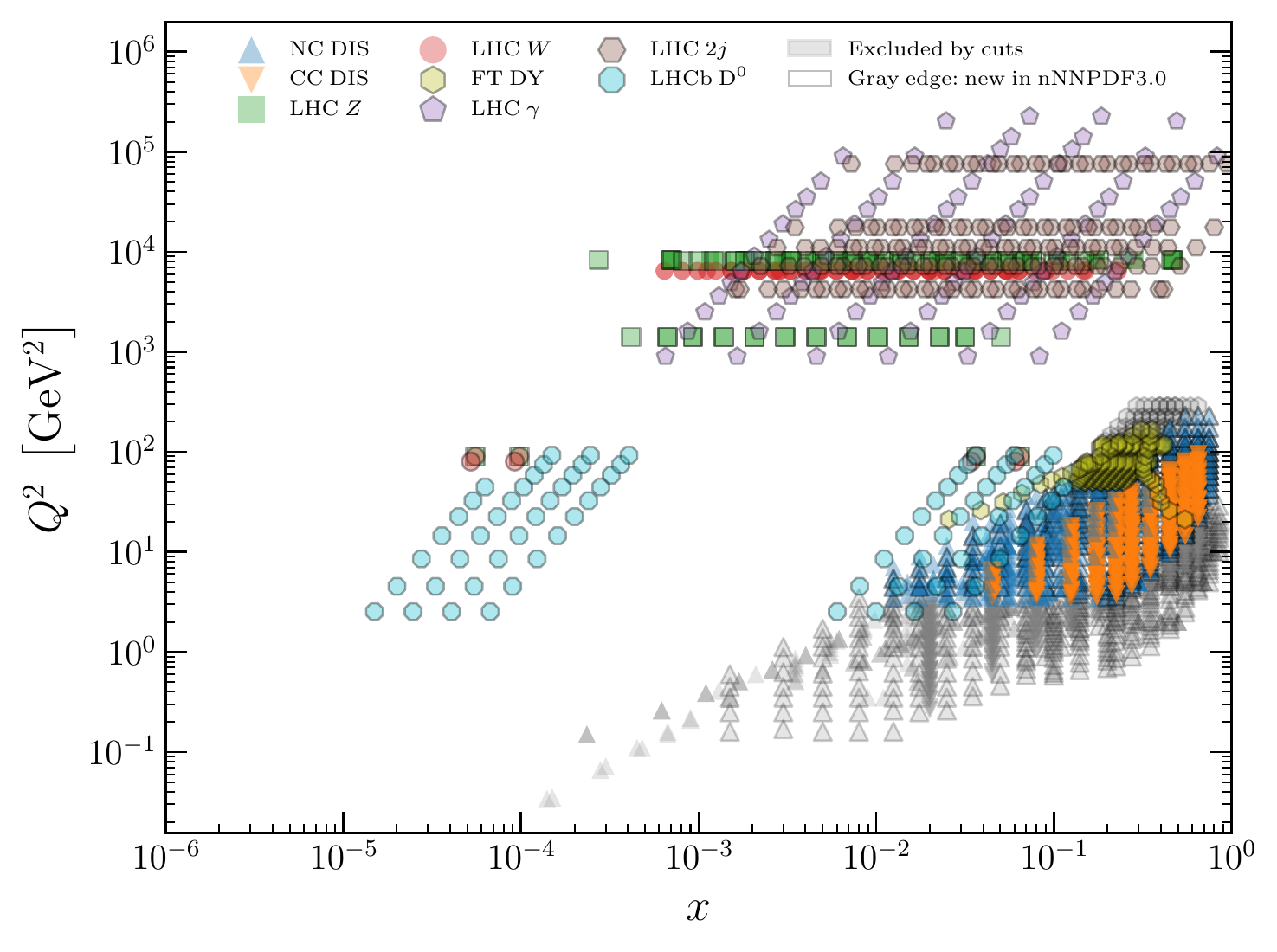}
  \caption{The kinematic coverage in the $(x,Q^2)$ plane of the nNNPDF3.0
    dataset. The evaluation of $x$ and $Q^2$ for the hadronic processes assumes
    LO kinematics. Data points are classified by process. Data points new in
    nNNPDF3.0 in comparison to nNNPDF2.0 are marked with a grey edge. Data
    points excluded by kinematic cuts are filled grey.}
  \label{fig:kinplot}
\end{figure}

The kinematic coverage in the $(x,Q^2)$ plane of the complete nNNPDF3.0 dataset
is displayed in Fig.~\ref{fig:kinplot}. For hadronic data, kinematic variables
are determined using leading order (LO) kinematics. Whenever an observable is
integrated over rapidity, the centre of the rapidity range is used to compute
the values of $x$. Data points are classified by process. Data points that are
new in nNNPDF3.0 (in comparison to nNNPDF2.0) are marked with a grey edge.

As customary, kinematic cuts are applied to the DIS structure function
measurements to remove data points that may be affected by large
non-perturbative or higher-twist corrections, namely
we require $Q^2\geq 3.5$~GeV$^2$ for the virtuality and $W^2\geq 12.5$~GeV$^2$
for the final-state invariant mass.
Cuts are also applied to the FNAL E605
measurement to remove data points close to the production threshold that
may be affected by large perturbative corrections. Namely we require
$\tau\leq 0.08$ and $|y/y_{\rm max}|\leq 0.663$, where $\tau=m^2/s$ and
$y_{\rm max}=-\frac{1}{2}\ln\tau$, with $m$ and $y$ the dilepton invariant mass
and rapidity and $\sqrt{s}$ the CoM energy of the collision. These cuts were
determined in~\cite{Bonvini:2015ira} and are also adopted in
NNPDF4.0~\cite{Ball:2021leu}. Data points excluded by kinematic cuts are
displayed in grey in Fig.~\ref{fig:kinplot}.

The total number of data points considered after applying these kinematic cuts
is $n_{\rm dat}=2188$; in comparison, the nNNPDF2.0 analysis
contained $n_{\rm dat}=1467$ points.
Of the new data points, 210 correspond to LHC measurements and the remaining to fixed-target data.
The kinematic coverage of the nNNPDF3.0 dataset is significantly
expanded in comparison to nNNPDF2.0, in particular at small $x$, where the LHCb
$D^0$-meson data covers values down to $x\simeq 10^{-5}$, and at high-$Q$, where
the ATLAS photon and CMS dijet data reaches values close to $Q\simeq 500$ GeV.

\subsection{General theory settings}
\label{sec:theory_settings}

The settings of the theoretical calculations adopted to describe the nNNPDF3.0
dataset follow those of the previous nNNPDF2.0
analysis~\cite{AbdulKhalek:2020yuc}.

Theoretical predictions are computed to next-to-leading order (NLO) accuracy
in the strong coupling $\alpha_s(Q)$. The strong coupling and (nuclear) PDFs
are defined in the \MSbar scheme, whereas heavy-flavour quarks are
defined in the on-shell scheme. The FONLL general-mass variable flavour number
scheme~\cite{Forte:2010ta} with $n_{f}^{\rm max} = 5$ (where $n_{f}^{\rm max}$ is
the maximum number of active flavours) is used to evaluate DIS structure
functions. Instead, for proton--nucleus collisions the zero-mass variable 
flavour number scheme is applied; the only exception being prompt $D$-meson
production  which is discussed in Sec.~\ref{subsubsec:DmesonProduction}.
The charm- and bottom-quark PDFs are evaluated perturbatively by applying
massive quark matching conditions. In the fit, the following input values are
used: $m_c=1.51$~GeV, $m_b=4.92$~GeV, and $\alpha_s(M_Z)=0.118$, respectively
for the charm and bottom quark masses, and for the strong coupling at a scale
equal to the $Z$-boson mass $M_Z$.

Predictions are made at LO in the electromagnetic
coupling, with the following input values for the on-shell gauge boson masses
(widths): $M_W=80.398$~GeV ($\Gamma_W=2.141$~GeV) and $M_Z=91.1876$~GeV
($\Gamma_Z=2.4952$~GeV). The $G_\mu$ scheme is used, with a value of the 
Fermi constant $G_F=1.1663787\ 10^{-5}$~GeV$^{-2}$.

The fitting procedure relies on the pre-computation of fast-interpolation grids
for both lepton--nucleus and proton--nucleus collisions. The {\sc\small FK}
table format, provided by {\sc\small APFELgrid}~\cite{Bertone:2016lga}, is used
for all fitted data. The format combines PDF and $\alpha_s$ evolution factors,
computed with {\sc\small APFEL}~\cite{Bertone:2013vaa}, with interpolated
weight tables, whose generation is process specific. For each of
the new LHC datasets included in nNNPDF3.0, this is detailed in the following.
For the datasets already part of nNNPDF2.0, the
set-up was detailed in~\cite{AbdulKhalek:2020yuc}.

\subsection{New LHC measurements and corresponding theory settings}
\label{sec:new_LHC_data}

The new LHC measurements included in nNNPDF3.0 are discussed in the following:
inclusive electroweak boson, prompt photon, dijet, and prompt $D^0$-meson
production. For inclusive electroweak boson production we consider data for
differential distributions obtained in pPb collisions. For all other processes,
differential distributions measured in pPb collisions are always normalised
to the corresponding distributions in pp collisions, measured at the same CoM
energy. These ratios take the schematic form
\begin{align}
  \label{eq:def_nuclear_modification_ratio_general}
  \frac{{\rm d} R_{\rm pPb}}{{\rm d}X} = \frac{ {\rm d}\sigma^{\rm pPb} }{{\rm d}X} \bigg{/} \frac{ {\rm d}\sigma^{\rm pp}}{{\rm d}X}\,,
\end{align}
where $X$ represents an arbitrary differential variable. The same form applies
to more (e.g. double) differential quantities.
The general rationale for applying this approach is that the LO predictions for prompt photon, dijet, and prompt $D$-meson production are $\mathcal{O}(\alpha_s)$.
As a consequence, the theoretical predictions for the absolute rates of these processes (at NLO QCD accuracy) are subject to uncertainties due to missing higher order effects which are typically in excess of the uncertainty related to nPDFs.
At the level of the ratio, uncertainties related to the overall normalisation of the distributions (e.g. the value of the coupling) cancel, while sensitivity to the nuclear modification of nPDFs is retained.
In Sect.~\ref{sec:results} it will be shown for prompt $D$-meson production (which has the largest relative theory uncertainties of the considered processes) that such observables ensure that nPDF uncertainties dominate over those due to scale uncertainties.
Notably, a shortcoming of this approach is that it becomes necessary to exclude the reference pp data from the proton baseline which enters the nPDF fit. 
The extension of the perturbative accuracy of the fit to NNLO QCD and/or including theoretical uncertainties (as in~\cite{NNPDF:2019ubu, NNPDF:2019vjt}) would allow one to consider absolute distributions instead of ratios for these
selected processes.
As a note, these ratios are constructed in the pp CoM frame. For collisions such
as pPb, they do not coincide with ratios constructed in the laboratory frame. 
The role of this asymmetry for the interpretation of pPb observables is
reviewed in App.~\ref{app:asymmetry}.

\subsubsection{Inclusive electroweak boson production}
\label{subsubsec:EWgauge}

The new datasets that we consider in this category are the following.
First we include the ALICE measurements of the fiducial cross-section for
$W$- and $Z$-boson production at $\sqrt{s}=5.02$~TeV in the muonic decay
channel~\cite{ALICE:2016rzo}. The data points cover the backward (Pb--going)
and forward (p--going) rapidity regions. 
The integrated luminosity is 5.81 and 5.03~nb$^{-1}$ in each case. 
Then we also include the related LHCb~\cite{ALICE:2020jff} and
ALICE~\cite{LHCb:2014jgh} measurements of the fiducial cross-section for
$Z$-boson production at $\sqrt{s}=5.02$~TeV and $\sqrt{s}=8.16$~TeV,
respectively. 
Backward and forward rapidity measurements are considered in both cases.
The integrated luminosities are, for ALICE, 8.40 and
12.7~nb$^{-1}$ in the Pb--going and in the p--going directions; and for LHCb,
1.6~nb$^{-1}$. 
Finally we include the CMS measurement of $Z$-boson production
in the dimuon decay channel at $\sqrt{s}=8.16$~TeV~\cite{CMS:2021ynu}.
In this latter case, differential cross-sections are presented with respect to
the rapidity and the invariant mass of the dimuon system, after being corrected
for acceptance effects.
The integrated luminosity is 173~nb$^{-1}$.

As was already the case for nNNPDF2.0, theoretical predictions for electroweak
gauge boson production are computed at NLO QCD accuracy with {\sc\small MCFM}
v6.8~\cite{Campbell:1999ah,Campbell:2011bn,Campbell:2015qma}. The
renormalisation and factorisation scales are set equal to the mass of the gauge
boson, for total cross-sections, and to the central value of the corresponding
invariant mass bin, in the case of the CMS differential
cross-sections~\cite{CMS:2021ynu}. 
Note that the choice of the scale is partly restricted by the grid generation procedure --- i.e. a fully dynamical event-by-event scale choice such as the
invariant mass of the muon pair is not accessible.
Experimental correlations are taken into account
whenever available, namely for the measurements
of~\cite{ALICE:2016rzo,LHCb:2014jgh,CMS:2021ynu}, otherwise statistical and
systematic uncertainties are added in quadrature.

\subsubsection{Prompt photon production}
\label{subsubsec:photon}

In this category nNNPDF3.0 includes the ATLAS measurement at
$\sqrt{s}=8.16$~TeV~\cite{Aaboud:2019tab}. Cross-sections for isolated
prompt photon production are presented in three pseudo-rapidity
$\eta_{\gamma}$ bins and then differentially in the photon transverse energy $E_T^\gamma$.
As discussed above, we only consider the ratio 
of the differential pPb cross-section normalised with respect to the reference pp results in the fit.
Notably, for prompt photon production this means that uncertainties due to the treatment of 
photon fragmentation and the choice of the value of the electromagnetic coupling are not important.
For completeness, in Sect.~\ref{sec:results} we also indicate how
the fit quality is deteriorated when the absolute 
cross-section is considered (without accounting for theory uncertainties). 
The ranges of the three photon pseudo-rapidity bins in the CoM frame
are $-2.83<\eta_{\gamma}<-2.02$, $-1.84<\eta_{\gamma}<0.91$ and
$1.09<\eta_{\gamma}<1.90$; the kinematic coverage in the photon transverse
energy is, for each of these, $20<E_T^\gamma<550$~GeV. The integrated luminosity
of this measurement is 165~nb$^{-1}$.

Theoretical predictions are computed at NLO QCD accuracy with {\sc\small MCFM} (v6.8)
following the calculational settings presented in~\cite{Campbell:2018wfu}.
The renormalisation and factorisation scales are set equal to the central value
of the photon transverse energy $E_T^\gamma$ for each bin.
Experimental correlations between transverse momentum and rapidity bins are
taken into account following the prescription provided in~\cite{Aaboud:2019tab}.

\subsubsection{Dijet production}
\label{subsubsec:dijets}

In this category we consider the CMS measurement at
$\sqrt{s}=5.02$~TeV~\cite{CMS:2018jpl}. The cross-section
is presented double differentially in the dijet average transverse momentum
$p_{T,{\rm dijet}}^{\rm ave}$, where $p_{T,{\rm dijet}}^{\rm ave} = \left(p_{T,1} + p_{T,2}\right)/2$,
and the dijet pseudo-rapidity $\eta_{\rm dijet}$. 
As in the case of prompt photon production, we consider only data for the ratio
of cross-sections obtained in  pPb with respect to that from pp collisions.
Again, for completeness, in Sect.~\ref{sec:results} we also show how the fit
quality deteriorates (without accounting for theory uncertainties) when the
absolute cross-section is considered.
The kinematic coverage in pseudo-rapidity and average transverse momentum is,
in the CoM frame, $-2.456<\eta<2.535$ and
$55~{\rm GeV}<p_{\rm T,dijet}^{\rm avg}<400$~GeV. The integrated
luminosity is 35 (27)~nb$^{-1}$ for the Pb--going (p--going) direction.

Theoretical predictions are computed at NLO QCD accuracy with
{\sc\small NLOjet++}~\cite{Nagy:2001fj}. We have verified that the independent
computation of~\cite{Gehrmann-DeRidder:2019ibf} is reproduced.
The renormalisation and factorisation scales are set equal to the dijet invariant mass $m_{jj}$, 
as in the dedicated pp study of~\cite{AbdulKhalek:2020jut}. Since experimental correlations for the
pPb to pp cross-section ratio are not provided,
statistical and systematic uncertainties are added in quadrature.

\subsubsection{Prompt $D$-meson production}
\label{subsubsec:DmesonProduction}

In this category we consider the LHCb measurements of prompt $D^0$-meson
production in pPb collisions at $\sqrt{s}=5.02$ TeV~\cite{LHCb:2017yua}.
The data is available both in the forward and backwards configurations, and is
presented as absolute cross-sections differential in the transverse momentum
$p^{D^0}_{\rm T}$ and rapidity $y^{D^0}$ of the $D^0$-mesons, namely
\be \label{eq:Rf}
\frac{d^2\sigma^{\rm pPb}}{dy^{D^0} dp^{D^0}_{\rm T}} \quad {\rm and} \quad \frac{d^2\sigma^{\rm Pbp}}{dy^{D^0} dp^{D^0}_{\rm T}}\,.
\ee
Following the discussion around Eq.~\eqref{eq:def_nuclear_modification_ratio_general}, we only consider data for cross-section ratios (which are obtained with
respect to reference pp data~\cite{LHCb:2016ikn}).
The double differential ratio for forward pPb measurements is
\begin{align}
  \label{eq:def_nuclear_modification_ratio}
R_{\rm pPb}(y^{D^0}, p^{D^0}_{\rm T})   = \left.\frac{d\sigma^{\rm pPb}(y^{D^0},p^{D^0}_{\rm T})}{dy^{D^0} dp^{D^0}_{\rm T}} \middle/ \frac{d\sigma^{\rm pp}(y^{D^0},p^{D^0}_{\rm T})}{dy^{D^0} dp^{D^0}_T}\right. \,,
\end{align}
where both pPb and pp cross-sections are given in the pp CoM frame (and $y^{D^0}$ is the $D^0$-meson rapidity in that frame).
The corresponding observable can be constructed for backwards Pbp collisions, which is denoted as $R_{\rm Pbp}$.

An additional ratio $R_{\rm fb}$ constructed from forward over backward cross-sections (defined in the pp CoM frame) can also be considered:
\be
\label{eq:Rfb}
R_{\rm fb}(y^{D^0}, p^{D^0}_{\rm T}) = \frac{d^2\sigma^{\rm pPb}(y^{D^0},p^{D^0}_{\rm T})}{dy^{D^0} dp^{D^0}_{\rm T}} \Bigg/
\frac{d^2\sigma^{\rm Pbp}(-|y^{D^0}|,p^{D^0}_{\rm T})}{dy^{D^0} dp^{D^0}_{\rm T}} \,.
\ee
This ratio benefits from a cancellation of a number of experimental systematics, and, as motivated in~\cite{Gauld:2015lxa}, provides sensitivity to nuclear modifications while the theoretical uncertainty from all other sources is reduced to a sub-leading level.

The LHCb measurements cover the rapidity range of $2.0 < y^{D^0} < 4.5$ for pp collisions, 
and the effective range of $1.5 < y^{D^0} < 4.0$ in the forward pPb collisions. 
The data on the $R_{\rm pPb}$ ratio Eq.~(\ref{eq:def_nuclear_modification_ratio}) is presented
in four rapidity bins in the region $2.0 < y^{D^0} < 4.0$, such that the coverage of
the forward pPb and the baseline pp measurements overlap.
For each of these bins in $y^{D^0}$, the coverage in the $D^0$-meson
transverse momentum is $0 < p^{D^0}_{\rm T} < 10$~GeV, adding up to a total of 37
data points. In contrast to pp measurements, for which different $D$ meson
species are detected, in the pPb case only $D^0$ mesons are reconstructed.
Because bin-by-bin correlations for systematic uncertainties are not available
for $R_{\rm pPb}$, we consider them as fully uncorrelated
and add them in quadrature with the statistical uncertainties.

Theoretical predictions for $D$-meson production in hadron-hadron collisions are computed at NLO QCD accuracy in a fixed-flavour number scheme with {\sc\small POWHEG}~\cite{Nason:2004rx,Frixione:2007vw,Alioli:2010xd} matched to {\sc\small Pythia8}~\cite{Sjostrand:2014zea}. The Monash 2013 Tune~\cite{Skands:2014pea} is used throughout.
Our computational set-up has been compared with that used in the calculations of charm production in pp collisions
from~\cite{Gauld:2016kpd}, finding good agreement. 
Furthermore, theoretical predictions for $R_{\rm pPb}$ and $R_{\rm Pbp}$ have been benchmarked against
the corresponding calculations used in the EPPS analysis~\cite{Eskola:2019bgf} (where the same set-up was considered).
It is relevant to mention that a comparison of how different theoretical approaches to describing $D^0$-meson production in pPb collisions impacts the extraction of nPDFs has been considered in~\cite{Eskola:2019bgf}.
This study demonstrated that (see Fig.~15 of~\cite{Eskola:2019bgf}), for suitably defined observables such as $R_{\rm fb}(y^{D^0}, p^{D^0}_{\rm T})$, consistent results are obtained between our chosen set-up ({\sc\small POWHEG+Pythia8}) and those in a general-mass variable flavour number scheme~\cite{Helenius:2018uul}.

The above discussion summarises how the various nuclear cross-section ratios for $D^0$-meson production are accounted for. 
However, it is also necessary to constrain the overall normalisation of the nPDFs,
and not just the size of the nuclear correction --- i.e. the normalisation of
the free nucleon PDFs must also be known.
To achieve this, we include the constraints from $D$-meson production in pp collisions at 7 TeV~\cite{LHCb:2013xam} and
13 TeV~\cite{LHCb:2015swx} into the proton PDF baseline which is used as a boundary condition for nNNPDF3.0. 
This is done following the analysis of~\cite{Gauld:2016kpd}, which considers LHCb $D$-meson data at the level of normalised differential cross-sections according to
\begin{align}
  \label{eq:normalized_xsec_pp}
  N^{AA'}_X(y^D ,p^D_{\rm T}) \equiv \left.\frac{d\sigma^{\rm AA'}\left(X\,\,\rm TeV\right)}{dy^D dp^D_{\rm T}}
  \middle/ \frac{d\sigma^{\rm AA'}\left(X\,\,\rm TeV\right)}{dy_{\rm ref}^D dp^D_T}\right. \,,
\end{align}
in terms of a reference rapidity bin $y_{\rm ref}^D$.
The main advantage of normalised observables such as Eq.~\eqref{eq:normalized_xsec_pp} is that scale uncertainties cancel to good approximation (since these depend mildly on rapidity) while some sensitivity to the PDFs is retained (as the rate of the change of the PDFs in $x$ is correlated with $y^D$).
This has been motivated in~\cite{Zenaiev:2015rfa} and subsequently in~\cite{Gauld:2015yia,Gauld:2016kpd}.
As in~\cite{Gauld:2016kpd} we consider pp measurements for the $\{D^0, D^+, D_s^+\}$ final states, adding up to a total of 79 and 126 data points for $N^{\rm pp}_7$ and $N^{\rm pp}_{13}$ respectively. 
To avoid double counting, the available data on $N^{\rm pp}_5$ is not included in the proton baseline, given that it will enter the nuclear PDF analysis through Eq.~\eqref{eq:def_nuclear_modification_ratio}.

Finally, we note that in contrast to all other considered scattering processes in this work, there is currently no public interface for the computation of fast interpolation tables for prompt $D$-meson production in pp or pPb collisions.
This complicates the inclusion of this data in the nNNPDF3.0 global analyses, which is realised by a multi-stage Bayesian reweighting procedure as detailed in Sect.~\ref{sec:RW} and summarised in Fig.~\ref{fig:strategy}.

\section{Analysis methodology}
\label{sec:fitting}

In this section we describe the fitting methodology that is adopted in
nNNPDF3.0. We start with an overview of the methodological aspects shared with
nNNPDF2.0. We then discuss the improvements in the free-proton baseline PDF used
as the $A=1$ boundary condition for the nuclear fit, and compare this baseline
to the one used in nNNPDF2.0. We proceed by describing hyperoptimisation,
the procedure adopted to automatically select the optimal set of model
hyperparameters such as the neural network architecture and the minimiser
learning rates. Finally we outline the strategy, based on Bayesian reweighting,
used in order to include the LHCb measurements of $D$-meson production in a
consistent manner both in the free-proton baseline PDFs and in the nPDFs.

\subsection{Methodology overview}
\label{sec:overview}

The fitting methodology adopted in nNNPDF3.0 closely follows the one used in
nNNPDF2.0. Here we summarise the methodological aspects common to the two
determinations. Additional methodological improvements will be discussed in
Sects.~\ref{sec:freeproton}-\ref{sec:hyperparameter_scan}.

\subsubsection{Parametrisation}
\label{subsubsec:parametrisation}

As in nNNPDF2.0, we parametrise six independent nPDF combinations in the
evolution basis. These are:
\bea
x\Sigma^{(p/A)}(x,Q_0) &=&x^{\alpha_\Sigma} (1-x)^{\beta_\Sigma} {\rm
NN}_\Sigma(x,A) \, , \nonumber \\
xT_3^{(p/A)}(x,Q_0) &=&x^{\alpha_{T_3}} (1-x)^{\beta_{T_3}} {\rm NN}_{T_3}(x,A)
\, , \nonumber \\
xT_8^{(p/A)}(x,Q_0) &=&x^{\alpha_{T_8}} (1-x)^{\beta_{T_8}} {\rm NN}_{T_8}(x,A)
\, , \label{eq:param2} \\
xV^{(p/A)}(x,Q_0) &=&B_{V}x^{\alpha_V} (1-x)^{\beta_V} {\rm NN}_V(x,A) \, ,
\nonumber \\
xV_3^{(p/A)}(x,Q_0) &=&B_{V_3}x^{\alpha_{V_3}} (1-x)^{\beta_{V_3}} {\rm
NN}_{V_3}(x,A) \nonumber\, ,  \\
xg^{(p/A)}(x,Q_0) &=&B_gx^{\alpha_g} (1-x)^{\beta_g} {\rm NN}_g(x,A) \, ,
\nonumber
\eea
where $f^{(p/A)}(x,Q_0)$, with $f=\Sigma,T_3,T_8,V,V_3,g$, denotes the nPDF of a
proton bound in a nucleus with atomic mass number $A$, see
App.~\ref{app:conventions} for the conventions adopted. The parametrisation
scale is $Q_0=1$~GeV, as in the free-proton baseline PDF set. The normalisation
coefficients $B_V$, $B_{V_3}$ and $B_g$ enforce the momentum
and valence sum rules, while the preprocessing exponents $\alpha_f$ and
$\beta_f$ are required to control the small- and large-$x$ behaviour of the
nPDFs, see Sect.~\ref{subsubsec:sumrules}. In Eq.~\eqref{eq:param2},
${\rm NN}_f(x,A)$
represents the value of the neuron in the output layer of the neural network
associated to each independent nPDF.
The input layer contains three neurons that take as input the values of the
momentum fraction $x$, $\ln(1/x)$, and the atomic mass number $A$, respectively.
The rest of the neural network architecture is determined through the
hyperoptimisation procedure described in Sect~\ref{sec:hyperparameter_scan}.
This is in contrast to nNNPDF2.0, where the optimal number of hidden layers
and neurons were determined by trial and error. The corresponding distributions
for bound neutrons are obtained from those of bound protons assuming isospin
symmetry. Under the isospin transformation, all quark and gluon combinations 
in Eq.~\eqref{eq:param2} are left invariant, except for:
\be
xT_3^{(p/A)}(x,Q_0)=-xT_3^{(n/A)}(x,Q_0) \, ,\quad
xV_3^{(p/A)}(x,Q_0)=-xV_3^{(n/A)}(x,Q_0) \, .
\ee

While the nNNPDF3.0 fits presented in this work are carried out in the
basis specified by Eq.~\eqref{eq:param2}, any other basis obtained as a linear 
combination of Eq.~\eqref{eq:param2} could be used. In principle, the
choice of any basis should lead to comparable nPDFs, within statistical
fluctuations, as demonstrated explicitly in the proton
case~\cite{Ball:2021leu}. In the following,
we will display the nNNPDF3.0 results in the flavour basis, given by
\be
\left\{xu^{(p/A)},\quad xd^{(p/A)},\quad x\bar{u}^{(p/A)},\quad
x\bar{d}^{(p/A)},\quad xs^{+(p/A)},\quad xg^{(p/A)}\right\} \, ,
\ee
see Sect.~3.1 of~\cite{Ball:2021leu} for the explicit relationship with the
evolution basis of Eq.~\eqref{eq:param2}. 

\subsubsection{Sum rules and preprocessing}
\label{subsubsec:sumrules}

Momentum and valence sum rules are enforced by requiring that the normalisation
coefficients $B_g$, $B_V$, and $B_{V_3}$ in Eq.~\eqref{eq:param2} take the
values:
\begin{align}
  B_g(A)
  &=
  \left(1 - \int_0^1 dx\, x\Sigma^{(p/A)}(x,Q_0)\right)
  \nonumber
  \Big/\left(\int_0^1 dx\, xg^{(p/A)}(x,Q_0)\right) \, ,\\
  B_V(A)
  &=
  3\Big/\left(\int_0^1 dx\, V^{(p/A)}(x,Q_0)\right)\label{eq:sumrules} \, ,\\
  B_{V_3}(A)
  &= 1\Big/\left(\int_0^1 dx\, V_3^{(p/A)}(x,Q_0)\right) \, .\nonumber
\end{align}
These coefficients must be determined for each value of $A$; perturbative
evolution ensures that, once enforced at the initial parametrisation scale
$Q_0$, momentum and valence sum rules are not violated for any $Q > Q_0$.

The preprocessing exponents $\alpha_f$ and $\beta_f$ in Eq.~\eqref{eq:param2}
facilitate the training process and are fitted simultaneously with the network
parameters. The exponents $\alpha_V$ and $\alpha_{V_3}$ are restricted
to lie in the range $[0,5]$ during the fit to ensure integrability
of the valence distributions. The other $\alpha_f$ exponents are
restricted to the range $[-1,5]$, consistently with the momentum sum rule
requirements, while the exponents $\beta_f$ lie in the range $[1,10]$.
While we do not explicitly impose integrability of $xT_3$ and $xT_8$,
consistently with the NNPDF3.1-like proton baseline, the fitted nPDFs turn out
to satisfy these integrability constraints anyway.
It is worth pointing out that a strategy 
to avoid preprocessing within the NNPDF methodology has been recently presented
in~\cite{Carrazza:2021yrg}. This strategy, which leads to consistent results
in the case of free-proton PDFs, may be applied to nPDFs in the future.

\subsubsection{The figure of merit}
\label{subsusbsec:figureofmerit}

The best-fit values of the parameters defining the nPDF parametrisation in
Eq.~\eqref{eq:param2} are determined by minimising a suitable figure of merit.
As in nNNPDF2.0, this figure of merit is an extended version of the $\chi^2$,
defined as
\be
\label{eq:chi2_fit}
\chi_{\rm fit}^2 = \chi_{\rm t_0}^2 + \kappa_{\rm pos}^2 + \kappa_{\rm BC}^2 \, . 
\ee
The first term, $\chi_{t_0}^2$, is the contribution from experimental data
\be
  \chi_{t_0}^2
  =
  \sum_{ij}^{n_{\rm dat}} (T_i - D_i)\,\,(\text{cov}_{t_0})_{ij}^{-1}\,\,(T_j - D_j)\, .
\ee
This is derived by maximising the likelihood of observing the data $D_i$
given a set of theory predictions $T_i$. The covariance matrix is constructed
according to the $t_0$ prescription~\cite{Ball:2009qv}, see
Eq.~(9) in~\cite{Ball:2012wy}.\footnote{Note that
  when quoting $\chi^2$ values we will always use
the experimental definition instead, Eq.~(8) in~\cite{Ball:2012wy}.}

The second term, $\kappa_{\rm pos}^2$,
ensures the positivity of physical cross-sections and is defined as
\be
  \label{eq:chi2_pos}
  \kappa_{\rm pos}^2
  =
  \lambda_{\rm pos} \sum_{l=1}^{n_{\rm pos}}\sum_{j=1}^{n_A} \sum_{i_l=1}^{n_{\rm dat}^{(l)}} {\rm max}\lp 0, -\mathcal{F}_{i_l}^{(l)}(A_j) \rp \, ,
\ee
where $l$ runs over the $n_{\rm pos}$ positivity observables
$\mathcal{F}^{(l)}$ (defined in Table~3.1 of~\cite{AbdulKhalek:2020yuc})
and each of the observables contain $n_{\rm dat}^{(l)}$ kinematic points that
are computed over all $n_A$ nuclei for which there are experimental data in the
fit. The Lagrange multiplier is fixed by trial and error to
$\lambda_{\rm pos} = 1000$. 

The third term, $\kappa_{\rm BC}^2$, ensures that, when taking the $A \to 1$
limit, the nNNPDF3.0 predictions reduce to those of a fixed free-proton
baseline in terms of both central values and uncertainties. The choice
of this baseline set differs from nNNPDF2.0 and is further discussed in
Sect.~\ref{sec:freeproton}. The $\kappa_{\rm BC}^2$ term is defined as
\be
\label{eq:chi2_BC}
\kappa_{\rm BC}^2
= \lambda_{\rm BC} \sum_{f}\sum_{j=1}^{n_x} \lp f^{(p/A)}(x_j,Q_0,A=1)
- f^{(p)}(x_j,Q_0) \rp^2 \, ,
\ee
where $f$ runs over the six independent nPDFs in the
evolution basis, and $j$ runs over $n_x$ points in $x$ as discussed in
Sect.~\ref{sec:freeproton}. The value of the Lagrange multiplier is fixed to
$\lambda_{\rm BC}=100$. In Eq.~\eqref{eq:chi2_BC}, for each replica in the
nNNPDF3.0 ensemble, we use a different random replica from the free-proton
baseline set. This way the free-proton PDF uncertainty is propagated into the
nPDF fit via the $A=1$ boundary condition.

In summary, in the present analysis the best-fit nPDFs are
determined by maximising the agreement between the theory predictions and the
corresponding
Monte Carlo replica of the experimental data, subject to the physical
constraints of cross-section positivity and the $A=1$ boundary condition,
together with methodological requirements such as cross-validation to prevent
overlearning.

\subsection{The free-proton boundary condition}
\label{sec:freeproton}

The neural network parametrisation of the nPDFs described by
Eq.~\eqref{eq:param2} is valid from $A=1$ (free proton) up to $A=208$ (lead).
This implies that the nNNPDF3.0 determination also contains a
determination of the free-proton PDFs (as $A\to1$). 
However, as discussed above, the PDFs in this $A=1$ limit are fixed
by means of the Lagrange multiplier defined in Eq.~\eqref{eq:chi2_BC}.
The central values and uncertainties of nNNPDF3.0 for $A=1$ hence reproduce
those of some external free-proton baseline PDFs, which effectively act as a
boundary condition to the nPDFs. This is a unique feature of the nNNPDF
methodology. We note that moderate deviations from this boundary condition may
appear as a consequence of the positivity constraints enforced through
Eq.~\eqref{eq:chi2_pos}. Distortions may also arise when nPDFs replicas are
reweighted with new data, as we will discuss in Sect.~\ref{sec:RW}.

In nNNPDF3.0, the free-proton baseline PDFs are chosen to be a variant of the
NNPDF3.1 NLO parton set~\cite{Ball:2017nwa} where the charm-quark PDF is
evaluated perturbatively  by applying massive quark matching conditions.
This baseline differs from that used in nNNPDF2.0 in the following respects.

\begin{itemize}
  
\item The free-proton baseline PDFs include all the datasets incorporated in the
  more recent NNPDF4.0 NLO analysis~\cite{Ball:2021leu}, except those involving
  nuclear targets (these are instead part of nNNPDF3.0), see in particular
  Tables 2.1--2.5 and App.~\ref{app:asymmetry} in~\cite{Ball:2021leu}. The
  extended NNPDF4.0 dataset provides improved constraints on the free-proton
  baseline PDFs in a wide range of $x$ for all quarks and the gluon. Note that,
  while our free-proton baseline PDFs are close to NNPDF4.0 insofar as the
  dataset is concerned, they are however based on the NNPDF3.1 fitting
  methodology. The reason being that the NNPDF4.0 methodology incorporates
  several modifications (see Sect.~3 in~\cite{Ball:2021leu} for a discussion)
  that are not part of the nNNPDF3.0 methodology. Among these, a strict
  requirement of PDF positivity. Those modifications do not alter the
  compatibility between PDF determinations obtained with the NNPDF3.1 or
  NNPDF4.0 methodologies, however they lead to generally smaller uncertainties in the latter case,
  see Sect.~8 in~\cite{Ball:2021leu}. 
  One could therefore expect a reduction of nPDF uncertainties for low-$A$ nuclei, and results consistent
  with those presented in the following, if the proton-only fit determined with the NNPDF4.0 methodology 
  was used as a boundary condition.
  
\item When implementing the free-proton boundary condition via
  Eq.~\eqref{eq:chi2_BC}, in nNNPDF3.0 we use a grid with $n_x=100$
  points, half of which are distributed logarithmically between
  $x_{\rm min}=10^{-6}$ and $x_{\rm mid}=0.1$ and the remaining half
  are linearly distributed between $x_{\rm mid}=0.1$ and $x_{\rm max}=0.7$.
  This is different from nNNPDF2.0, where the grid had 60 points, 10 of which
  were logarithmically spaced between $x_{\rm min}=10^{-3}$ and
  $x_{\rm mid}=0.1$ and the remaining 50 were linearly spaced between
  $x_{\rm mid}=0.1$ and $x_{\rm max}=0.7$. This extension in the $x$ range is
  necessary to account for the kinematic coverage of the LHCb $D$-meson
  production measurements in pp and pPb collisions, see also
  Fig.~\ref{fig:kinplot}.
 
\end{itemize}

\begin{figure}[!t]
  \centering
  \includegraphics[width=\textwidth]{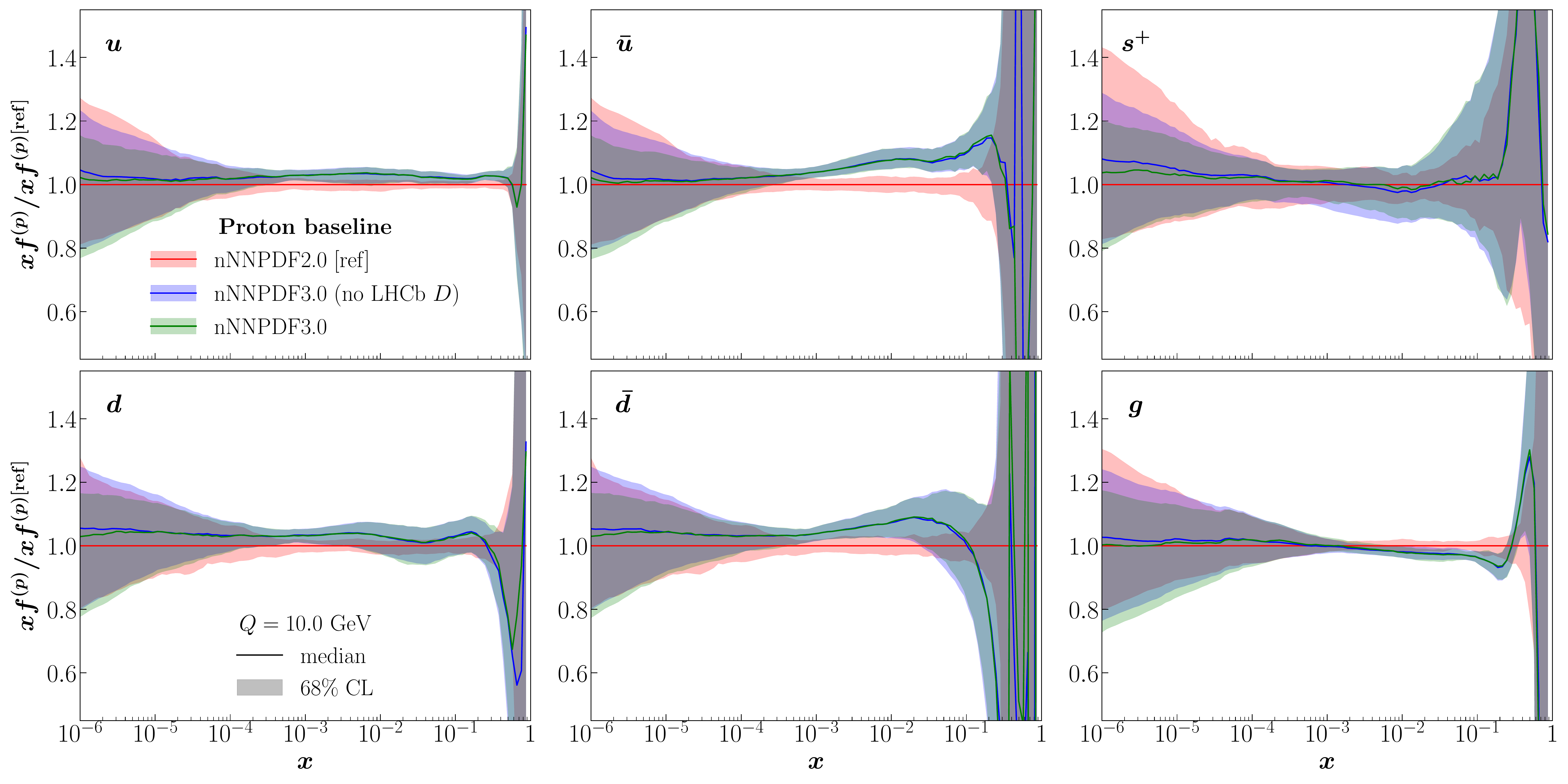}\\
  \caption{Comparison between the free-proton baseline PDFs used in nNNPDF2.0
    and in nNNPDF3.0, without and with the LHCb $D$-meson data. These PDFs are
    used to enforce the $A=1$ boundary condition in the corresponding nPDF
    fits by means of Eq.~(\ref{eq:chi2_BC}). Results are normalised to the
    nNNPDF2.0 free-proton baseline PDFs and are shown at $Q=10$~GeV in the
    extended range of $x$ for which the boundary condition is enforced.}
  \label{fig:freeproton}
\end{figure}

Furthermore, we produce two variants of these free-proton baseline PDFs,
respectively with and without the LHCb $D$-meson production data in pp
collisions at 7 and 13~TeV, see Sect.~\ref{subsubsec:DmesonProduction}.
The two variants
are used consistently with variants of the nPDF fit in which LHCb $D^0$-meson
production data in pPb collisions are included or not. These two variants are
compared (and normalised) to the free-proton baseline PDFs used in nNNPDF2.0 in 
Fig.~\ref{fig:freeproton}. We show the PDFs at $Q=10$~GeV in the
same extended range of $x$ for which the boundary condition
Eq.~\eqref{eq:chi2_BC} is enforced, that is between $x=10^{-6}$ and $x=0.7$.

Differences between the three free--proton baseline PDFs displayed in
Fig.~\ref{fig:freeproton} arise from differences in the fitted dataset: the
nNNPDF3.0 baselines do not include nuclear data that were instead part of the
nNNPDF2.0 baseline; and conversely the nNNPDF3.0 baselines benefit from the
extended NNPDF4.0 dataset, which was not part of the nNNPDF2.0 baseline.
The effect of these differences are: an increase of the up and down quark and
anti-quark central values in the region $10^{-3} \lsim x \lsim 10^{-1}$;
a slight increase of the PDF uncertainties in the same region, in particular
for the down quark and antiquark PDFs and for the total strangeness;
and a suppression of the gluon central value for $10^{-3} \lsim x \lsim 10^{-1}$
followed by an enhancement at larger values of $x$, accompanied by an
uncertainty reduction in the same region.

The datasets responsible for each of the effects observed in the nNNPDF3.0
baseline PDFs have been identified in the NNPDF4.0 analysis~\cite{Ball:2021leu}
and in related
studies~\cite{Ball:2018twp,Ball:2020xqw,AbdulKhalek:2020jut,Faura:2020oom}.
The enhancement of the central values of the up quark and antiquark PDFs
in the region $10^{-3} \lsim x \lsim 10^{-1}$ is a consequence of the new
measurements of inclusive and associated DY production from ATLAS, CMS,
and LHCb. The increase of uncertainties for the down quark and antiquark
PDFs and for the total strangeness are due to the removal of the deuteron DIS
and DY cross-sections. This piece of information is however not lost since it
is subsequently included in the nuclear fit. The variation of the central value
and the reduction of the uncertainty of the gluon PDF are a consequence
of the new ATLAS and CMS dijet cross-sections at 7~TeV.

On the other hand, a comparison between the two nNNPDF3.0 free-proton baseline
PDFs, with and without the pp 7 and 13~TeV LHCb $D$-meson production data,
reveals that central values remain mostly unchanged. This fact indicates that
the data is described reasonably well even if this is not included in the fit.
Uncertainties are unaffected for $x\gsim 10^{-4}$, while they are reduced
at smaller values of $x$, in particular for the sea quark and gluon PDFs.

It should be finally noted that several datasets and processes constrain both
the free-proton baseline PDFs and the nPDFs to a similar degree of precision.
There is therefore some interplay between the two. The free-proton baseline
PDFs will directly constrain the low-$A$ nPDFs, and indirectly the nPDFs
corresponding to higher values of $A$.

\subsection{Hyperparameter optimisation}
\label{sec:hyperparameter_scan}

A common challenge in training neural-network based models is the choice of
the hyperparameters of the model itself. These include, for instance, the
architecture and activation functions of the neural network, the optimisation
algorithm and learning rates. The choice of hyperparameters affect the
performance of the model and of its training. Hyperparameters can be tuned by
trial and error, however this is computationally inefficient and may leave
unexplored relevant regions of the hyperparameter space. A heuristic approach
designed to address this
problem more effectively is hyperparameter scan, or hyperparameter optimisation
(or hyperoptimisation in short). It consists in finding the best combination of
hyperparameters through an iterative search of the hyperparameter space
following a specific optimisation algorithm. In the context of NNPDF fits,
this approach has been proposed in~\cite{Carrazza:2019mzf} and has been
used in recent free-proton PDF fits~\cite{Ball:2021leu}.

Hyperoptimisation is realised as follows. First a figure of merit (also known
as loss function) to minimise and a search domain in the hyperparameter space
are defined. Here we define the loss function as the average of the training
and validation $\chi^2$,
\be
\label{eq:hyperopt_loss}
L_{\rm hyperopt} = \frac{1}{2}\lp \chi^2_{\rm tr} + \chi^2_{\rm val}\rp \, ,
\ee
see~\cite{Ball:2021leu} for alternative choices. We carry out the
hyperparameter scan for various subsets of Monte Carlo data replicas
and check that the results converge to a unique combination of hyperparameters.
During the initialisation process, the loss function is evaluated for a few
random sets of hyperparameters. Based on the results from these searches, the
optimisation algorithm constructs models in which the hyperparameters that
reduce the loss function are selected. The models are then updated during the
trials based on historical observations and subsequently define new sets of
hyperparameters to test. In this analysis, we implement the tree-structured
Parzen Estimator (TPE), also known as Kernel Density Estimator
(KDE)~\cite{10.1214/aoms/1177728190,10.1214/aoms/1177704472}, as optimisation
algorithm for hyperparameter tuning. The TPE selects the most promising sets of
hyperparameters to evaluate the loss function by constructing a
probabilistic model based on previous trials, and has been proven to
outperform significantly any random or grid searches.

\begin{figure}[!t]
  \centering
  \includegraphics[width=0.495\textwidth]{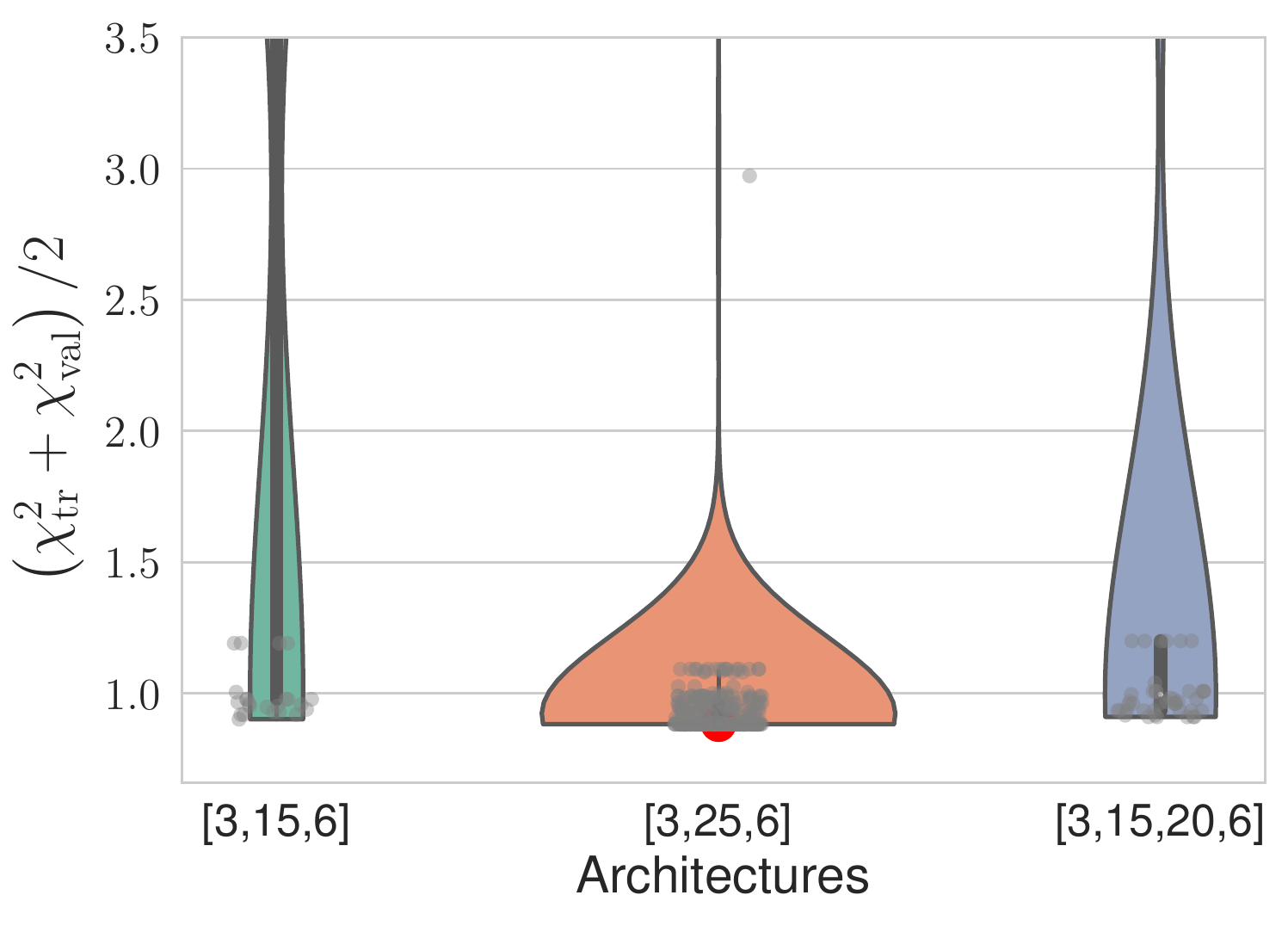}
  \includegraphics[width=0.495\textwidth]{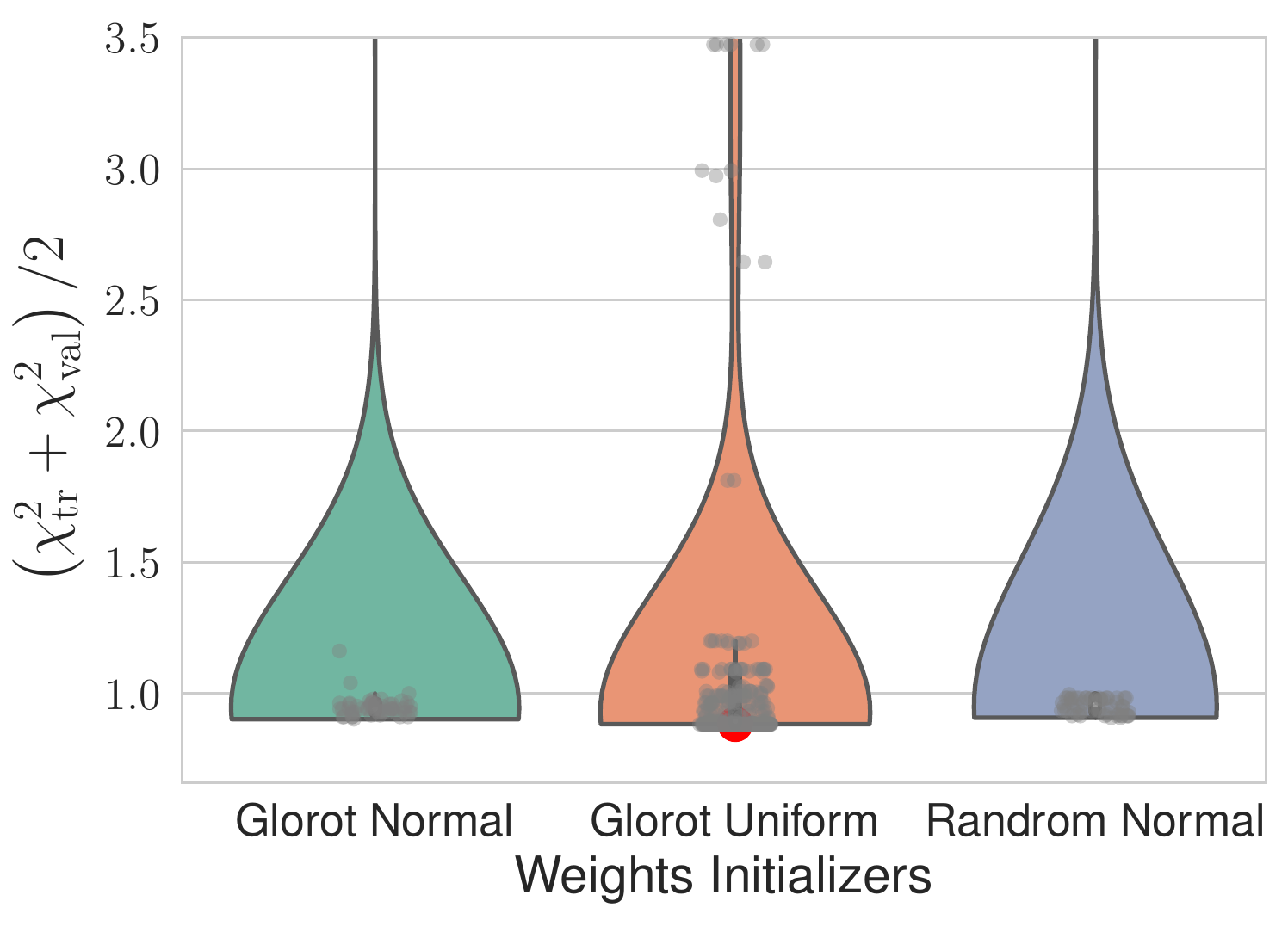}\\
  \includegraphics[width=0.495\textwidth]{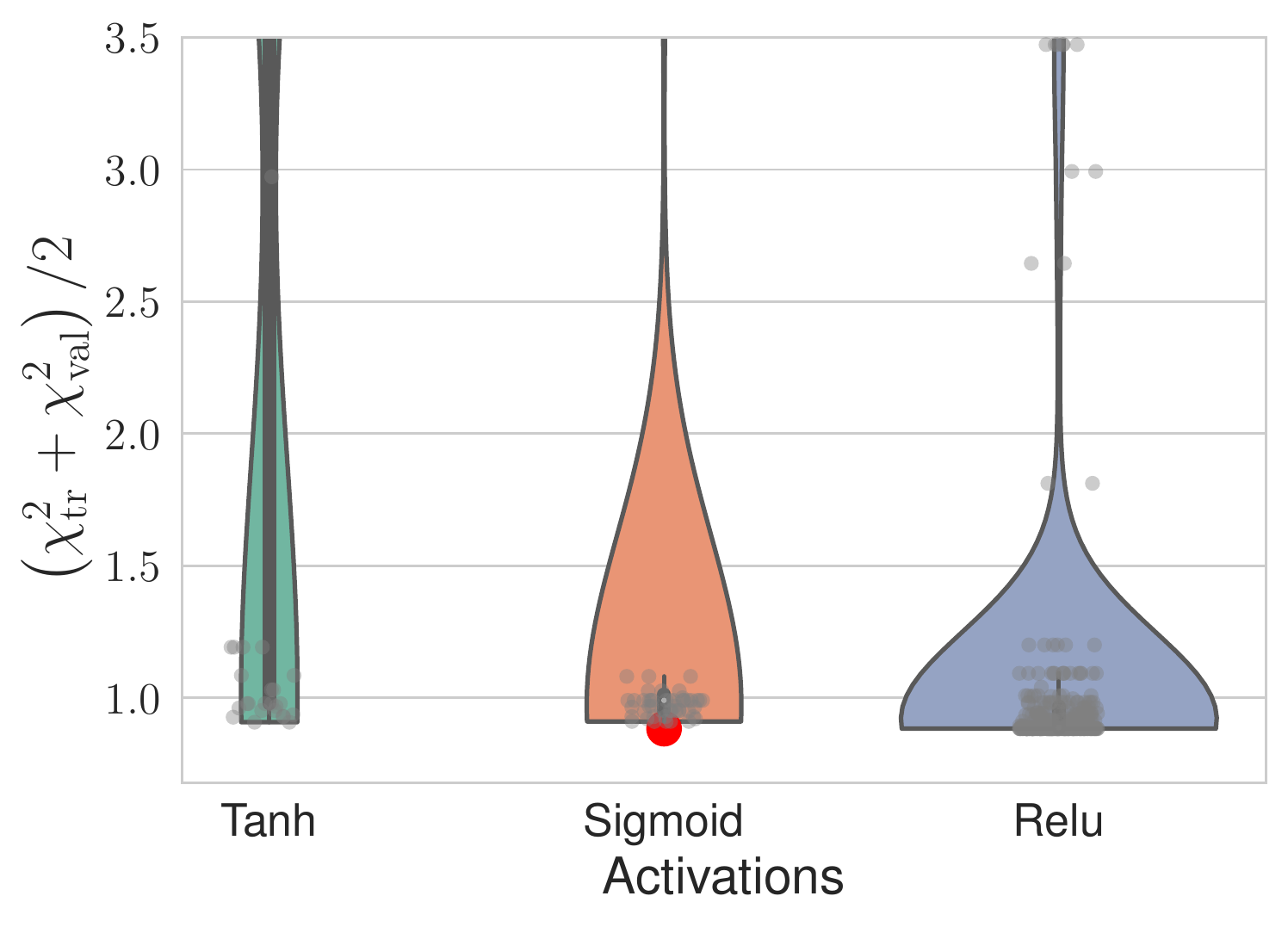}
  \includegraphics[width=0.495\textwidth]{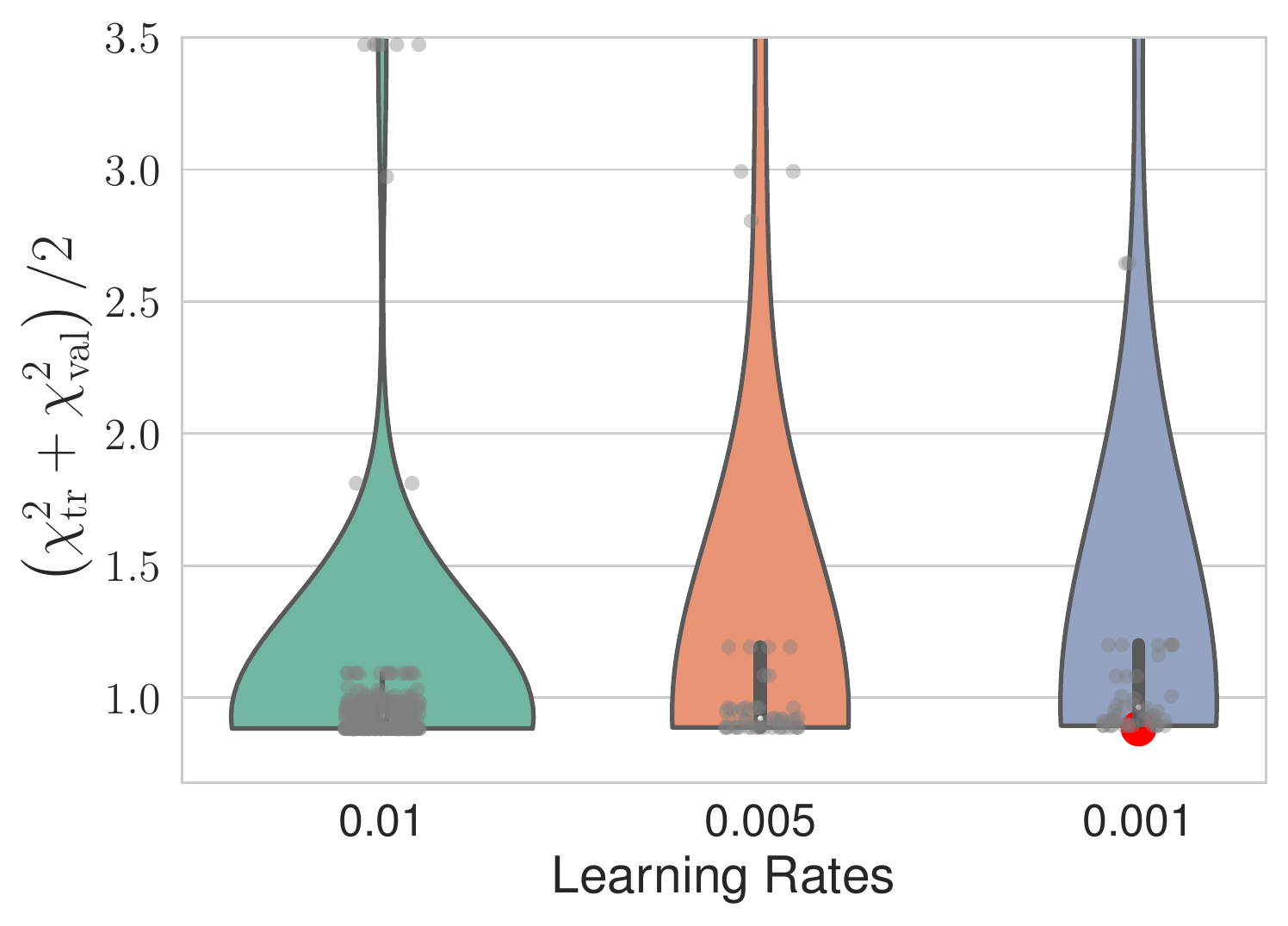}\\
  \caption{Graphical representation of a hyperparameter scan for
    representative hyperparameters produced with $1000$ trial
    searches using the TPE  algorithm.
    The values of the hyperparameter are represented on the $x$-axis
    while the hyperopt loss function is in the $y$-axis.
    In addition to the outcome of individual trials we 
    display a reconstruction of the probability distribution
    by means of the KDE method.
    The red dots indicate the values of the hyperparameters
    used in the nNNPDF2.0 analysis.
  }
  \label{fig:hyperscan}
\end{figure}

Figure~\ref{fig:hyperscan} is a graphical representation of a hyperparameter
scan obtained with $1000$ trial searches using the TPE algorithm. We show the
results for the architecture, weight initialisation method, activation
function, and learning rates of the optimiser. The values of the
hyperparameters are represented on the $x$-axis while the loss function,
Eq.~\eqref{eq:hyperopt_loss}, is reported on the $y$-axis. In addition to the
outcome of individual trials, we also display a reconstruction of the
probability distribution by means of the KDE method. The red dots indicate the
values of the best hyperparameters used in nNNPDF2.0, which were determined by
means of a trial and error selection procedure.

Parameters that exhibit denser tails in the KDE distributions of
Fig.~\ref{fig:hyperscan} are considered better choices as they yield more
stable trainings. For instance, one observes that a network with one single
hidden layer and $25$ nodes significantly outperforms a network with two hidden
layers. Indeed, not only the chosen architecture leads to the smallest value of
the loss function but also to a more stable behaviour. Similar patterns are
observed for the activation function and learning rate. On the other hand, no
clear preference is seen for the initialisation of the weights. Despite the
fact that the \texttt{Glorot Uniform} initialisation leads to the smallest
value of the objective function, the \texttt{Glorot Normal} one appears to yield
slightly better stability with more gray points concentrated at the tail and
hence a fatter distribution.

\begin{table}[!t]
  \centering
  \footnotesize
  \renewcommand{\arraystretch}{1.35}
\begin{tabularx}{\textwidth}{lC{6.6cm}C{6.0cm}}
  \toprule
  & nNNPDF3.0 & nNNPDF2.0 \\
  \midrule
  Architecture & $\left[3,25,6\right]$ & $\left[3,25,6\right]$ \\
  \midrule
  Weight initialisation & \texttt{Glorot Normal} & \texttt{Glorot Uniform}
  \\
  \midrule
  Bias initialisation & \texttt{Zeros} & \texttt{Zeros} \\
  \midrule
  Activation function & \textsc{ReLU} & \textsc{Sigmoid} \\
  \midrule
  Learning rate & $10^{-2}$ & $10^{-3}$ \\
  \midrule
  Optimiser & \textsc{Adam} & \textsc{Adam} \\
  \bottomrule
\end{tabularx}

    \vspace{0.3cm}
  \caption{The neural-network hyperparameters used in nNNPDF3.0, determined by
    means of the automated hyperoptimisation algorithm, compared with its
    counterparts determined from trial and error selection in nNNPDF2.0.}
  \label{tab:hyperparameter}
\end{table}

Following this hyperparameter optimisation process, we have determined
the baseline hyperparameters to be used in the nNNPDF3.0 analysis. We
list them in Table~\ref{tab:hyperparameter}.
For  reference, we also show the hyperparameters that
were chosen in nNNPDF2.0 by means of a trial and error selection procedure.
Remarkably, the automatically-selected optimal hyperparameters in nNNPDF3.0
coincide in many case with those selected by trial and error in nNNPDF2.0,
in particular for the neural network architecture, the initialisation of the
weight, and the optimiser. In terms of the initialisation of the weights,
the \texttt{Glorot Normal} and \texttt{Glorot Uniform} strategies exhibit
similar training behaviours. The main differences between the hyperparameters
of the nNNPDF2.0 and nNNPDF3.0 methodologies hence lie in the activation
function and the learning rate, where for the latter it is found that faster
convergence is achieved with a larger step size. We recall that situations
where the model could converge to a suboptimal solution are avoided thanks to 
the Adaptive Momentum Stochastic Gradient Descent (\textsc{Adam}) optimiser that
dynamically adjusts the learning rate.

All in all, the differences between the previous
and new model hyperparameters listed in Table~\ref{tab:hyperparameter}
are found to be rather moderate,
confirming the general validity of the choices that were adopted
in the nNNPDF2.0 analysis.

\subsection{The LHCb $D$-meson data and PDF reweighting}
\label{sec:RW}

As mentioned previously, the constraints on nNNPDF3.0 from the datasets
described in Sect.~\ref{sec:expdata} are accounted for by means of the
experimental data contribution, $\chi^2_{t_0}$, to the cost function in
Eq.~\eqref{eq:chi2_fit} used for the neural network training. The only
exceptions are the LHCb measurements of $D$-meson production discussed in 
Sect.~\ref{subsubsec:DmesonProduction}. The impact of these measurements is
instead determined by means of Bayesian
reweighting~\cite{Ball:2010gb,Ball:2011gg}.
The reason is that interpolation tables, that combine PDF and $\alpha_s$
evolution factors with weight tables for the hadronic matrix elements
(see Sect.~\ref{sec:theory_settings}), need to be pre-computed to allow for a
fast determination  of theoretical predictions.
The efficient computation of these theoretical predictions is critical for the
fit, as they must be evaluated a large number of times,  $\mathcal{O}(10^5)$,
as part of the minimisation procedure. However, no interface is currently
publicly available to generate interpolation tables in the format 
required to include LHCb $D$-meson data in the free-proton and nuclear
nNNPDF3.0 fits. Bayesian reweighting then provides a suitable alternative to
account for the impact of these measurements in the nPDF determination, since
the corresponding theory predictions must be evaluated only once per each of
the prior replicas.

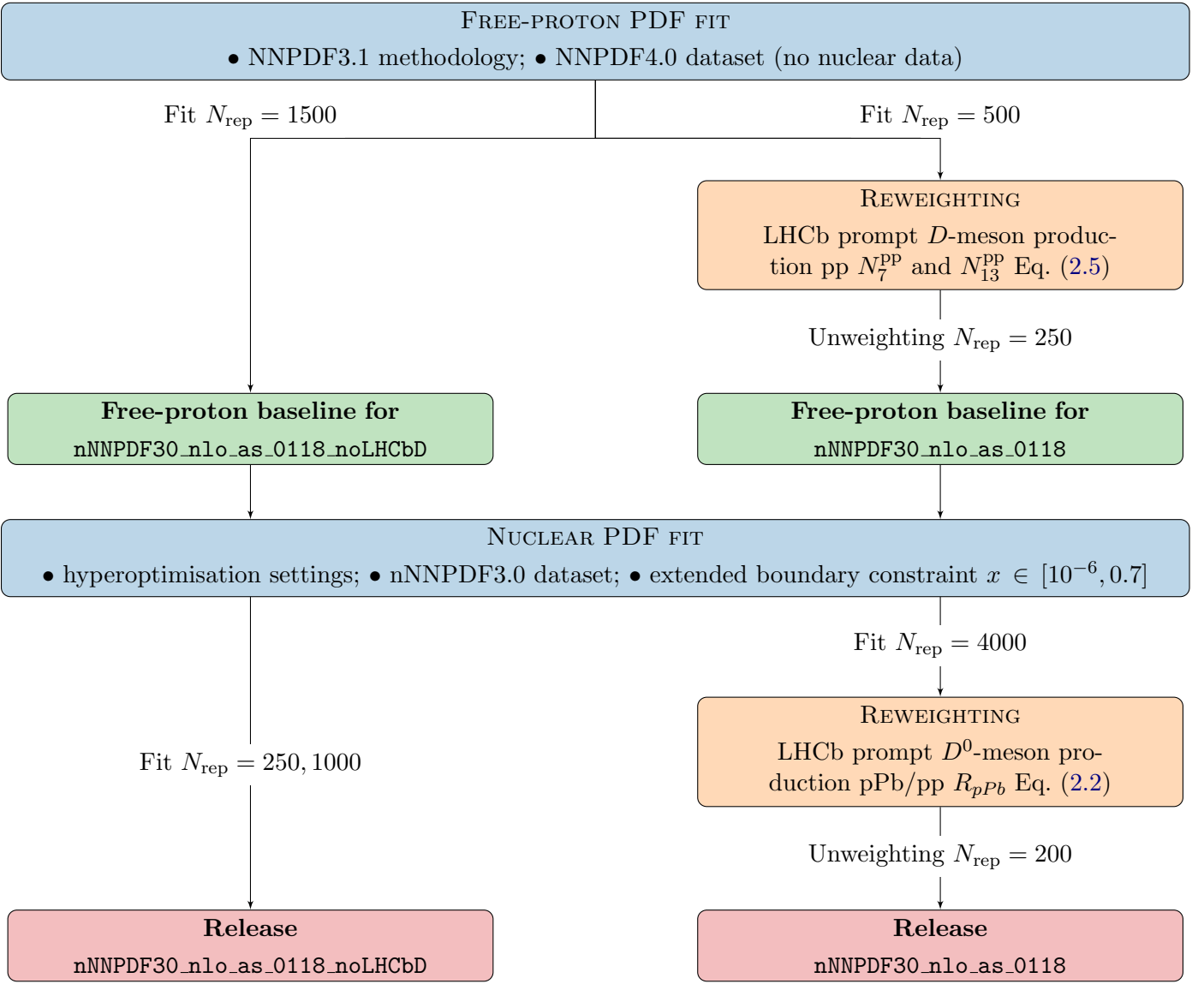
\begin{figure}[!t]
  \begin{center}
\begin{tikzpicture}
  \node [block] (proton) {{\sc Free-proton PDF fit}\\
    \vspace{0.1cm}
    $\bullet$ NNPDF3.1 methodology; $\bullet$ NNPDF4.0 dataset (no nuclear data)};
  \node [block1, below of=proton, node distance=3cm, xshift=5.3cm] (rw-proton) {{\sc Reweighting}\\
    \vspace{0.1cm}
    LHCb prompt $D$-meson production pp $N^{\rm pp}_7$ and $N^{\rm pp}_{13}$~Eq.~\eqref{eq:normalized_xsec_pp}};
  \node [block2, below of=proton, node distance=6cm, xshift=5.3cm] (proton-baseline-LHCb)
        {{\bf Free-proton baseline for}\\ \vspace{0.1cm}{\tt nNNPDF30\_nlo\_as\_0118}};
        \node [block2, below of=proton, node distance=6cm, xshift=-5.3cm] (proton-baseline)
        {{\bf Free-proton baseline for}\\ \vspace{0.1cm}{\tt nNNPDF30\_nlo\_as\_0118\_noLHCbD}};
    \node [block, below of=proton, node distance=8cm] (nuclear-fit) {{\sc Nuclear PDF fit}\\
      \vspace{0.1cm}
      $\bullet$ hyperoptimisation settings; $\bullet$ nNNPDF3.0 dataset; $\bullet$ extended boundary constraint $x\in[10^{-6},0.7]$};
    \node [block1, below of=proton, node distance=11cm, xshift=5.3cm] (rw-nuclear) {{\sc Reweighting}\\
      \vspace{0.1cm}
      LHCb prompt $D^0$-meson production pPb/pp $R_{pPb}$~Eq.~\eqref{eq:Rf}};
    \node [block3, below of=proton, node distance=14cm, xshift=5.3cm] (nuclear-LHCb)
          {{\bf Release}\\ \vspace{0.1cm}{\tt nNNPDF30\_nlo\_as\_0118}};
    \node [block3, below of=proton, node distance=14cm, xshift=-5.3cm] (nuclear)
          {{\bf Release}\\ \vspace{0.1cm}{\tt nNNPDF30\_nlo\_as\_0118\_noLHCbD}};
   \draw [line] (proton) |- (5cm,-1.5cm) -| node[above=0.,fill=white] {Fit $N_{\rm rep}=500$} (rw-proton);
   \draw [line] (proton) |- (5cm,-1.5cm) -| node[above=0.,fill=white] {Fit $N_{\rm rep}=1500$} (proton-baseline);     
   \draw [line] (rw-proton) -- node[above=-0.3,fill=white] {Unweighting $N_{\rm rep}=250$}  (proton-baseline-LHCb);
   \draw [line] (proton-baseline) -- (-5.3,-7.4);
   \draw [line] (proton-baseline-LHCb) -- (5.3,-7.4);
   \draw [line] (-5.3,-8.6) -- node[above=-0.5,fill=white] {Fit $N_{\rm rep}=250,1000$} (nuclear);      
   \draw [line] (5.3,-8.6) -- node[above=-0.3,fill=white] {Fit $N_{\rm rep}=4000$} (rw-nuclear);
   \draw [line] (rw-nuclear) -- node[above=-0.3,fill=white] {Unweighting $N_{\rm rep}=200$} (nuclear-LHCb);
\end{tikzpicture}
\end{center}
  \caption{Schematic representation of the fitting strategy used to construct
    the nNNPDF3.0 determination. The starting point is a dedicated proton
    global fit based on the NNPDF3.1 methodology but with the NNPDF4.0
    proton-only dataset. This proton PDF fit is then reweighted with the LHCb
    $D$ meson production data in pp collisions, which upon unweighting results
    into the proton PDF baseline to be used for nNNPDF3.0 (via the $A=1$
    boundary condition). Then to assemble nNNPDF3.0 we start from the nNNPDF2.0
    dataset, augment it with the NNPDF4.0 deuteron data and the new pPb LHC
    cross-sections, and produce a global nPDF fit with the hyperoptimised 
    methodology. Finally this is reweighted by the LHCb $D^0$ meson production
    measurements in pPb collisions, and upon unweighting we obtain the final
    nNNPDF3.0 fits.}
  \label{fig:strategy}
\end{figure}

Our strategy therefore combines fitting and reweighting procedures, as
illustrated in the right branch of the flowchart in Fig.~\ref{fig:strategy}.
We describe the various steps of this strategy in turn.

\begin{itemize}

\item The first step concerns the free-proton PDF baseline. A variant of the
  NNPDF3.1 NLO fit is constructed as described in Sect.~\ref{sec:freeproton},
  which is then reweighted with the LHCb $D$-meson measurements of
  $N^{\rm pp}_7$ and $N^{\rm pp}_{13}$, see Eq.~(\ref{eq:normalized_xsec_pp}) and
  the discussion in Sect.~\ref{subsubsec:DmesonProduction}. The NNPDF3.1 fit
  variant is composed of $N_{\rm rep}=500$ replicas; after reweighting one ends
  up with $N_{\rm eff}=250$ effective replicas. Table~\ref{tab:chi2_N7N13}
  collects the values of $\chi^2/n_{\rm dat}$ for the LHCb $D$-meson measurements
  of $N^{\rm pp}_7$ and $N^{\rm pp}_{13}$ before and after reweighting.
  We observe that the dataset is relatively well described already before
  reweighting; the improvement after reweighting is therefore noticeable,
  although not dramatic. As shown in Fig.~\ref{fig:freeproton}, the impact of
  the LHCb $N^{\rm pp}_7$ and $N^{\rm pp}_{13}$ data on
  the free-proton baseline PDFs consists of a reduction of uncertainties in the
  small-$x$ region; central values are left mostly unaffected.
  
\begin{table}[!t]
  \centering
  \footnotesize
  \renewcommand{\arraystretch}{1.35}
\begin{tabularx}{\textwidth}{XC{3.5cm}C{3.5cm}}
  \toprule
  $\chi^2/n_{\rm dat}$  &   $N^{\rm pp}_7$   & $N^{\rm pp}_{13}$  \\
  \midrule
  Prior        &  0.81   &  1.06             \\
  \midrule
  Reweighted   &  0.76   &  0.91           \\
  \bottomrule
\end{tabularx}

  \vspace{0.2cm}
  \caption{Values of $\chi^2/n_{\rm dat}$ for the LHCb $D$-meson measurements of
    $N^{\rm pp}_7$ and $N^{\rm pp}_{13}$ before and after reweighting.}
  \label{tab:chi2_N7N13}
\end{table}

\item The second step consists in producing the nNNPDF3.0 prior fit. This is
  based on the dataset described in Sect.~\ref{sec:expdata} (except
  LHCb pPb $D^0$-meson data) and makes use of the free-proton PDF set
  determined at the end of the previous step. The nNNPDF3.0 prior fit is made
  of $N_{\rm rep}=4000$ replicas. Such a large number of replicas ensures
  sufficiently high statistics for the subsequent reweighting of this nNNPDF3.0
  prior fit with the LHCb $D$-meson pPb data.

\item The third step is the reweighting of the nNNPDF3.0 prior fit with the
  LHCb measurements of $D^0$-meson production in pPb collisions. As discussed
  in Sect.~\ref{subsubsec:DmesonProduction} we use the ratio of pPb to pp
  spectra, in the forward region, Eq.~\eqref{eq:def_nuclear_modification_ratio},
  by default. The stability of our results if instead the forward-to-backward
  ratio measurements Eq.~\eqref{eq:Rfb} are used is quantified in
  Sect.~\ref{sec:stability}. After reweighting, we end up with 
  $N_{\rm eff}=200$~(500) effective replicas when reweighting with
  $R_{\rm pPb}$~($R_{\rm fb}$). A satisfactory description of the LHCb $D^0$-meson
  data in pPb collisions is achieved, as will be discussed in
  Sect.~\ref{sec:D0_reweighting}. Our final nNNPDF3.0 set is
  constructed after unweighting~\cite{Ball:2011gg} and contains $N_{\rm rep}=200$
  replicas. This is released in the usual {\sc\small LHAPDF} format for the
  relevant values of $A$, listed in Sect.~\ref{subsection:delivery}.

\end{itemize}  

In addition to this baseline nNNPDF3.0 fit, as indicated on the left
branch of the flowchart in Fig.~\ref{fig:strategy}, we also produce and release
a variant without any LHCb $D$-meson data, neither in the free-proton baseline
nor in the nuclear fit. For completeness, we list here the specific procedure
adopted to construct this variant.

\begin{itemize}

\item The first step concerns again the free-proton PDF baseline.
  This is the  variant of the NNPDF3.1 NLO fit constructed as described
  in Sect.~\ref{sec:freeproton} and made of $N_{\rm rep}=1500$ replicas.

\item The second step consists in producing the variant of the nNNPDF3.0 fit
  from the datasets described in Sect.~\ref{sec:expdata}, except the LHCb
  $D$-meson cross-sections. We produce two ensembles with $N_{\rm rep}=250$
  and $N_{\rm rep}=1000$ replicas, which we also release in the usual
  {\sc\small LHAPDF} format for the relevant values of $A$.
  In Sects.~\ref{sec:nnpdf30_noLHCb} and~\ref{sec:D0_reweighting} we will
  compare the nNNPDF3.0 baseline fit and the variant without any LHCb
  $D$-meson data.

\end{itemize}

We have explicitly verified the validity of the reweighting procedure when
applied to nPDFs (see App.~\ref{app:RWex} for more details), by comparing the
outcome of a fit where a subset of the CMS dijet measurements from pPb
collisions are included either by a fit or by reweighting.
We have confirmed that the two methodologies lead to compatible
results, in particular that nPDF central values are similarly shifted and
uncertainties are similarly reduced.
Nevertheless, we remark that the reweighting procedure has some inherent
limitations as compared to a fit.
First of all, reweighting is only expected to reproduce the outcome of a
fit provided that statistics, i.e. the number of effective replicas,
is sufficiently high.
Second, the results obtained by means of reweighting may differ from
those obtained after a fit in those cases where the figure of merit used to
compute the weights is different from that used for fit minimisation.
For instance, if the $\chi^2$ function used for reweighting does not account
for the cross-section positivity and the $A=1$ free-proton boundary condition
constraints as in Eq.~\eqref{eq:chi2_fit}.

\section{Results}
\label{sec:results}

In this section we present the main results of this work, namely the
nNNPDF3.0 global analysis of nuclear PDFs.
First of all, we discuss the key features of the variant of nNNPDF3.0 without the LHCb 
$D$-meson data, and compare it with the nNNPDF2.0 reference. 
Second, we describe the outcome of the reweighting  of a
nNNPDF3.0 prior
set with the  LHCb $D$-meson data, which defines the nNNPDF3.0 default
determination, and the resulting constraints on the nuclear modification factors.
Third, we study the goodness-of-fit to the new datasets
incorporated in the present analysis and carry out
representative comparisons with experimental data.
Fourth, we study the $A$-dependence of our results and assess
the local statistical significance of nuclear modifications.
Finally, we compare the nNNPDF3.0 determination with two other
global analyses of nuclear PDFs, EPPS16 and nCTEQ15WZ+SIH.

The stability of nNNPDF3.0  with respect to methodological 
and dataset variations is then studied in Sect.~\ref{sec:stability}, while its implications for 
the ultra high-energy neutrino-nucleus interaction cross-sections are quantified in Sect.~\ref{sec:uheneut}.
Furthermore, representative comparisons between the
predictions from nNNPDF3.0
and experimental data from pPb collisions 
can be found in App.~\ref{app:datacomp}.

\subsection{The nNNPDF3.0 (no LHCb $D$) fit}
\label{sec:nnpdf30_noLHCb}

We present first the main features of the nNNPDF3.0
variant that excludes the LHCb $D$ meson data (from both pp and
pPb collisions), following the strategy
indicated in Sect.~\ref{sec:fitting}.
In the following, this variant  is denoted as nNNPDF3.0 (no LHCb $D$).
This fit differs
from nNNPDF2.0 due to three main factors: {\it i)} the significant number of new datasets
involving D, Cu,
and Pb targets, {\it ii)} the improved treatment of $A=1$ free-proton PDF boundary condition,
and {\it iii)} the automated optimisation of the model hyperparameters.

\begin{figure}[!t]
  \centering
   \includegraphics[width=\textwidth]{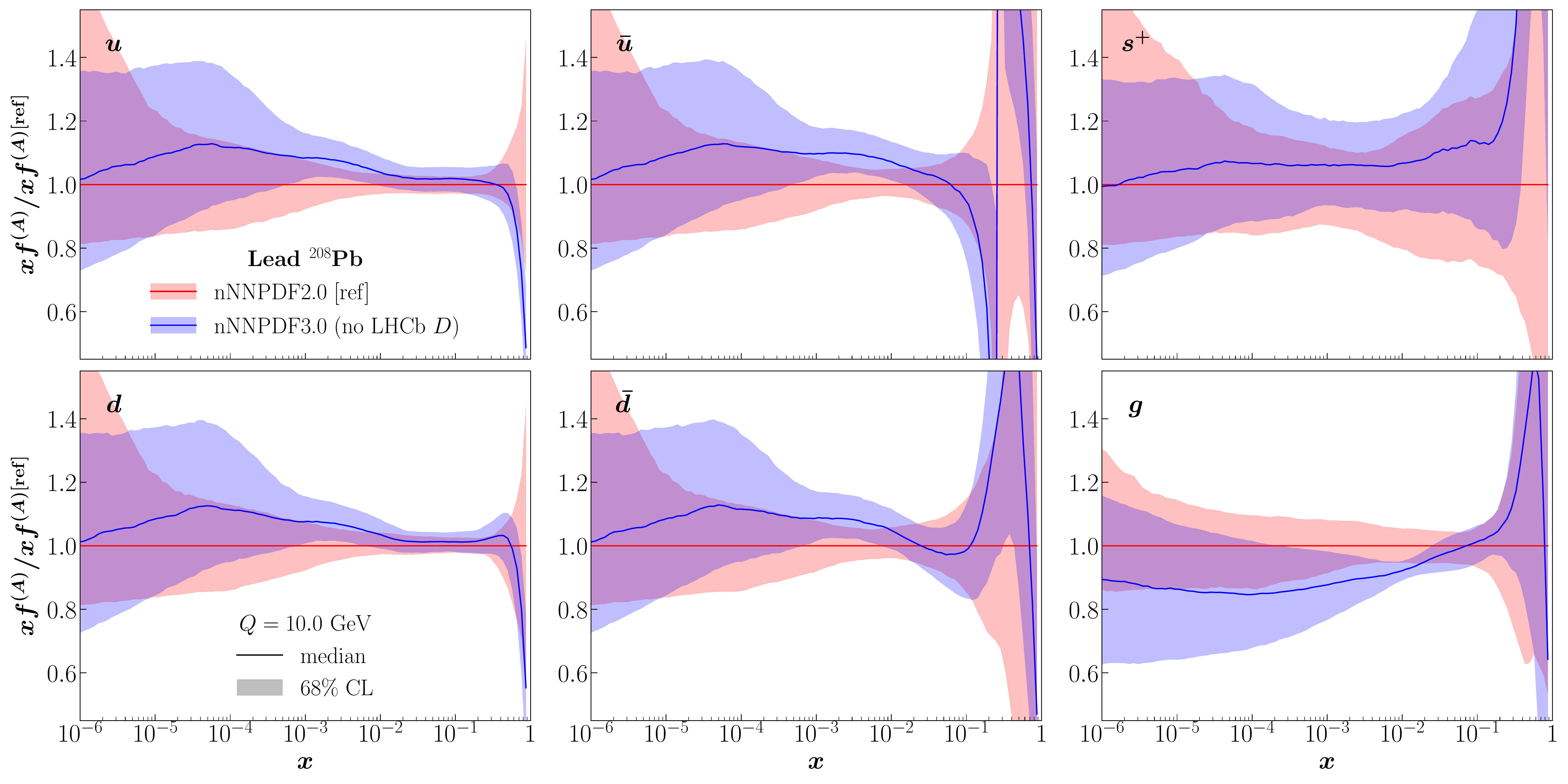}
   \caption{\small Comparison between the nNNPDF2.0 and nNNPDF3.0 (no LHCb $D$) fits
     for the lead PDFs at $Q=10$ GeV, normalised to the central value
     of the nNNPDF2.0 reference.
     The uncertainty bands indicate the 68\% CL intervals.
   }
  \label{fig:nNNPDF30vs20_Pb208_a_RationPDFs_Q10}
\end{figure}

\begin{figure}[!t]
  \centering
   \includegraphics[width=\textwidth]{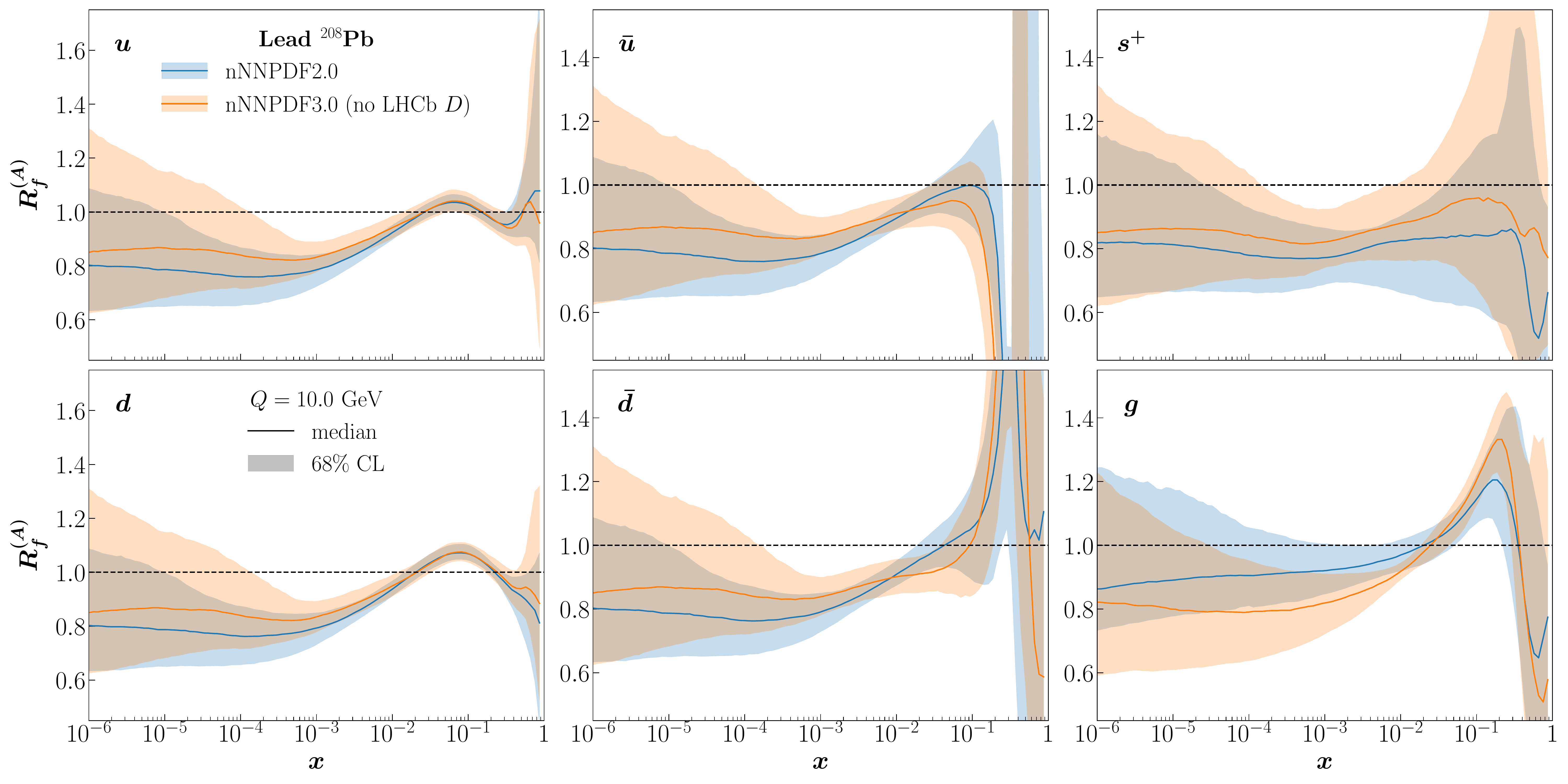}
   \caption{\small Same as Fig.~\ref{fig:nNNPDF30vs20_Pb208_a_RationPDFs_Q10}
     now in terms of the nuclear modification ratios
$R_f^{(A)}(x,Q^2)$.
   }
  \label{fig:nNNPDF30vs20_Pb208_a_NuclRationPDFs_Q10}
\end{figure}

The comparison between the nNNPDF2.0 and nNNPDF3.0 (no LHCb $D$) fits
is presented in Fig.~\ref{fig:nNNPDF30vs20_Pb208_a_RationPDFs_Q10}
at the level of lead PDFs and in Fig.~\ref{fig:nNNPDF30vs20_Pb208_a_NuclRationPDFs_Q10}
at the level of nuclear modification ratios, defined
as
  \be
  \label{eq:nuclear_modification_ratios}
  R_f^{(A)}(x,Q) \equiv \frac{ f^{(N/A)}(x,Q)}{\frac{Z}{A} f^{(p)}(x,Q) + \frac{(A-Z)}{A}f^{(n)}(x,Q)}\, ,
  \ee
  where $f^{(N/A)}$, $f^{(p)}$, and $f^{(n)}$ indicate the PDFs of the average nucleon $N$
  bound in a nucleus with $Z$ protons and $A-Z$ neutrons, the free-proton, and the free-neutron
  PDFs respectively,
  see App.~\ref{app:conventions} for an overview of the conventions
  and notation used throughout this work.
In both cases, the results display the 68\% CL uncertainties and are evaluated
at $Q=10$ GeV.

First of all,
the two determinations are found to be consistent within
uncertainties for all the nPDF flavours in the full range of $x$
for which experimental data is available.
The qualitative behaviour of the nuclear modification ratios defined in
Eq.~(\ref{eq:nuclear_modification_ratios})
is similar in the two determinations, with the strength of the small-$x$ shadowing
being reduced (increased) in the quark (gluon) nPDFs in nNNPDF3.0 (no LHCb $D$) as
compared to the nNNPDF2.0 analysis.
From Fig.~\ref{fig:nNNPDF30vs20_Pb208_a_NuclRationPDFs_Q10}, one also observes
how
the large-$x$ behaviour of the nuclear modification factors for the quark PDFs
is similar in the two fits,
while for the gluon one finds an increase in the strength of anti-shadowing
peaking at $x\simeq 0.2$.
These differences in $R_g^{(A)}$ reported between nNNPDF3.0 (no LHCb $D$)
and nNNPDF2.0
can be traced back to the constraints
provided by the CMS dijet cross-sections in pPb, which will be further discussed
in Sect.~\ref{sec:impact_dijet}.
One also finds that uncertainties in the small-$x$ region, $x\lsim 10^{-3}$,
where neither of the two fits includes direct constraints, are increased in nNNPDF3.0.
This is a consequence of the improved implementation of the $A=1$ proton
PDF boundary condition discussed in Sect.~\ref{sec:freeproton}, as will
be further studied in Sect.~\ref{sec:nnnpdf20_reloaded}.

Furthermore, in the region where the bulk of experimental data  on nuclear targets lies,
$x\gsim 10^{-3}$, the uncertainties on the quark nPDFs of lead are also
basically unchanged between the two analyses.
The impact of the new LHC $W$ and $Z$ production measurements in nNNPDF3.0
is mostly visible for the up and down anti-quark PDFs,
both in terms of a shift in the central values and of a moderate reduction
of the nPDF uncertainties.
The increase in the central value of the total strangeness is related
to the inclusion of deuteron and copper cross-sections in nNNPDF3.0 together
with the improved $A=1$ boundary condition,
as will be demonstrated in Sect.~\ref{sec:nnnpdf20_reloaded}.

\subsection{The nNNPDF3.0 determination}
\label{sec:D0_reweighting}

The nNNPDF3.0~(no LHCb~$D$) fit presented in the previous section is
the starting point to quantify the constraints
on the proton and nuclear PDFs provided by LHCb $D$-meson production data
by means of Bayesian reweighting.
As discussed in Sect.~\ref{sec:RW}, the first step is to produce the nNNPDF3.0 prior fit,
which coincides with
nNNPDF3.0~(no LHCb~$D$) with the only difference being that the proton PDF boundary
condition now accounts for the constraints provided by the LHCb $D$-meson data in pp collisions at 7 and 13 TeV.
The differences and similarities between the proton PDF boundary conditions used for
the nNNPDF3.0 and nNNPDF3.0 (no LHCb $D$) fits and their nNNPDF2.0 counterpart
were studied in Fig.~\ref{fig:freeproton}.
Subsequently, the LHCb data for $R_{\rm pPb}$ in the forward region
is added to this prior nPDF set using reweighting.

\begin{figure}[!t]
    \centering
     \includegraphics[width=0.95\textwidth]{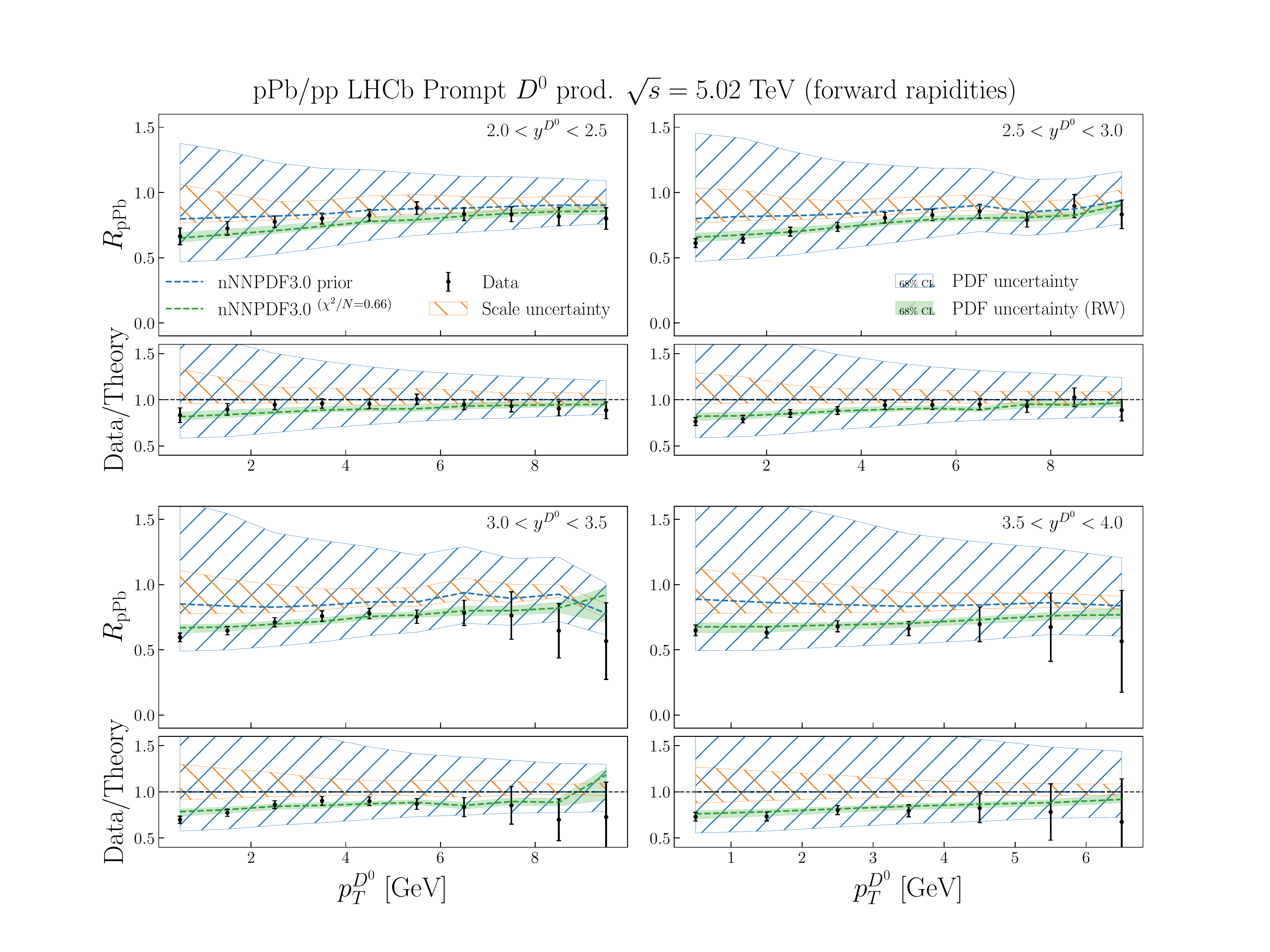}
     \caption{\small Comparison between the LHCb data on $D^0$-meson production
       from pPb collisions
       in the forward region and the corresponding theoretical predictions based on the
       nNNPDF3.0 prior set described in  Sect.~\ref{sec:RW}.
       The ratio between $D^0$-meson  spectra in pPb and pp collisions, $R_{\rm pPb}$ in
       Eq.~(\ref{eq:def_nuclear_modification_ratio}), is presented in four bins
       in $D^0$-meson rapidity $y^{D^0}$ as a function of the transverse momentum $p_T^{D^0}$.
       We display separately the PDF and scale uncertainty bands, and the bottom panels
     show the ratios to the central value of the theory prediction based on the prior.}
    \label{fig:D0_pPb_forward}
\end{figure}

Fig.~\ref{fig:D0_pPb_forward} displays the
comparison between the LHCb  data for $R_{\rm pPb}$, Eq.~(\ref{eq:def_nuclear_modification_ratio}),
for $D^0$-meson production in pPb collisions (relative to that in pp collisions)
in the forward region, and the corresponding theoretical predictions based on this
nNNPDF3.0 prior set.
The LHCb measurements are
presented in four bins
in $D^0$-meson rapidity $y^{D^0}$ as a function of the transverse momentum $p_T^{D^0}$,
and we display separately the PDF and scale uncertainty bands, and the bottom panels
show the ratios to the central value of the theory prediction.

\begin{figure}[!t]
  \centering
  \includegraphics[width=0.95\textwidth]{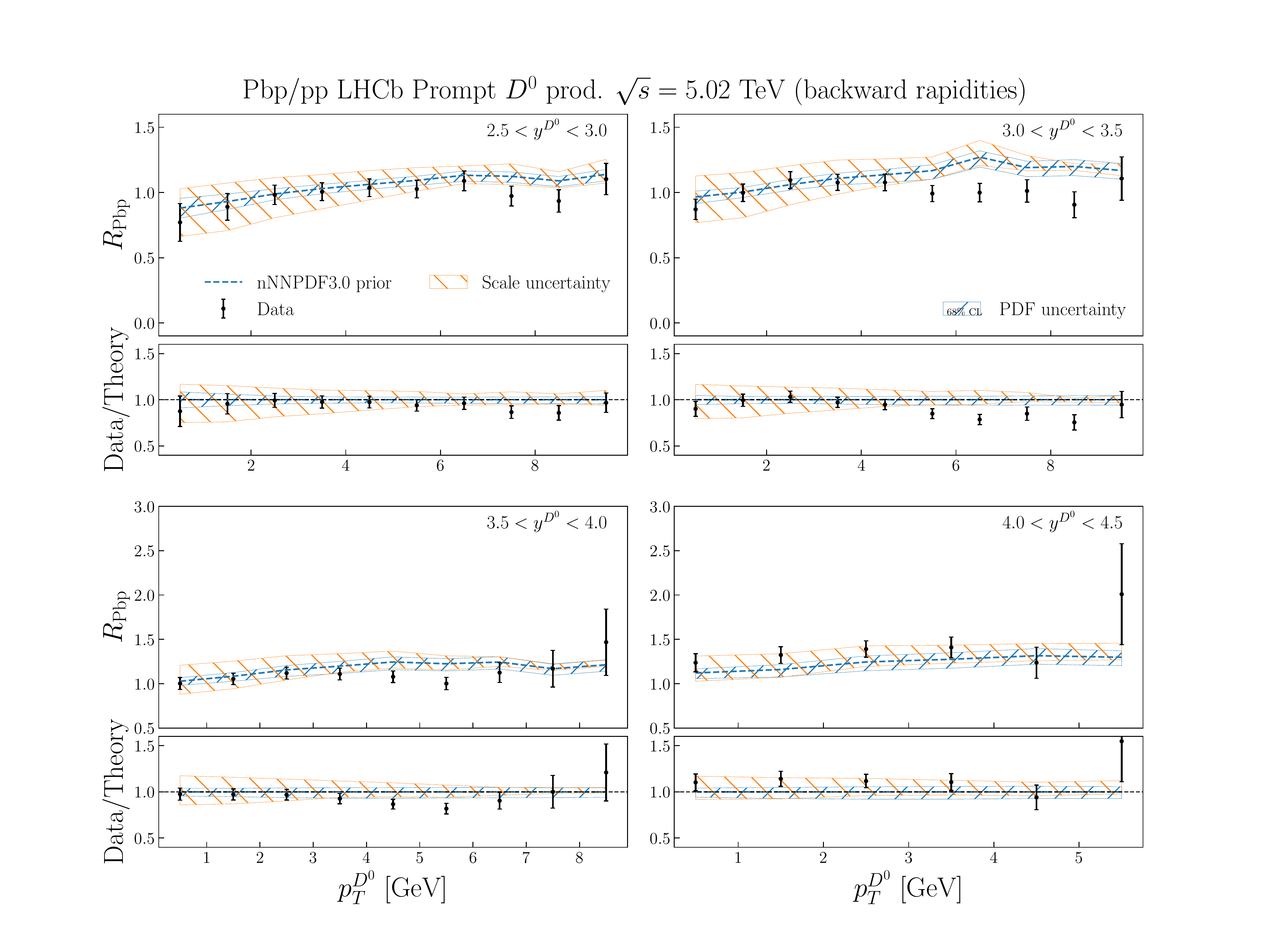}
  \caption{Same as Fig.~\ref{fig:D0_pPb_forward}, now
    comparing the theory predictions based on the nNNPDF3.0 prior fit
    with the backward rapidity bins
    of the LHCb measurement of $D^0$-meson production in Pbp collisions.
    While the predictions are consistent with the LHCb data,
    uncertainties due to MHOs now are the dominant source of theory error (as opposed to the pPb forward data), 
    hence this dataset is not amenable for inclusion in nNNPDF3.0.
  }
  \label{fig:D0_pPb_backward}
\end{figure}

From Fig.~\ref{fig:D0_pPb_forward} one can observe how PDF uncertainties of the
prior (that does not yet contain $R_{\rm pPb}$ $D^0$-meson data) are very
large, and completely dominate over the uncertainties due to missing higher
order (MHOs), for the whole kinematic range for which the LHCb measurements are
available.
The uncertainties due to MHOs (or scale uncertainties) are evaluated here by independently varying the factorisation and renormalisation scales around the nominal scale $\mu = E_{T}^c$ with the constraint $1/2 \leq \mu_F/\mu_R \leq 2$, and correlating those scales choices between numerator and denominator of the ratio observable defined in Eq.~\eqref{eq:def_nuclear_modification_ratio}.
Furthermore, these PDF uncertainties are also much larger than the experimental
errors, especially for the bins in the low $p_T^{D^0}$ region which dominate the sensitivity
to the small-$x$ nPDFs of lead.
Within these large PDF uncertainties, the predictions based on the nNNPDF3.0
prior fit agree well with the LHCb measurements.
This feature makes the LHCb forward $R_{\rm pPb}$ data amenable
to inclusion in a nPDF analysis, as opposed to the situation
with the corresponding measurements in the backward
region, shown in Fig.~\ref{fig:D0_pPb_backward}, where uncertainties due to MHOs
are larger than both PDF and experimental uncertainties.
Because of this, the LHCb backward $R_{\rm Pbp}$ data are not further considered in the nNNPDF3.0 analysis.
Considering  the low $p_T^{D^0}$ region of the $R_{\rm pPb}$ measurements,
one finds that the LHCb data prefer a smaller
central value than the central prediction for the nNNPDF3.0 prior
fit, indicating than a stronger shadowing at small-$x$ is being favoured.
Indeed, as we show next, once the LHCb $D$-meson constraints are included
via reweighting, the significance of small-$x$ shadowing in nNNPDF3.0 markedly increases.

\begin{figure}[!t]
  \centering
\includegraphics[width=\textwidth]{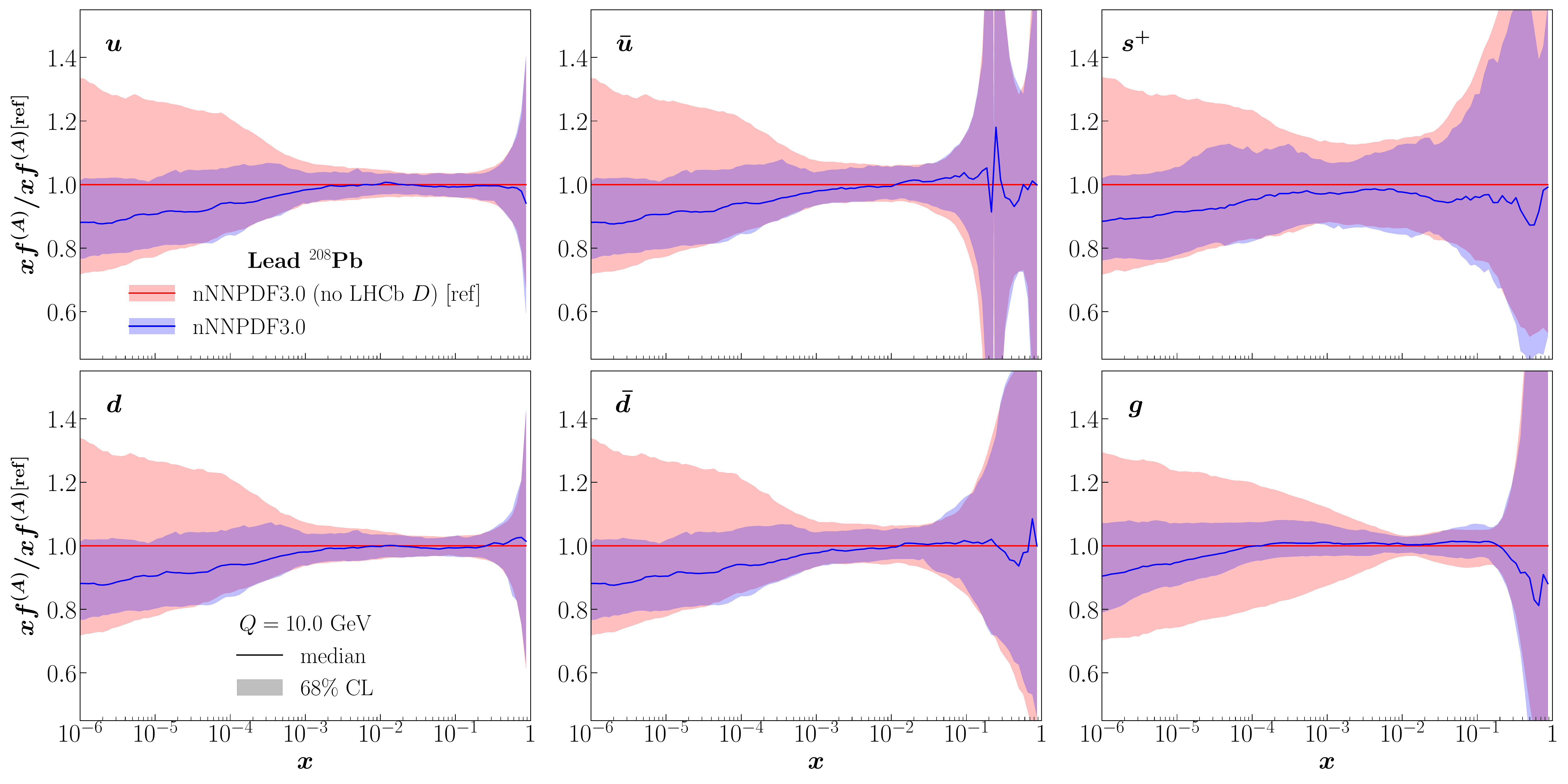}
  \caption{\label{fig:nNNPDF30_Pb208_a_RationPDFs_Q10}  Comparison of the nPDFs
of lead nuclei at $Q=10$ GeV between nNNPDF3.0~(no LHCb~$D$) and nNNPDF3.0,
normalised to the central value of the former.}
\end{figure}

The comparison between the  nPDFs of
lead nuclei at $Q=10$ GeV for the nNNPDF3.0 prior fit and
the corresponding reweighted results,
normalised to the central value of the former, is shown in
Fig.~\ref{fig:nNNPDF30_Pb208_a_RationPDFs_Q10}.
This comparison quantifies the impact of the LHCb $D$-meson production
measurements when added to the nNNPDF3.0 prior fit.
The reweighted nPDFs of lead nuclei are consistent with those of the prior, and display
a clear uncertainty reduction for $x\lsim 10^{-2}$ (for the gluon)
and $x\lsim 10^{-3}$ (for the sea quarks).
For instance, in the case of the gluon the nPDF uncertainties
are reduced by around a factor three for $x\simeq 10^{-4}$, highlighting
the constraining power of these LHCb measurements.
In terms of the central values, that of the gluon is mostly left
unchanged as compared to the prior, while for the sea quarks
one gets a suppression of up to a few percent.
We note that the reweighting procedure also affects the proton baseline,
given that weights are applied to each replica including the full $A$-dependence
of the parametrisation.

\begin{figure}[!t]
  \centering
\includegraphics[width=\textwidth]{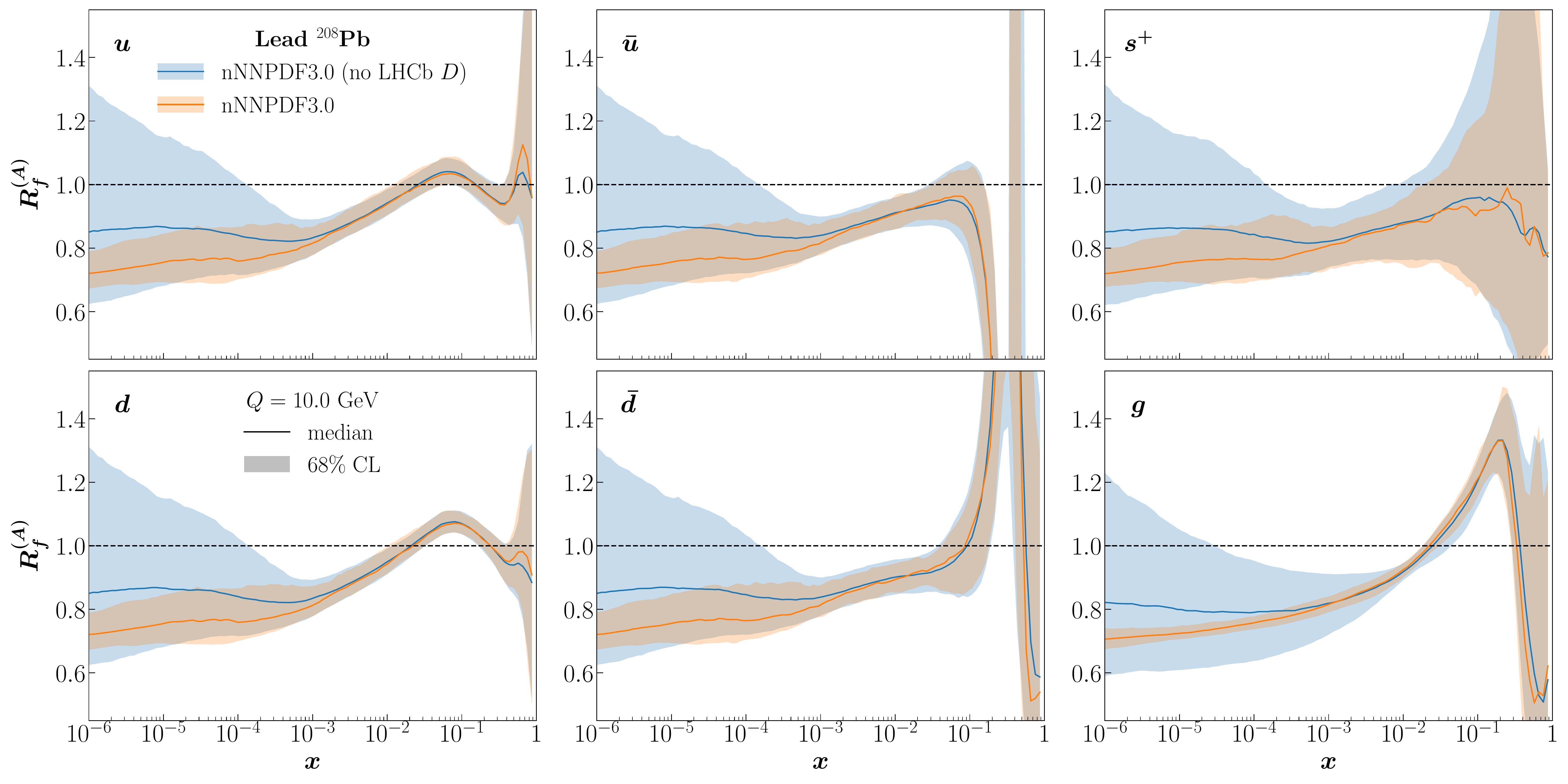}
\caption{Same as Fig.~\ref{fig:nNNPDF30_Pb208_a_RationPDFs_Q10} now
  presented in terms of the terms of the nuclear modification ratios $R_f^{(A)}(x,Q)$.
\label{fig:nNNPDF30_RA_Pb208_a_NuclearRationPDFs_Q10}
}
\end{figure}

The same comparison as Fig.~\ref{fig:nNNPDF30_Pb208_a_RationPDFs_Q10}
is displayed now in terms of the nuclear modification ratios $R_f^{(A)}(x,Q)$
in Fig.~\ref{fig:nNNPDF30_RA_Pb208_a_NuclearRationPDFs_Q10}.
As discussed in Sect.~\ref{sec:RW}, in the present analysis we consider in a coherent
manner the constraints of the LHCb $D$-meson data both on the proton
and nuclear PDFs while keeping track of their correlations,
and hence the impact on the ratios $R_f^{(A)}$ is in general expected
to be more marked as compared to that restricted to the lead PDFs.
Indeed, considering first the nuclear modification ratio for the gluon,
we find that the LHCb $D^0$-meson measurements in pPb collisions bring in
an enhanced shadowing for $x\lsim 10^{-4}$ together with an associated reduction
of the PDF uncertainties in this region by up to a factor five.
Hence the LHCb data constrain $R_g$ more than it does the absolute
lead PDFs in Fig.~\ref{fig:nNNPDF30_Pb208_a_RationPDFs_Q10}, demonstrating the
importance of accounting for the correlations between proton and lead PDFs.
In the case of the sea quark PDFs, the enhanced shadowing for  $x\lsim 10^{-3}$
and the corresponding uncertainty reduction is qualitatively similar to that
observed at the lead PDF level.
The preference of the LHCb $D$-meson production
measurements for a strong small-$x$ shadowing of the quark and gluon
PDFs of lead is in agreement with related studies of
the same process in the literature~\cite{Eskola:2019bgf,Eskola:2021nhw,Kusina:2017gkz}.

Whenever the nuclear ratios deviate from unity, $R^{(A)}_f(x,Q)\ne 1$,
the fit results favour non-zero nuclear modifications of the free-proton
PDFs.
However, such non-zero nuclear modifications will not be significant unless
the associated nPDF uncertainties are small enough.
In order to quantify the
local statistical significance of the nuclear modifications,
it is useful to evaluate the pull on $R^{(A)}_f(x,Q)$ defined as
\be
\label{eq:pull_global}
P\lc R_f^{(A)}\rc (x,Q) \equiv \frac{\lp R^{(A)}_f(x,Q)-1 \rp}{\delta R^{(A)}_f(x,Q)} \, ,
\ee
where  $\delta R^{(A)}_f(x,Q)$ indicates the 68\% CL uncertainties associated
to the nuclear modification ratio for the $f$-th flavour.
Values of these pulls such that $|P|\lsim 1$ indicate consistency with no nuclear
modifications at the 68\% CL, while $|P|\gsim 3$ corresponds
to a local statistical significance of nuclear modifications at the $3\sigma$ level,
the usually adopted threshold for evidence, in units of the nPDF uncertainty.

\begin{figure}[!t]
  \centering
\includegraphics[width=\textwidth]{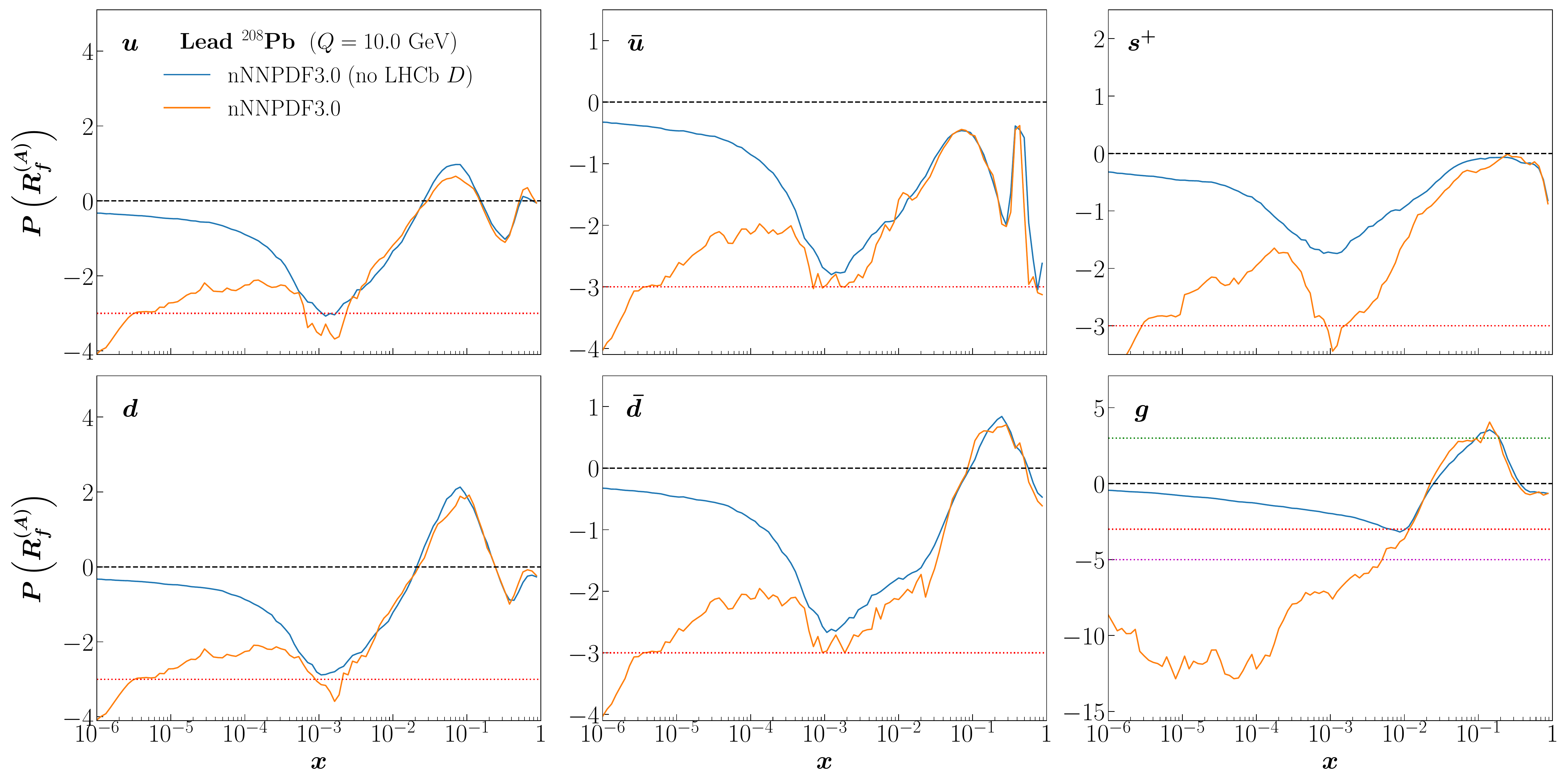}
\caption{Same as Fig.~\ref{fig:nNNPDF30_RA_Pb208_a_NuclearRationPDFs_Q10} now presented
  in terms of the pulls defined in Eq.~(\ref{eq:pull_global}).
  The dotted horizontal lines indicate the threshold for which
  nuclear modifications differ from zero at the $3\sigma~(5\sigma)$ level.
  \label{fig:nNNPDF30_RA_Pb208_a_NuclearRatio_pullnPDFs_Q10}
}
\end{figure}

These pulls are displayed in Fig.~\ref{fig:nNNPDF30_RA_Pb208_a_NuclearRatio_pullnPDFs_Q10}
for both nNNPDF3.0 and the prior fit at $Q=10$ GeV, where
 dotted horizontal lines indicate the threshold for which
nuclear modifications differ from zero at the $3\sigma$~($5\sigma$) level.
In the case of the quarks, the LHCb $D$-meson data enhances
the pulls in the region $x\simeq 10^{-3}$,
leading to a strong evidence for small-$x$ shadowing in the quark sector.
At larger values of $x$, the pull for anti-shadowing reaches
between the $1\sigma$
the $2\sigma$ level for up and down quarks and the down antiquark,
while for $\bar{u}$ it is absent.
The significance of the EMC effect remains at the $1\sigma$ level
of the up and down quarks.
Considering next the pull on the gluon modification ratio, we observe
how the LHCb $D$-meson measurements markedly increase both the significance
and the extension
of shadowing in the small-$x$ region.
Once the LHCb constraints are accounted for, one finds that nNNPDF3.0
favours a marked and statistically significant shadowing of
the small-$x$ gluon nPDF of lead in the region $ x \le 10^{-2}$.
To the best of our knowledge, this is the first time that such a strong
significance for gluon shadowing in heavy nuclei has been reported.

\subsection{Fit quality and comparison with data}
\label{sec:fit_quality}

We now turn to discuss the fit quality in the nNNPDF3.0 analysis,
for both the variants with and without the LHCb $D$-meson data,
and present representative comparisons between NLO QCD predictions
and the corresponding experimental data.
Table~\ref{tab:chi2baseline_1} reports the values of the $\chi^2$ per data point
for the DIS datasets that enter nNNPDF3.0.
For each dataset we indicate its name, reference, the nuclear
species involved, the number of data points, and the values
of $\chi^2/n_{\rm dat}$ obtained both with nNNPDF3.0 (no LHCb $D$)
and with nNNPDF3.0.
Datasets labelled with {\bf (*)} are new in nNNPDF3.0 as compared
to its predecessor nNNPDF2.0.
Table~\ref{tab:chi2baseline_2} displays the same information
as Table~\ref{tab:chi2baseline_1} now for the fixed-target
and LHC DY production datasets, the CMS dijet cross-sections, and the
ATLAS direct photon production measurements.
The last row of the table indicates the values corresponding to the global dataset.
Values indicated within brackets ($\lc~ \rc $) correspond to datasets
that are not part of the nNNPDF3.0 baseline, and whose $\chi^2$ values
are reported only for comparison purposes.
Finally in Table~\ref{tab:rw_details} we report the details of the reweighting and 
unweighting procedures, including the number of effective replicas $N_{\text{eff}}$, 
the number of replicas in the unweighted set $N_{\rm unweight}$, the number of data points included 
and the values of the $\chi^2$ before and after reweighting.

\begin{table}[!t]
  \centering
  \footnotesize
  \footnotesize
    \renewcommand{\arraystretch}{1.35}
\begin{tabularx}{\textwidth}{lC{2.0cm}C{1.3cm}C{3.9cm}C{2.8cm}}
  \toprule
 \multicolumn{3}{c}{} & nNNPDF3.0~(no LHCb~$D$) & nNNPDF3.0\\ \hline
Dataset & $A$ &$n_{\rm dat}$ & $\chi^ 2/n_{\rm dat}$ & $\chi^ 2/n_{\rm dat}$ \\ \hline
NMC 96~\cite{Arneodo:1996kd,Arneodo:1996qe} {\bf (*)} & $^1$p/$^2$D & 123 & 0.97 & 1.01\\
SLAC 91~\cite{Whitlow:1991uw} {\bf (*)} & $^2$D & 38 & 1.18 & 1.24\\
BCDMS 89~\cite{Benvenuti:1989rh} {\bf (*)} & $^2$D & 250 & 1.28 & 1.24\\ \hline
SLAC E-139~\cite{Gomez:1993ri} & $^4$He/$^2$D & 3 & 0.57 & 0.58\\
NMC 90, re.~\cite{Amaudruz:1995tq} & $^4$He/$^2$D & 13 & 1.15 & 1.16\\ \hline
NMC 95~\cite{Arneodo:1995cs} & $^6$Li/$^2$D & 12 & 1.10 & 1.06\\ \hline
SLAC E-139~\cite{Gomez:1993ri} & $^9$Be/$^2$D & 3 & 1.10& 1.17\\
NMC 96~\cite{Arneodo:1996rv} & $^9$Be/$^{12}$C & 14 & 0.30 & 0.29\\ \hline
EMC 88, EMC 90~\cite{Ashman:1988bf,Arneodo:1989sy} & $^{12}$C/$^{2}$D & 12 & 1.19  & 1.18\\
SLAC E-139~\cite{Gomez:1993ri} & $^{12}$C/$^{2}$D & 2 & 0.30 & 0.34\\
NMC 95, NMC 95, re.~\cite{Amaudruz:1995tq,Arneodo:1995cs} & $^{12}$C/$^{2}$D & 26 & 2.42 & 2.23\\
FNAL E665~\cite{Adams:1995is} & $^{12}$C/$^{2}$D & 3 & 0.76 & 0.79\\
NMC 95, re.~\cite{Amaudruz:1995tq} & $^{12}$C/$^{6}$Li & 9 & 1.00 & 1.00\\ \hline
BCDMS 85~\cite{Alde:1990im} & $^{14}$N/$^{2}$D & 9 & 2.14 & 2.06\\ \hline
SLAC E-139~\cite{Gomez:1993ri} & $^{27}$Al/$^{2}$D & 3 & 0.20 & 0.15\\
NMC 96~\cite{Arneodo:1996rv} & $^{27}$Al/$^{12}$C & 14 & 0.36  & 0.33\\
SLAC E-139~\cite{Gomez:1993ri} & $^{40}$Al/$^{2}$D & 2 & 0.87  & 0.88\\
NMC 95, re.~\cite{Amaudruz:1995tq} & $^{40}$Al/$^{2}$D & 12 & 1.63 & 1.49\\
EMC 90~\cite{Arneodo:1989sy} & $^{40}$Al/$^{2}$D & 3 & 1.68  & 1.64\\
FNAL E665~\cite{Adams:1995is} & $^{40}$Al/$^{2}$D & 3 & 0.90  & 0.96\\ \hline
NMC 95, re.~\cite{Amaudruz:1995tq} & $^{40}$Ca/$^{6}$Li & 9 & 0.19 & 0.20\\
NMC 96~\cite{Arneodo:1996rv} & $^{40}$Ca/$^{12}$C & 23 & 0.52 & 0.51\\ \hline
EMC 87~\cite{Aubert:1987da} & $^{56}$Fe/$^{2}$D & 58 & 0.73 & 0.71\\
SLAC E-139~\cite{Gomez:1993ri} & $^{56}$Fe/$^{2}$D & 8 & 1.68 & 1.60\\
NMC 96~\cite{Arneodo:1996rv} & $^{56}$Fe/$^{12}$C & 14 & 0.78 & 0.76\\
BCDMS 85, BCDMS 87~\cite{Alde:1990im,Benvenuti:1987az} & $^{56}$Fe/$^{2}$D & 16 & 1.46 & 1.31\\ \hline
EMC 88, EMC 93~\cite{Ashman:1988bf,Ashman:1992kv} & $^{64}$Cu/$^{2}$D & 27 & 0.61& 0.62\\ \hline
SLAC E-139~\cite{Gomez:1993ri} & $^{108}$Ag/$^{2}$D & 2 & 0.55& 0.60\\ \hline
EMC 88~\cite{Ashman:1988bf} & $^{119}$Sn/$^{2}$D & 8 & 2.23  & 2.14\\
NMC 96, $Q^2$ dependence~\cite{Arneodo:1996ru} & $^{119}$Sn/$^{12}$C & 119 & 0.65& 0.64\\ \hline
FNAL E665~\cite{Adams:1992vm} & $^{131}$Xe/$^{2}$D & 4 & 0.35 & 0.4\\ \hline
SLAC E-139~\cite{Gomez:1993ri} & $^{197}$Au/$^{2}$D & 3 & 0.91 & 0.87\\ \hline
FNAL E665~\cite{Adams:1995is} & $^{208}$Pb/$^{2}$D & 3 & 2.12 & 2.16\\
NMC 96~\cite{Arneodo:1996ru} & $^{208}$Pb/$^{12}$C & 14 & 0.97 & 0.96\\ \thickhline
\textbf{Total NC DIS} &  & 862 & 1.05 & 1.03 \\ \hline \hline
NuTeV $\nu$~\cite{Goncharov:2001qe} & $^{56}$Fe & 39 & 0.52 & 0.48\\
NuTeV $\bar{\nu}$~\cite{Goncharov:2001qe} & $^{56}$Fe & 37 & 1.59 & 1.31\\ \hline
CHORUS $\nu$~\cite{Onengut:2005kv} & $^{208}$Pb & 423 & 1.04 & 1.03\\
CHORUS $\bar{\nu}$~\cite{Onengut:2005kv} & $^{208}$Pb & 423 & 1.03 & 1.01\\ \thickhline
\textbf{Total CC DIS} &  & 922 & 1.04 & 1.01 \\ \bottomrule
\end{tabularx}
\vspace{0.3cm}

  \caption{\small The values of the $\chi^2$ per data point
    for the DIS datasets that enter nNNPDF3.0.
    For each dataset we indicate its name, reference, the nuclear
    species involved, the number of data points, and the values
    of $\chi^2/n_{\rm dat}$ obtained both with the nNNPDF3.0 prior fit
    (without the LHCb $D$-meson cross-sections) and with nNNPDF3.0.
    Datasets labelled with {\bf (*)} are new in nNNPDF3.0 as compared
    to its predecessor nNNPDF2.0.
    The datasets are separated into neutral-current (upper) and charged-current
    (bottom part) structure functions.
  }
  \label{tab:chi2baseline_1}
\end{table}

\begin{table}[!t]
  \centering
  \footnotesize
  \footnotesize
    \renewcommand{\arraystretch}{1.35}
\begin{tabularx}{\textwidth}{lC{1.5cm}C{1.0cm}C{4.0cm}C{2.7cm}}
  \toprule
 \multicolumn{3}{c}{} & nNNPDF3.0 (no LHCb~$D$) & nNNPDF3.0\\ \hline
 Dataset & $A$ &$n_{\rm dat}$ & $\chi^ 2/n_{\rm dat}$ & $\chi^ 2/n_{\rm dat}$ \\ \hline
   FNAL Drell-Yan E605~\cite{Moreno:1990sf} {\bf (*)} & $^{64}$Cu & 85 & 0.82 & 0.85\\ \hline
   FNAL Drell-Yan E886~\cite{Webb:2003ps,Webb:2003bj,Towell:2001nh} {\bf (*)} &
   $^{2}$d/$^{1}$p & 15 & 1.04 & 1.16\\ \hline
ATLAS Z $\sqrt{s}=5.02$ TeV~\cite{Aad:2015gta} & $^{208}$Pb & 14 & 0.91  & 0.93\\
CMS Z $\sqrt{s}=5.02$ TeV~\cite{Khachatryan:2015pzs} & $^{208}$Pb & 12 & 0.6 & 0.6\\
CMS W$^-$ $\sqrt{s}=5.02$ TeV~\cite{Khachatryan:2015hha} & $^{208}$Pb & 10 & 1.02 & 1.07\\
CMS W$^+$ $\sqrt{s}=5.02$ TeV~\cite{Khachatryan:2015hha} & $^{208}$Pb & 10 & 1.11 & 1.08\\
CMS W$^-$ $\sqrt{s}=8.16$ TeV~\cite{Sirunyan:2019dox} & $^{208}$Pb & 24 & 0.72 & 0.73\\
CMS W$^+$ $\sqrt{s}=8.16$ TeV~\cite{Sirunyan:2019dox} & $^{208}$Pb & 24 & 0.77 & 0.8\\
ALICE Z $\sqrt{s}=5.02$ TeV~\cite{ALICE:2016rzo} {\bf (*)} & $^{208}$Pb & 2 & 0.14 & 0.14\\
ALICE W$^-$ $\sqrt{s}=5.02$ TeV~\cite{ALICE:2016rzo} {\bf (*)} & $^{208}$Pb & 2 & 0.18 & 0.18\\
ALICE W$^+$ $\sqrt{s}=5.02$ TeV~\cite{ALICE:2016rzo} {\bf (*)} & $^{208}$Pb & 2 & 2.55 & 2.54\\
LHCb Z $\sqrt{s}=5.02$ TeV~\cite{LHCb:2017yua} {\bf (*)} & $^{208}$Pb & 2 & 0.9 & 0.9\\
CMS Z $\sqrt{s}=8.16$ TeV~\cite{CMS:2021ynu} {\bf (*)} & $^{208}$Pb & 36 & 2.49 & 2.49\\
ALICE Z $\sqrt{s}=8.16$ TeV~\cite{ALICE:2020jff} {\bf (*)} & $^{208}$Pb & 2 & 0.02 & 0.03\\
   \midrule
   \textbf{Total Drell-Yan} &  & 240 & 1.08 & 1.11 \\ \hline \hline
CMS dijet pPb $\sqrt{s}=5.02$ TeV~\cite{CMS:2018jpl} & $^{208}$Pb & 85 & [13.6] & [13.96]\\
CMS dijet pPb/pp $\sqrt{s}=5.02$ TeV~\cite{CMS:2018jpl} {\bf (*)} & $^{208}$Pb & 84 & 1.81 & 1.75\\ \thickhline
ATLAS photon pPb $\sqrt{s}=8.16$ TeV~\cite{Aaboud:2019tab} & $^{208}$Pb & 46 & [3.33] & [3.21]\\ \hline
ATLAS photon pPb/pp $\sqrt{s}=8.16$  TeV~\cite{Aaboud:2019tab} {\bf (*)} & $^{208}$Pb & 43 & 1.03& 1.03\\
   \midrule
   \midrule
   \textbf{Total dataset} &  &{\bf 2151} & {\bf 1.11} & {\bf 1.09} \\ \bottomrule
\end{tabularx}
\vspace{0.3cm}

  \caption{Same as Table~\ref{tab:chi2baseline_1} now for the fixed-target
    and LHC DY production datasets, the CMS dijet cross-sections, and the
    ATLAS direct photon production measurements.
    The last row indicates the values corresponding to the global dataset.
    Values indicated within brackets ($\lc~ \rc $) correspond to datasets
    that are not part of the nNNPDF3.0 baseline, and are reported only
    for comparison purposes.
    See Table~\ref{tab:rw_details} for the corresponding $\chi^2$ values
    for the LHCb $D^0$-meson forward data.
  }
  \label{tab:chi2baseline_2}
\end{table}

\begin{table}[!t]
  \centering
  \footnotesize
  \centering
\small
      \renewcommand{\arraystretch}{1.45}
\begin{tabularx}{0.8\textwidth}{C{2.5cm}C{2.5cm}C{2.5cm}C{2.5cm}C{2.5cm}}
    \toprule
    $N_{\text{eff}}$  &   $N_{\text{unweight}}$   & $n_{\text{dat}}$ & $\chi^2_{\rm prior}$ & $\chi^2_{\text{rw}}$  \\
    \midrule
    185      &  200   &  37 & 32.16 & 0.66             \\
\bottomrule
\end{tabularx}
\vspace{0.3cm}

  \caption{The number of effective $N_{\text{eff}}$ and unweighted $N_{\text{unweight}}$
    replicas associated to the inclusion of the LHCb $R_{{\rm pPb}}$ data on
    the nNNPDF3.0 prior fit via reweighting, together with the number of data points
    and the values of the $\chi^2$ for this dataset before and after reweighting.
  }
  \label{tab:rw_details}
\end{table}

Several interesting observations can be derived from the results
presented in Tables~\ref{tab:chi2baseline_1} and~\ref{tab:chi2baseline_2}.
First of all, the global fit $\chi^2/n_{\rm dat}$ values are
satisfactory, with $\chi^2/n_{\rm dat}\simeq 1.10$ for both nNNPDF3.0 variants.
Actually, the values obtained for the nNNPDF3.0 fits
with and without the constraints from the LHCb $D$-meson data are very similar
in all cases.
This observation is explained because, as discussed above,
the LHCb $D$-meson data constraints are restricted to $x\lsim 10^{-3}$
where there is little overlap with other datasets entering the nuclear
fit, as also highlighted by the kinematic plot of Fig.~\ref{fig:kinplot}.
Given that one is combining $n_{\rm dat}=2151$ data points
from 54 datasets corresponding to 7 different processes, such
a satisfactory fit quality is a reassuring, non-trivial consistency test
of the reliability of the QCD factorisation framework
when applied to nuclear collisions.
Likewise, a very good description of the nuclear DIS structure function data is achieved,
with $\chi^2/n_{\rm dat}\simeq 1.02$~(1.04) for 1784 data points
for nNNPDF3.0 (its variant without LHCb $D$ data).
This fit quality is similar to that reported for nNNPDF2.0,
see~\cite{AbdulKhalek:2020yuc,AbdulKhalek:2019mzd}
for a discussion of the somewhat higher $\chi^2$ values
obtained for a few of the DIS datasets.

\begin{figure}[!t]
  \centering
   \includegraphics[width=0.85\textwidth]{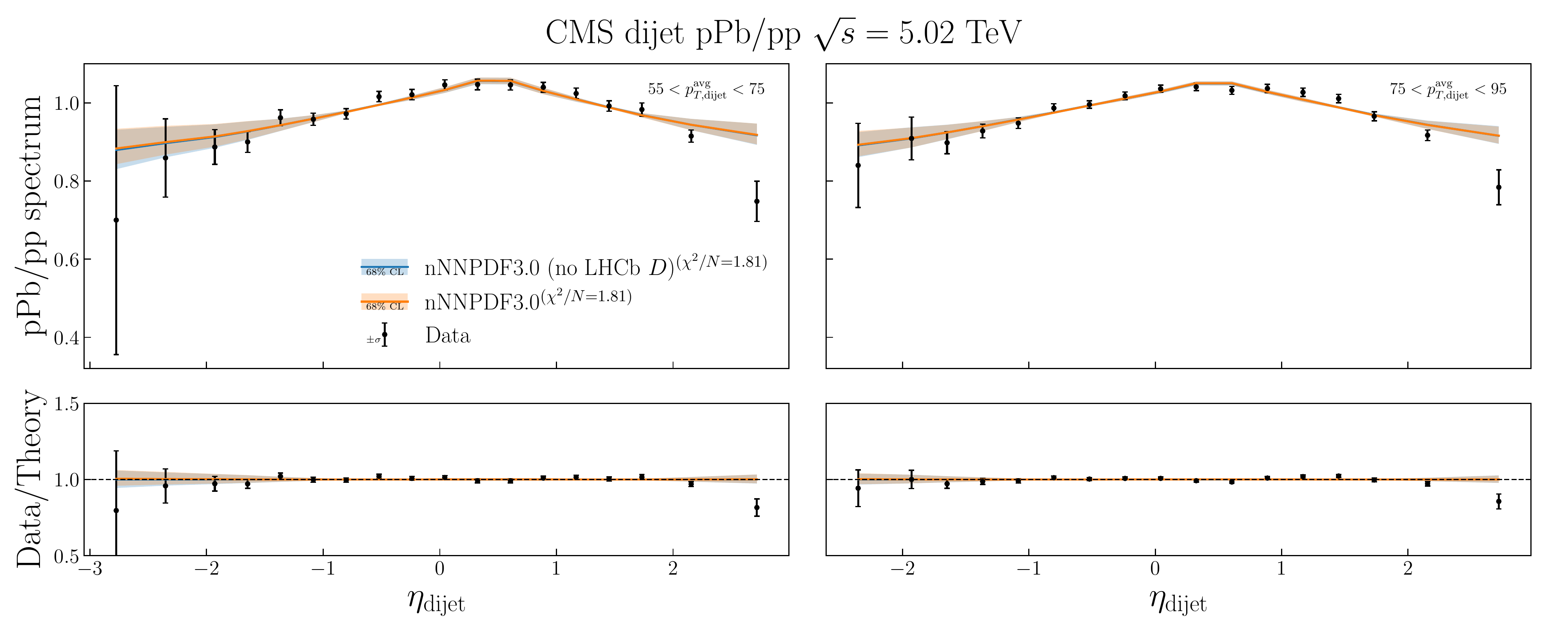}
   \includegraphics[width=0.85\textwidth]{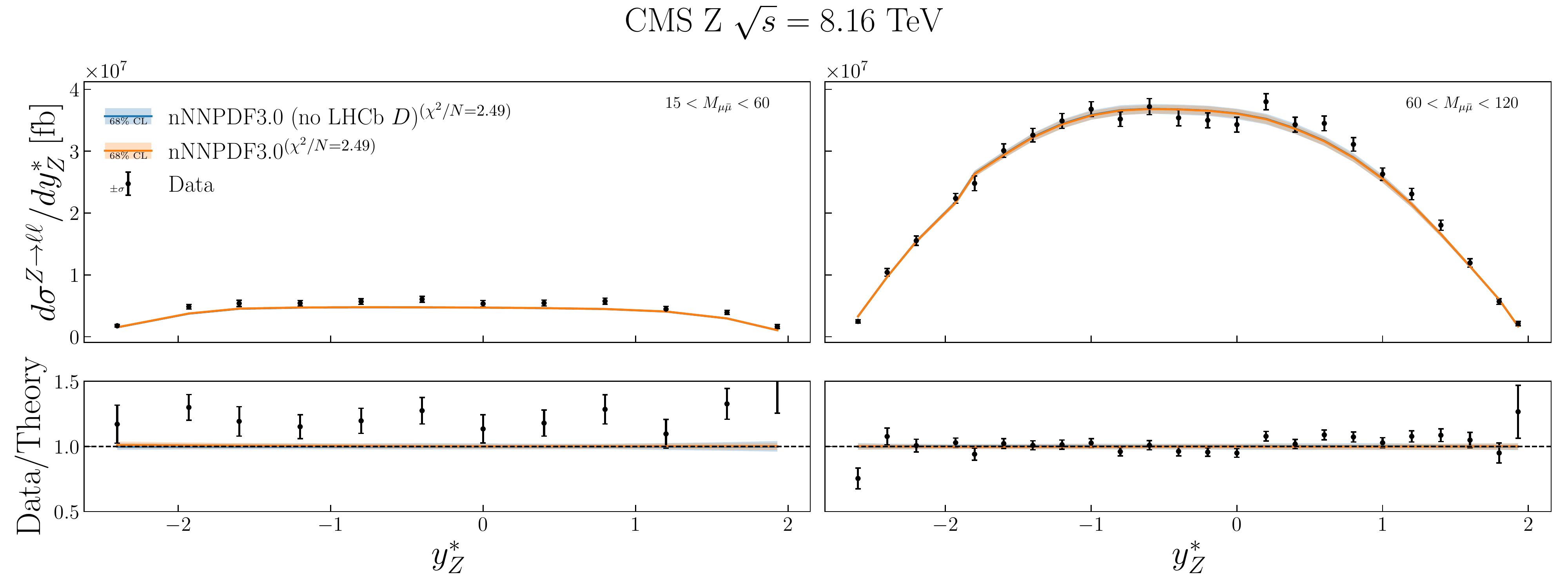}
   \caption{\small Comparison between the NLO QCD theory predictions
     based on the nNNPDF3.0 and nNNPDF3.0~(no LHCb~$D$) fits with the corresponding
     experimental data for the first two $p_{T,{\rm dijet}}^{\rm avg}$ bins
     of the CMS $\sqrt{s}=5.02$ TeV dijet production measurement (upper)
     and for the two dimuon invariant mass bins
     from the CMS $\sqrt{s}=8.16$ $Z$ production measurement (bottom panels).}
  \label{fig:data_vs_theory_lhc}
\end{figure}

Concerning the new LHC gauge boson production datasets from pPb collisions
added to nNNPDF3.0, a satisfactory fit quality is obtained for all of them
except for the ALICE $W^+$ at 5.02 TeV (2 points) and CMS $Z$ at 8.16 TeV (36
points) measurements.
From the data versus theory comparisons reported in Fig.~\ref{fig:data_vs_theory_lhc}
(see also App.~\ref{app:datacomp})
for the case of the CMS $Z$ dataset,  in the low dimuon
invariant mass bin with $15\le M_{\mu\bar\mu} \le 60$ GeV the NLO QCD theory
predictions undershoot the data by around 10\% to 20\%.
This shift should be reduced by the addition of the NNLO QCD corrections~\cite{Anastasiou:2003yy},
which may be relatively large in this region.
We have verified that the nNNPDF3.0 fit results are unaffected if this
low invariant mass bin is removed.
For the on-peak invariant mass bin, with $60\le M_{\mu\bar\mu} \le 120$ GeV
and 
for which the NNLO QCD corrections are relatively small,
a good description of the experimental data is obtained except for the left-most
rapidity bin, where the cross-section is very small.

Concerning the CMS measurements of dijet production at 5.02 TeV and the ATLAS ones of isolated
photon production at 8.16 TeV, in both cases presented as ratio between the pPb
and pp spectra, one finds a fit quality of $\chi^2/n_{\rm dat}\simeq 1.8$ and 1.0 respectively.
In the case of the CMS dijets, inspection of the comparison
between the fit results and data in the bottom panels of Fig.~\ref{fig:data_vs_theory_lhc} (see
App.~\ref{app:datacomp} for the rest of the bins)
reveals that in general one has a good agreement with the exception of one or
two bins in the forward (proton-going) direction.
In particular, the most forward bin systematically undershoots the NLO QCD theory prediction.
We have verified that if the two most forward rapidity bins are removed
in each $p_{T,{\rm dijet}}^{\rm avg}$ bin, the fit quality improves to
$\chi^2/n_{\rm dat}\simeq 1.3$ without any noticeable
impact on the fit results.
Hence, it is neither justified nor necessary to apply ad-hoc kinematic cuts in the
dijet rapidity,
and one can conclude that a satisfactory description of this dataset is obtained.
In the case of the ATLAS isolated photon measurements, good agreement
between data and theory is obtained for the whole range of $E_T^\gamma$ and $\eta^\gamma$
covered by the data.

As indicated by Table~\ref{tab:chi2baseline_2}, when the
theory predictions based on nNNPDF3.0 are compared to
the absolute pPb spectra for dijet production and isolated photon production,
much worse $\chi^2/n_{\rm dat}$ values are found, 13.6 and 3.3 for each dataset respectively.
As discussed in Sect.~\ref{sec:new_LHC_data}, the prediction of the absolute cross-section rates
at NLO QCD accuracy suffers from large uncertainties due to MHO effects.
Dedicated studies of these two processes at NNLO QCD accuracy (such as those in~\cite{AbdulKhalek:2020jut,Campbell:2018wfu}) will be required to determine if this data can be reliably included in an nPDF fit.

\subsection{$A$-dependence of nuclear modifications}

\begin{figure}[!t]
  \centering
\includegraphics[width=\textwidth]{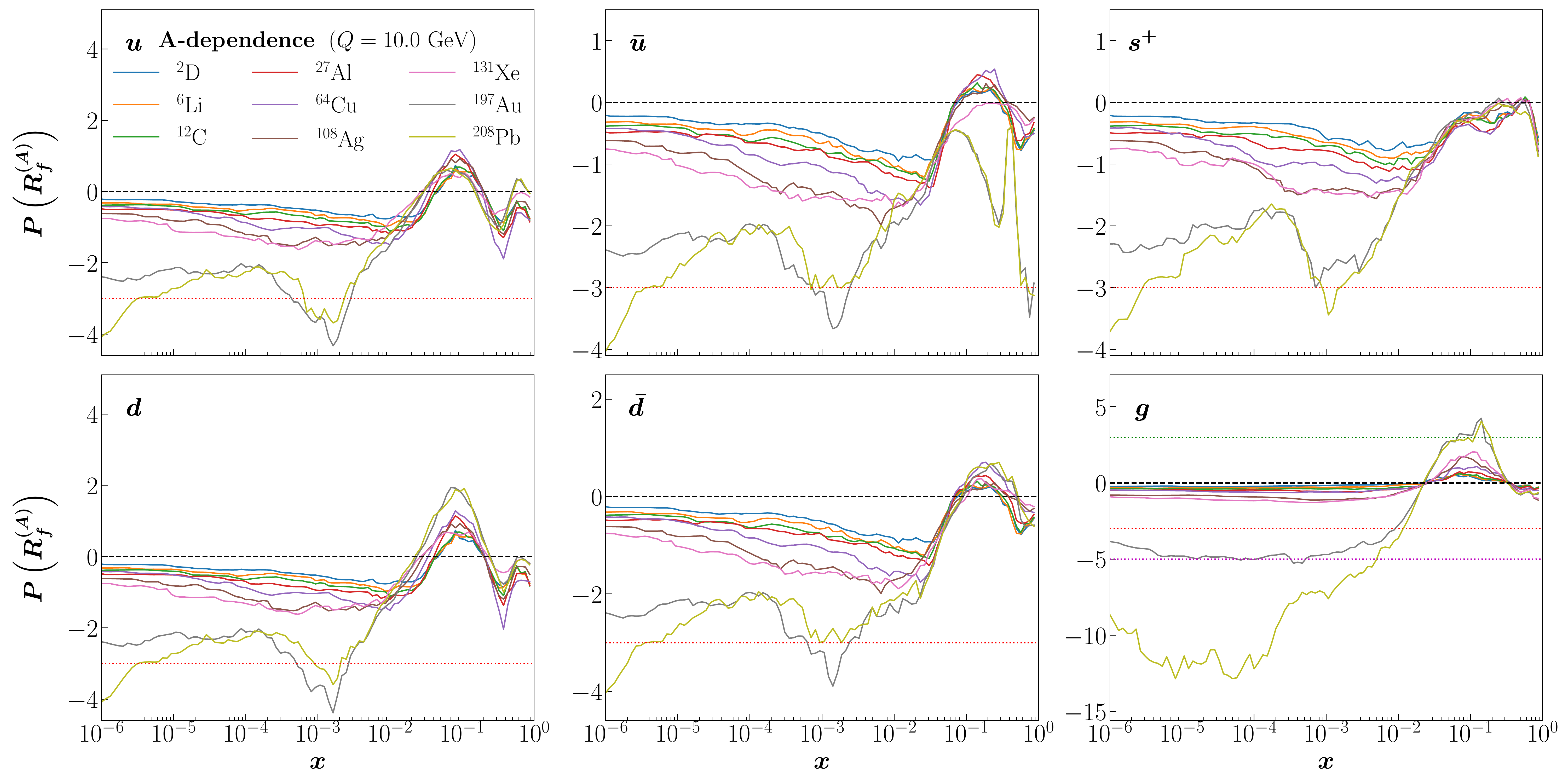}
\caption{\label{fig:Adep_pull} The dependence with the atomic mass number $A$
  of the pulls defined in Eq.~(\ref{eq:pull_global}) in nNNPDF3.0 for a range
  of nuclei from deuterium ($A=2$) up to lead ($A=208$).
  Recall from Eq.~(\ref{eq:pull_global}) that nuclear modifications associated
  to the different numbers of protons and neutrons have already been accounted for.
}
\end{figure}

The results discussed so far have focused on the nuclear modifications
of lead, which is the nuclear species
for which hard-scattering data from the LHC is available.
Here we study how these nuclear modifications depend on the atomic mass number.
Fig.~\ref{fig:Adep_pull} displays
the dependence with the atomic mass number $A$
of the pulls defined in Eq.~(\ref{eq:pull_global}) for the nNNPDF3.0 global
analysis.
These pulls are displayed for $Q=10$ GeV as a function of $x$ for a range
of nuclei from deuterium ($A=2$) up to lead ($A=208$).
Recall that nuclear modifications associated
to different numbers of protons and neutrons have already been accounted for.
We note that a small value of the pull in Fig.~\ref{fig:Adep_pull}
does not necessarily imply that nuclear corrections for such value of $A$ are small,
it can also mean that nPDF uncertainties are relatively large (for example due
to lack of direct experimental constraints) as compared to other nuclei.

In the case of the gluon nuclear modifications, one can observe how the pulls are small
and similar for light nuclear, all the way up to $^{27}$Al.
The pulls become somewhat larger as $A$ is increased up to $^{131}$Xe, favoring shadowing and anti-shadowing
at small and large-$x$ respectively.
However, it is only for the heavier nuclei, $^{197}$Au and $^{208}$Pb, for which the pulls
reach the $3\sigma$ level, providing evidence for
shadowing and anti-shadowing
at small and large-$x$ respectively.
The large impact of the LHCb $D^0$-meson $R_{\rm pPb}$ data on $R_g$ for Pb is clearly visible,
and indirectly also constrains the nuclear modifications of Au.
The fact that the absolute pulls increase with $A$ arises from the combination of two factors:
nuclear effects are known to become more important for heavier nuclei,
and that the heavier nuclei benefit from the nPDF uncertainty reduction provided
by the LHC measurements from pPb collisions.

A similar picture is observed for the pulls associated to the quark and anti-quark nuclear
modification ratios.
One difference is that in this case one observes nuclear modifications
with associated pulls at the $1\sigma$ level already for the light nuclei,
from deuteron onwards, e.g. with anti-shadowing at $x\simeq 0.1$.
As was also the case for the gluon, the significance of small-$x$ shadowing in the quark sector
increases with the atomic mass number $A$, and reaches the $3\sigma$ level for
the heavier nuclei at $x\simeq 10^{-3}$.
For smaller values of $x$, nPDF errors grow up and this significance is washed out
except for the heavier nuclei considered.
Interestingly, the strong dependence with $A$ observed for small-$x$ shadowing is less clear
for larger-$x$ phenomena such as anti-shadowing and the EMC effect in the quark
sector.
In the specific case of the down quark, the pulls in the anti-shadowing region reach
the $2\sigma$ level for
$x\simeq 0.1$ only when considering the heavier nuclei, but such trend is not observed
in the case of its up quark counterpart.

\begin{figure}[!t]
  \centering
\includegraphics[width=\textwidth]{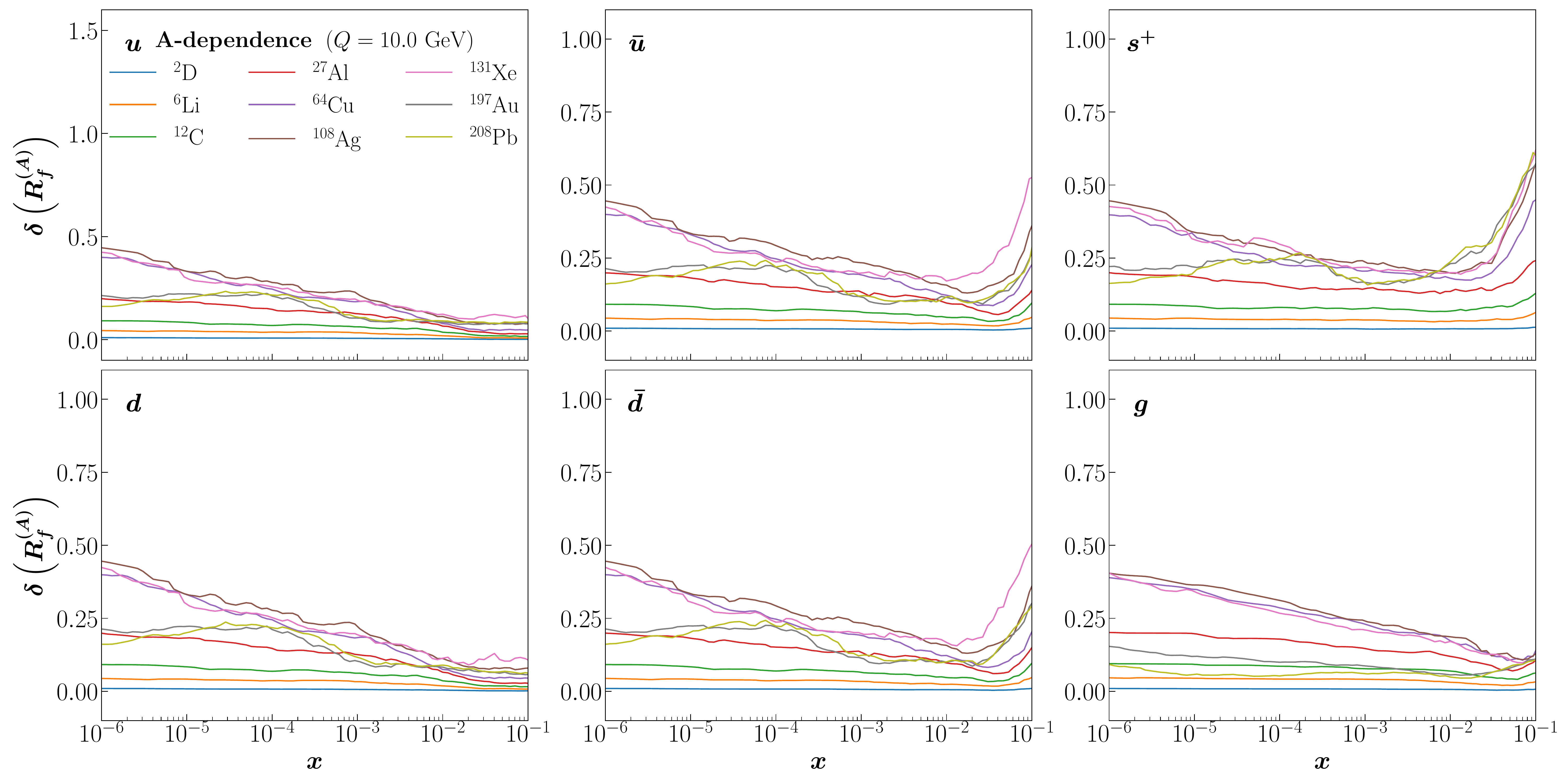}
\caption{\label{fig:Adep_relativeerror} The relative 68\% CL uncertainties on the nuclear
  modification ratios, $\delta ( R_f^{(A)})$, evaluated at $Q=10$ GeV  for the same
  nuclei as those presented in Fig.~\ref{fig:Adep_pull}.
}
\end{figure}

Then in Fig.~\ref{fig:Adep_relativeerror}
we present the relative 68\% CL uncertainties on the nuclear
modification ratios, $\delta ( R_f^{(A)})$, evaluated at $Q=10$ GeV for the same
nuclei as those presented in Fig.~\ref{fig:Adep_pull}.
For the lighter nuclei, these uncertainties are the smallest
for the deuteron and then increase monotonically up to $^{27}$Al.
The fact that nuclear modifications with low-$A$ nuclei are well constrained follows
not only from the free-proton boundary condition but also from the large amount
of data taken on a deuteron target used in nNNPDF3.0.
Even so, already for relatively light nuclei as $^{27}$Al the
uncertainties have become quite large, demonstrating how
the $A$-dependence of our parametrisation is not over-constraining.

For heavier nuclei, the results for $\delta (R_f^{(A)})$ depend on the flavour.
For the gluon, we see that uncertainties keep increasing with $A$ until we get to $^{131}$Xe,
but then $^{197}$Au and $^{208}$Pb are better constrained 
as compared to lighter nuclei.
This is a direct consequence of the strong constraints imposed by the CMS dijet and LHCb $D$-meson data
on the nuclear modifications of the gluon PDF.
Remarkably, the PDF uncertainties for $A=208$ (lead) are smaller than those of light nuclei
such as $A=12$ (carbon) for $x\lsim 10^{-2}$.
A similar picture applies for the up and down quarks and anti-quarks,
where for example the uncertainties in $^{208}$Pb (and $^{197}$Au) are clearly
reduced as compared to $^{131}$Xe for the entire $x$ region considered.
In this case, the origin of this improvement can be traced back to the information provided
by the LHC measurements of weak gauge boson production in pPb collisions.
This trend is however absent for the PDF uncertainties in the case of
the total strangeness, most likely since the fit does not contain direct constraints
on the nuclear modifications of strangeness in heavy nuclei.

\begin{figure}[!t]
  \centering
  \includegraphics[scale=0.30]{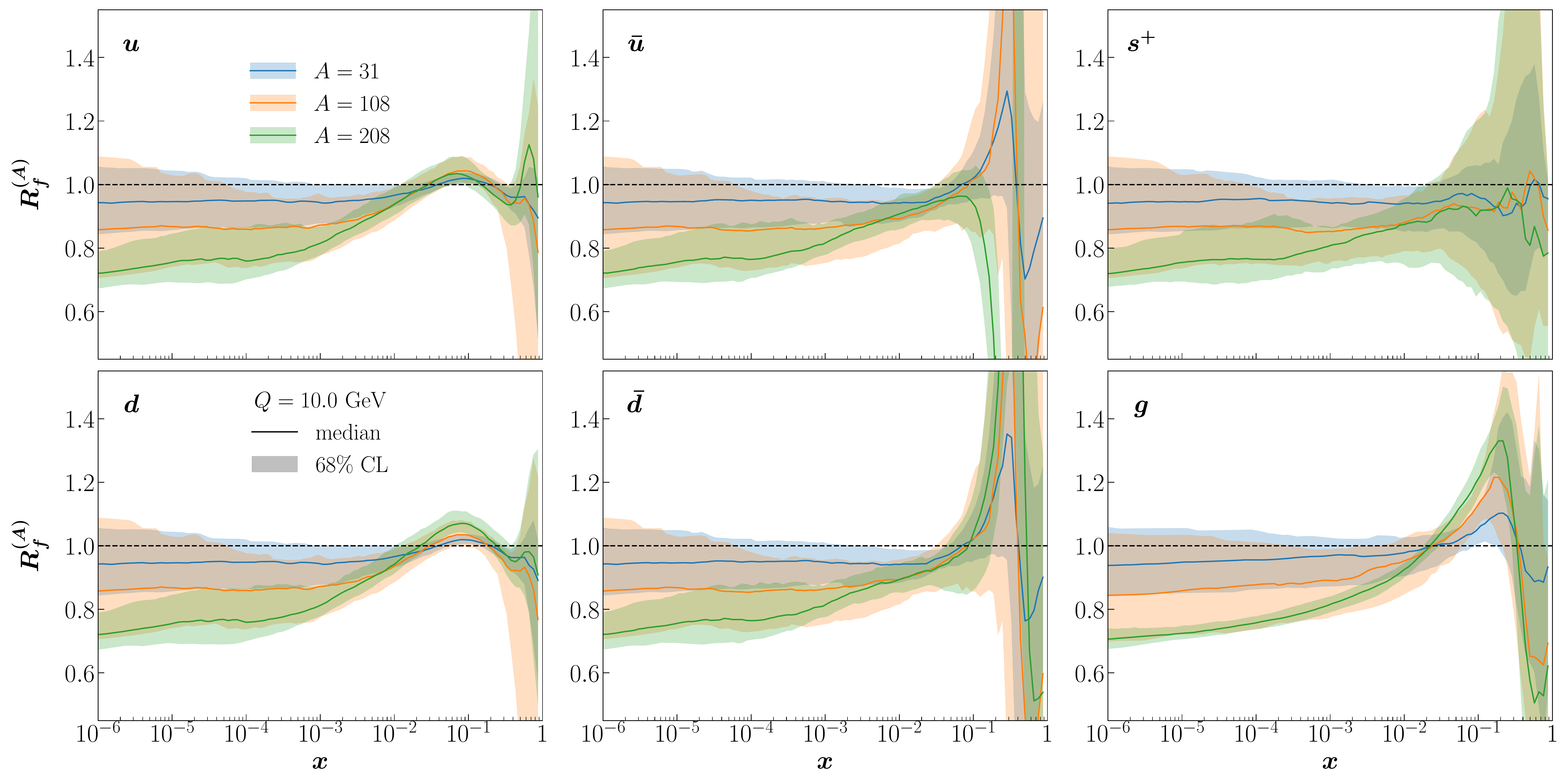}
  \caption{The nuclear modification factors $R_f^{(A)}(x,Q)$ at
    $Q=10$ GeV for $A=208$ (lead), $A=108$ (silver)
    and $A=31$ (the mean atomic mass number of nuclei typically 
    encountered by ultra-high-energy neutrinos propagating through the Earth).}
    \label{fig:NucleiComparison}
\end{figure}

A complementary picture of the $A$-dependence of our results is provided by the
nuclear modification factors $R_f^{(A)}(x,Q)$ for different values of $A$.
Figure~\ref{fig:NucleiComparison} compares $R_f^{(A)}$ for three representative
values of $A$: $A=208$ (lead), $A=108$ (silver) and $A=31$ (the mean atomic mass number of nuclei typically 
encountered by ultra-high-energy neutrinos propagating through the Earth). While nuclear modifications are well constrained
for lead, given the abundance of LHC data included in the fit for this nucleus,
only a handful of points (corresponding specifically to DIS structure functions)
are available for silver, and none for $A=31$. From this comparison one
observes how the uncertainties on $R_f^{(A)}$ are largest for $A=108$, due to
the scarcity of experimental data and to the fact that this value of $A$ is far
from both $A=1$, constrained by the proton boundary condition, and $A=208$,
for which most of the heavier nuclear data is available. In particular, for
$A=108$ the small-$x$ gluon is consistent with no nuclear modifications within
uncertainties. The predictions for $A=31$ are obtained from the outcome of the
neural network parametrisation, Eq.~\eqref{eq:param2} trained to other values
of $A$, and for this nuclear species $R_f^{(A)}$ turns out to be consistent with
unity at the 68\% confidence level.

\begin{figure}[!t]
  \centering
  \includegraphics[width=\textwidth]{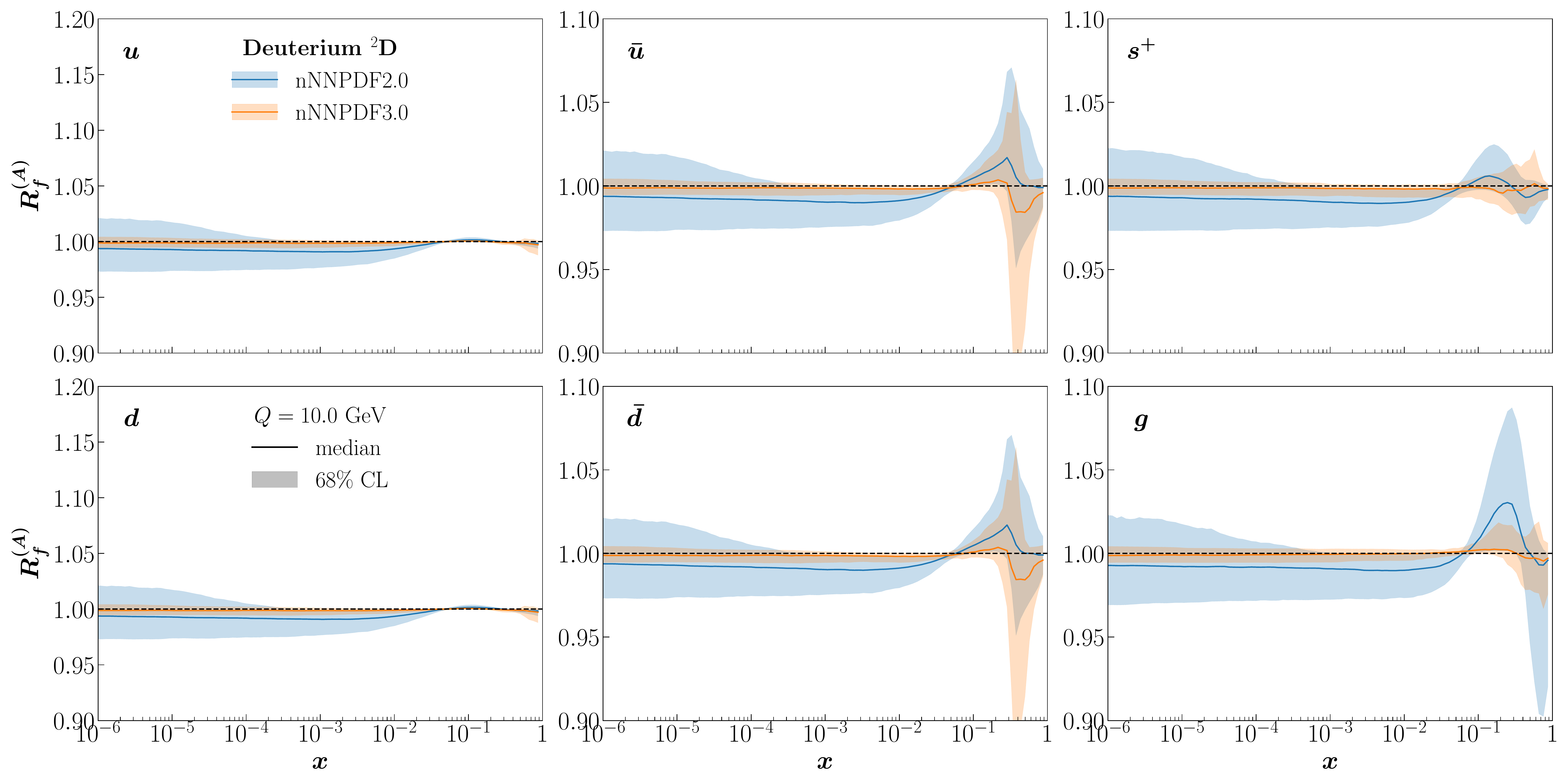}
  \caption{\small The nuclear modification ratios in nNNPDF3.0 for deuterium ($A=2$)
    compared to nNNPDF2.0.
  }
  \label{tab:nuclear_ratios_deuterium}
\end{figure}

A particularly interesting case is that of the nuclear modification factors associated
to the deuteron, $A=2$.
The reason is that the treatment of deuteron data has been improved as compared
to our previous analysis: in nNNPDF2.0, deuteron DIS and DY cross-sections were part
of the proton baseline (where they were included assuming no nuclear effects), while
now instead the free-proton baseline does not contain any deuteron data, which
is instead considered entirely at the level of the nuclear fit.
Hence, one expects the nuclear corrections associated to deuteron nuclei to be better constrained
in nNNPDF3.0, since these are now being directly determined from the data rather than indirectly via
the boundary condition.
Fig.~\ref{tab:nuclear_ratios_deuterium} displays
the nuclear modification ratios in nNNPDF3.0 for deuterium nuclei ($A=2$),
compared to those of nNNPDF2.0.
While in both cases the deuteron nuclear modification ratios agree with unity within uncertainties,
the nNNPDF3.0 predictions are indeed rather more precise than those of its
predecessor.
From these results, deuteron corrections for the light quarks
appear to be constrained to be at the $\lsim 1\%$ level in the
kinematic region
where deuteron experimental data is available, with somewhat larger uncertainties
in the case of large-$x$ antiquarks.

\subsection{Comparison with other global nPDF analyses}
\label{sec:comparison_global_npdfs}

Several groups have presented determinations
of nuclear PDFs~\cite{Eskola:2016oht, Duwentaster:2021ioo,Kusina:2016fxy,Kusina:2020lyz,Kovarik:2015cma,AbdulKhalek:2019mzd,deFlorian:2003qf,deFlorian:2011fp,Eskola:2008ca,Eskola:2009uj,Eskola:2021nhw,Walt:2019slu,Helenius:2021tof}, which differ in terms of the input dataset, fitting methodology, and/or theoretical settings.
Here we compare the nNNPDF3.0 results with two other recent nuclear PDF analyses\footnote{
  While the new EPPS21 set has been presented in~\cite{Eskola:2021nhw}, the corresponding
{\sc\small LHAPDF} grids are not yet publicly available.} based
on global datasets, namely EPPS16~\cite{Eskola:2016oht} and nCTEQ15WZ+SIH~\cite{Duwentaster:2021ioo}.
The EPPS16 study considers fixed-target DIS and DY cross-sections
on nuclear targets complemented with LHC data on gauge boson and dijet production and
with RHIC measurements.
Follow up studies based on the EPPS16 framework have focused
on the nPDF constraints provided by $D$-meson~\cite{Eskola:2019bgf}, dijet~\cite{Eskola:2019dui},
and fixed-target large-$x$ DIS~\cite{Paukkunen:2020rnb} data.
nCTEQ15WZ+SIH is the most recent nPDF analysis from the nCTEQ collaboration, building
upon previous results first based on DIS, fixed-target DY, and RHIC
data on nuclear targets~\cite{Kovarik:2015cma}, then extended to gauge boson production in pPb
collisions~\cite{Kusina:2016fxy,Kusina:2020lyz}, and recently
to single inclusive hadron (SIH) production from RHIC and the LHC~\cite{Duwentaster:2021ioo}.
Furthermore, a nCTEQ15 variant studying the nPDF constraints from low-$Q^2$
DIS structure functions at JLab has also been presented~\cite{Segarra:2020gtj}.

\begin{figure}[!t]
  \centering
  \includegraphics[width=\textwidth]{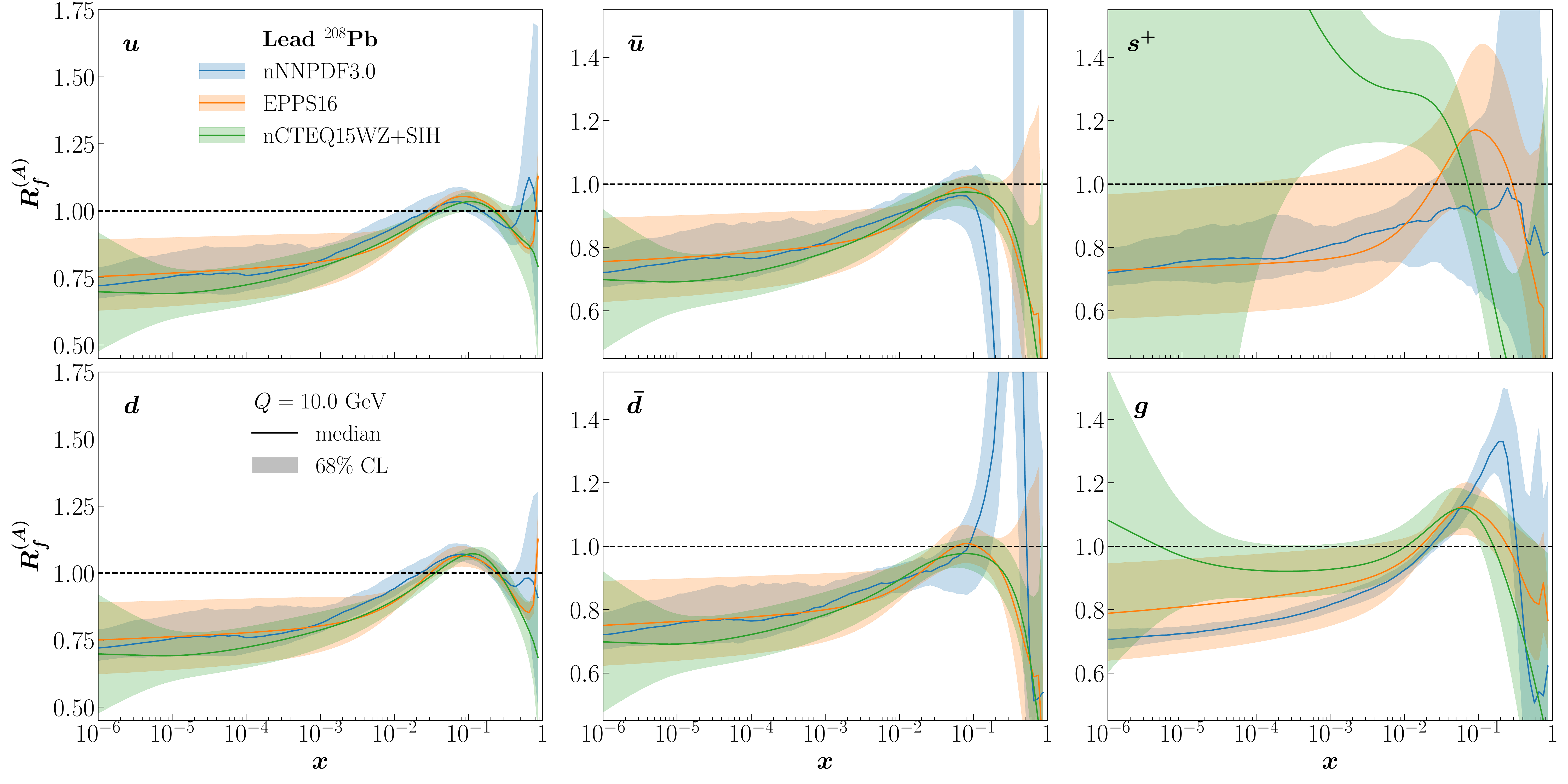}
     \caption{\small The nNNPDF3.0 predictions for the nuclear modification ratios in lead
    at $Q=10$ GeV, compared to the corresponding results from
    the EPPS16
    and nCTEQWZ+SIH global analyses.
    The PDF uncertainty bands correspond in all cases to 68\% CL intervals.
  }
\label{fig:comparison_npdfs_global}
\end{figure}

\begin{figure}[!t]
  \centering
  \includegraphics[width=\textwidth]{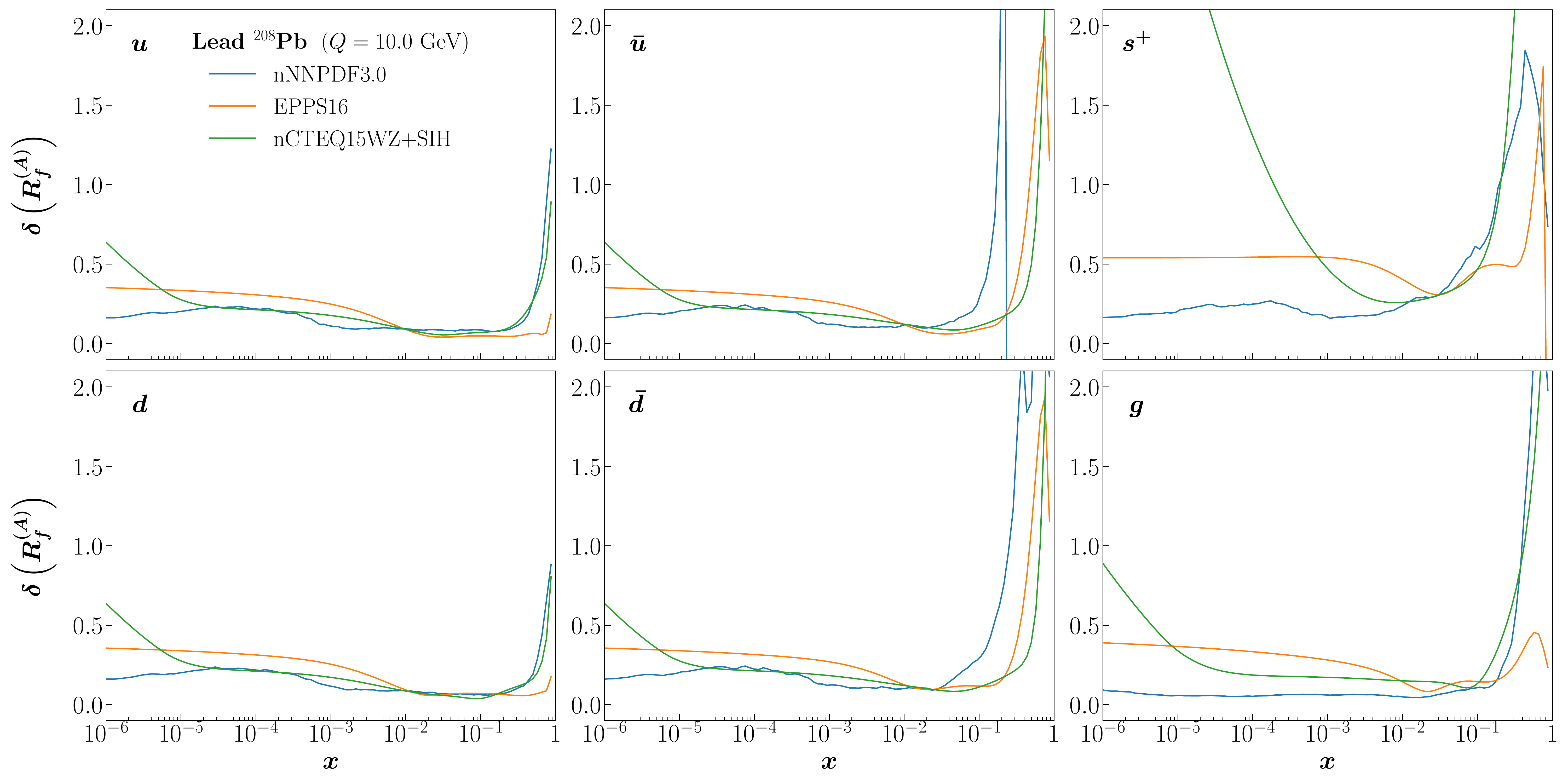}
  \caption{\small Same as Fig.~\ref{fig:comparison_npdfs_global}, now comparing
    the relative nPDF uncertainties associated to $R^{(A)}_f(x,Q)$.
  }
\label{fig:comparison_npdfs_global_2}
\end{figure}

The nNNPDF3.0 predictions for the nuclear modification ratios $R^{(A)}_f(x,Q)$ are compared with its
counterparts from the EPPS16 and nCTEQ analyses in
Fig.~\ref{fig:comparison_npdfs_global}.
We display the results for the lead PDFs at $Q=10$ GeV, where the uncertainty bands
correspond in all cases to the 68\% CL intervals.
We note that the EPPS16 and nCTEQ15WZ+SIH Hessian
uncertainties are provided as 90\% CL, hence we rescale them to obtain 68\% CL bands.
To ease the interpretation of this comparison, the relative 68\% CL
uncertainties associated to these nuclear modification ratios
are plotted separately in Fig.~\ref{fig:comparison_npdfs_global_2}.

In general one finds reasonable agreement between the results of the three global analyses, but also some
differences both in terms of central values and uncertainties.
Concerning the gluon nuclear modifications, nNNPDF3.0 favours both a stronger
shadowing at small-$x$ and a more intense anti-shadowing at large-$x$ as compared to the other two groups.
Except for the region around $x\simeq 0.2$, where the nNNPDF3.0  nuclear
ratio is somewhat higher, the nNNPDF3.0 predictions
agree within uncertainties in the full $x$ range with EPPS16, while
for $x\lsim 10^{-3}$ the nCTEQ prediction for $R_g$ is instead higher
and consistent with no gluon shadowing.
In terms of the nPDF uncertainties on $R_g$, these turn out to be similar between the three groups
in the region where the bulk of the data lies, $x\gsim 10^{-2}$, while at smaller $x$
those of EPPS16 become larger than those of nNNPDF3.0,
the latter result being explained by the strong constraints provided by the LHCb $D$-meson
measurements in this kinematic region.

In the case of the nuclear modifications associated to the up and down quarks,
good agreement is found both in terms
of the central values and of the PDF uncertainties among the three groups
for the whole range of $x$.
The good agreement between the three fits is, as expected,
also present for the up and down antiquarks for $x\lsim 10^{-2}$,
where the valence contribution is negligible.
On the other hand, for $x\gsim 0.1$ the predictions for $R_{\bar{u}}$
and $R_{\bar{d}}$ from
nNNPDF3.0 are rather different as compared to the other two groups,
and furthermore their uncertainties are also significantly larger in this region.
We remark that the
experimental constraints on the large-$x$ nuclear antiquarks
are limited, and hence the methodological assumptions play a bigger role.

The largest differences between the three groups are observed for the strange
PDF: while nNNPDF3.0 and EPPS16 favour small-$x$ shadowing along the lines of
the up and down quark sea, nCTEQ displays
a positive nuclear correction of up to 50\% for $x\lsim 0.1$ followed by a strong
suppression at larger $x$.
It is unclear what the origin of this difference is, especially since EPPS16 and nCTEQ share
the same free-proton PDF baseline.

\begin{figure}[!t]
  \centering
  \includegraphics[width=0.95\textwidth]{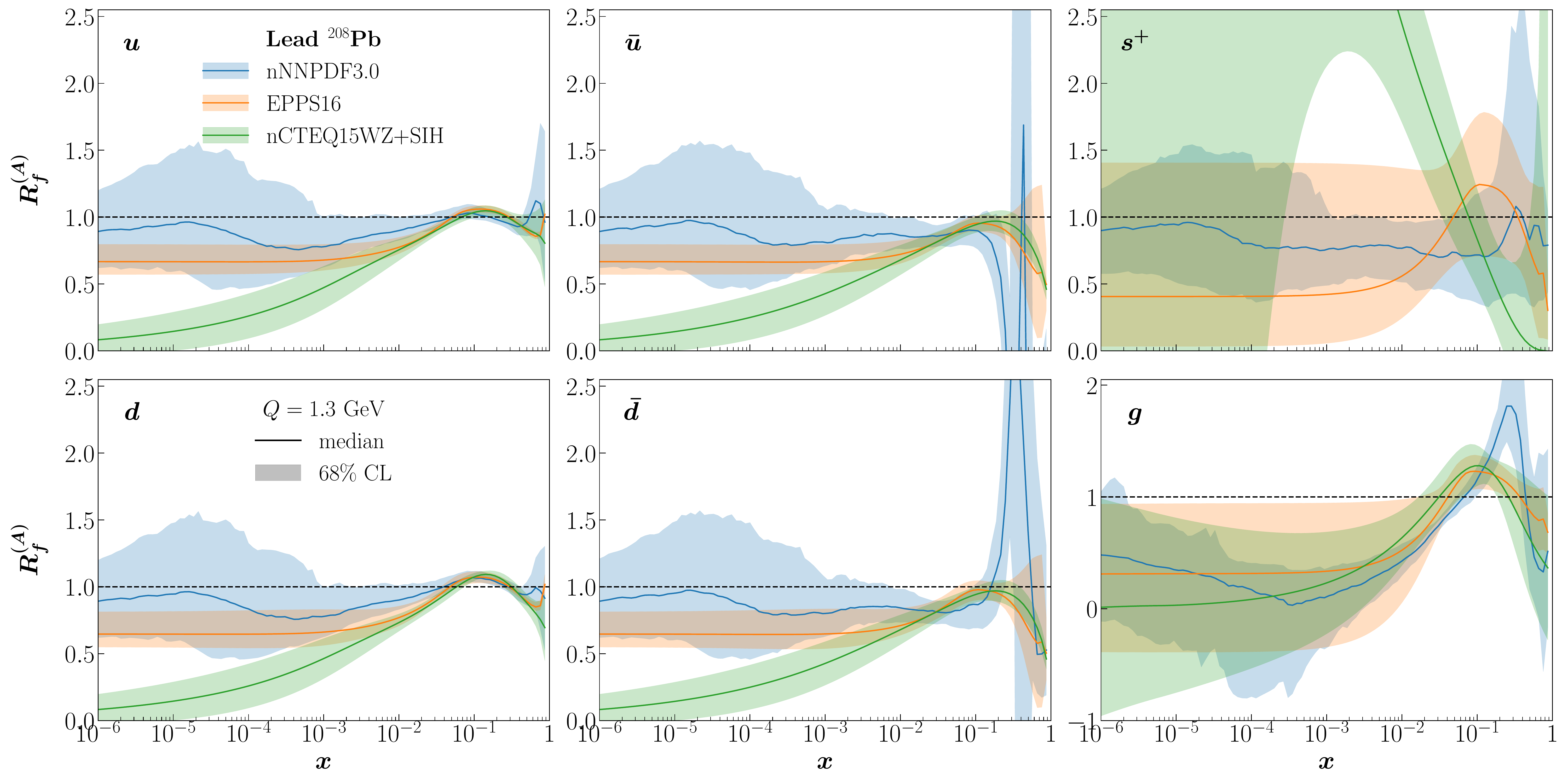}
  \includegraphics[width=0.95\textwidth]{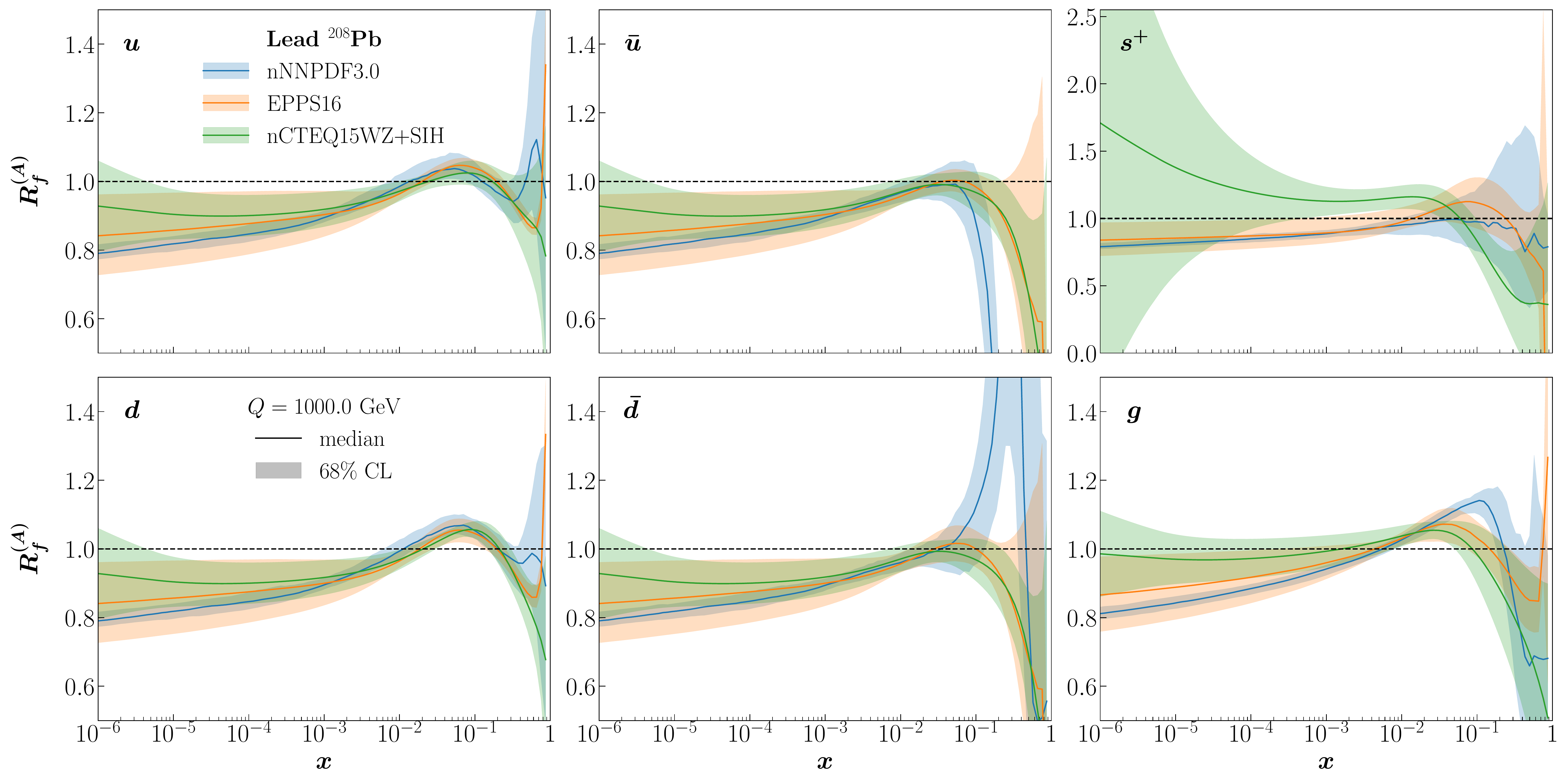}
  \caption{Same as Fig.~\ref{fig:comparison_npdfs_global} now displayed
    at $Q=1.3$~GeV (the lowest
    common scale among the three nPDF sets) and at $Q=1$~TeV in the upper
    and lower panels respectively.}
  \label{fig:nPDF_comparison_global_lowhighQ}
\end{figure}

It should be noted that, due to DGLAP evolution, the comparison of nuclear
modification factors across various groups may be subject to a different
interpretation if it were carried out at other values of $Q$. In general,
DGLAP evolution effects tend to smoothen out differences
present at medium- and small-$x$ as $Q$ is increased. To highlight this point, 
we display in Fig.~\ref{fig:nPDF_comparison_global_lowhighQ} the
same comparison as in Fig.~\ref{fig:comparison_npdfs_global} for
both the lowest scale $Q=1.3$~GeV, common to all nPDF sets, and
for a very high scale, $Q=1$~TeV. One sees that the PDF uncertainties in the
low-$x$ nuclear modification factors are large at $Q=1.3$~GeV, while they are
markedly reduced at large energy scales $Q=1$~TeV.
As expected, for $x\lsim 10^{-2}$ the differences between groups
are reduced as the scale $Q$ is increased due to DGLAP evolution.
Nevertheless, differences remain up to $Q=1$~TeV for the poorly known large-$x$
antiquark and strange nuclear modification ratios. Interestingly, the evidence
for small-$x$ quark and gluon shadowing and for large-$x$ gluon anti-shadowing
found in nNNPDF3.0 at $Q=10$~GeV persists up to the high scale considered here.

In summary, despite the several differences in input dataset, fitting methodology,
and theory calculations adopted, a reasonably consistent
picture emerges from the comparison between the three global analyses.
Especially remarkable is the agreement for the up and down quark nuclear ratios,
as well as for the corresponding antiquark ratios for $x\lsim 0.1$.
The main disagreements between the three groups are related to the gluon, in particular
concerning the strength of small-$x$ shadowing and large-$x$ anti-shadowing,
the strangeness nuclear modifications,
and the behaviour of the nuclear antiquarks at large $x$.
We note that for all these cases the choice of free-proton PDF baseline may
play a role, e.g. the gluon and strange PDFs already exhibit discrepancies
at the level of proton PDF fits.

\section{Stability analysis}
\label{sec:stability}

Here we present 
a number of studies assessing the stability of the nNNPDF3.0 results
and studying the impact of specific datasets and methodological choices.
First of all, we present the variant of nNNPDF2.0 used as starting point of the present
analysis, differing from the published version in methodological improvements
related to hyperparameter optimisation
and to the implementation of the proton boundary condition.
Second, we quantify the impact of those nuclear datasets that have been moved
from the proton baseline to the nuclear PDF analysis as discussed in Sect.~\ref{sec:expdata}.
Third, we study the constraints provided
by the CMS dijet cross-sections from pPb collisions
on the gluon nPDF.
Finally, we demonstrate the stability of nNNPDF3.0 upon two variations in the
treatment of the LHCb $D^0$-meson measurements: we apply different cuts on the
$D^0$-meson transverse momentum (i.e. restricting that data set to larger transverse momentum values); 
and we replace the measurements for the forward pPb-to-pp cross-section ratio with those for the forward-to-backward
ratio.

\subsection{nNNPDF2.0 reloaded}
\label{sec:nnnpdf20_reloaded}

We consider first a variant of the  nNNPDF2.0 analysis denoted
by nNNPDF2.0r (where `r' stands for  `reloaded' set).
This variant differs from the published nNNPDF2.0 set for the
methodological improvements described in Sect.~\ref{sec:fitting},
that we also summarise as follows.

First, the range in $x$ for which the proton boundary condition
is imposed has been lowered from $x_{\rm min}=10^{-3}$ to $x_{\rm min}=10^{-6}$,
motivated by the extension of
the kinematic coverage that is provided by the nNNPDF3.0
dataset, in particular
due to the LHCb $D^0$-meson production cross-section as shown in Fig.~\ref{fig:kinplot}.
Since these measurements are also included in the proton baseline, ensuring
that the free-proton
boundary condition is satisfied down to $x_{\rm min}=10^{-6}$ becomes necessary.

\begin{figure}[!t]
    \centering
    \includegraphics[width=1.0\textwidth]{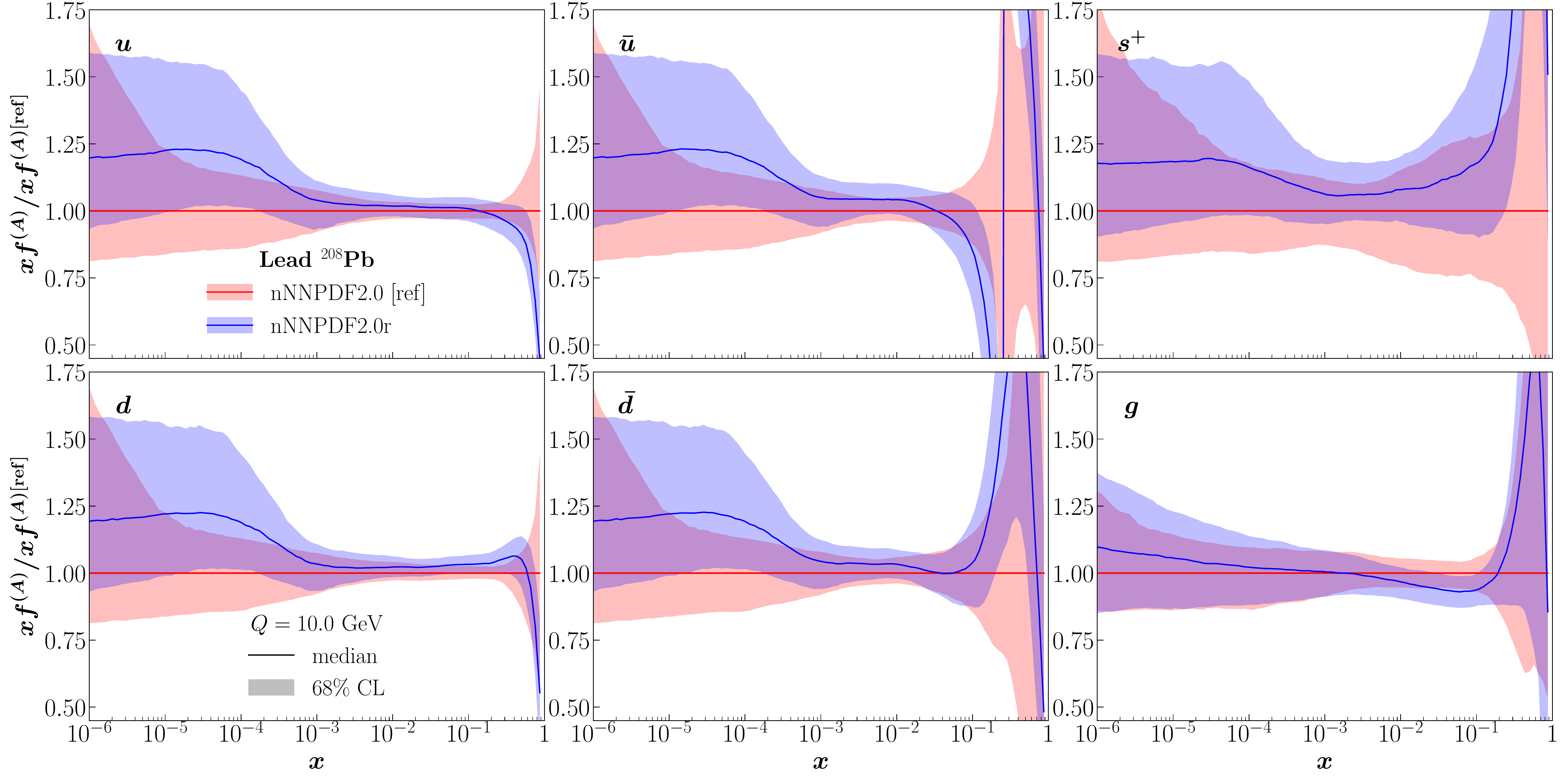}
    \caption{\small Comparison of nNNPDF2.0 with the nNNPDF2.0r variant.
      Results are shown for the lead PDFs at $Q=10$ GeV normalised to the central value of nNNPDF2.0,
      and the uncertainty bands represent the 68\% CL intervals.
    }
    \label{fig:nNNPDF20r}
\end{figure}

Second, the proton PDF baseline itself has been improved
as compared to the one used in nNNPDF2.0.
There, a variant of NNPDF3.1 with the heavy nuclear datasets
(taken on iron and lead targets) removed was adopted.
The new proton baseline in nNNPDF3.0 is also based on the NNPDF3.1 fitting
methodology but it is now extended to include
all the new datasets from pp collisions
considered in NNPDF4.0~\cite{Ball:2021leu}.
Specifically, this new proton baseline
includes the datasets labelled with {\bf (*)} in Tables 2.1--2.5 of~\cite{Ball:2021leu}
(with the exception of those included via reweighting),
see also the discussion in App.~B of~\cite{Ball:2021leu} for more details.
These datasets correspond, among others, to new measurements of inclusive and associated production
of gauge bosons, single and top-quark pair production, and jet and photon
production from ATLAS, CMS and LHCb.
Furthermore, no deuteron or copper datasets are considered
in this proton PDF baseline, since these enter the nuclear fit as discussed below.
Crucially, the resulting proton PDF baseline contains
the most updated measurements available from pp collisions for the same
processes that are considered in the corresponding nuclear PDF analysis.

The third methodological
improvement consists on the hyperparameter settings determined
by means of the optimisation procedure described in Sect.~\ref{sec:hyperparameter_scan}.
Interestingly, the optimal hyperparameters turn out to be very close to those
found manually in the nNNPDF2.0 analysis, with the only moderate differences for the activation
function, the weight initialization, and the learning rate of the SGD minimiser.

The aggregate impact of these various improvements 
is illustrated in Fig.~\ref{fig:nNNPDF20r}, which compares the lead PDFs in
nNNPDF2.0 and nNNPDF2.0r at $Q=10$ GeV normalised to the central value of the former.
One observes how in all cases the two fits agree within the corresponding
68\% CL uncertainties, except for the up antiquark at very large-$x$.
We also find that the PDF uncertainties in the region $x\lsim 10^{-3}$
in nNNPDF2.0r are increased as compared
to the published variant.
This result implies that the nPDF uncertainties in nNNPDF2.0 were somewhat
underestimated there, due to imposing the $A=1$ limit in a restricted
region of $x$.
While differences are in general moderate in the data region,
they can be marked in the 
extrapolation regions at small- and large-$x$
where there are limited  experimental constraints.
Overall, good consistency between the two fits is found.

\subsection{The impact of the deuteron and copper NNPDF3.1 datasets}

As discussed in Sects.~\ref{sec:expdata} and~\ref{sec:fitting}, the nNNPDF3.0
analysis contains a number of datasets taken on deuteron and copper
targets that previously were accounted for by means of the proton
PDF boundary condition.
Specifically, these consist on the NMC~\cite{NewMuon:1996uwk} deuteron
to proton DIS structure functions; the SLAC~\cite{Whitlow:1991uw} and
BCDMS~\cite{BCDMS:1989qop} deuteron structure functions; and the
E866~\cite{NuSea:2001idv} fixed-target DY deuteron to proton cross-section ratio.
Furthermore, this category also includes the fixed-target DY measurement performed
on Cu by E605~\cite{Nagarajan:1991zz}.
Here we ascertain the impact of these deuteron and copper
datasets by adding them on top of the nNNPDF2.0r fit.

\begin{figure}[!t]
    \centering
    \includegraphics[width=1.0\textwidth]{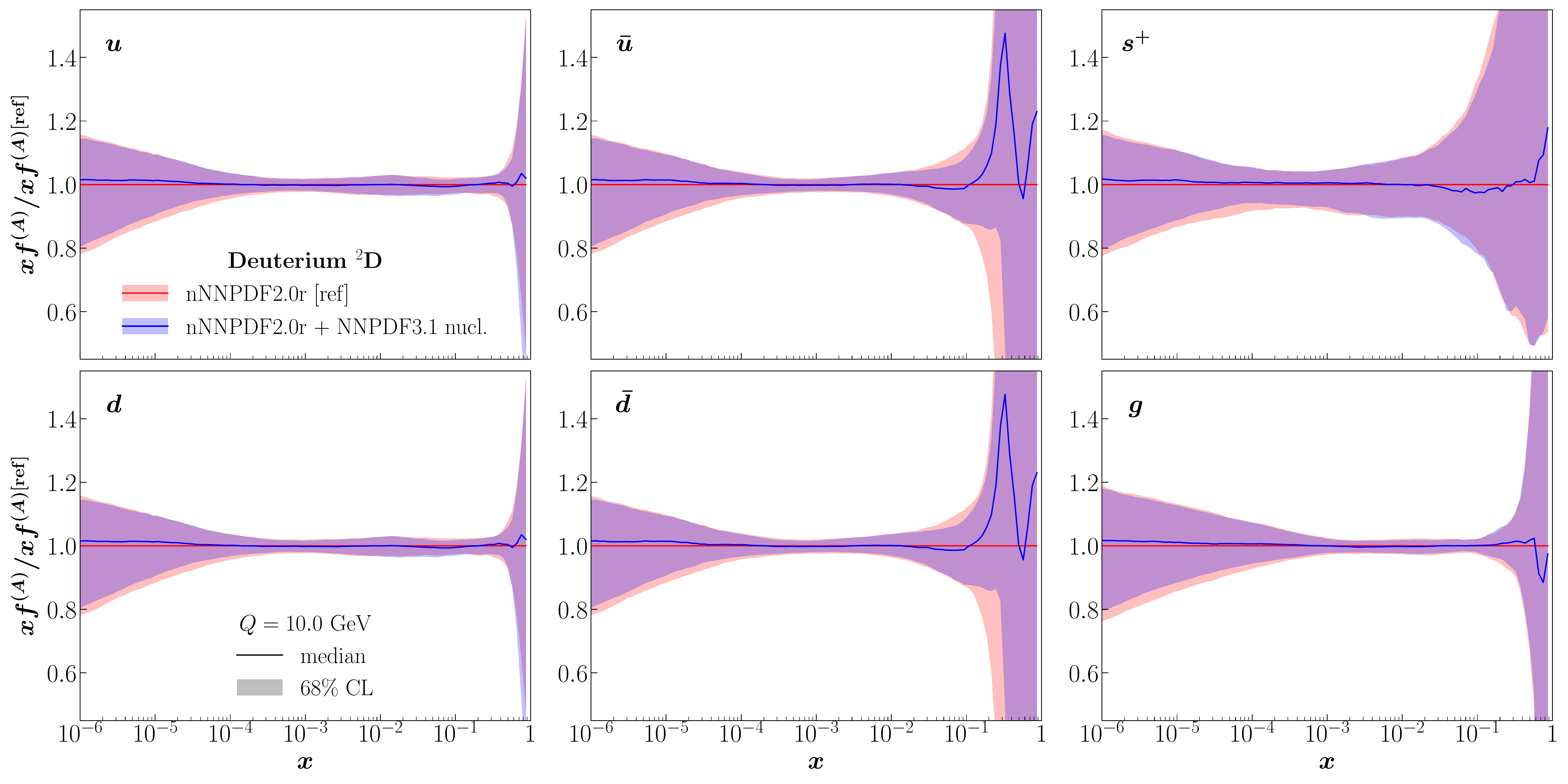}
    \caption{\small  Comparison of nNNPDF2.0r
      with the same fit where the deuteron and copper datasets from
      NNPDF3.1 have been included, see text for more details.
      Results are shown for the deuterium PDFs ($A=2$)
      at $Q=10$ GeV normalised to the central value of nNNPDF2.0r,
      and the uncertainty bands represent the 68\% CL intervals. }
    \label{fig:nNNPDF20r_nuclNNPDF3_D2}
\end{figure}

\begin{figure}[!t]
    \centering
    \includegraphics[width=1.0\textwidth]{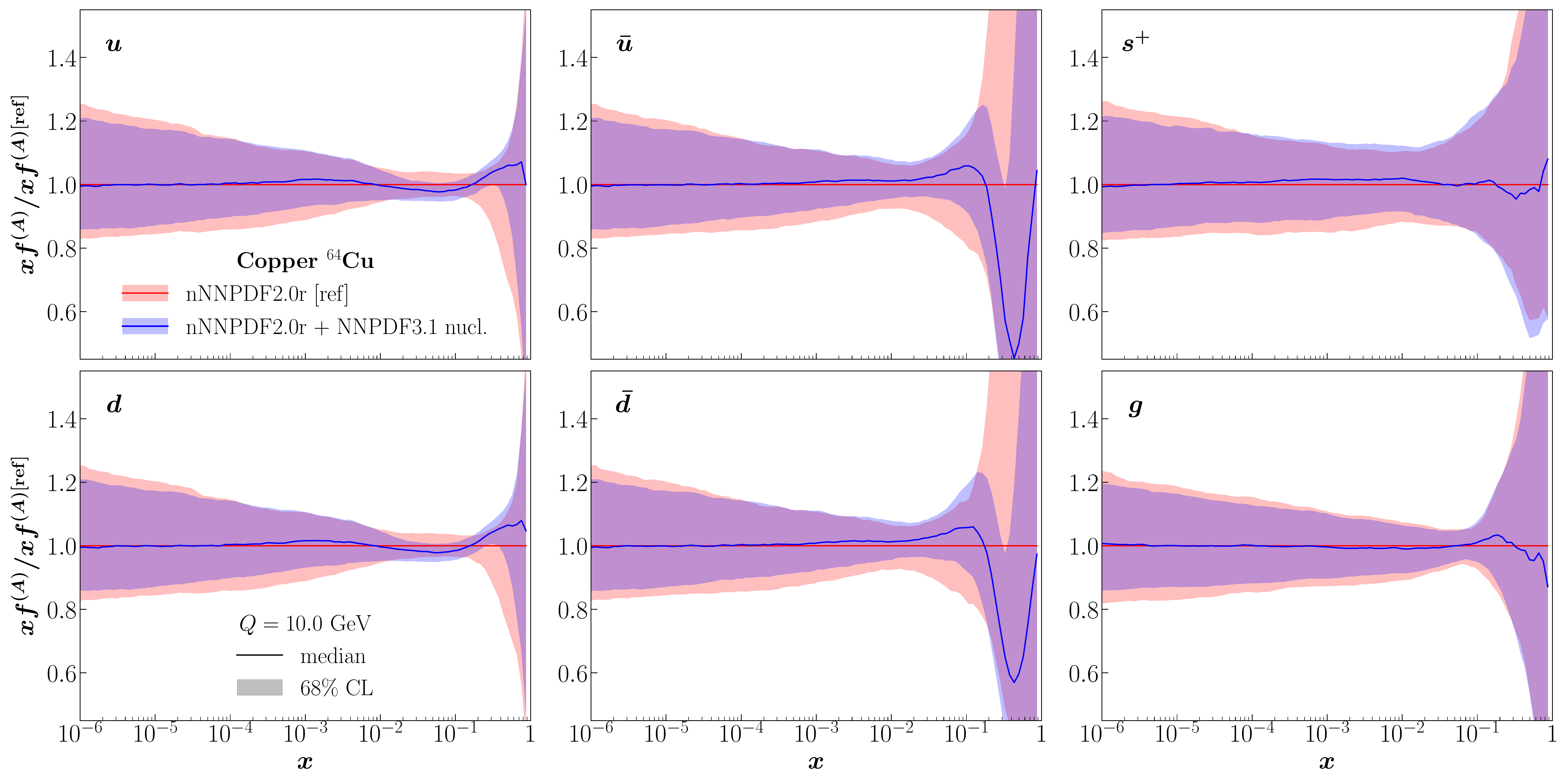}
    \caption{\small Same as Fig.~\ref{fig:nNNPDF20r_nuclNNPDF3_D2} for
    the nPDFs corresponding to copper ($A=64$) nuclei.}
    \label{fig:nNNPDF20r_nuclNNPDF3_Cu64}
\end{figure}

Figs.~\ref{fig:nNNPDF20r_nuclNNPDF3_D2} and~\ref{fig:nNNPDF20r_nuclNNPDF3_Cu64}
compare nNNPDF2.0r with the same fit where the deuteron and copper datasets from
NNPDF3.1 have been included, showing the nPDFs of
deuterium ($A=2$) and copper ($A=64$) nuclei respectively.
Considering the results for $A=2$, one finds good stability,
with the impact of the deuteron fixed-target data most visible for the large-$x$
up and down antiquarks (which are identical for this isoscalar nucleus),
an effect which is consistent with the smallness of nuclear effects in 
 in deuterium.
More marked effects are found at the level of copper nuclei,
Fig.~\ref{fig:nNNPDF20r_nuclNNPDF3_Cu64},
where the proton-copper DY cross-sections from E605 lead
to a reduction of the nPDF uncertainties, in particular in the region
with $x\gsim 10^{-2}$ where this data has kinematical coverage.
The appreciable impact of these measurements on the nPDFs
is consistent with the fact that direct constraints
on the Cu nPDFs are limited to a few DIS structure function data points.
Furthermore, while the central value of the up, down, and strange quarks
is mostly unaffected (as is the case for the gluon), the large-$x$ antiquarks
are suppressed following the inclusion of the E605 data.

\subsection{Impact of dijet production}
\label{sec:impact_dijet}

Among the new datasets that enter the nNNPDF3.0
determination and discussed in Sect.~\ref{sec:expdata},  the CMS measurement of dijet production
in pPb collisions at 5.02 TeV~\cite{CMS:2018jpl} is one of those carrying
the most information on the nuclear PDFs.
In our analysis we consider the ratio of pPb-to-pp dijet
spectra, double-differential in
the  dijet
average transverse momentum $p_{T,{\rm dijet}}^{\rm ave}$ and the
dijet pseudo-rapidity $\eta_{\rm dijet}$.
As indicated in Fig.~\ref{fig:kinplot}, this measurement
covers a range in $x$ between $10^{-3}$ and 1
and in $Q^2$ between 400 GeV$^2$ and $10^5$ GeV$^2$. 
Since jet production in hadronic collisions is dominated in this kinematic region
by quark-gluon scattering~\cite{Rojo:2014kta}, the CMS measurement provides direct constraints
on the nuclear modifications of the gluon PDF for $x\gsim 10^{-3}$.
We also point out that the ATLAS and CMS measurements of dijet cross-sections in
pp collisions at 7 TeV are already accounted for by means of the free-proton
PDF boundary condition.

\begin{figure}[!t]
    \centering
    \includegraphics[width=0.8\textwidth]{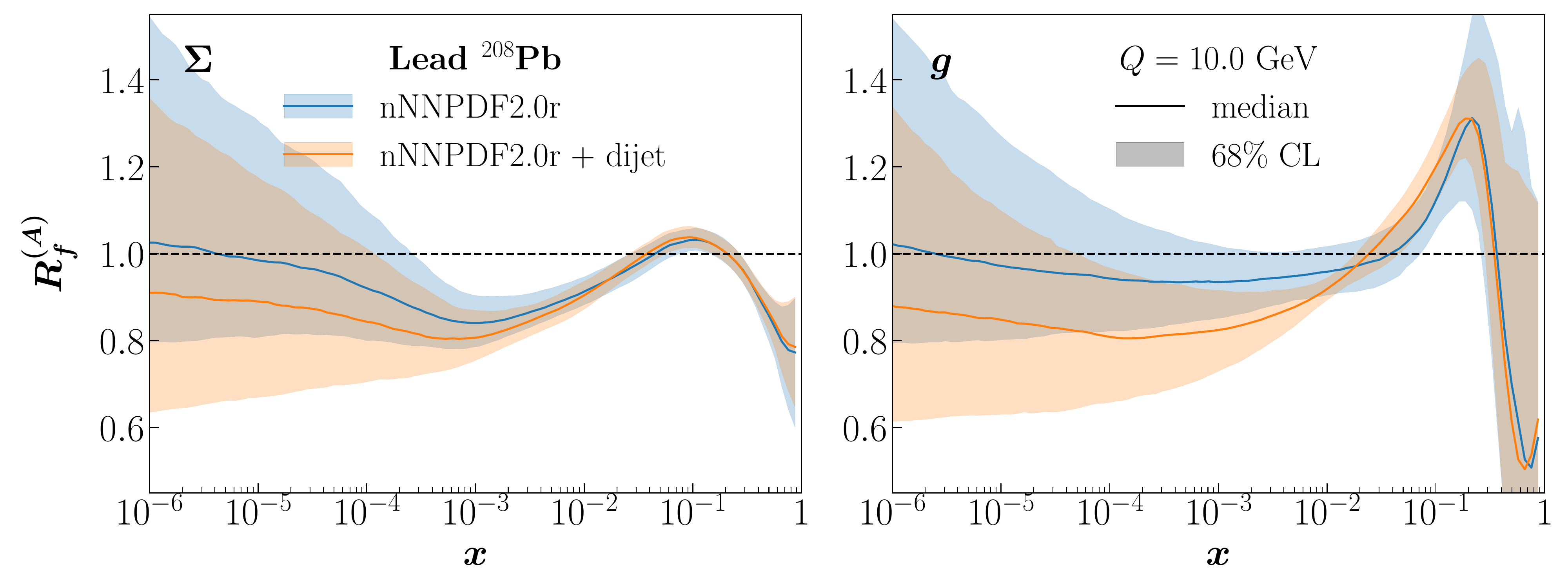}
    \includegraphics[width=0.8\textwidth]{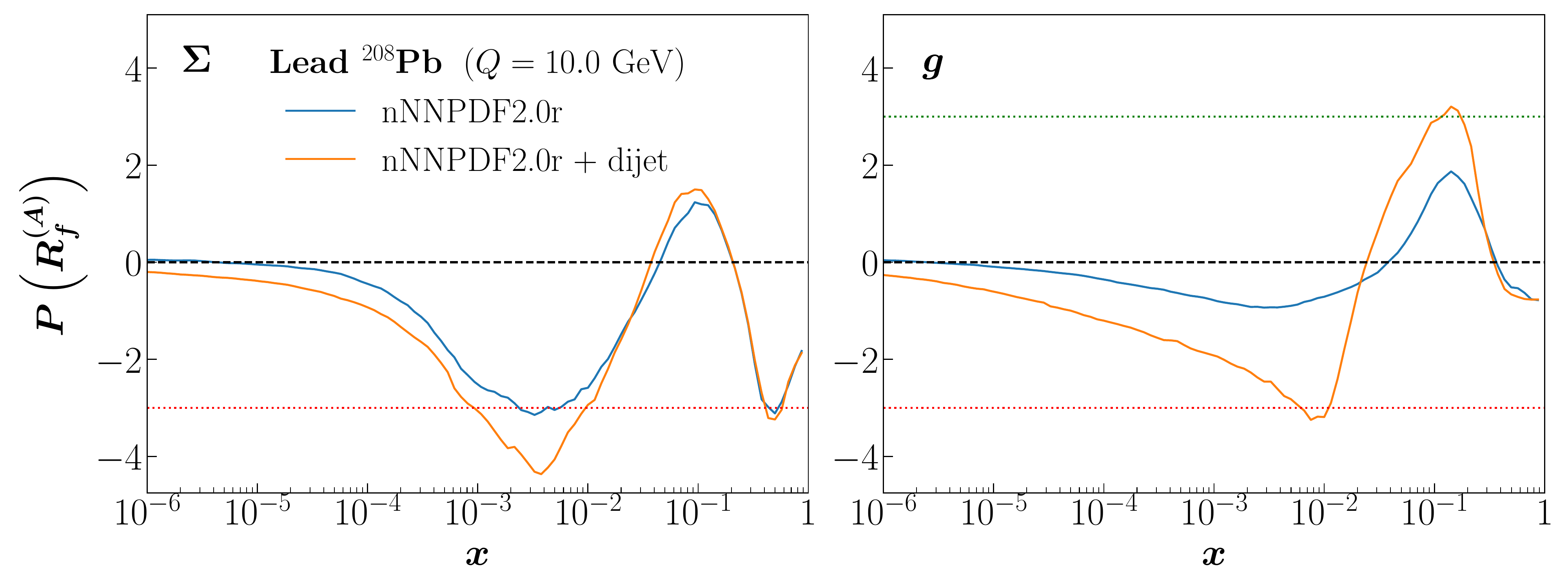}
    \caption{\small The impact of the CMS dijet pPb-to-pp ratio
      measurements at 5.02 TeV on the quark singlet (left)
      and gluon (right panel) nPDFs of lead. We display results for the nuclear modification factors
      $R^{(A)}(x,Q)$ at $Q=10$ GeV, comparing the nNNPDF2.0r fit with the variant including
      the CMS dijet data (upper)
      as well as for the corresponding pulls defined in Eq.~(\ref{eq:pull_global})
      (bottom panels).
      The dotted horizontal lines indicate the $3\sigma$ thresholds.
    }
    \label{fig:nNNPDF20r_dijets_RA}
\end{figure}

Here we present a variant of nNNPDF3.0 where the CMS dijet cross-section ratio
is the only measurement added on top of the NNPDF2.0r baseline fit defined
in Sect.~\ref{sec:nnnpdf20_reloaded}.
Fig.~\ref{fig:nNNPDF20r_dijets_RA} displays the impact of these CMS dijet measurements
on the quark singlet and gluon  nPDFs of lead when added to nNNPDF2.0r.
Results are presented for  the nuclear modification factor $R^{(A)}(x,Q)$ at $Q=10$ GeV
as well as for the corresponding pulls defined in Eq.~(\ref{eq:pull_global}).
For the quark singlet nPDF, the impact of the dijet data is moderate
and restricted to the small-$x$ region, where a stronger shadowing
is favored.
In the case of the gluon nPDF, it is found that the dijet measurements significantly
reduce the uncertainties for $10^{-3}\lsim x \lsim 0.4$.
For smaller values of $x$, the nPDF uncertainty is unaffected
but the central 
value of $R^{(A)}$ remains suppressed as compared to the nNNPDF2.0r reference,
which in turn enhances the significance of small-$x$ gluon shadowing.
This comparison confirms that the CMS dijet cross-sections prefer a strong small-$x$ shadowing
of the lead gluon nPDF, a feature also reported in~\cite{Eskola:2019dui}.
Furthermore,  since the central value of the gluon
around $x\simeq 0.2$ is unchanged but the uncertainties
are almost halved, the CMS dijet data also enhances the fit preference for a strong gluon
anti-shadowing in the large-$x$ region.

The impact of the CMS dijet measurements on the nPDF fit is also illustrated
by the comparison of the pulls before and after its inclusion, displayed in the bottom
panels of Fig.~\ref{fig:nNNPDF20r_dijets_RA}.
For the quark singlet, the dijet data enhances the pull for $x\lsim 10^{-2}$,
strengthening the evidence for quark sector shadowing in this region.
For $x\lsim 5\times 10^{-3}$, outside the region covered by the CMS dijet data,
uncertainties grow rapidly and the fit results are consistent
with no nuclear modifications at the 68\% CL.
Considering the gluon nPDF, one finds that while the baseline fit
is consistent with no small-$x$ anti-shadowing within uncertainties,
once the CMS dijets are accounted for this significance reaches
the $3\sigma$ level, peaking at $x\simeq 10^{-2}$.
The constraints provided by the CMS dijets are also visible for the large-$x$
anti-shadowing, whose significance increases from the $2\sigma$
to the $3\sigma$ level.
All in all, these comparisons illustrate
how the CMS dijet measurements are instrumental in nNNPDF3.0 to pin down
the modifications of the gluon PDF.

\subsection{Impact of kinematic cuts on the $D^0$-meson transverse momentum}
\label{sec:cut_impact}

As mentioned in Sect.~\ref{sec:expdata}, no kinematic cuts are applied to the
transverse momentum of the $D^0$ meson, $p_T^{D^0}$, in the analysis of the LHCb
$D^0$-meson data entering our baseline fits. We assess the stability of our
results upon this choice by repeating the reweighting of the prior fit
displayed in Fig.~\ref{fig:nNNPDF30_RA_Pb208_a_NuclearRationPDFs_Q10}
after applying a cut on $p_T^{D^0}$, for different values of the cut.
We specifically consider the following cases: $p_T^{D^0}>2$~GeV, $p_T^{D^0}>3$~GeV, and
$p_T^{D^0}>5$~GeV. For each of them, Table~\ref{tab:rw_pcut} collects the
same statistical estimators reported in Table~\ref{tab:rw_details}: the number
of effective replicas $N_{\rm eff}$, the number of replicas in the unweighted
set $N_{\rm unweight}$, the number of data points, and the values of the $\chi^2$
per data point before and after reweighting. We note that, as the cut on
$p_T^{D^0}$ is increased, the value of the prior $\chi^2$ decreases, the number of
effective replicas increases, and the value of the $\chi^2$ after
reweighting remains similar irrespective of the $p_T^{D^0}$ cut.

\begin{table}[!t]
  \centering
  \footnotesize
  \small
\renewcommand{\arraystretch}{1.45}
\begin{tabularx}{\textwidth}{XC{2.5cm}C{2.5cm}C{3.5cm}C{2.5cm}C{2.5cm}}
  \toprule
  $p_T^D$ cut &
  $N_{\text{eff}}$  &   $N_{\text{unweight}}$   & $n_{\text{dat}}$ & $\chi^2_{\rm prior}$ & $\chi^2_{\text{rw}}$  \\
  \midrule
  no $p_T^D$ cut & 185 & 200 & 37  & 32.2 & 0.66 \\
  $p_T^D>2$~GeV  & 330 & 200 & 29  & 11.8 & 0.53 \\
  $p_T^D>3$~GeV  & 443 & 200 & 25  & 6.7  & 0.54 \\
  $p_T^D>5$~GeV  & 776 & 200 & 17  & 2.8  & 0.57 \\
  \bottomrule
\end{tabularx}

  \caption{Same as Table~\ref{tab:rw_details}, now for varying cuts on the
    transverse momentum of the $D^0$ meson $p_T^{D^0}$.
  }
  \label{tab:rw_pcut}
\end{table}

\begin{figure}[!t]
  \centering
  \includegraphics[scale=0.29]{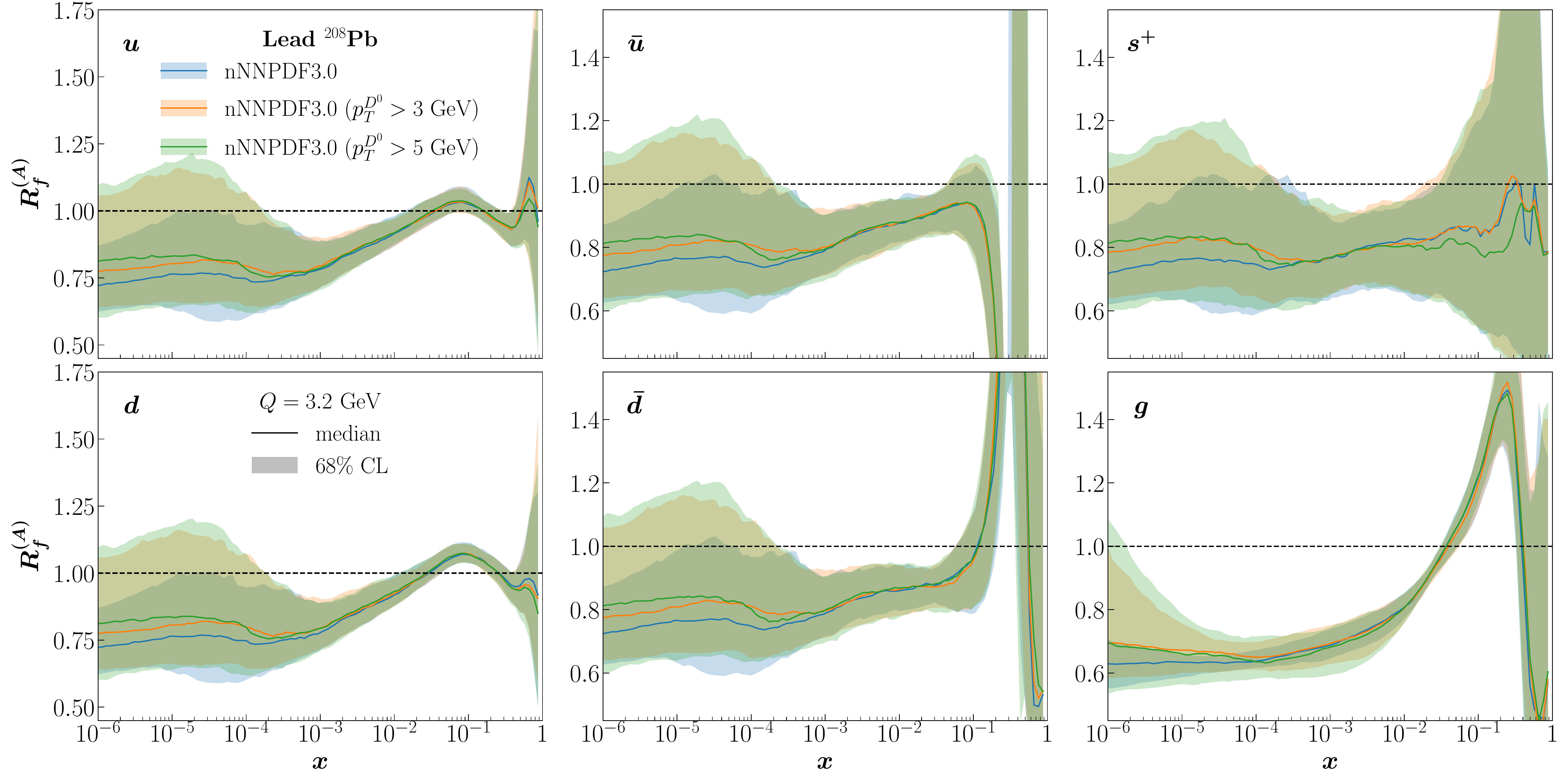}
  \caption{Same as Fig.~\ref{fig:nNNPDF30_RA_Pb208_a_NuclearRationPDFs_Q10},
    now at $Q^2=10$ GeV$^2$,  
    comparing the baseline nNNPDF3.0 determination with the
    variants where kinematic cuts of $p_T^{D^0} \ge 3$~GeV
    and $p_T^{D^0} \ge 5$~GeV are applied to the LHCb $D^0$-meson cross-section data,
    see also Table~\ref{tab:rw_pcut}.
  \label{rw:Rf_cuts_LHCbDmeson}}
\end{figure}

Figure~\ref{rw:Rf_cuts_LHCbDmeson} displays the resulting PDF nuclear
modification factors for the nNNPDF3.0 variants carried out with kinematic
cuts of $p_T^{D^0} \ge 3$~GeV and  $p_T^{D^0} \ge 5$~GeV at $Q^2=10$ GeV$^2$.
The effect of introducing a $p_T^{D^0}$ cut is moderate, though generally leads an
increase in the resultant nPDF uncertainties. The more restrictive the cut, the larger the
uncertainty increase. For instance, in the region $x\simeq 10^{-4}$,
the nPDF errors on the gluon nuclear modification factor increase by around
a factor of two for $p_T^{D^0} \ge 5$~GeV. 
However, even in the conservative case, where the restriction of $p_T^{D^0} \ge 5$~GeV
is applied to the LHCb $D^0$-meson data, the reduction of nPDF uncertainties is substantial
as compared to the no LHCb $D^0$-meson scenario 
(see for example the comparison in Fig.~\eqref{fig:nNNPDF30_RA_Pb208_a_NuclearRationPDFs_Q10})
This fact highlights how even only a subset of the LHCb data still imposes significant constraints on the
small-$x$ nPDFs. 
In addition, the consistency of the results obtained when restricting the data with a varying $p_T^{D^0}$ cut 
also indicate that the nNNPDF3.0 determination is robust upon introduction and variation of a cut
on $p_T^{D^0}$.
%
%

\subsection{Constraints from the $D^0$-meson forward-to-backward ratio}
\label{sec:Rfb_impact}

In Sect.~\ref{sec:D0_reweighting} we assessed the impact of the $D^0$-meson production measurements
from LHCb in nNNPDF3.0 in terms of the ratio between pPb and pp spectra
in the forward region defined in  Eq.~\eqref{eq:def_nuclear_modification_ratio}.
An alternative observable to assess the impact of the LHCb $D^0$-meson data
in the nPDF fit is provided by the ratio of forward-to-backward
measurements defined in Eq.~(\ref{eq:Rfb}).
Since in the nucleon-nucleon CoM frame LHCb measurements cover the range
$1.5 < y^{D^0} < 4.0$ and $2.5 < y^{D^0} < 5.0$ for pPb (forward) and pPb (backwards) collisions respectively,
the forward-to-backward ratio $R_{\rm fb}$ is provided
in the three overlapping rapidity bins covering the region  $2.5 < y^{D^0} < 4.0$,
adding up to  a total of 27 data points. 

\begin{figure}[!t]
  \centering
  \includegraphics[width=\textwidth]{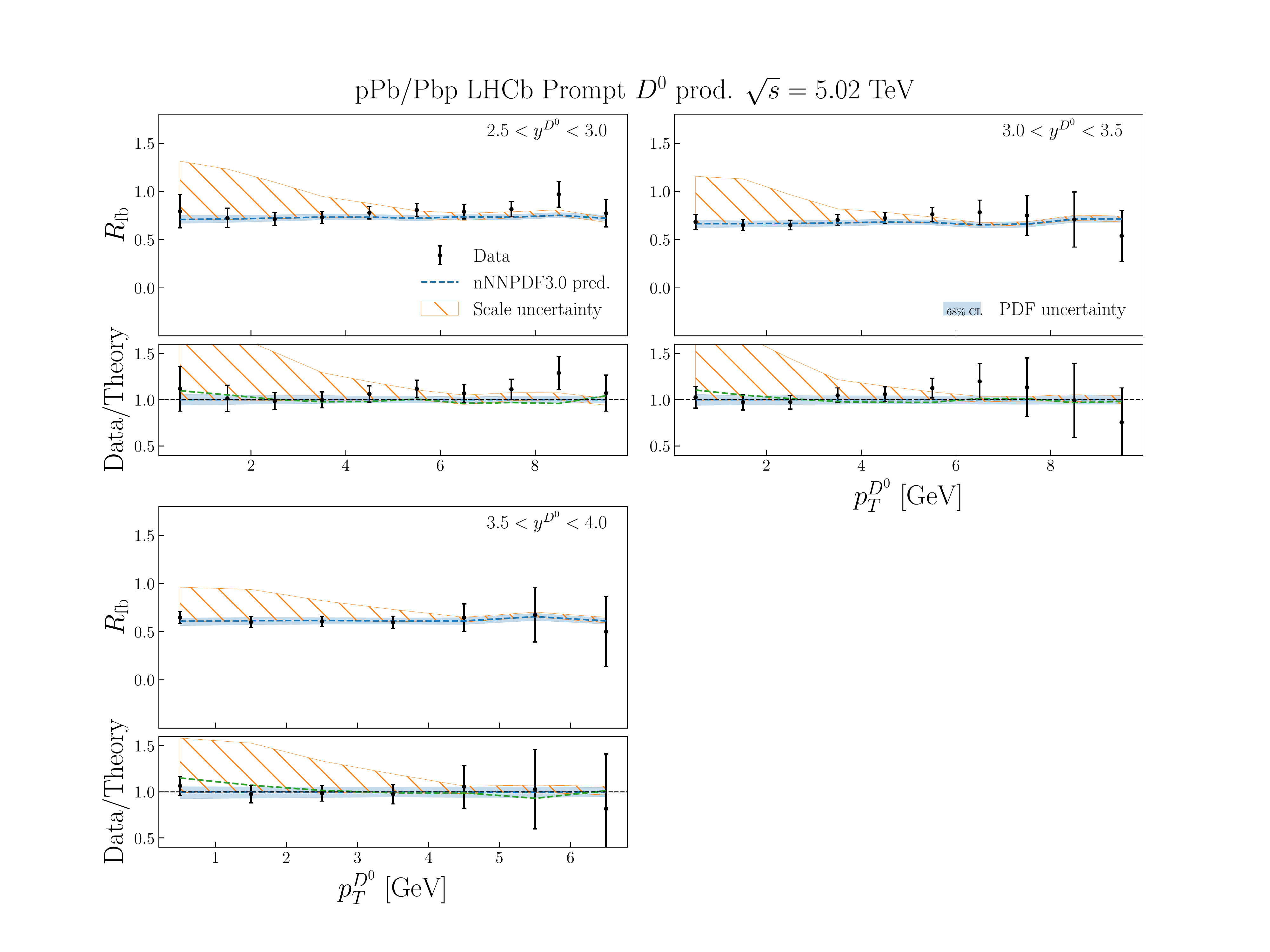}
  \caption{Comparison between the LHCb data on the ratio of forward-to-backward
    measurements $R_{\rm fb}$ and the corresponding theoretical predictions based on
    nNNPDF3.0 (which includes instead the data on $R_{\rm pPb}$).
    The $R_{\rm fb}$ ratio is available in the three rapidity
    bins for which the forward (pPb) and backward (Pbp) measurements
    overlap.
    We display separately the PDF and scale uncertainties.
  }
  \label{fig:Rfp_data_th_nNNPDF3.0}
\end{figure}

First of all, in Fig.~\ref{fig:Rfp_data_th_nNNPDF3.0} we display theoretical predictions
for $R_{\rm fb}$ in these three rapidity bins computed using nNNPDF3.0 as input
(which includes the constraints provided by $R_{\rm fb}$).
This comparison accounts for both uncertainties due to PDFs and MHOs,
and demonstrates that predictions based on nNNPDF3.0 also provide a satisfactory description 
of the $R_{\rm fb}$ data, which were not included in the fit.
Hence, the information on the nPDFs provided by the two LHCb observables
is consistent.

\begin{figure}[!t]
    \centering
    \includegraphics[width=\textwidth]{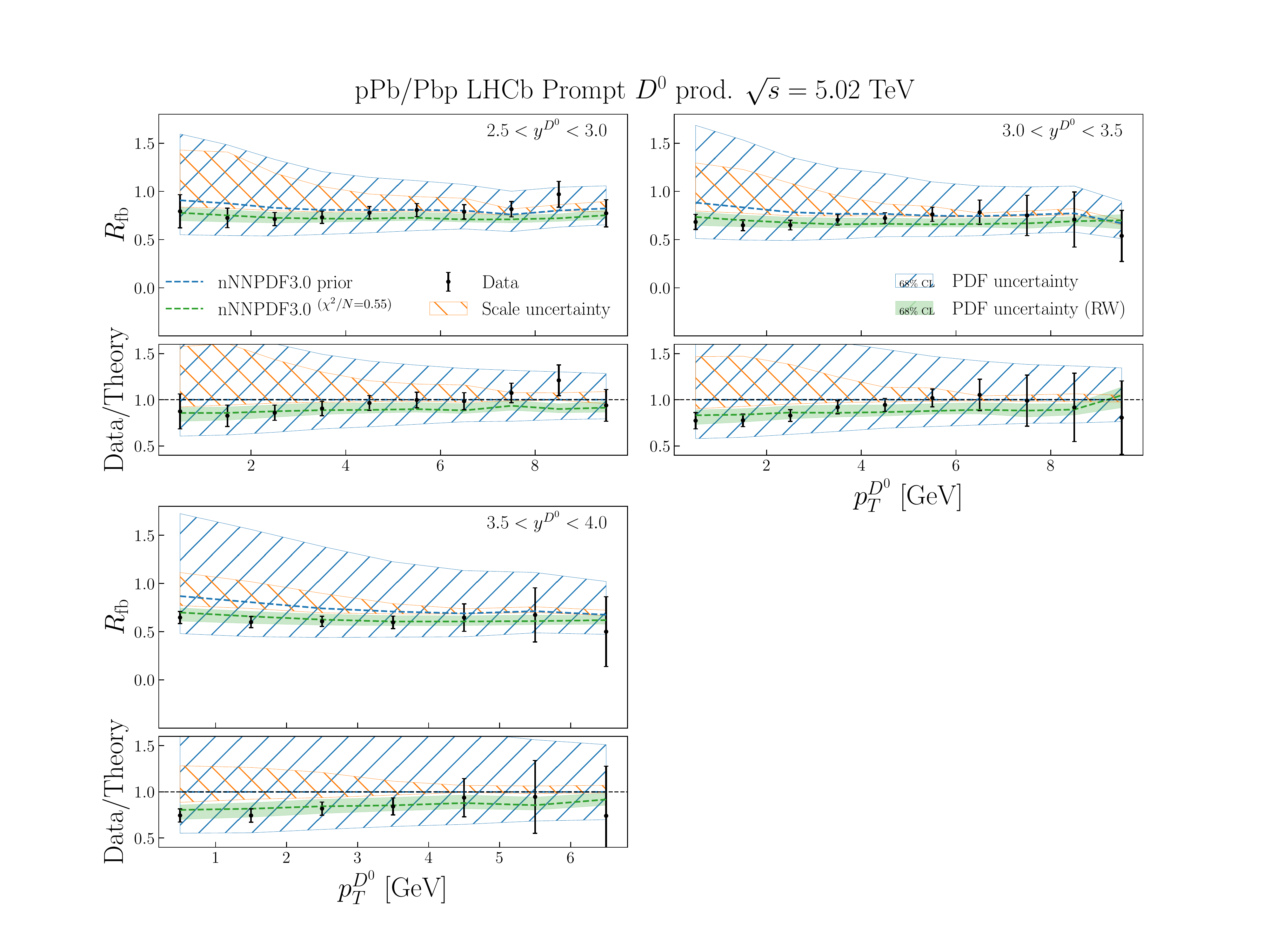}
    \caption{Same as Fig.~\ref{fig:D0_pPb_forward} now
      for the case in which the LHCb data on the forward-to-backward ratio
      $R_{\rm fb}$ defined in Eq.~(\ref{eq:Rfb}) is added to the prior set
      by reweighting.
    }
    \label{fig:Rfp_data_th}
\end{figure}

As an additional check of the stability of our results, we have studied the impact
of the inclusion of the $R_{\rm fb}$ data on the nNNPDF3.0 prior described in  Sect.~\ref{sec:RW}.
Fig.~\ref{fig:Rfp_data_th} displays the theoretical
predictions for the nNNPDF3.0 prior set (with the LHCb $D$-meson data accounted for only
via the pp baseline) compared to the corresponding LHCb measurements
of the forward-backward ratio $R_{\rm fb}$
as well as to the result of including this dataset in the fit
by means of Bayesian reweighting.
As in the case of Fig.~\ref{fig:D0_pPb_forward},
we display separately the uncertainties due to PDFs and MHOs for the prediction
based on the prior fit.
In the same manner as for the forward Pb-to-pp ratio Eq.~\eqref{eq:def_nuclear_modification_ratio},
also in $R_{\rm fb}$ the PDF uncertainties dominate both
over scale uncertainties (partially cancelling out in the ratio)
and over the experimental errors.
This assessment indicates that the LHCb measurements
of $R_{\rm fb}$ are also suitable to be included in nNNPDF3.0 via reweighting.
As was the case for the $R_{\rm pPb}$ data, the theoretical
predictions for $R_{\rm fb}$ display a significantly reduced PDF
uncertainty once this dataset has been added to the prior via reweighting.
In Table~\ref{tab:rw_Rfb_details} we report the details of the reweighting and unweighting procedures.

\begin{table}[!t]
  \centering
  \footnotesize
  \centering
\small
\renewcommand{\arraystretch}{1.45}
\begin{tabularx}{0.7\textwidth}{C{2.5cm}C{2.5cm}C{2.5cm}C{2.5cm}}
    \toprule
    $N_{\text{eff}}$  &   $n_{\text{dat}}$ & $\chi_{\rm prior}^2$ & $\chi^2_{\text{rw}}$  \\
    \midrule
    509      &   27 & 11.63 & 0.55             \\
\bottomrule
\end{tabularx}
\vspace{0.3cm}

  \caption{Same as Table~\ref{tab:rw_details}, now for the case
    of the  inclusion of of the LHCb data on $R_{\rm fb}$ to the prior
    fit by means of reweighting.
  }
  \label{tab:rw_Rfb_details}
\end{table}

\begin{figure}[!t]
  \centering
  \includegraphics[width=\textwidth]{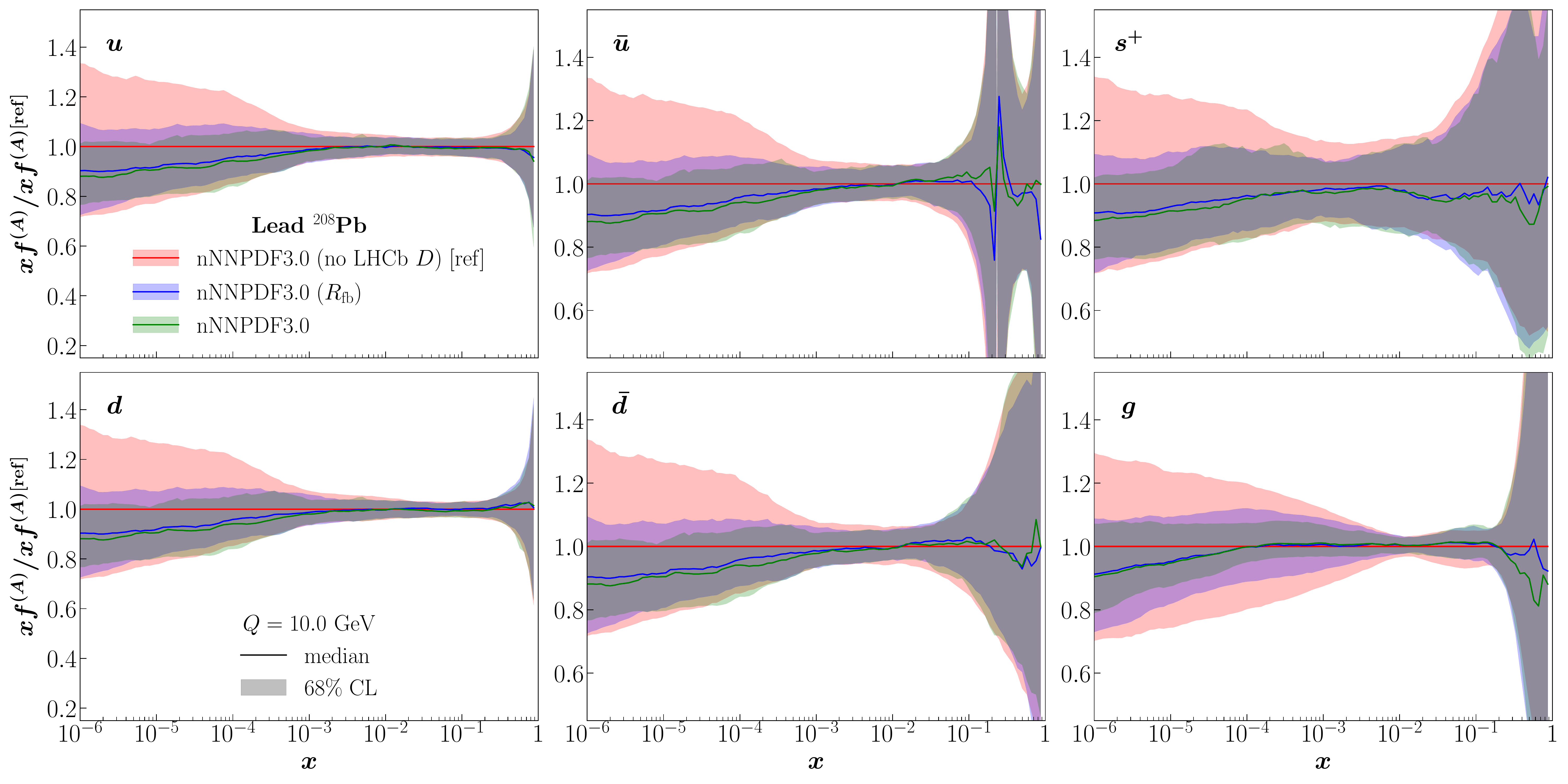}
  \caption{\small The impact of the $D^0$-meson production data from pPb collisions on
    the lead PDF at $Q=10$ GeV.
    We compare the nNNPDF3.0 prior set
    with the outcome of the reweighting with the LHCb $D^0$-meson pPb data
    using either the forward pPb-to-pp ratio $R_{\rm pPb}$ ,
    Eq.~(\ref{eq:def_nuclear_modification_ratio}), or
    the forward-to-backward ratio $R_{\rm fb}$, Eq.~(\ref{eq:Rfb}).
    Results are presented normalised to the central value of the nNNPDF3.0 prior.
  } 
  \label{fig:D0_impact_lead_npdfratios}
\end{figure}

\begin{figure}[!t]
  \centering
  \includegraphics[width=\textwidth]{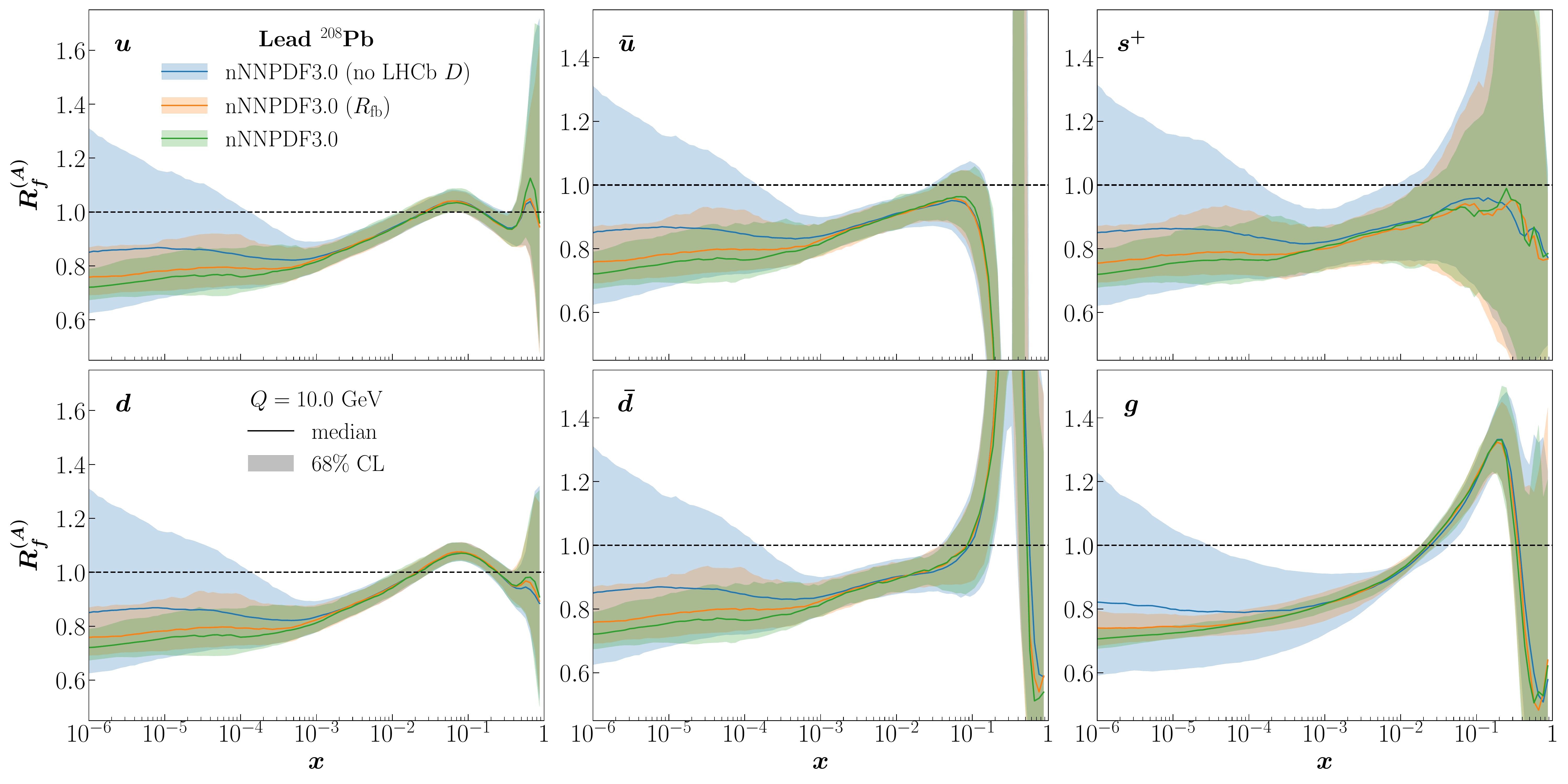}
  \caption{\small Same as Fig.~\ref{fig:D0_impact_lead_npdfratios}
  for the nuclear modification factors $R_f^{(A)}(x,Q^2)$.}
\label{fig:D0_impact_lead_nuclearmodifications}
\end{figure}

Figs.~\ref{fig:D0_impact_lead_npdfratios} and~\ref{fig:D0_impact_lead_nuclearmodifications}
then compare the impact on the
nNNPDF3.0 prior fit of the  LHCb $D^0$-meson pPb data
when accounted for 
using either the forward pPb-to-pp ratio $R_{\rm pPb}$,
Eq.~(\ref{eq:def_nuclear_modification_ratio}),  or
the forward-to-backward ratio $R_{\rm fb}$, Eq.~(\ref{eq:Rfb}), at the level
of the lead PDFs and of the  nuclear modification factors respectively.
One finds that the outcome of incuding the LHCb $D^0$-meson data using
either $R_{\rm pPb}$  or $R_{\rm fb}$ is fully compatible,
with the former leading to a somewhat larger reduction
of the PDF uncertainties and hence justifying our baseline choice
in the nNNPDF3.0 dataset.
The obtained central values are also similar for the two observables,
especially in the case of the gluon nuclear modification factor.
We note that,
as already remarked in Sect.~\ref{sec:D0_reweighting},
the impact of the LHCb $D$-meson data is substantial
in the nuclear ratio, which accounts for the correlations
between the proton and lead PDF uncertainties.
These results demonstrate that the impact of the LHCb $D$-meson
data in the nNNPDF3.0 determination is robust upon variations of the experimental
observable included in the fit.

\section{(Ultra-)High-energy neutrino-nucleus interactions}
\label{sec:uheneut}

As motivated in Sect.~\ref{sec:introduction}, precise and accurate knowledge of nPDFs is crucial for predicting the absolute rate of ultra-high-energy (UHE) neutrino--nucleon DIS interactions.
This information is important for the interpretation of UHE neutrino events observed at large-volume based neutrino detectors such as IceCube~\cite{IceCube:2006tjp} and KM3NeT~\cite{KM3Net:2016zxf}, which can in turn provide vital information on the rates of atmospheric and cosmic neutrino production mechanisms.
At yet higher energies (beyond the PeV), proposed large volume detectors such as GRAND~\cite{GRAND:2018iaj} or POEMMA~\cite{Olinto:2017xbi} may also contribute to our knowledge of the cosmic neutrino flux for energies up to $10^9$~GeV (see~\cite{Denton:2020jft} for a feasibility study).
In those cases, knowledge of nuclear corrections is either necessary for describing neutrino-matter interactions within the detector volume (e.g. water), or required to describe the attenuation rate of neutrinos as they travel through Earth towards the various detector~\cite{Garcia:2020jwr}.
The uncertainty related to the magnitude of these nuclear corrections is the dominant source of theoretical uncertainty in describing UHE neutrino-matter interactions (see Fig.~5 of~\cite{Gauld:2019pgt} for a breakdown of the various uncertainties).

We are now in a position to present updated predictions for UHE neutrino-nucleus cross-sections based on nNNPDF3.0.
Our calculational settings follow~\cite{Bertone:2018dse} for the neutrino-nucleon DIS cross-section and use {\sc\small APFEL}~\cite{Bertone:2013vaa} to evaluate the structure functions for both charged- and neutral-current interactions.
These calculations are performed in the FONLL general-mass variable flavour number scheme (consistently with the nNNPDF3.0 fit), but differ from the calculation presented in~\cite{Bertone:2018dse} through the inclusion of two-quark mass contributions to the charged-current DIS structure function.
The latter contributions are necessary to account for fixed-order corrections of the form $\alpha_s \ln[m_b] m_t^2$ when following the FONLL implementation~\cite{Ball:2011mu}, and become numerically relevant in the region of $E_{\nu} \geq 10^{6}~\GeV$, where $E_\nu$ is the energy of the neutrino (see the discussion in App.~B.3 of~\cite{Garcia:2020jwr}). These additional corrections are evaluated with an independent code developed in~\cite{Gauld:2021zmq}. 

In the following, we present results separately for inclusive cross-sections in charged- and neutral-current DIS as a function of $E_{\nu}$. 
Predictions are provided assuming an isoscalar nuclear target with
nuclear mass numbers of $A = 1, 16$, and $31$.
These choices are representative of a free nucleon, and oxygen nucleon, and the average atomic mass number $\langle A \rangle$ encountered by a neutrino traversing Earth respectively.
Those values of $A$ are hence relevant for describing neutrino-matter interactions within a detector volume composed of ${\rm H}_{2}{\rm O}$ molecules (e.g. IceCube, KM3NeT), as well as neutrinos traveling through the Earth.
Notably, here we focus only on the dominant neutrino-nucleon DIS contribution to the cross-section for which the presented nPDFs are relevant. 
There are additional (in)elastic resonant and coherent scattering contributions~\cite{Seckel:1997kk,Alikhanov:2015kla,Gauld:2019pgt,Zhou:2019vxt,Zhou:2019frk} which must also be included to achieve percent-level accurate predictions (see~\cite{Garcia:2020jwr} for a summary).

\begin{figure}[!t]
    \centering
    \includegraphics[width=0.49\textwidth]{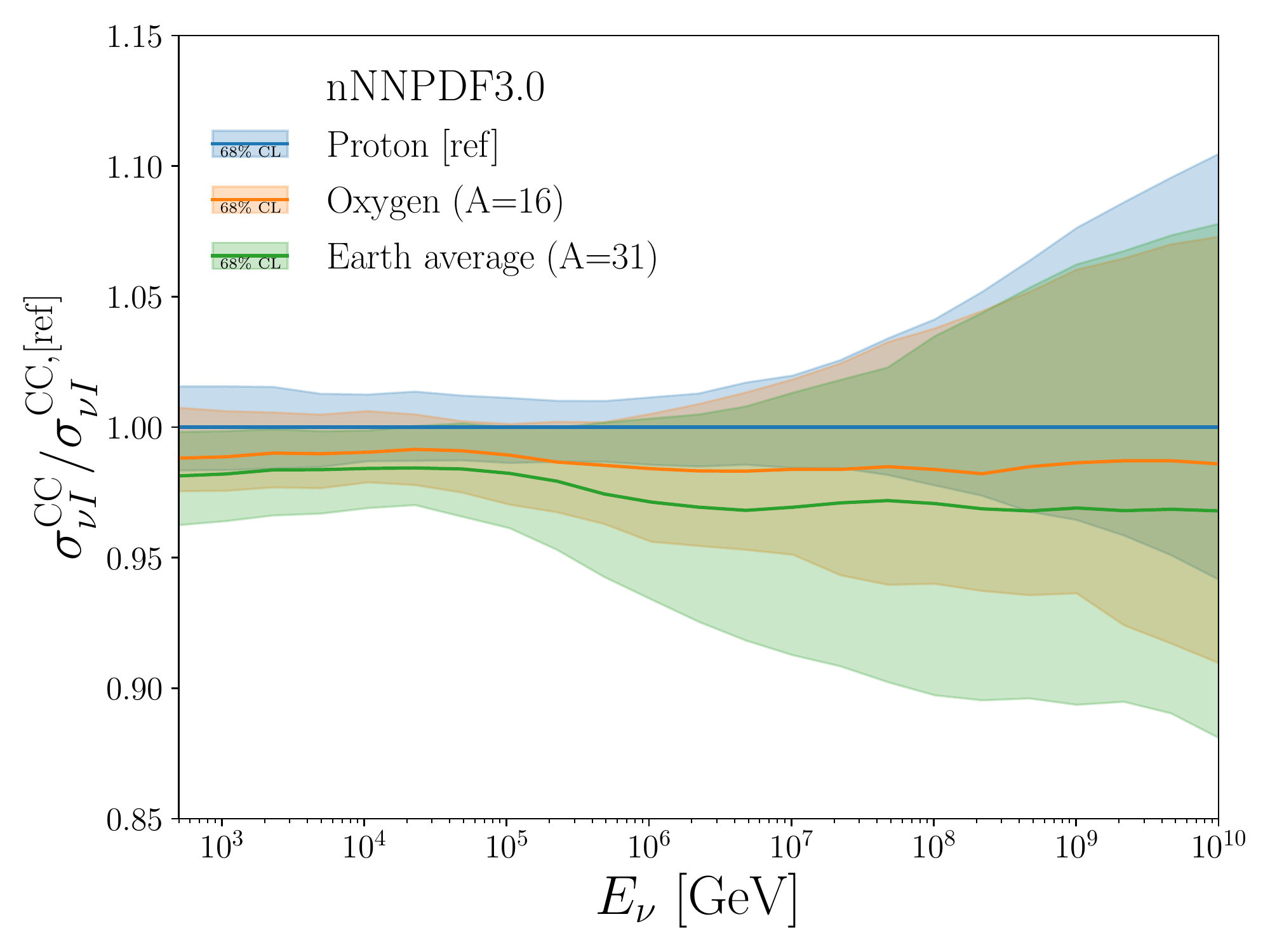}
    \includegraphics[width=0.49\textwidth]{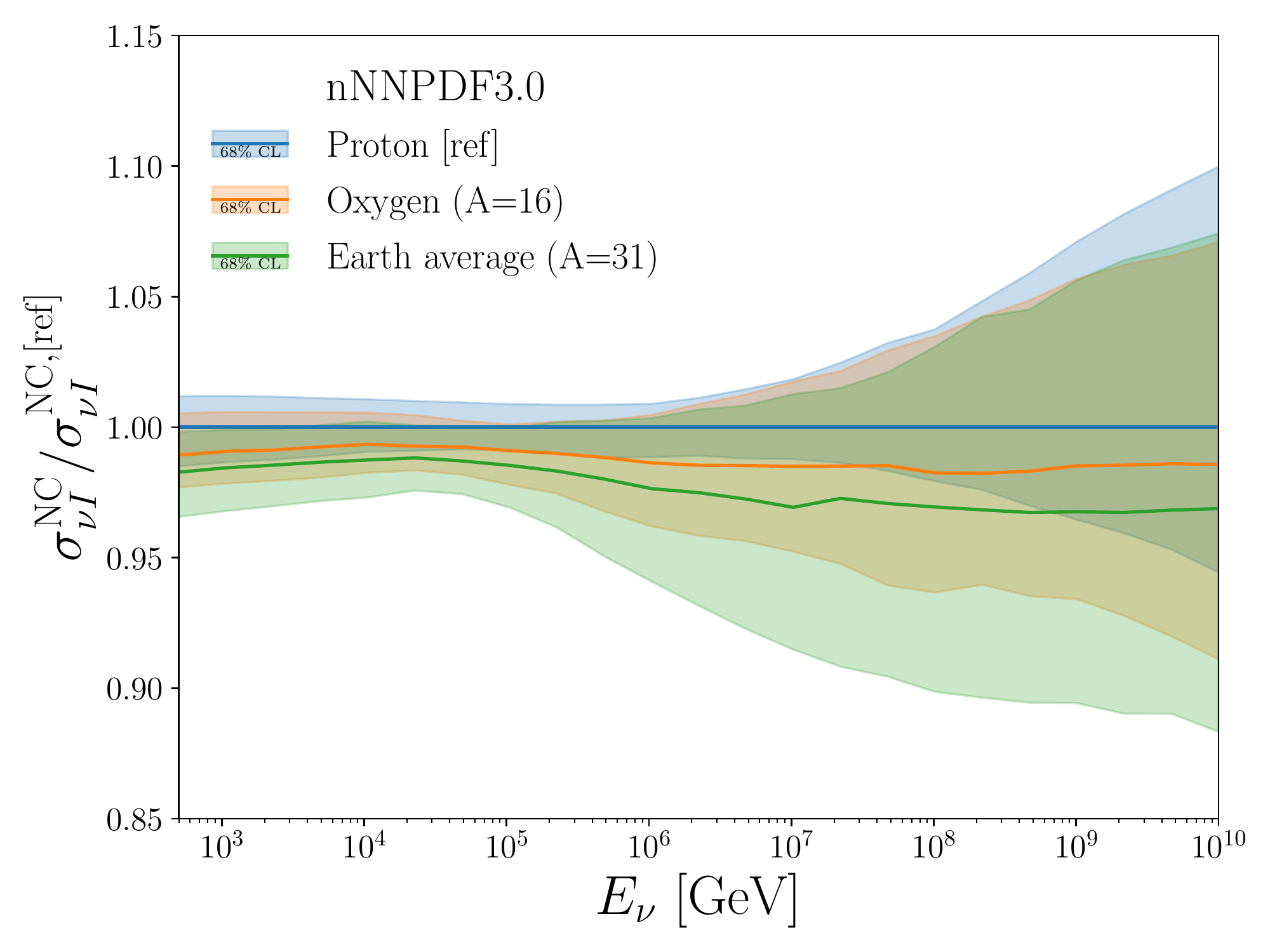}    
    \caption{\small Comparison of the inclusive CC (left) and NC (right) neutrino-nucleon cross-section as a function of incident neutrino energy.
    Predictions are shown for isoscalar nuclear target with $A = 1, 16, 31$,
    where the uncertainty bands represent the nPDF 68\% CL intervals.
    The prediction are normalised with respect to the $A=1$ central value.
    }
    \label{fig:UHE_Avariation}
\end{figure}

In Fig.~\ref{fig:UHE_Avariation} the results for the charged- (left) and neutral-current (right) cross-sections are shown for
predictions based on nNNPDF3.0 for $A = 1, 16, 31$.
The central value and uncertainty band represent the median and 68\% CL interval respectively, and in each case the cross-sections are shown normalised with respect to the $A=1$ central value.
At low $E_{\nu}$ values, the nuclear corrections act to lower the inclusive cross-section by $1~(2)\%$ for $A=16~(31)$, which are corrections that are of the same magnitude as the $A=1$ PDF uncertainties (i.e. those in the absence of nuclear corrections).
In the PeV ($10^{6}$~GeV) energy regime, a kinematic region currently accessible by neutrino telescopes, the nuclear corrections remain negative and amount to $\simeq 2~(3)\%$ for $A=16~(31)$.
Even at the  highest energies, the overall PDF uncertainties remain below $10\%$ and the nuclear-induced suppression
of the central values is at most 4\%.
Note that unlike the previous determinations~\cite{Bertone:2018dse,Gauld:2019pgt,Garcia:2020jwr}, the nuclear dependence of these cross-sections is computed directly without the need to factorise free-nucleon PDF and nPDF effects.

\begin{figure}[!t]
    \centering
    \includegraphics[width=0.49\textwidth]{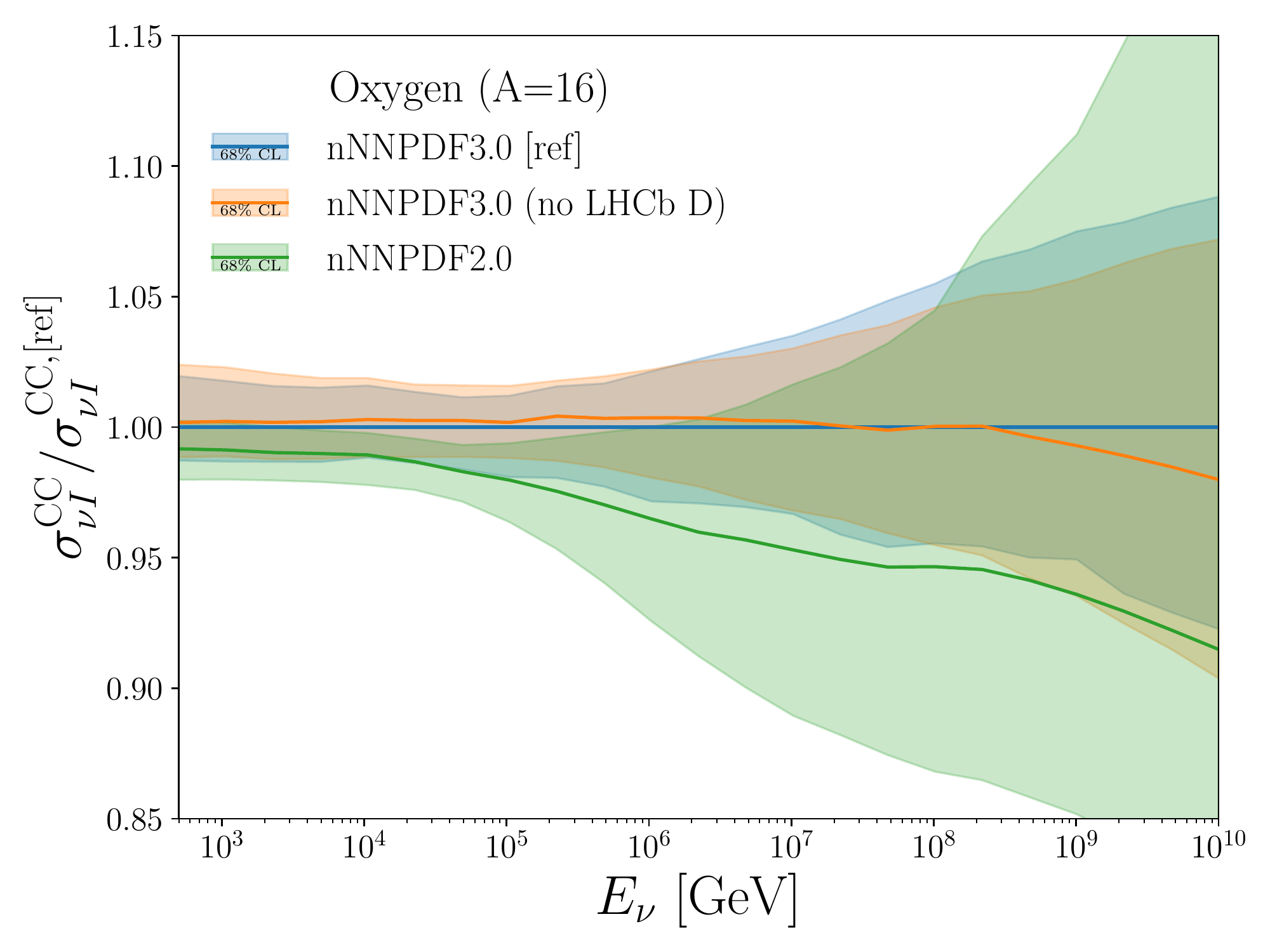}
    \includegraphics[width=0.49\textwidth]{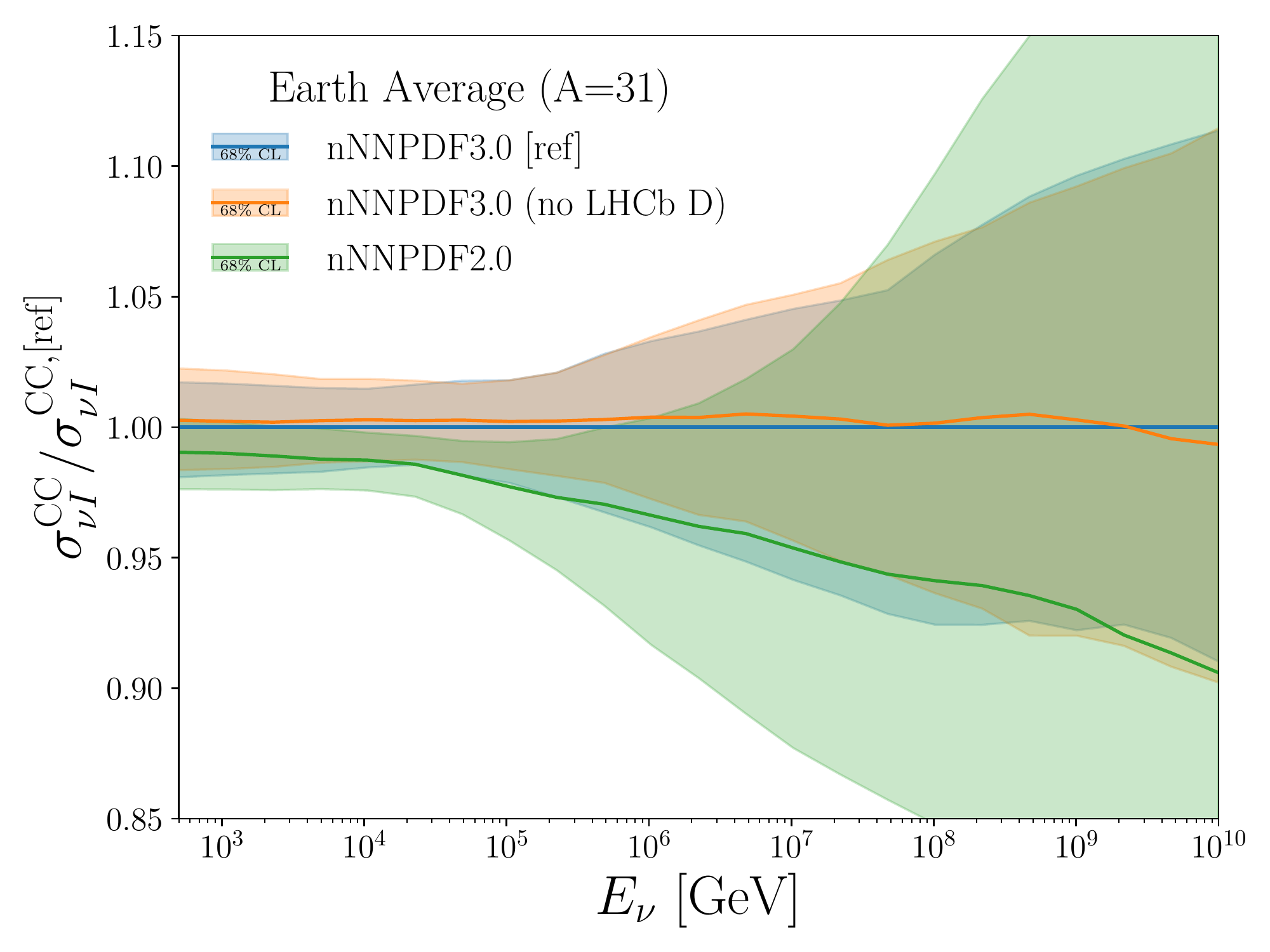}
    \caption{\small Comparison of the inclusive CC neutrino-nucleon cross-section as a function of incident neutrino energy for different nPDF sets.
    Predictions are shown for an isoscalar nuclear target with $A = 16$ (left) and $A = 31$ (right), and the uncertainty bands represent the nPDF 68\% CL intervals.
    Each prediction is shown normalised with respect to the nNNPDF3.0 central value.    
    }
    \label{fig:UHE_PDFvariation}
\end{figure}

In Fig.~\ref{fig:UHE_PDFvariation} we display results for the absolute cross-sections in charged-current DIS predicted for $A=16$ (left) and $A=31$ (right).
Predictions are obtained with nNNPDF3.0, nNNPDF3.0 (no LHCb $D$), nNNPDF2.0, and in each case the central value and uncertainties represent the median and 68\% CL intervals. For comparison, these predictions have been normalised to the central value of the nNNPDF3.0 result.
Overall, these predictions demonstrate the improvement with respect to the previous determination of nPDFs (nNNPDF2.0).
This improvement is a result of the knowledge of the quark and gluon PDFs at small-$x$ values, which is driven by the inclusion of both dijet and the LHCb $D$-meson data.
In practice, the inclusion of LHCb $D$-meson data in nNNPDF3.0 is found to have only a moderate impact on these predictions.
This is because the nuclear corrections for the PDFs of low mass nuclei (i.e. $A = 16, 31$) at small-$x$ values are found to be small, combined with the fact that the boundary condition for the nPDF fit (i.e. the proton baseline) is similar whether the LHCb $D$-meson data is included or not (see Fig.~\ref{fig:freeproton}).

Overall, the results presented here demonstrate the important role played by collider physics data in improving our understanding of scattering processes which are essential to the study of both atmospheric and astrophysical neutrinos.

\section{Delivery, usage, summary and outlook}
\label{sec:summary}

The nNNPDF3.0 analysis presented in this paper has led to the determination of
two nPDF sets, obtained with and without the LHCb $D$-meson production
data respectively. In comparison to the previous nNNPDF2.0
analysis, they both benefit from an extended dataset
and from a more sophisticated fitting methodology. We list 
the nPDF grid files that are made available in the
{\sc\small LHAPDF6} format in  Sect.~\ref{subsection:delivery},
provide prescriptions for their usage in Sect.~\ref{subsection:usage},
and present a summary of the main features of 
nNNPDF3.0 and of possible future developments in
Sect.~\ref{subsection:summary}.

\subsection{Delivery}
\label{subsection:delivery}

The nNNPDF3.0 parton sets are made available as interpolation grids in the
{\sc\small LHAPDF6} format~\cite{Buckley:2014ana} for all
phenomenologically relevant nuclei from
$A=1$ to $A=208$.
Grid files are available both for the nPDFs of bound protons,
$f^{(p/A)}(x,Q^2)$, and for bound average nucleons, $f^{(N/A)}(x,Q^2)$, see
the conventions described in App.~\ref{app:conventions}.
Each of these sets is composed by
$N_{\rm rep}=200$~(250) correlated replicas
for the fits with~(without) the LHCb $D$-meson data.
In addition, for the nNNPDF3.0 variant without the LHCb
$D$-meson data, we also make available high-statistics grids with
$N_{\rm rep}=1000$ replicas.

The nPDF sets that include the LHCb $D$-meson production measurements are named:

\begin{center}
  \renewcommand{\arraystretch}{1.15}
\begin{longtable}{ll}
   $f^{(N/A)}(x,Q^2)\qquad\qquad\qquad  $ &  $ f^{(p/A)}(x,Q^2)\qquad\qquad \qquad$ \\
{\tt nNNPDF30\_nlo\_as\_0118\_p}     &  {\tt nNNPDF30\_nlo\_as\_0118\_p} \\
{\tt nNNPDF30\_nlo\_as\_0118\_A2\_Z1}     &  {\tt nNNPDF30\_nlo\_as\_0118\_p\_D2}\\
{\tt nNNPDF30\_nlo\_as\_0118\_A4\_Z2}    &  {\tt nNNPDF30\_nlo\_as\_0118\_p\_He4}\\
{\tt nNNPDF30\_nlo\_as\_0118\_A6\_Z3}    &  {\tt nNNPDF30\_nlo\_as\_0118\_p\_Li6}\\
{\tt nNNPDF30\_nlo\_as\_0118\_A9\_Z4}    &  {\tt nNNPDF30\_nlo\_as\_0118\_p\_Be9}\\
{\tt nNNPDF30\_nlo\_as\_0118\_A12\_Z6}    &  {\tt nNNPDF30\_nlo\_as\_0118\_p\_C12}\\
{\tt nNNPDF30\_nlo\_as\_0118\_A14\_Z7}    &  {\tt nNNPDF30\_nlo\_as\_0118\_p\_N14}\\
{\color{blue}{\tt nNNPDF30\_nlo\_as\_0118\_A16\_Z8}}    &
{\color{blue}{\tt nNNPDF30\_nlo\_as\_0118\_p\_O16}}\\
{\tt nNNPDF30\_nlo\_as\_0118\_A27\_Z13}   &  {\tt nNNPDF30\_nlo\_as\_0118\_p\_Al27}\\
{\color{blue}{\tt nNNPDF30\_nlo\_as\_0118\_A31\_Z15} }  &
{\color{blue}{\tt nNNPDF30\_nlo\_as\_0118\_p\_A31}}\\
{\tt nNNPDF30\_nlo\_as\_0118\_A40\_Z20}   &  {\tt nNNPDF30\_nlo\_as\_0118\_p\_Ca40}\\
{\tt nNNPDF30\_nlo\_as\_0118\_A56\_Z26}   &  {\tt nNNPDF30\_nlo\_as\_0118\_p\_Fe56}\\
{\tt nNNPDF30\_nlo\_as\_0118\_A64\_Z29}   &  {\tt nNNPDF30\_nlo\_as\_0118\_p\_Cu64}\\
{\tt nNNPDF30\_nlo\_as\_0118\_A108\_Z54}  &  {\tt nNNPDF30\_nlo\_as\_0118\_p\_Ag108}\\
{\tt nNNPDF30\_nlo\_as\_0118\_A119\_Z59}  &  {\tt nNNPDF30\_nlo\_as\_0118\_p\_Sn119}\\
{\tt nNNPDF30\_nlo\_as\_0118\_A131\_Z54}  &  {\tt nNNPDF30\_nlo\_as\_0118\_p\_Xe131}\\
{\color{blue}{\tt nNNPDF30\_nlo\_as\_0118\_A184\_Z74}}  &
{\color{blue}{\tt nNNPDF30\_nlo\_as\_0118\_p\_W184}}\\
{\tt nNNPDF30\_nlo\_as\_0118\_A197\_Z79}  &  {\tt nNNPDF30\_nlo\_as\_0118\_p\_Au197}\\
{\tt nNNPDF30\_nlo\_as\_0118\_A208\_Z82}  &  {\tt nNNPDF30\_nlo\_as\_0118\_p\_Pb208}\\
\end{longtable}
\end{center}
where each of these grid files contains $N_{\rm rep}=200$ replicas, fully
correlated among the different values of $A$.
The names indicated in blue correspond to nuclear species
for which direct experimental constraints are not available, and which are
made available (via the continuous $A$ parameterisation of the nPDFs) because of their phenomenological relevance.

The nPDF sets that do not include the LHCb $D$-meson production measurements
are named:
\begin{center}
  \renewcommand{\arraystretch}{1.15}
  \begin{longtable}{ll}
   $f^{(N/A)}(x,Q^2)\qquad\qquad\qquad\qquad \qquad\qquad\qquad\quad  $ &  $ f^{(p/A)}(x,Q^2)\qquad\qquad \qquad$ \\
{\tt nNNPDF30\_nlo\_as\_0118\_noLHCbD\_p}     &  {\tt nNNPDF30\_nlo\_as\_0118\_noLHCbD\_p} \\
{\tt nNNPDF30\_nlo\_as\_0118\_noLHCbD\_A2\_Z1}     &  {\tt nNNPDF30\_nlo\_as\_0118\_noLHCbD\_p\_D2}\\
{\tt nNNPDF30\_nlo\_as\_0118\_noLHCbD\_A4\_Z2}    &  {\tt nNNPDF30\_nlo\_as\_0118\_noLHCbD\_p\_He4}\\
{\tt nNNPDF30\_nlo\_as\_0118\_noLHCbD\_A6\_Z3}    &  {\tt nNNPDF30\_nlo\_as\_0118\_noLHCbD\_p\_Li6}\\
{\tt nNNPDF30\_nlo\_as\_0118\_noLHCbD\_A9\_Z4}    &  {\tt nNNPDF30\_nlo\_as\_0118\_noLHCbD\_p\_Be9}\\
{\tt nNNPDF30\_nlo\_as\_0118\_noLHCbD\_A12\_Z6}    &  {\tt nNNPDF30\_nlo\_as\_0118\_noLHCbD\_p\_C12}\\
{\tt nNNPDF30\_nlo\_as\_0118\_noLHCbD\_A14\_Z7}    &  {\tt nNNPDF30\_nlo\_as\_0118\_noLHCbD\_p\_N14}\\
{\color{blue}{\tt nNNPDF30\_nlo\_as\_0118\_noLHCbD\_A16\_Z8}}    &
{\color{blue}{\tt nNNPDF30\_nlo\_as\_0118\_noLHCbD\_p\_O16}}\\
{\tt nNNPDF30\_nlo\_as\_0118\_noLHCbD\_A27\_Z13}   &  {\tt nNNPDF30\_nlo\_as\_0118\_noLHCbD\_p\_Al27}\\
{\color{blue}{\tt nNNPDF30\_nlo\_as\_0118\_noLHCbD\_A31\_Z15} }  &
{\color{blue}{\tt nNNPDF30\_nlo\_as\_0118\_noLHCbD\_p\_A31}}\\
{\tt nNNPDF30\_nlo\_as\_0118\_noLHCbD\_A40\_Z20}   &  {\tt nNNPDF30\_nlo\_as\_0118\_noLHCbD\_p\_Ca40}\\
{\tt nNNPDF30\_nlo\_as\_0118\_noLHCbD\_A56\_Z26}   &  {\tt nNNPDF30\_nlo\_as\_0118\_noLHCbD\_p\_Fe56}\\
{\tt nNNPDF30\_nlo\_as\_0118\_noLHCbD\_A64\_Z29}   &  {\tt nNNPDF30\_nlo\_as\_0118\_noLHCbD\_p\_Cu64}\\
{\tt nNNPDF30\_nlo\_as\_0118\_noLHCbD\_A108\_Z54}  &  {\tt nNNPDF30\_nlo\_as\_0118\_noLHCbD\_p\_Ag108}\\
{\tt nNNPDF30\_nlo\_as\_0118\_noLHCbD\_A119\_Z59}  &  {\tt nNNPDF30\_nlo\_as\_0118\_noLHCbD\_p\_Sn119}\\
{\tt nNNPDF30\_nlo\_as\_0118\_noLHCbD\_A131\_Z54}  &  {\tt nNNPDF30\_nlo\_as\_0118\_noLHCbD\_p\_Xe131}\\
{\color{blue}{\tt nNNPDF30\_nlo\_as\_0118\_noLHCbD\_A184\_Z74}}  &
{\color{blue}{\tt nNNPDF30\_nlo\_as\_0118\_noLHCbD\_p\_W184}}\\
{\tt nNNPDF30\_nlo\_as\_0118\_noLHCbD\_A197\_Z79}  &  {\tt nNNPDF30\_nlo\_as\_0118\_noLHCbD\_p\_Au197}\\
{\tt nNNPDF30\_nlo\_as\_0118\_noLHCbD\_A208\_Z82}  &  {\tt nNNPDF30\_nlo\_as\_0118\_noLHCbD\_p\_Pb208}\\
  \end{longtable}
  \end{center}
with $N_{\rm rep}=250$ replicas each. The corresponding sets with
$N_{\rm rep}=1000$ replicas have the same names with an additional suffix
{\tt \_1000}. The nPDF sets indicated in blue are as above.

The nPDF sets are also available on the NNPDF Collaboration website:
\begin{center}
\url{http://nnpdf.mi.infn.it/for-users/nnnpdf3-0/}.
\end{center}

\subsection{Usage}
\label{subsection:usage}

As discussed in previous nNNPDF studies~\cite{AbdulKhalek:2019mzd,AbdulKhalek:2020yuc}, the recommended usage
of the nNNPDF3.0 sets is given by the following prescription.
Consider a general nPDF-depending quantity, indicated
schematically by
\be
\mathcal{F}\lc f_{i_1}^{(p/A_1)}(x_1,Q_1),f_{i_2}^{(p/A_2)}(x_2,Q_2), \ldots\rc \, .
\ee
This quantity could represent e.g. a nPDF, $\mathcal{F}= f_g^{(p/A)}(x,Q)$, a nuclear
ratio, $\mathcal{F}= f_g^{(p/A)}(x,Q)/f_g^{(p)}(x,Q)$, a proton-lead LHC 
cross-section, or a UHE cross-section.
To evaluate the most-likely value and uncertainty for $\mathcal{F}$ based
on nNNPDF3.0, first one evaluates this quantity for the $N_{\rm rep}$ replicas composing this set:
\be
\label{eq:cl_calculation}
\mathcal{F}^{(k=1)},\mathcal{F}^{(k=2)},\mathcal{F}^{(k=3)},\ldots,\mathcal{F}^{(k=N_{\rm rep})} \, ,
\ee
recalling that within a given set, different values of $A$ are fully correlated 
(including $A=1$, the free-proton baseline).
Next, order the elements of Eq.~(\ref{eq:cl_calculation}) in ascending order
and remove symmetrically $(100-X)\%$ of the replicas
with the highest and lowest values.
The resulting interval defines the $X\%$ confidence level
for this quantity, given the nPDF set used in the calculation.
For instance, a 68\% CL interval (corresponding to a 1-$\sigma$ 
interval for a Gaussian distribution) is obtained by 
keeping the central 68\% replicas by removing the 
lowest 16\% and highest 16\% of the (ordered) replicas.
The best-fit value for the quantity $\mathcal{F}$ is taken 
to be the median evaluated over all of the replicas.

The rationale for estimating the nPDF 
uncertainties as CL intervals, as opposed to the variance, 
is that nNNPDF3.0 probability distributions are not always well 
described by a Gaussian approximation.
We also note that implementations of this procedure are available 
in most numerical libraries. For example in {\tt NumPy} if {\tt  F} 
corresponds to the (unordered) array containing the $N_{\rm rep}$ replicas 
of $\mathcal{F}$, one can compute the lower and upper limits 
of the 68\% CL interval with {\tt high = np.nanpercentile(F,84)} and {\tt low = np.nanpercentile(F,16)}.

\subsection{Summary and outlook}
\label{subsection:summary}

The nNNPDF3.0 analysis presented in this work is based on an extensive
set of measurements using nuclear probes. These include in
particular pPb LHC data for dijet, isolated photon and $D^0$-meson production.
A key aspect of nNNPDF3.0 is the coherent treatment of the experimental input
in both the proton, deuteron, and nuclear PDFs. This ensures its consistent
theoretical and methodological treatment throughout the fitting procedure.

Overall we find an excellent compatibility between the constraints provided by
the data already in nNNPDF2.0 and the new data in nNNPDF3.0. The new data
significantly improves the precision with which nuclear modification factors
are determined.
In particular, we have established strong evidence of deviations from the
free-proton baseline for small-$x$ shadowing in lead nuclei both in the quark
and in the gluon sector, as well as for gluon anti-shadowing at large-$x$ 
also in the case of lead.
Furthermore, we have studied the dependence of the nuclear
modification factors on the atomic number $A$, assessed the robustness of
nNNPDF3.0 with respect to variations in the input dataset and the fitting
methodology, and outlined the impact on nPDFs of specific processes,
in particular of dijet and $D$-meson production cross-sections.

As a representative phenomenological application of nNNPDF3.0,
we have presented updated predictions for the ultra-high-energy
neutrino-nucleus scattering cross-sections for different nuclear targets.
We have considered $A=16$, relevant for neutrino scattering
off water and ice targets, and $A=31$, required for the calculations of
neutrino flux attenuation due to their interactions with matter within Earth
before being detected. In both cases the significance of nuclear modification
effects, in particular small-$x$ shadowing, are enhanced in comparison to
nNNPDF2.0, and exhibit substantially reduced theory uncertainties.

The results presented in this work could be expanded in several theoretical,
experimental, and methodological aspects.
Concerning theory, the accuracy of
the nPDF determination could be improved by including NNLO QCD corrections in
the solution of the DGLAP equations and in the hard-scattering cross-sections.
One could also account for missing higher-order corrections as
additional correlated uncertainties, {\it e.g.} by means of the method developed
in~\cite{NNPDF:2019ubu,NNPDF:2019vjt}. Even if the fit quality of nNNPDF3.0
is overall very good, the precision of forthcoming LHC measurements is
expected to increase with the higher luminosity achieved in Run III.
Higher-order corrections may therefore become essential to describe future data,
especially for absolute spectra in processes such as dijet and isolated
photon production for which the NLO calculations are currently unsatisfactory.
The inclusion of higher-order corrections (and theory uncertainties) is likely relevant for the
description of $D$-meson data, which has a large impact on nPDFs in the region of small-$x$.
This was already indicated by the study~\cite{Eskola:2019bgf}, which showed that the treatment 
of $\mathcal{O}(\alpha_s^4)$ terms in $D$-meson production has a small but 
noticeable impact on the extracted value of the gluon PDF.

Concerning experimental data, additional analyses of LHC Run
II pPb collision measurements are expected to become available, as are new
analyses of LHC Run III. In particular, the upcoming pO and OO
LHC runs~\cite{Brewer:2021kiv} will constrain the nuclear modification factors
for a much lower value of $A$ than pPb collisions. One such example is
dijet production~\cite{Paakkinen:2021jjp}. In addition, nuclear
structure function measurements in fixed-target DIS at
JLab~\cite{Paukkunen:2020rnb} may be used to improve the determination of
nuclear modifications at very large values of $x$. In the longer term,
nPDFs will be probed at the EIC~\cite{AbdulKhalek:2021gbh,Anderle:2021wcy,
  Khalek:2021ulf}, by means of GeV-scale lepton scattering on light and heavy
nuclei, and at the FPF~\cite{Anchordoqui:2021ghd}, by means of TeV-scale
neutrino scattering on heavy nuclear targets.

Finally, concerning methodology, one may consider integrating more coherently
the free-proton PDF boundary condition with the $A$-dependent nPDFs.
Given that, in general, proton-nucleon collision constrain both proton and
nuclear PDFs, and that both will be probed with similar precision,
this separation appears to be artificial and  undesirable. A more sophisticated
approach
should aim to determining proton, deuteron, and nuclear PDFs simultaneously
within a single QCD analysis. Such an approach will bypass the need to carry
out proton and nuclear fits separately, and to use one as input to the other.
Such a program, which is being developed, e.g. for the simultaneous
determination of (polarised) PDFs and fragmentation
functions~\cite{Ethier:2017zbq,Sato:2019yez,Moffat:2021dji}, represents
a major milestone to fully exploit a much wider range of future measurements.

\subsection*{Acknowledgments}
We are grateful to Jake~Ethier and Gijs van Weelden for previous contributions relevant to this work.
We thank Kari Eskola, Petja Paakkinen, and Hannu Paukkunen for discussions
and information about the EPPS16 analysis and for providing benchmark
numbers for their dijet and $D$ meson production calculations.
We thank Pit Duwentaster and Fred Olness for information
concerning the nCTEQ15 analyses.
R.~A.~K. and J.~R. are (partially) supported by the Dutch Research Council (NWO).
T.~G. is supported by NWO  via an ENW-KLEIN-2 project.
E.~R.~N. is supported by the U.K. Science and
Technology Facility Council (STFC) grant ST/P000630/1.
The research of T.~R. has been partially supported by an ASDI grant
of The Netherlands eScience Center.

\appendix
\section{Notation and conventions}
\label{app:conventions}

Throughout this work we adopt the following conventions.

\begin{itemize} 
\item The laboratory frame refers to the reference frame of
  the asymmetric pPb collisions as they take place at LHC, see also the discussion in
  App.~\ref{app:asymmetry}.
\item The nucleon--nucleon (NN) or CoM frame refers to
  the reference frame where the two colliding nucleons (one from the proton beam
  and another from the lead nucleus) have a vanishing total three-momentum, $\vec{p}_{\rm tot}
  = \vec{p}_{\rm p} + \vec{p}_{\rm N/Pb} =0$, where $\vec{p}_{\rm N/Pb}$ indicates the average
  linear momentum carried by a nucleon N (proton or neutron) within the lead nucleus.
  \item pPb (Pbp) collisions indicate collisions where the proton beam circulates in the
  positive (negative) $z$-direction in the laboratory frame, corresponding to forward
  (backward) rapidity regions in this reference frame.
  Hence in pPb (Pbp) collisions the region $z >0$ ($z<0$) corresponds to the proton--going
  direction, and $z <0$ ($z>0$) to the lead--going direction instead. 
\item The nPDFs $f^{(A)}(x,Q)$ of a nucleus with atomic charge $Z$ and
  atomic mass number $A$,
  following the convention used in  nNNPDF2.0, are defined as:
    \begin{equation}
      \label{app:nPDF_def}
    f^{(A)}(x,Q) = A f^{(N/A)}(x,Q) = \left( Z f^{(p/A)}(x,Q) + (A-Z)f^{(n/A)}(x,Q) \right) \, ,
    \end{equation}
    such that the nPDF $f^{(N/A)}(x,Q)$ corresponding to
    the average bound nucleon $N$ within the nucleus $A$ is given by
     \begin{equation}
      \label{app:nPDF_def_average}
    f^{(N/A)}(x,Q) = \frac{ Z}{A} f^{(p/A)}(x,Q) + \frac{(A-Z)}{A}f^{(n/A)}(x,Q) \, ,
     \end{equation}
     where $f$ is a quark or the gluon. For $(Z,A)=(1,1)$ one reproduces
     the free-proton PDFs.
     $f^{(p/A)}$ and $f^{(n/A)}(x,Q)$ denote the nPDFs of bound protons and
     neutrons, respectively, within a nucleus with atomic mass number $A$,
     and are related to each other via isospin symmetry e.g. $u^{(p/A)}=d^{(n/A)}$.

   \item Genuine nuclear effects on the nPDFs correspond to the case where
     these differ from their free-nucleon counterparts once isospin effects are accounted
     for, that is,
     \be
     R_f^{(A)}(x,Q) \equiv \frac{ f^{(N/A)}(x,Q)}{\frac{ Z}{A} f^{(p)}(x,Q) + \frac{(A-Z)}{A}f^{(n)}(x,Q)} \ne 1 \, ,
     \ee
     where  $f^{(p)}$ and $f^{(n)}$ are the free proton and neutron PDFs, again
     related to each other by isospin symmetry.
     
\end{itemize}

\section{Reference frames in asymmetric pPb collisions }
\label{app:asymmetry}

An important difference between pp and pPb collisions at the
LHC is that the latter are asymmetric because the energy per nucleon
of the lead projectile is smaller as compared to the proton one.
This implies that the final state of the collision will be boosted
along the proton--going direction, and hence that the laboratory
frame does not coincide with the CoM frame in these
collisions.
In this appendix we review the treatment of asymmetric pPb collisions
and the role played by the transformation between the
laboratory and the center-of-mass frame in the global nPDF analysis.

\paragraph{Boost between laboratory and CoM frames.}
When accelerated in opposite directions at the LHC, the proton and lead
beams are required
to have an equal magnetic rigidity\footnote{Defined as  $p/Z$, 
the particle's linear momentum $p$ normalised by its total electric charge $Z$.}~\cite{Albacete:2013ei}.
This implies that the energy of lead in the laboratory frame, $E^{\rm lab}_{\rm Pb}$,
must be related to that of the colliding protons $E^{\rm lab}_{\rm p}$
by  $E^{\rm lab}_{\rm Pb} = Z E^{\rm lab}_{\rm p}$, where the atomic number for lead is $Z=82$.
In the laboratory frame we can express
the four-momenta of the proton  $p_{\rm p}^{\mu}$ and of the average nucleon $N$ in lead,
denoted as $p_{\rm N/Pb}^{\mu}$, by
\bea
p_{\rm p}^{\mu} &=& \lp  E_p, 0, 0, E_p\rp \nonumber \, , \\
p_{\rm N/Pb}^{\mu} &=& \lp \frac{Z}{A} E_p, 0, 0, -\frac{Z}{A} E_p\rp \, , 
\eea
where the atomic mass number for lead is  $A=208$, we neglect nucleon mass effects, 
the beams collide along the $z$-direction, and the positive direction coincides with the proton direction of motion.
The total four-momentum and CoM energy of the nucleon--nucleon collision is given by
\be
\label{eq:app_ptot}
p_{\rm tot}^{\mu} = p_{\rm p}^{\mu} + p_{\rm N/Pb}^{\mu} = \lp E_p+\frac{Z}{A} E_p, 0, 0, E_p-\frac{Z}{A} E_p\rp \, , 
\ee
\be
\label{eq:app_sNN}
\sqrt{s_{\rm NN}} = \sqrt{ p_{\rm tot}^2} = 2\sqrt{\frac{Z}{A}}E_p \simeq 1.2558E_p \, .
\ee
Since $\sqrt{s_{\rm NN}}$ is a Lorentz invariant, its value is the same in
any reference frame and hence Eq.~(\ref{eq:app_sNN}) holds both in the laboratory
and in the (NN) CoM frame. 

Eq.~(\ref{eq:app_ptot}), together with momentum conservation, implies that in the laboratory frame the
final state of the pPb collision will move with a non-zero momentum $p_z>0$
along the proton direction.
Therefore, the value of $p_z$ of the measured final state particles (and
of their rapidity) will be different
from that expected in the CoM frame, where $p_{\rm z,tot}=0$ vanishes
by construction.
Accounting for this asymmetry is necessary for the interpretation
of hard-scattering cross-sections in pPb collisions.
Indeed, the rapidity $y$ of a particle with energy $E$ and linear momentum in the beam direction $p_z$
is defined as
\begin{equation} \label{eq:rapidity_def}
  y =  \frac{1}{2}\ln\left({\frac{E+p_{z}}{E-p_{z}}}\right) \, ,
\end{equation}
and depends on the specific reference frame (though rapidity differences  do not).
For a collision taking place in the CoM frame,
one has that the final state is characterised by $p_z^{\rm tot}=0$ and hence $y=0$.
However, in the laboratory frame instead, using the four-vector in Eq.~(\ref{eq:app_ptot})
one finds that the rapidity of the collision final-state in the laboratory frame is
\be
y^{\rm lab} = \frac{1}{2}\ln \frac{A}{Z} \simeq 0.46541 \, ,
\ee
reflecting how after the collision the system keeps moving in the proton--going direction.
Hence, a rapidity boost arises, $\Delta y \equiv y^{\rm lab} -y = 0.46541 $, between particle
kinematics as measured in the laboratory frame and those predicted in the
CoM frame that needs to be accounted for:
\bea
\label{eq:rapdist}
y^{\rm lab} &=& y + \Delta y~({\rm from~CoM~frame~to~lab~frame}) \, , \\
y &=& y^{\rm lab} - \Delta y~({\rm from~lab~frame~to~CoM~frame}) \, . \nonumber
\eea
This rapidity shift $\Delta y $ between the two reference frames is
invariant under longitudinal Lorentz boosts and therefore it is not affected by the partonic kinematics
as discussed below.

\paragraph{Partonic kinematics.}
The general factorised expression for hard-scattering cross-sections
in pPb collisions can be written as
\begin{equation} \label{eq:DY_xsec}
  \frac{d\sigma_{{\rm pPb}}}{dQ^2dy} =  \sum_{a,b} \int_{x_1}^1 d\xi_{1} \int_{x_2}^1 d\xi_{2}~~f_a^{(\rm{p})}(\xi_1,Q^2)f_b^{(\rm{Pb})}(\xi_2,Q^2)\frac{d\hat{\sigma}_{ab}}{dQ^2d\hat{y}}\left(\frac{x_1}{\xi_1},\frac{x_2}{\xi_2},Q^2\right) \, ,
\end{equation}
where $y$~($\hat{y}$) denotes the rapidity of the hadronic (partonic) final state, $Q^2$ is the hard
scale that makes the process perturbative,
$f_a^{(\rm{p})}$ and $f_b^{(\rm{Pb})}$ indicate the PDFs of the proton and of the lead nucleus
respectively (see App.~\ref{app:conventions} for the adopted conventions), and $a,b$
are partonic indices.
We have omitted the dependence on
the factorisation and renormalisation scales for simplicity.
Neglecting hadron and parton mass effects, and assuming the CoM
reference frame,
one can relate the incoming partonic $\hat{p}_i^{\mu}$ and hadronic
$p_i^\mu$ four-momenta as
\begin{equation}
\hat{p}_i^{\mu} = x_ip_i^\mu = x_i E(1,0,0,\pm 1),\qquad i={\rm p, N/Pb} \, , \label{eq:partonic_momentum}
\end{equation}
where $x_i$ is the momentum fraction carried by the colliding
parton from nucleon $i$, $p_i^z = +x_iE$~($-x_iE$) for $i=p$ ($i={\rm N/Pb}$), and the nucleon energy $E$
does not depend on $i$ in this reference frame.
One can then evaluate the hadronic and partonic collision energies in this reference frame as usual:
\begin{align}
    \sqrt{s} &= \sqrt{(p_1+p_2)^2}  = 2E  \, ,\\
    \sqrt{\hat{s}} &= \sqrt{(\hat{p}_1+\hat{p}_2)^2} =
    \sqrt{x_1x_2}\sqrt{s} \, .
    \label{eq:partonic_CoM_energy}
\end{align}
Even for collisions in this CoM frame,
the final state particles
will exhibit a non-trivial rapidity distribution related to the distribution
in momentum fractions $x_1$ and $x_2$ dictated by the PDFs.
Indeed, the rapidity of the produced final state
assuming colliding partons
carrying momentum fractions $x_1$ and $x_2$ is
\begin{equation}  \label{eq:CoM_rap}
  y = 
  \frac{1}{2}\ln\left({\frac{\hat{E}+\hat{p}_{z}}{\hat{E}-\hat{p}_{z}}}\right)=
  \frac{1}{2}\ln\left({\frac{E(x_1+x_2)+E(x_1-x_2)}{E(x_1+x_2)-E(x_1-x_2)}}\right)
  = \frac{1}{2}\ln\left({\frac{x_1}{x_2}}\right) \, .
\end{equation}
Note that the distribution in rapidity $y$ hence follows closely
the distribution in $x_1$ and $x_2$ dictated by the PDFs
through Eq.~(\ref{eq:DY_xsec}).
This distribution will be in general asymmetric due to the
difference between the proton and nuclear PDFs.
Even in the absence of
genuine nuclear effects, nuclear PDFs are different from the proton
ones due to isospin effects.

For a $2\to1$ type process such as Drell-Yan, Eq.~\eqref{eq:CoM_rap} can be re-arranged, assuming leading-order kinematics, to write the momentum fractions $x_1$ and $x_2$ in terms of the rapidity of the reconstructed gauge-boson as 
\begin{equation}
x_{1(2)} = \sqrt{\frac{Q^2}{s}}e^{(-)y}\,,
\end{equation}
with $Q^2 = m_{V} = \sqrt{\hat{s}}$.
For $2\to2$ type processes a similar relation in terms of out-going particle kinematics may be written
\begin{equation} \label{eq:x12_def2to2}
x_{1(2)} = \frac{1}{\sqrt{s}} \left( E_{T,3} e^{(-)y_3} + E_{T,4} e^{(-)y_4} \right) \,,
\end{equation}
where the labels `3' and `4' refer to the outgoing particles.
For the proton--nucleus scattering processes considered  in this work, assuming leading-order kinematics, we have
\begin{equation}
E_{T,3} = E_{T,4} = \begin{cases}
    p_{T}^{j} \quad &\text{Dijet production} \\
    E_T^{\gamma}\quad &\text{Photon production} \\
    \sqrt{(p_{T}^{c})^2 +m_c^2}\quad &c\text{-quark production}
    \end{cases}
\end{equation}
This information, in combination with the approximation $y_3 \approx y_4$, is used to
estimate the kinematic coverage
of the available experiment data which are displayed in Fig.~\ref{fig:kinplot}.

\paragraph{Implications for pPb collisions.}
The relations in  Eq.~(\ref{eq:rapdist})
make it possible to relate cross-sections measured
in the laboratory frame with those predicted in the
CoM frame and vice-versa by means of a rapidity shift
\bea
\label{app:rap_matching}
\frac{d\sigma(Q^2,y)}{dQ^2dy}\Bigg|_{\rm lab} &=&
\frac{d\sigma(Q^2,y-\Delta y)}{dQ^2dy}\Bigg|_{\rm NN} \, ,
\\  \frac{d\sigma(Q^2,y)}{dQ^2dy}\Bigg|_{\rm NN} &=&
\frac{d\sigma(Q^2,y+\Delta y)}{dQ^2dy}\Bigg|_{\rm lab} \, .
\eea
These relations are particularly useful when evaluating theoretical
cross-sections for hard-probes in pPb collisions.
The reason is that most available codes assume
symmetric NN collisions, but the data
is in many cases presented in the laboratory frame.
As indicated by Eq.~(\ref{app:rap_matching}),
it suffices to shift either the rapidity bins or
equivalently the cross-section histograms 
by $\Delta y = 0.4651$ before or after the computation
respectively.
We also note that switching the (p) and (Pb) superscripts in Eq.~(\ref{eq:DY_xsec}) is 
equivalent to switching the minus and plus signs in Eq.~(\ref{eq:partonic_momentum}), i.e. switching the proton and lead beams.
    
It is also worth mentioning that given pPb (Pbp) collisions, the data measured in the forward (backward) rapidity regions constrain the large-$x$ region of the proton PDFs and the small-$x$ region of the lead PDFs, see Eq.~(\ref{eq:x12_def2to2}), and vice-versa in Pbp (pPb) collisions.
The ATLAS and CMS detectors cover the central rapidity region and hence provide access to both forward and backward final states.
The asymmetric configuration of the ALICE and LHCb detectors imply that final states are only detected in the forward rapidity region~\cite{HADJIDAKIS20211} and hence recording data from both pPb and Pbp collisions is instrumental to maximise the coverage in Bjorken-$x$ for the lead PDF.

\section{Comparison between nNNPDF3.0 and experimental data}
\label{app:datacomp}

In this appendix we present representative
comparisons between the NLO QCD theoretical predictions
obtained from nNNPDF3.0 and the corresponding
experimental measurements.
In particular, we focus on the LHC pPb datasets,
with the exception of the LHCb $D^0$-meson measurements that have
been already discussed in Sects.~\ref{sec:results}
and~\ref{sec:stability}.
We describe first of all the data versus theory comparisons for those datasets
which are part of nNNPDF3.0, and then for completeness for those
excluded from the nNNPDF3.0 baseline
for the reasons discussed in Sect.~\ref{sec:expdata}.

\begin{figure}[!t]
  \centering
\includegraphics[width=0.8\textwidth]{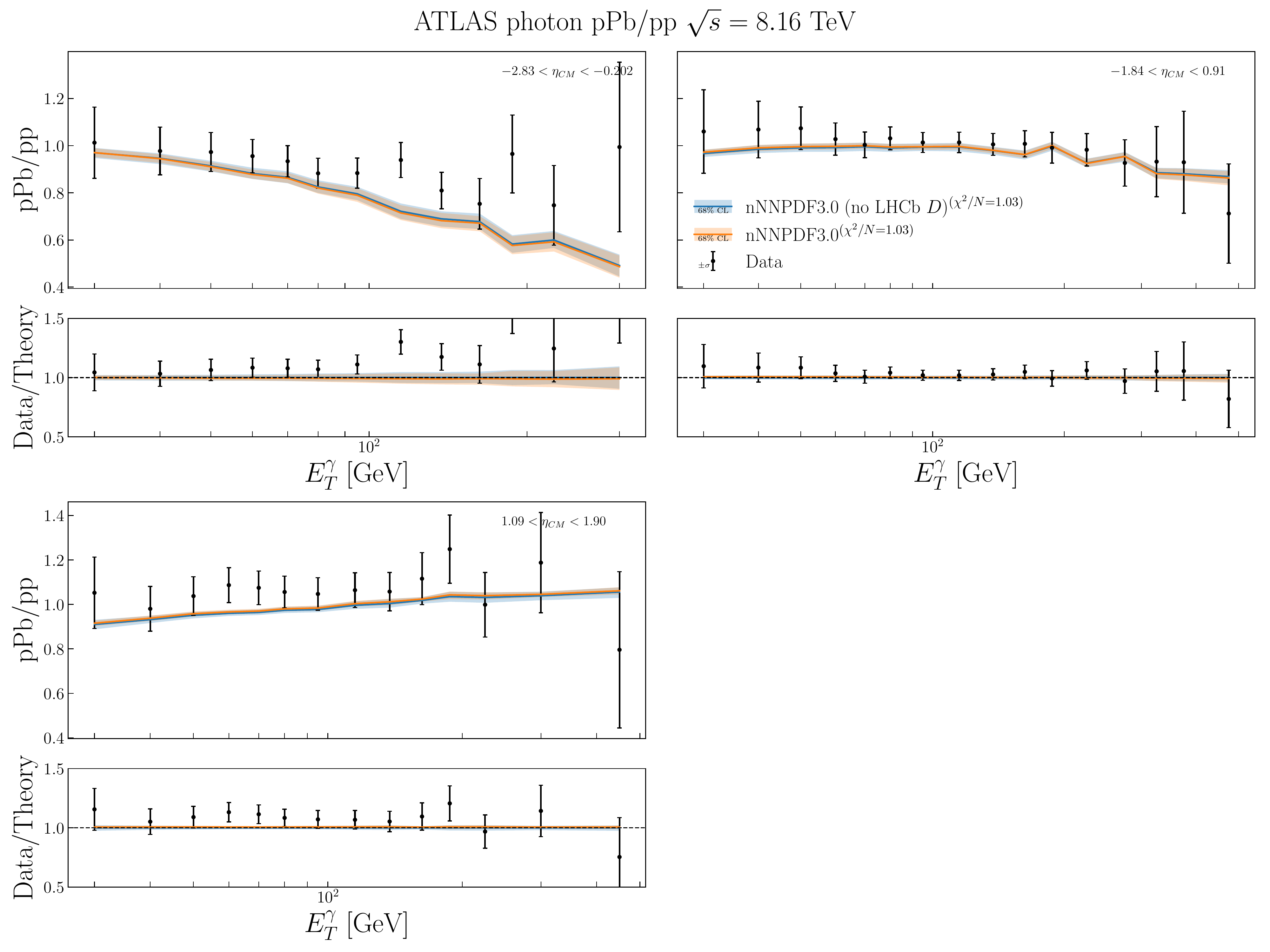}
\caption{\label{fig:data_vs_theory_ATLASphotons} Comparison
  between the theoretical predictions based on both nNNPDF3.0
  and the variant without the LHCb $D$-meson cross-sections
  and the ATLAS measurements of the ratio of isolated photon
  production spectra between pPb and pp collisions at $\sqrt{s}=8.16$ TeV.
  Results are presented differential in the photon transverse energy
  $E_T^\gamma$ for three photon pseudo-rapidity $\eta_{CM}$ bins
  in the CoM frame.
  The band in the theory prediction indicates the PDF uncertainty,
  and the bottom panels display the ratio to the central theory.
  No scale uncertainties are considered in this comparison.
  The value of the $\chi^2$ to this dataset for both nNNPDF3.0 is also indicated
  in the legend.
}
\end{figure}

\paragraph{Comparison with datasets included in nNNPDF3.0.}
To begin with, Fig.~\ref{fig:data_vs_theory_ATLASphotons}
displays a comparison
  between the theoretical predictions based on both nNNPDF3.0
  and the variant without the LHCb $D$-meson cross-sections
  and the ATLAS measurements of the ratio of isolated photon
  production spectra between pPb and pp collisions at $\sqrt{s}=8.16$ TeV.
  Results are presented differential in the photon transverse energy
  $E_T^\gamma$ for three photon pseudo-rapidity $\eta_{CM}$ bins
  in the CoM frame.
  The band in the theory prediction indicates the PDF uncertainty,
  and the bottom panels display the ratio to the central theory.
  No uncertainty due to MHOs is considered in this comparison.
  The value of the $\chi^2$ to this dataset for both nNNPDF3.0 is also indicated
  in the legend.
  We note that the same format will be used for the rest of the plots in this appendix.
  This comparison confirms that
  a good description of this dataset is achieved by the NLO QCD calculation
in the three rapidity bins available.

\begin{figure}[!t]
  \centering
  \includegraphics[width=0.49\textwidth]{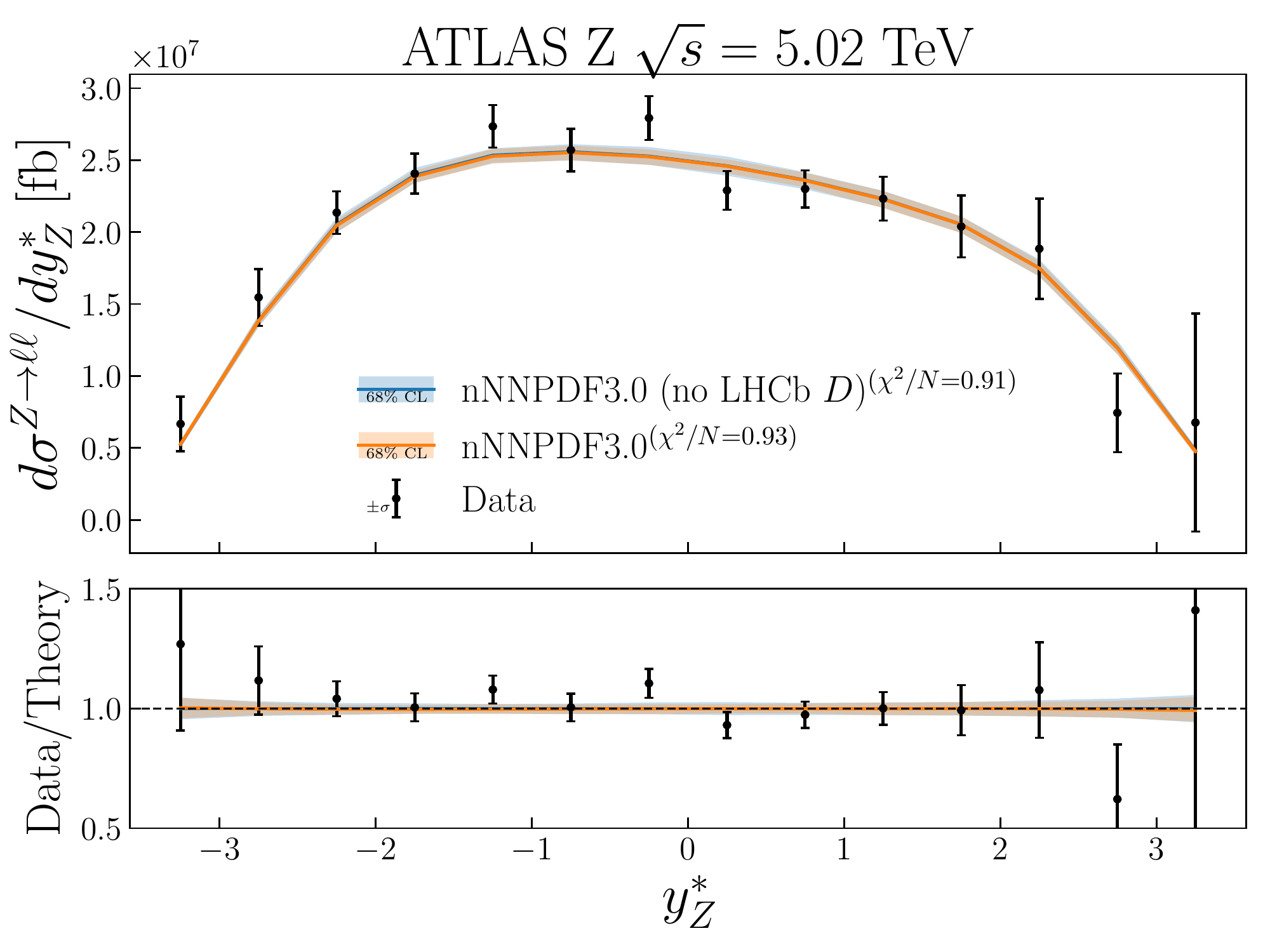}
  \includegraphics[width=0.49\textwidth]{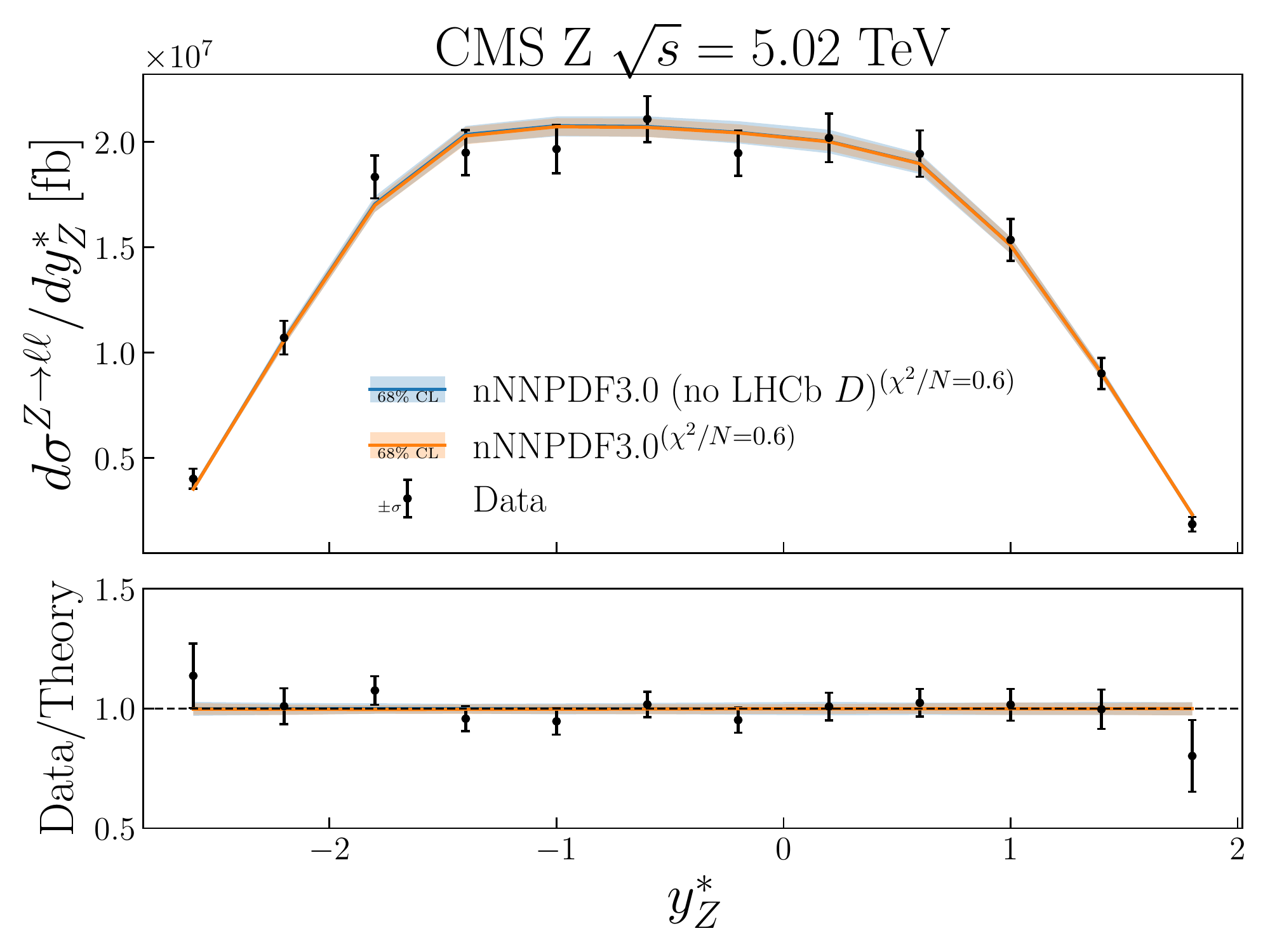}
  \includegraphics[width=0.49\textwidth]{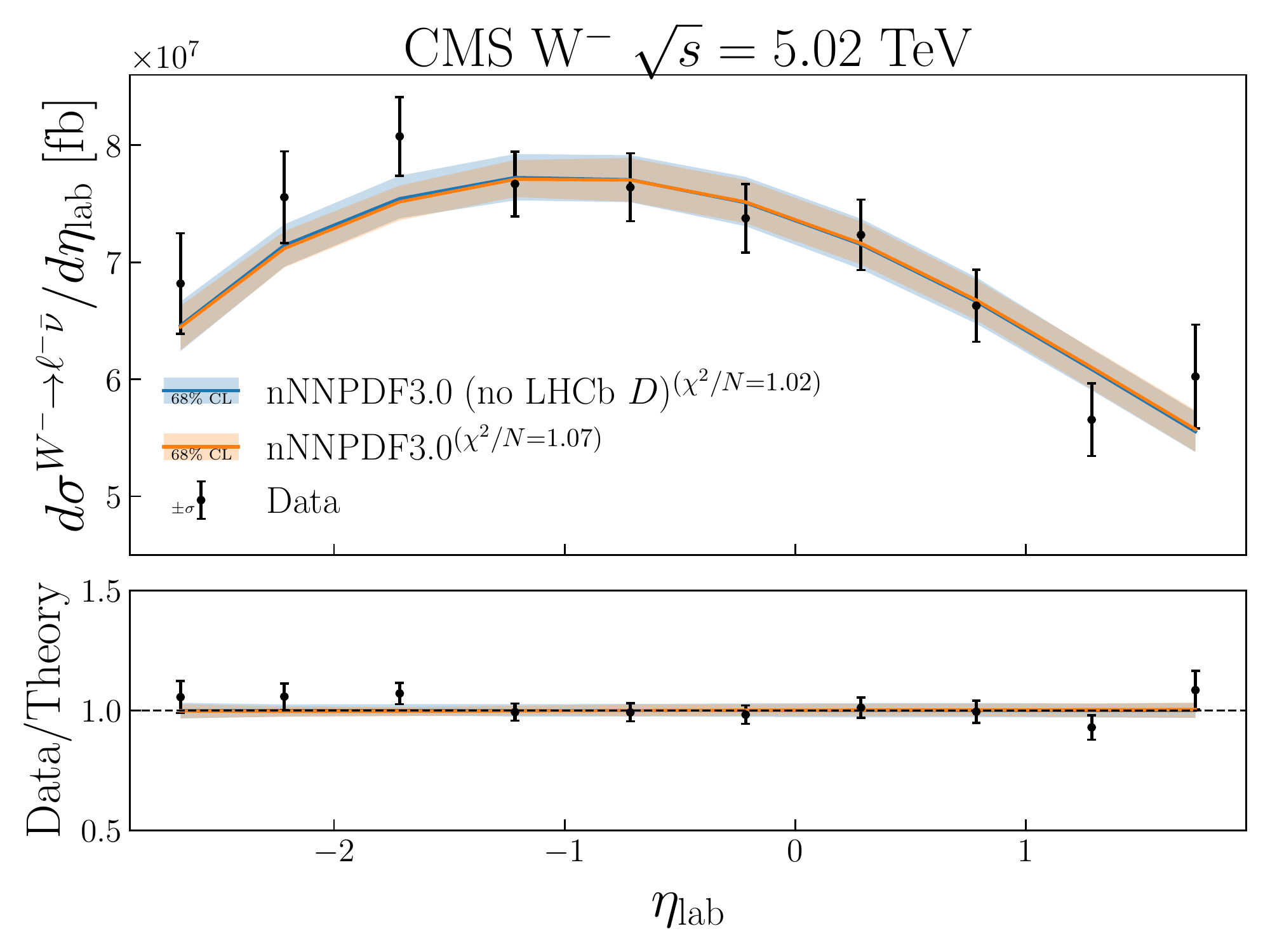}
  \includegraphics[width=0.49\textwidth]{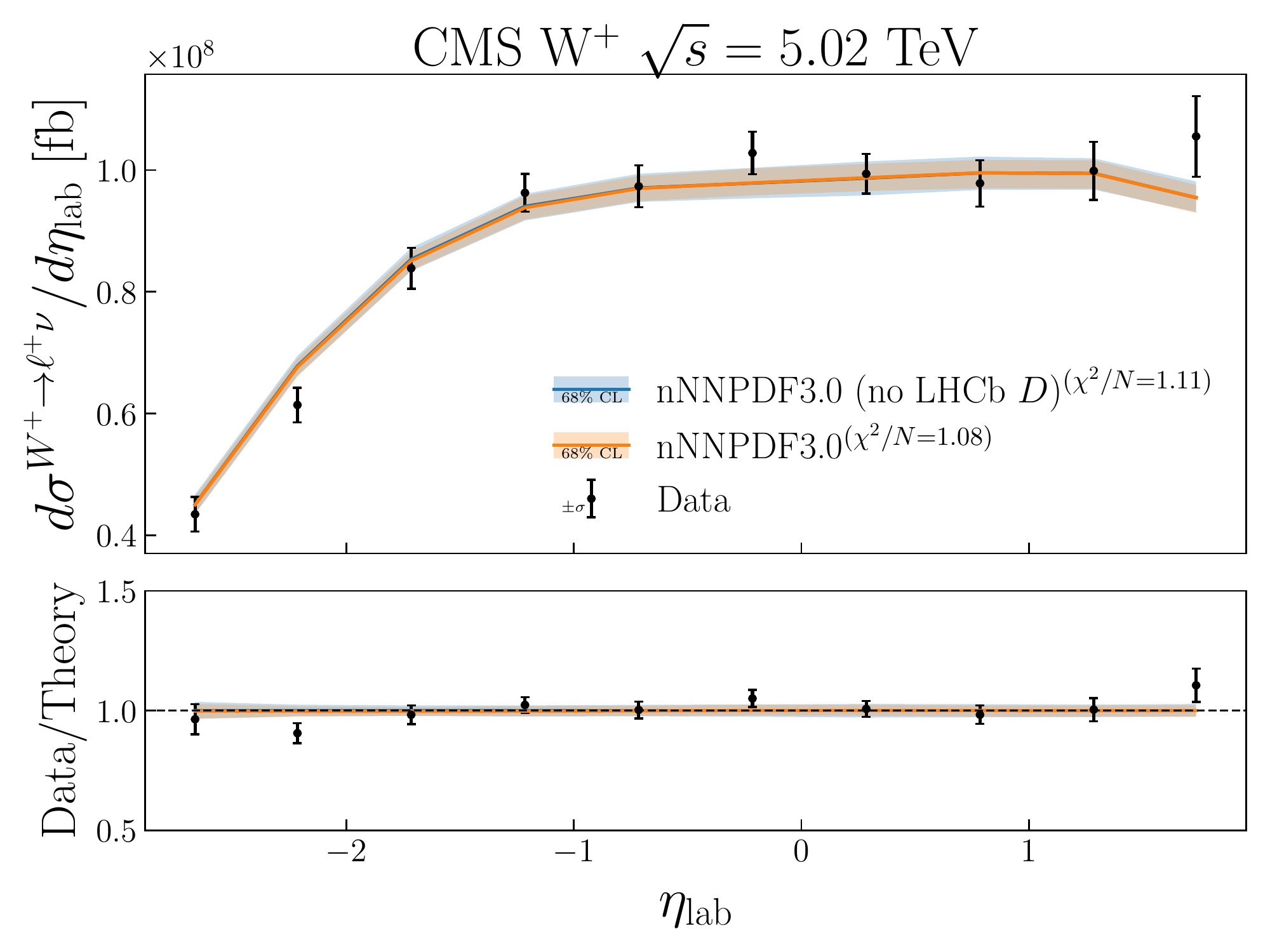}
  \includegraphics[width=0.49\textwidth]{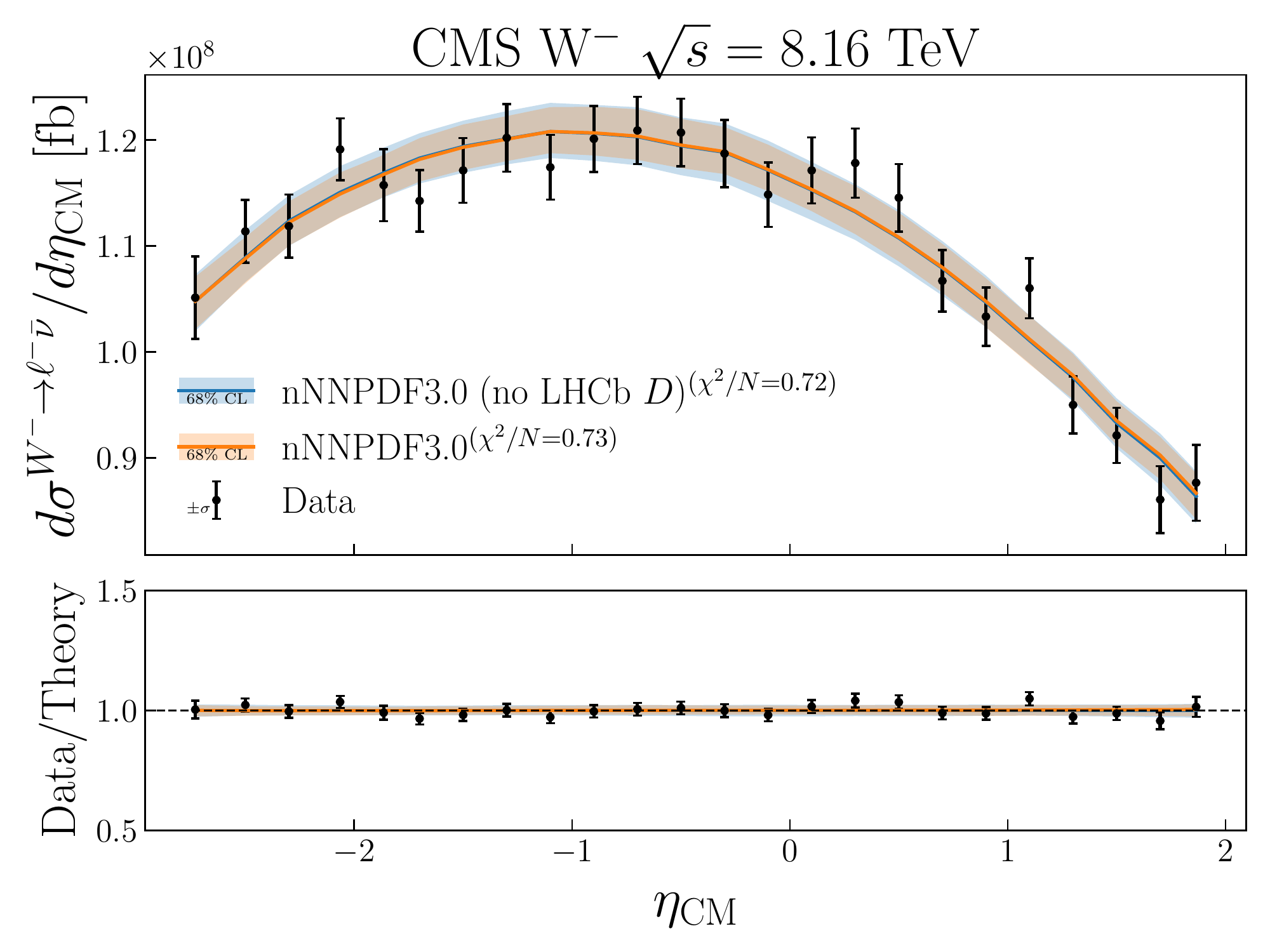}
   \includegraphics[width=0.49\textwidth]{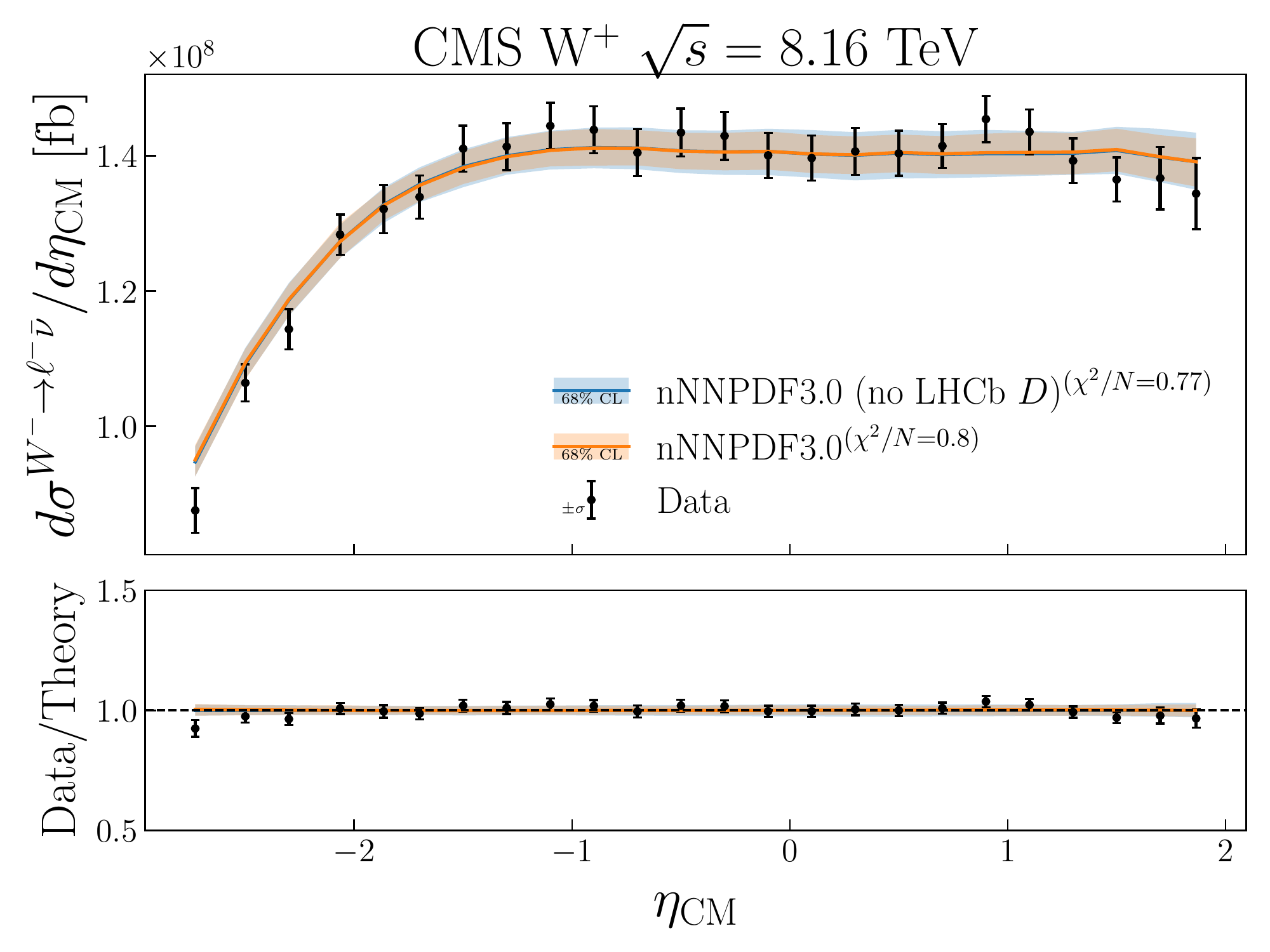}
   \caption{\label{fig:data_vs_theory_LHCewk1}
     Same as Fig.~\ref{fig:data_vs_theory_ATLASphotons}
    for LHC datasets on gauge boson production in pPb collisions, specifically
    for the ATLAS and CMS $Z$ production measurements at 5.02 TeV,
    and the charged lepton rapidity distributions for $W^+$ and $W^-$
  collisions from CMS at 5.02 TeV and 8.16 TeV.}
\end{figure}

\begin{figure}[!t]
  \centering
  \includegraphics[width=0.49\textwidth]{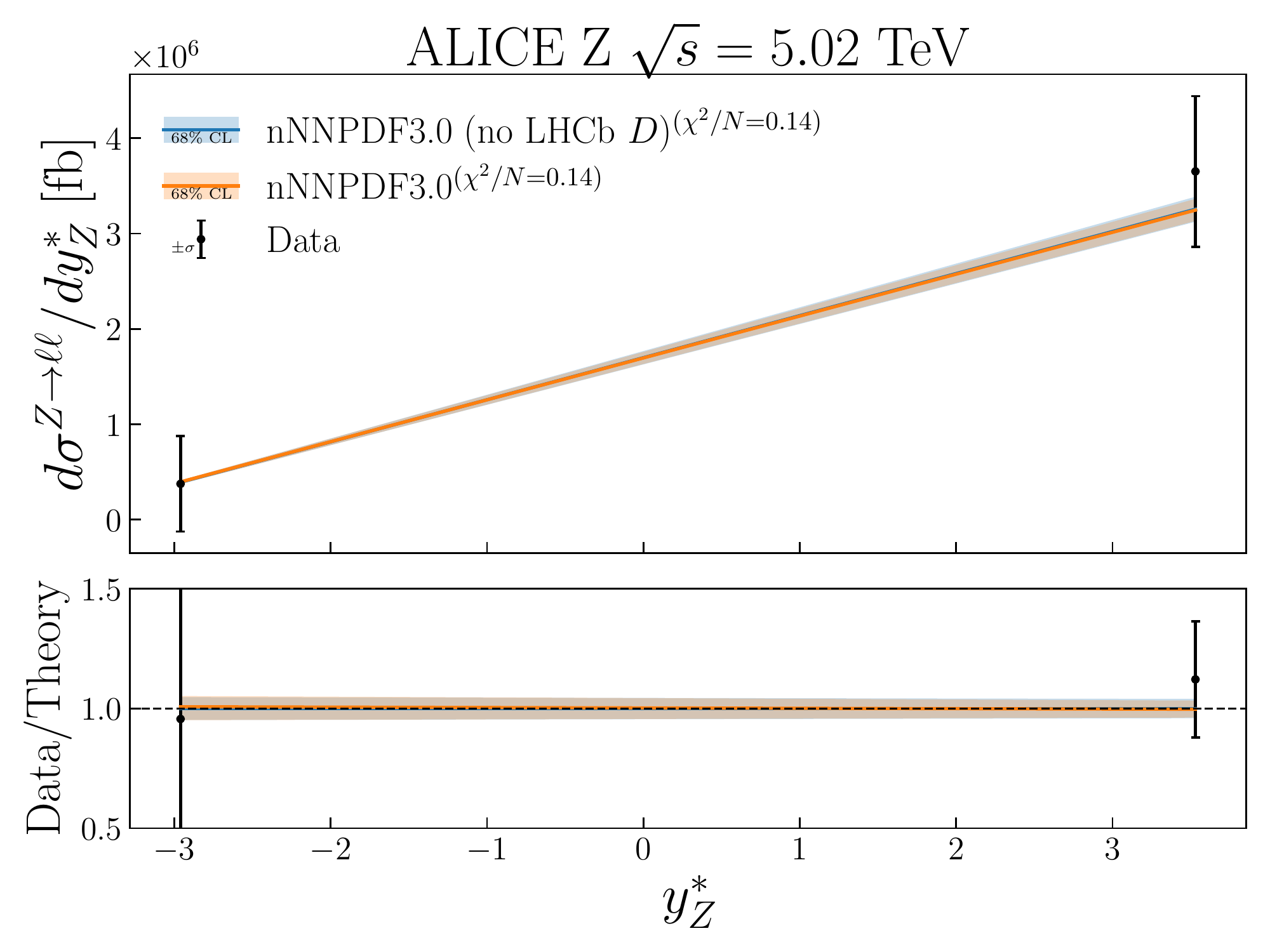}
  \includegraphics[width=0.49\textwidth]{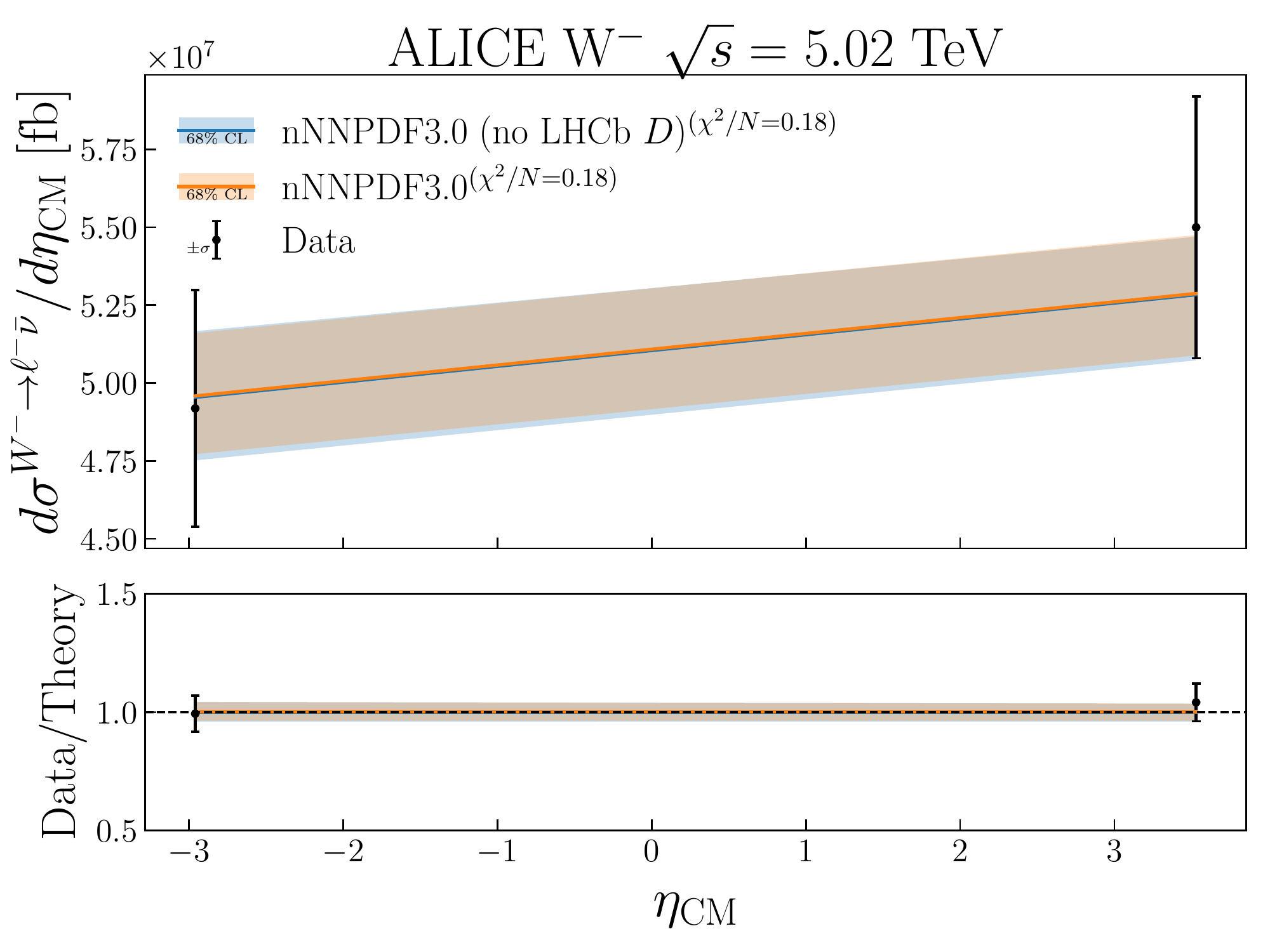}
   \includegraphics[width=0.49\textwidth]{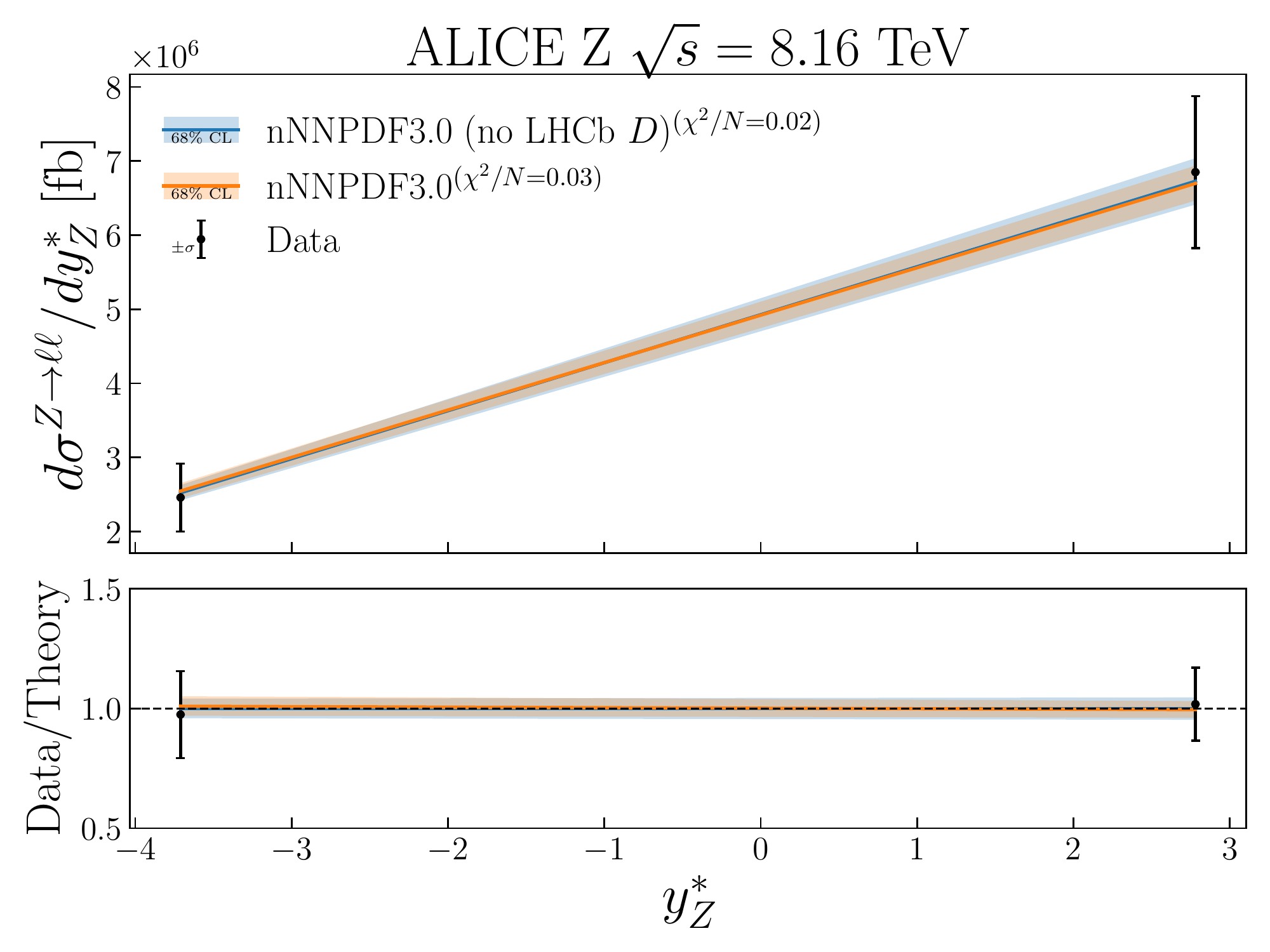}
  \includegraphics[width=0.49\textwidth]{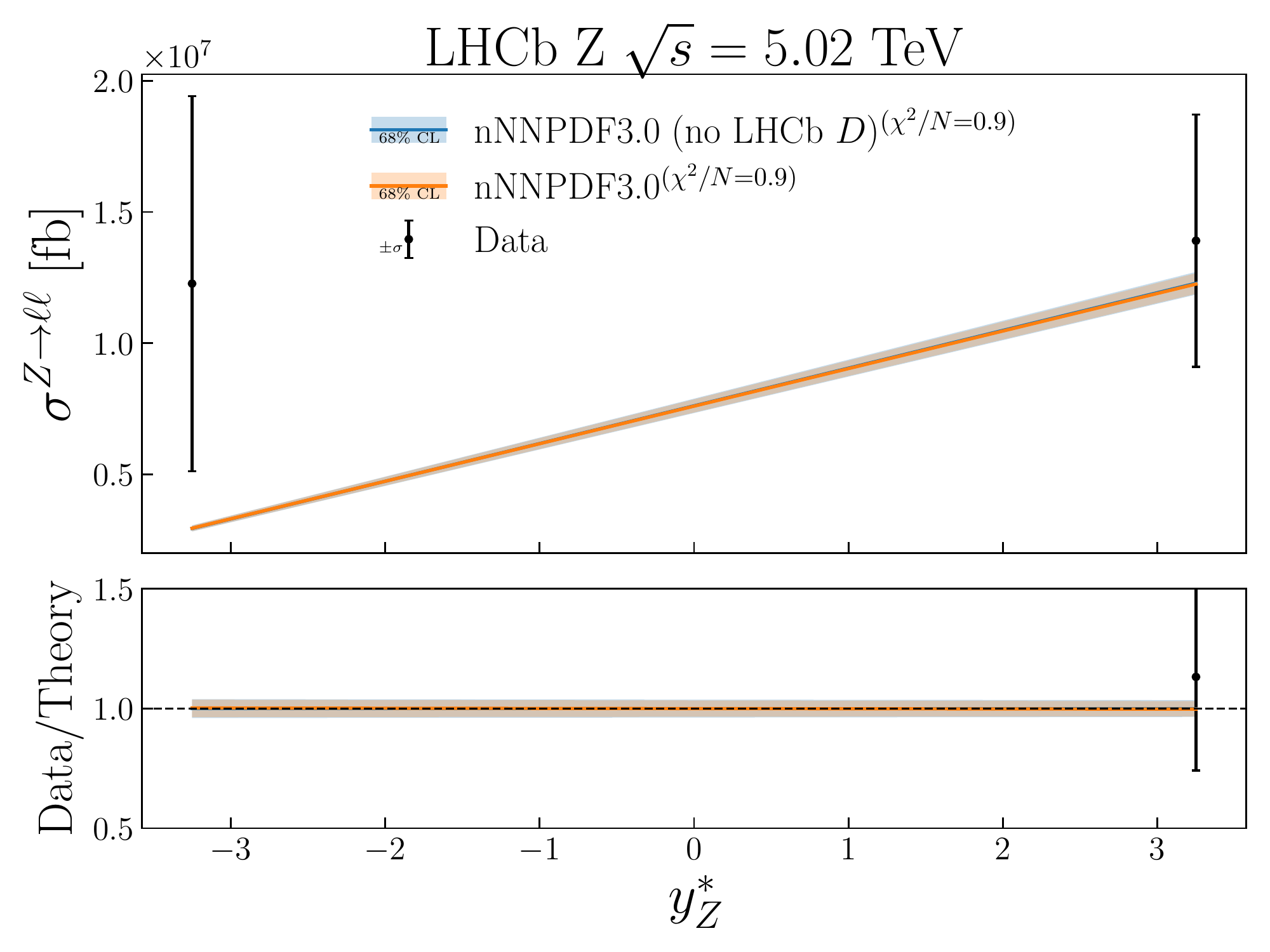}
 \includegraphics[width=\textwidth]{plots/50_nCMS_pPb_Z_8TEV.pdf}
  \caption{\label{fig:data_vs_theory_LHCewk2} Same as Fig.~\ref{fig:data_vs_theory_ATLASphotons}
    for other LHC datasets on gauge boson production in pPb collisions,
    in particular the ALICE and LHCb forward and backward measurements
    of $W$ and $Z$ production at 5.02 TeV and 8.16 TeV
    and the differential measurements of $Z$ production at 8.16  TeV from
  CMS in two bins of the dimuon invariant mass $M_{\mu\bar\mu}$.}
\end{figure}

Then in Figs.~\ref{fig:data_vs_theory_LHCewk1}
and~\ref{fig:data_vs_theory_LHCewk2} we display
similar comparisons as those in Fig.~\ref{fig:data_vs_theory_ATLASphotons}
for LHC datasets on gauge boson production in pPb collisions.
Specifically, we show in turn
 the ATLAS and CMS $Z$ production measurements at 5.02 TeV;
the charged lepton rapidity distributions for $W^+$ and $W^-$
collisions from CMS at 5.02 TeV and 8.16 TeV;
the ALICE and LHCb forward and backward measurements
of $W$ and $Z$ production at 5.02 TeV and 8.16 TeV;
and the differential measurements of $Z$ production at 8.16 from
CMS in two bins of the dimuon invariant mass $M_{\mu\bar\mu}$.

From this comparison, one can observe that in general there is very good agreement
between the NLO QCD predictions based on nNNPDF3.0 and the corresponding
experimental measurements.
As expected, given that their differences are localised at small-$x$,
the predictions based on the variant without the LHCb $D$-meson
data are very similar to those from the baseline fit.
The only pPb gauge boson production dataset for which
the quality of the data description is somewhat unsatisfactory
are the CMS differential measurements
of  $Z$ production at 8.16  TeV.
From the bottom plots of Fig.~\ref{fig:data_vs_theory_LHCewk2},
we see however that for the on-peak region,
with $60\le M_{\mu\bar\mu}\le 120~{\rm GeV}$, there is excellent
agreement between theory and data for all rapidity bins
except for the most backward ones, where the data undershoots
theory by about $3\sigma$.
Concerning the off-peak invariant mass region, $15\le M_{\mu\bar\mu}\le 60~{\rm GeV}$,
the theory undershoots the data by a factor which is more or less rapidity
independent and that could be explained by the absence of NNLO QCD corrections,
which are known to be non-negligible in this kinematic region.

\begin{figure}[!t]
  \centering
  \includegraphics[width=\textwidth]{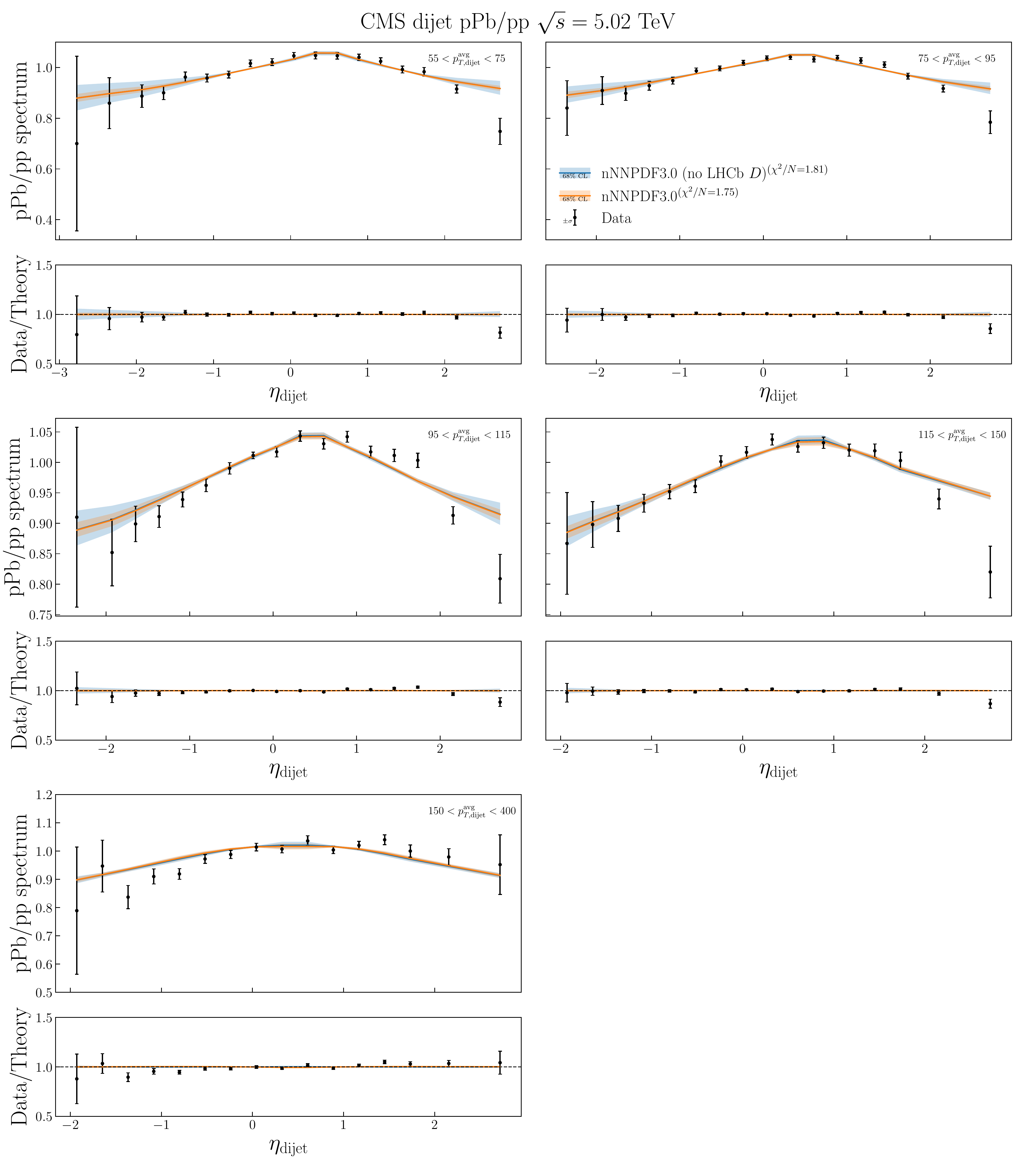}
    \caption{\label{fig:data_vs_theory_cmsdijets} Same as Fig.~\ref{fig:data_vs_theory_ATLASphotons}
      for the CMS measurements of dijet production in proton-lead collisions at 5.02 TeV,
    presented in terms of the ratio between pPb and pp spectra.}
\end{figure}

Finally, Fig.~\ref{fig:data_vs_theory_cmsdijets} compares
the NLO QCD calculations based on nNNPDF3.0 with the CMS measurements
of dijet production at  5.02 TeV presented in terms of the ratio between pPb and pp spectra.
These measurements are presented as a function of the dijet rapidity $\eta_{\rm dijet}$
in five bins of the dijet average transverse momentum $p_{T,{\rm dijet}}^{\rm avg}$.
We find how in general there is good agreement between the CMS data
and the theory calculations for most of the range in $\eta_{\rm dijet}$
and $p_{T,{\rm dijet}}^{\rm avg}$ covered.
One difference with the LHC electroweak measurements is that for dijet production
one can appreciate how the nNNPDF3.0 predictions exhibit reduced
PDF uncertainties, especially for the forward and backward bins,
as compared to the variant without the LHC $D$-meson cross-sections included.
The only noticeable discrepancy arises for the two most forward bins in $\eta_{\rm dijet}$
for the first four $p_{T,{\rm dijet}}^{\rm avg}$ bins, where the CMS data markedly
undershoots the theory predictions.
We have verified that in a fit where these forward bins are removed, the fit
quality to the CMS dijets is improved down to $\chi^2/n_{\rm dat}$ without any appreciable
change at the level of the nPDFs themselves.

\section{Reweighting validation}
\label{app:RWex}

In this appendix we assess the performance of the Bayesian reweighting
method, by comparing it with the direct inclusion of the same dataset in the prior nPDF set.
As discussed in Sect.~\ref{sec:RW}, in general
the results of including a new dataset on the prior
nPDF fit will not be identical in both methods.
One reason is that the figure of merit used
in the reweighting procedure is $\chi_{\rm t_0}^2 $ in Eq.~(\ref{eq:chi2_fit}),
and hence it does not account for the theoretical
constraints that enter the full $\chi_{\rm fit}^2$ in the fit, namely
the $A=1$ free-proton PDF boundary condition and the positivity
of cross-sections.
Furthermore, Bayesian reweighting assumes some degree of compatibility
between the prior fit and the new data, such that the $\chi^2$
evaluated over the unweighted replicas for the new dataset
follows (approximately) a  $\chi^2$-like distribution.
This condition may not be satisfied in the case of 
e.g. internal inconsistencies of the added dataset.
Within a direct fit this problem is avoided,
since stopping is based on look-back cross-validation
rather than in reaching target $\chi^2$ values.
Finally, finite-sample effects may also lead to small differences,
since formally reweighting and refitting coincide only
in the $N_{\rm rep},~N_{\rm eff}\to \infty$ limit.

\begin{figure}[!t]
  \centering
\includegraphics[width=0.5\textwidth]{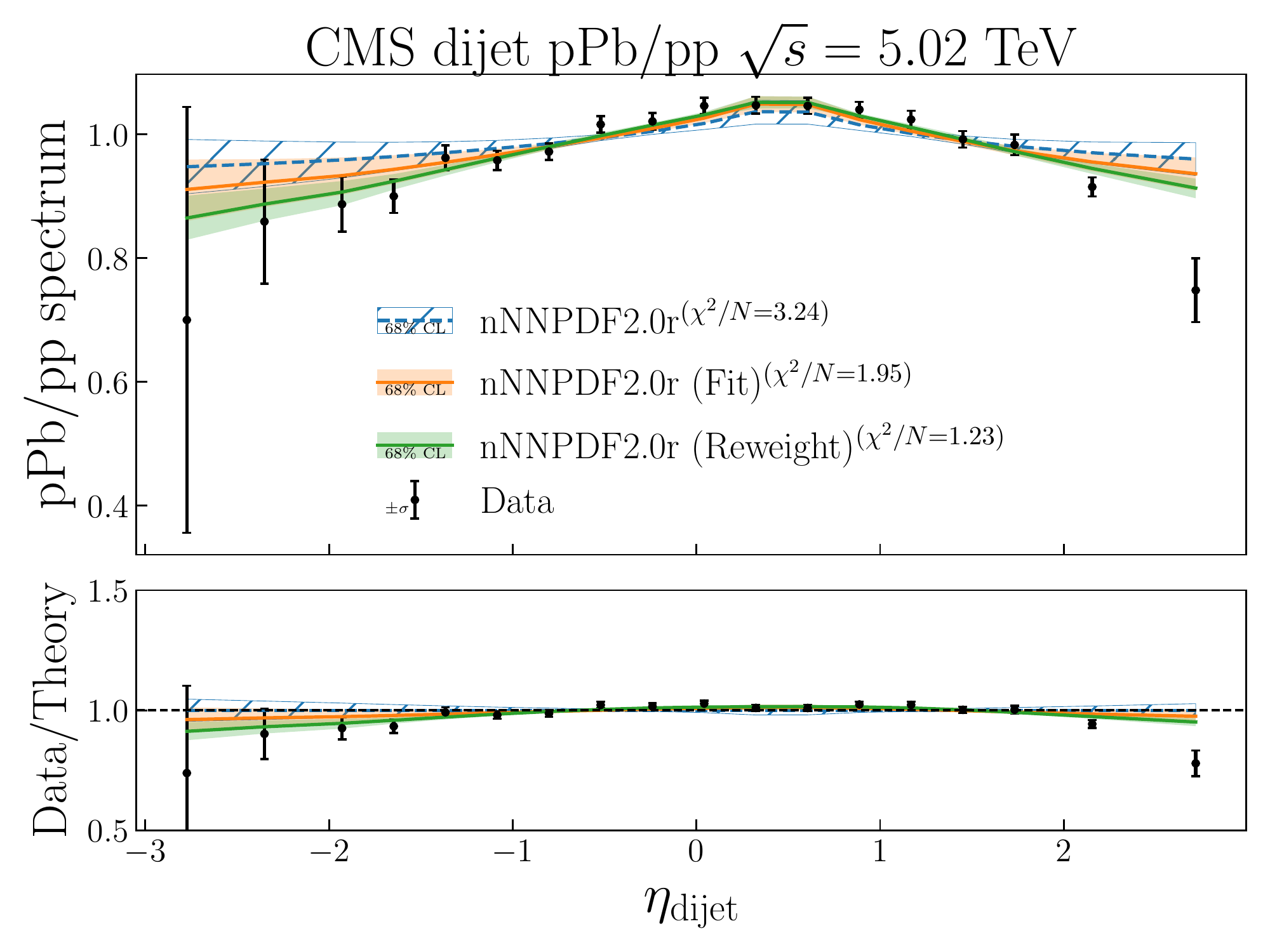}
\caption{\label{fig:data_vs_theory_RW_dijet} Comparison
  between the theoretical predictions based on nNNPDF2.0r
  and on two variants where the first bin ($55<p_{T,{\rm dijet}}^{\rm avg}<75~{\rm GeV}$)
  of the CMS dijet ratio
  data has been added either via direct inclusion in the fit or
  by means of Bayesian reweighting.
  The band in the theory prediction indicates the PDF uncertainty,
  while the bottom panels display the ratio to the central theory prediction.
  No estimate of uncertainties due to MHOs are considered in this comparison.
  We also indicate in the legend the corresponding values
  for the $\chi^2/n_{\rm dat}$ in each case.
}
\end{figure}

For this validation exercise, we consider nNNPDF2.0r
as the prior nuclear PDF set and the first bin
of the  CMS dijet pPb/pp ratio measurements
($55<p_{T,{\rm dijet}}^{\rm avg}<75~{\rm GeV}$) as the new dataset
to be added either via direct inclusion in the fit
or by means of Bayesian reweighting.
Fig.~\ref{fig:data_vs_theory_RW_dijet} displays the comparison 
between the theoretical predictions based on nNNPDF2.0r 
and in the two associated variants where this dataset 
has been added either via direct inclusion in the fit or by means of reweighting.
The band in the theory prediction indicates the PDF uncertainty, 
while the bottom panels display the ratio to the central theory prediction.
We also indicate in the legend the corresponding values for the $\chi^2/n_{\rm dat}$ in each case.

The nNNPDF2.0r prior describes this dataset (recall that only
the first CMS dijet bin is considered here) rather poorly,
with $\chi^2/n_{\rm dat}=3.24$ for $n_{\rm dat}=18$.
The fit quality is markedly improved once this dataset
is included in the fit either by direct inclusion or via reweighting,
with $\chi^2/n_{\rm dat}=1.95$ and 1.23 respectively.
In both cases, we observe how the theoretical predictions
move closer to the data, favouring large-$x$ anti-shadowing  (central
rapidities) and small-$x$ shadowing (forward and backward rapidities) respectively for the gluon nPDF.
The reweighted predictions agree within uncertainties with the direct
fit results for all the $\eta_{\rm dijet}$ bins considered.
For the most forward and backward rapidity bins the reweighted predictions
prefer smaller values for the cross-section ratio (i.e. a stronger suppression of the nPDFs),
providing a slightly better description of the data.
This may be explained by the fact that the evaluation of the $\chi^2$ in the reweighting 
procedure is not constrained by the free proton boundary condition constraint.
Consequently, the proton boundary condition can be modified in the reweighting procedure, 
which provides extra flexibility when describing the ratio observable.

\begin{figure}[!t]
  \centering
  \includegraphics[width=0.95\textwidth]{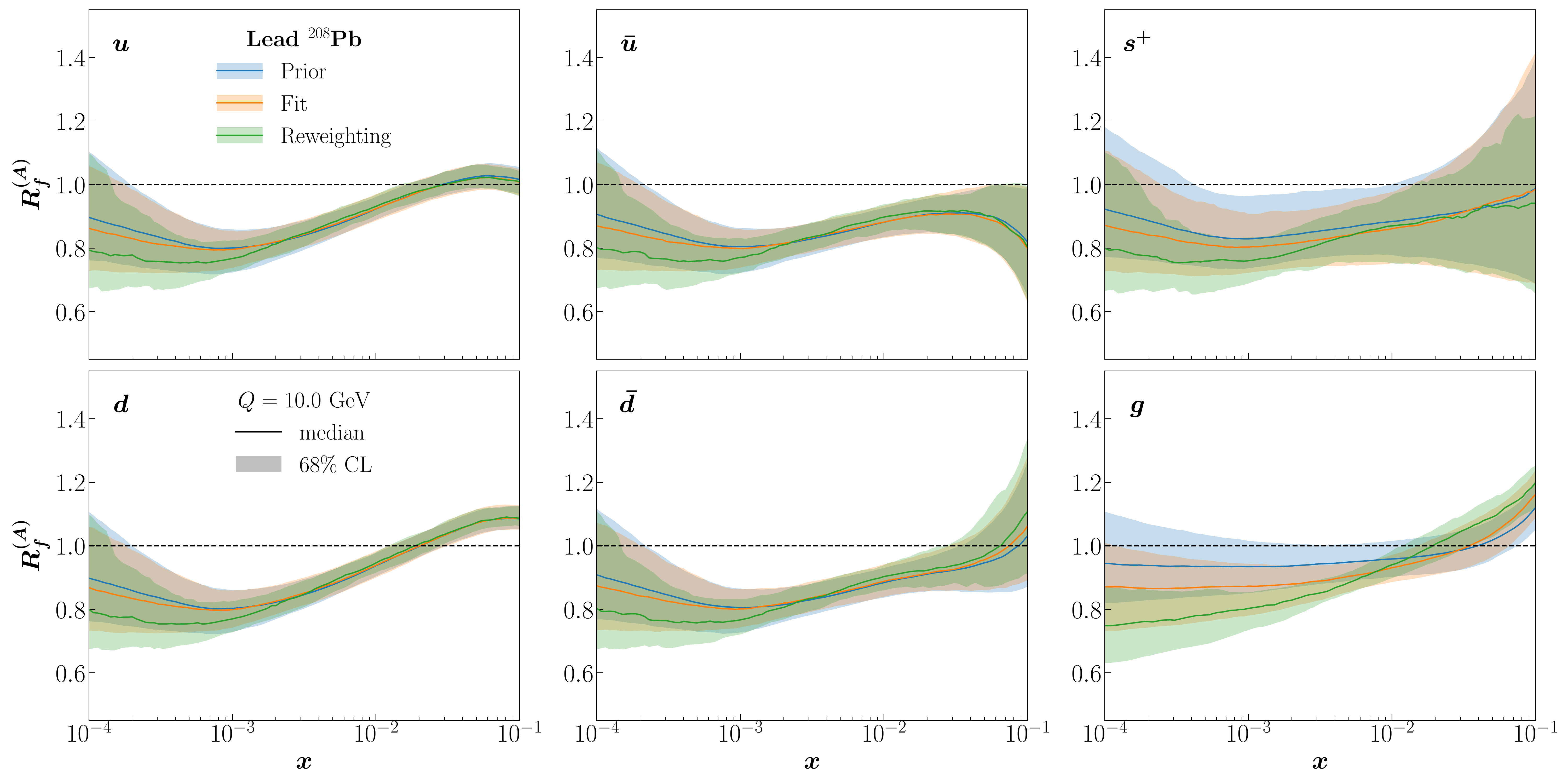}
  \includegraphics[width=0.95\textwidth]{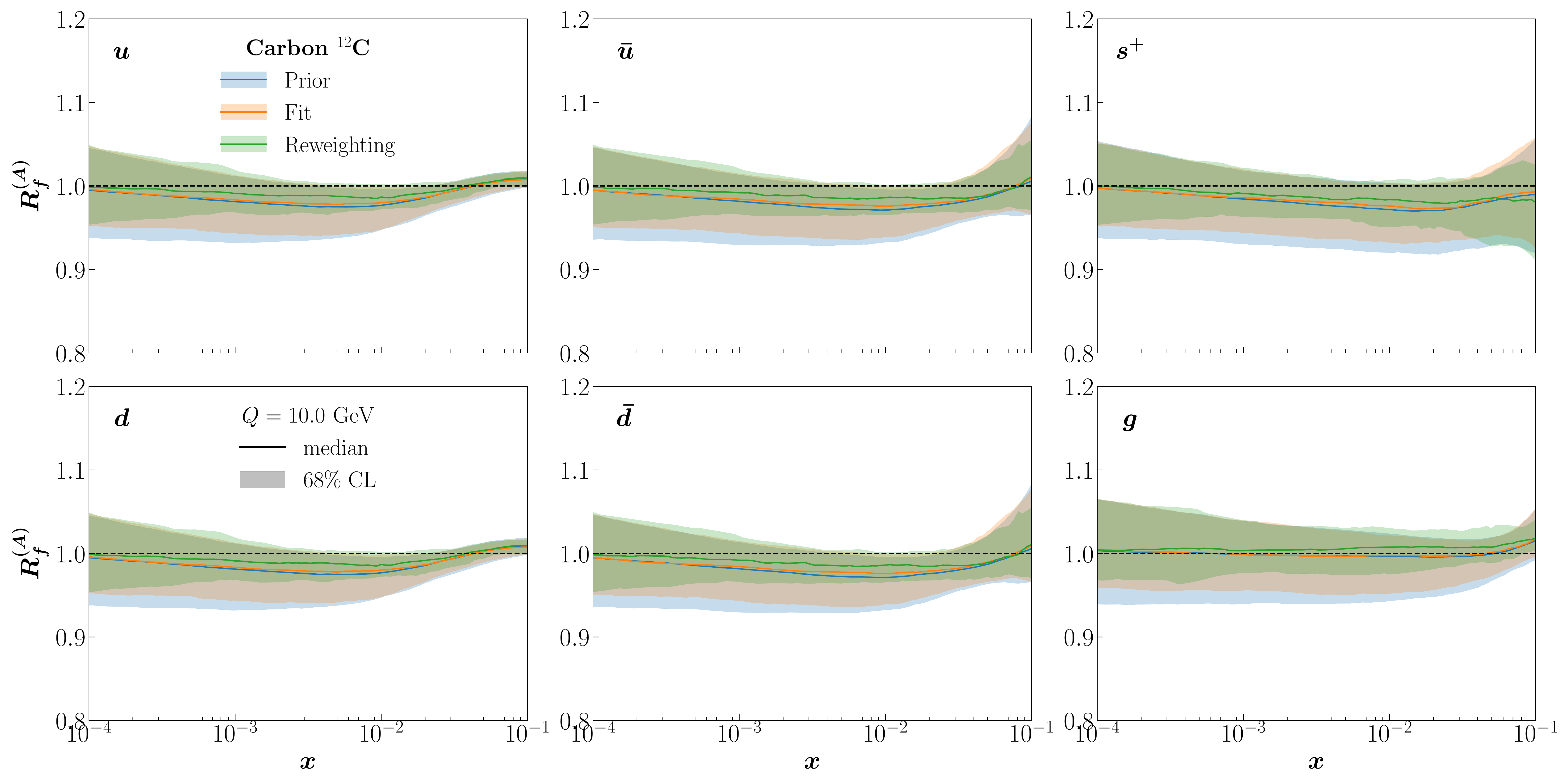}
  \caption{\label{fig:RW_validation}
    The nuclear modification ratios $R_f^{(A)}(x,Q^2)$ in nNNPDF2.0r,
    compared to the outcome of adding to nNNPDF2.0r the first
    $p_{T,{\rm dijet}}^{\rm avg}$ bin of the CMS dijet measurements either
    via direct inclusion in the fit of via Bayesian reweighting.
    We show the results for $^{208}$Pb (top)
    and $^{12}$C (bottom panels).
  }
\end{figure}

Fig.~\ref{fig:RW_validation} then displays the
nuclear modification ratios $R_f^{(A)}(x,Q^2)$ in nNNPDF2.0r
compared to those associated to the variant fits where the CMS dijet data
is added to the prior either via direct inclusion in the fit of via Bayesian reweighting.
From top to bottom, we show the results for lead ($^{208}$Pb)
and then for carbon ($^{12}$C) nuclei.
Recall that the reweighting procedure determines the weight
of each replica taking into account the information contained
in the new dataset, and hence it modifies
the predictions for the nuclear modifications also
for different values of $A$ as compared to that
for which the data is provided ($A=208$ in this case).

In the case of the nuclear modifications of lead, we observe
that in the kinematic region for which the CMS
data provides sensitivity ($ 10^{-3}\lsim x \lsim 0.1$)
the reweighted results reproduce the qualitative behaviour
of the direct fit predictions.
In particular, for the gluon nuclear ratio the reweighting
procedure correctly identifies the small-$x$ suppression 
and the large-$x$ enhancement as compared to the prior. 
As observed in Fig.~\ref{fig:RW_validation}, reweighting
also tends to somewhat overestimate the impact of the new dataset
as compared to direct inclusion in the fit, at least for
this specific case.
Hence this analysis confirms that reweighting
captures the main features of the information provided by
the new dataset on the nPDFs when the same value of $A$
is considered.
However, in the case of a low-$A$ nuclei such as carbon ($A=12$),
reweighting appears to induce a reduction
of the nPDF uncertainty absent in the direct fit.
As discussed above, this effect
may be attributed to the absence of the $A=1$ boundary condition
in the evaluation of the $\chi^2$ value in the reweighting procedure.

We emphasise that the specific dataset considered in the exercise
displayed in Figs.~\ref{fig:data_vs_theory_RW_dijet} and~\ref{fig:RW_validation}
represents an extreme case, in the sense that the CMS dijet measurement
is one of the datasets carrying most information in the nNNPDF3.0 fit.

\bibliography{nNNPDF30}

\providecommand{\href}[2]{#2}\begingroup\raggedright\begin{thebibliography}{100}

\bibitem{Rojo:2019uip}
J.~Rojo, {\it {The Partonic Content of Nucleons and Nuclei}},  {\em Oxford
  Encyclopedia of Physics} (10, 2019)
  [\href{http://arxiv.org/abs/1910.03408}{{\tt arXiv:1910.03408}}].

\bibitem{Kovarik:2019xvh}
K.~Kova\v{r}\'\i{}k, P.~M. Nadolsky, and D.~E. Soper, {\it {Hadronic structure
  in high-energy collisions}},  {\em Rev. Mod. Phys.} {\bf 92} (2020), no.~4
  045003, [\href{http://arxiv.org/abs/1905.06957}{{\tt arXiv:1905.06957}}].

\bibitem{Ethier:2020way}
J.~J. Ethier and E.~R. Nocera, {\it {Parton Distributions in Nucleons and
  Nuclei}},  {\em Ann. Rev. Nucl. Part. Sci.} (2020), no.~70 1--34,
  [\href{http://arxiv.org/abs/2001.07722}{{\tt arXiv:2001.07722}}].

\bibitem{Fischer:2014wfa}
W.~Fischer and J.~M. Jowett, {\it {Ion Colliders}},  {\em Rev. Accel. Sci.
  Tech.} {\bf 7} (2014) 49--76.

\bibitem{Jowett:2018yqk}
J.~Jowett, {\it {Colliding Heavy Ions in the LHC}},  in {\em {9th International
  Particle Accelerator Conference}}, 6, 2018.

\bibitem{Brewer:2021kiv}
J.~Brewer, A.~Mazeliauskas, and W.~van~der Schee, {\it {Opportunities of OO and
  $p$O collisions at the LHC}},  in {\em {Opportunities of OO and pO collisions
  at the LHC}}, 3, 2021.
\newblock \href{http://arxiv.org/abs/2103.01939}{{\tt arXiv:2103.01939}}.

\bibitem{Hadjidakis:2018ifr}
C.~Hadjidakis et~al., {\it {A fixed-target programme at the LHC: Physics case
  and projected performances for heavy-ion, hadron, spin and astroparticle
  studies}},  {\em Phys. Rept.} {\bf 911} (2021) 1--83,
  [\href{http://arxiv.org/abs/1807.00603}{{\tt arXiv:1807.00603}}].

\bibitem{Cooper-Sarkar:2011jtt}
A.~Cooper-Sarkar, P.~Mertsch, and S.~Sarkar, {\it {The high energy neutrino
  cross-section in the Standard Model and its uncertainty}},  {\em JHEP} {\bf
  08} (2011) 042, [\href{http://arxiv.org/abs/1106.3723}{{\tt
  arXiv:1106.3723}}].

\bibitem{Bertone:2018dse}
V.~Bertone, R.~Gauld, and J.~Rojo, {\it {Neutrino Telescopes as QCD
  Microscopes}},  {\em JHEP} {\bf 01} (2019) 217,
  [\href{http://arxiv.org/abs/1808.02034}{{\tt arXiv:1808.02034}}].

\bibitem{Garcia:2020jwr}
A.~Garcia, R.~Gauld, A.~Heijboer, and J.~Rojo, {\it {Complete predictions for
  high-energy neutrino propagation in matter}},  {\em JCAP} {\bf 09} (2020)
  025, [\href{http://arxiv.org/abs/2004.04756}{{\tt arXiv:2004.04756}}].

\bibitem{Connolly:2011vc}
A.~Connolly, R.~S. Thorne, and D.~Waters, {\it {Calculation of High Energy
  Neutrino-Nucleon Cross Sections and Uncertainties Using the MSTW Parton
  Distribution Functions and Implications for Future Experiments}},  {\em Phys.
  Rev. D} {\bf 83} (2011) 113009, [\href{http://arxiv.org/abs/1102.0691}{{\tt
  arXiv:1102.0691}}].

\bibitem{Gauld:2015yia}
R.~Gauld, J.~Rojo, L.~Rottoli, and J.~Talbert, {\it {Charm production in the
  forward region: constraints on the small-x gluon and backgrounds for neutrino
  astronomy}},  {\em JHEP} {\bf 11} (2015) 009,
  [\href{http://arxiv.org/abs/1506.08025}{{\tt arXiv:1506.08025}}].

\bibitem{Garzelli:2016xmx}
{\bf PROSA} Collaboration, M.~V. Garzelli, S.~Moch, O.~Zenaiev,
  A.~Cooper-Sarkar, A.~Geiser, K.~Lipka, R.~Placakyte, and G.~Sigl, {\it
  {Prompt neutrino fluxes in the atmosphere with PROSA parton distribution
  functions}},  {\em JHEP} {\bf 05} (2017) 004,
  [\href{http://arxiv.org/abs/1611.03815}{{\tt arXiv:1611.03815}}].

\bibitem{Zenaiev:2019ktw}
{\bf PROSA} Collaboration, O.~Zenaiev, M.~V. Garzelli, K.~Lipka, S.~O. Moch,
  A.~Cooper-Sarkar, F.~Olness, A.~Geiser, and G.~Sigl, {\it {Improved
  constraints on parton distributions using LHCb, ALICE and HERA heavy-flavour
  measurements and implications for the predictions for prompt
  atmospheric-neutrino fluxes}},  {\em JHEP} {\bf 04} (2020) 118,
  [\href{http://arxiv.org/abs/1911.13164}{{\tt arXiv:1911.13164}}].

\bibitem{IceCube:2006tjp}
{\bf IceCube} Collaboration, A.~Achterberg et~al., {\it {First Year Performance
  of The IceCube Neutrino Telescope}},  {\em Astropart. Phys.} {\bf 26} (2006)
  155--173, [\href{http://arxiv.org/abs/astro-ph/0604450}{{\tt
  astro-ph/0604450}}].

\bibitem{KM3Net:2016zxf}
{\bf KM3Net} Collaboration, S.~Adrian-Martinez et~al., {\it {Letter of intent
  for KM3NeT 2.0}},  {\em J. Phys. G} {\bf 43} (2016), no.~8 084001,
  [\href{http://arxiv.org/abs/1601.07459}{{\tt arXiv:1601.07459}}].

\bibitem{AbdulKhalek:2021gbh}
R.~Abdul~Khalek et~al., {\it {Science Requirements and Detector Concepts for
  the Electron-Ion Collider: EIC Yellow Report}},
  \href{http://arxiv.org/abs/2103.05419}{{\tt arXiv:2103.05419}}.

\bibitem{Anderle:2021wcy}
D.~P. Anderle et~al., {\it {Electron-ion collider in China}},  {\em Front.
  Phys. (Beijing)} {\bf 16} (2021), no.~6 64701,
  [\href{http://arxiv.org/abs/2102.09222}{{\tt arXiv:2102.09222}}].

\bibitem{Khalek:2021ulf}
R.~A. Khalek, J.~J. Ethier, E.~R. Nocera, and J.~Rojo, {\it {Self-consistent
  determination of proton and nuclear PDFs at the Electron Ion Collider}},
  {\em Phys. Rev. D} {\bf 103} (2021), no.~9 096005,
  [\href{http://arxiv.org/abs/2102.00018}{{\tt arXiv:2102.00018}}].

\bibitem{Anchordoqui:2021ghd}
L.~A. Anchordoqui et~al., {\it {The Forward Physics Facility: Sites,
  Experiments, and Physics Potential}},
  \href{http://arxiv.org/abs/2109.10905}{{\tt arXiv:2109.10905}}.

\bibitem{Gelis:2010nm}
F.~Gelis, E.~Iancu, J.~Jalilian-Marian, and R.~Venugopalan, {\it {The Color
  Glass Condensate}},  {\em Ann. Rev. Nucl. Part. Sci.} {\bf 60} (2010)
  463--489, [\href{http://arxiv.org/abs/1002.0333}{{\tt arXiv:1002.0333}}].

\bibitem{Ball:2018twp}
{\bf NNPDF} Collaboration, R.~D. Ball, E.~R. Nocera, and R.~L. Pearson, {\it
  {Nuclear Uncertainties in the Determination of Proton PDFs}},  {\em Eur.
  Phys. J.} {\bf C79} (2019), no.~3 282,
  [\href{http://arxiv.org/abs/1812.09074}{{\tt arXiv:1812.09074}}].

\bibitem{Ball:2020xqw}
R.~D. Ball, E.~R. Nocera, and R.~L. Pearson, {\it {Deuteron Uncertainties in
  the Determination of Proton PDFs}},  {\em Eur. Phys. J. C} {\bf 81} (2021),
  no.~1 37, [\href{http://arxiv.org/abs/2011.00009}{{\tt arXiv:2011.00009}}].

\bibitem{Ball:2021leu}
R.~D. Ball et~al., {\it {The Path to Proton Structure at One-Percent
  Accuracy}},  \href{http://arxiv.org/abs/2109.02653}{{\tt arXiv:2109.02653}}.

\bibitem{AbdulKhalek:2019mzd}
{\bf NNPDF} Collaboration, R.~Abdul~Khalek, J.~J. Ethier, and J.~Rojo, {\it
  {Nuclear parton distributions from lepton-nucleus scattering and the impact
  of an electron-ion collider}},  {\em Eur. Phys. J.} {\bf C79} (2019), no.~6
  471, [\href{http://arxiv.org/abs/1904.00018}{{\tt arXiv:1904.00018}}].

\bibitem{AbdulKhalek:2020yuc}
R.~Abdul~Khalek, J.~J. Ethier, J.~Rojo, and G.~van Weelden, {\it {nNNPDF2.0:
  quark flavor separation in nuclei from LHC data}},  {\em JHEP} {\bf 09}
  (2020) 183, [\href{http://arxiv.org/abs/2006.14629}{{\tt arXiv:2006.14629}}].

\bibitem{CMS:2018jpl}
{\bf CMS} Collaboration, A.~M. Sirunyan et~al., {\it {Constraining gluon
  distributions in nuclei using dijets in proton-proton and proton-lead
  collisions at $\sqrt{s_{_\mathrm{NN}}} =$ 5.02 TeV}},  {\em Phys. Rev. Lett.}
  {\bf 121} (2018), no.~6 062002, [\href{http://arxiv.org/abs/1805.04736}{{\tt
  arXiv:1805.04736}}].

\bibitem{LHCb:2017yua}
{\bf LHCb} Collaboration, R.~Aaij et~al., {\it {Study of prompt D$^{0}$ meson
  production in $p$Pb collisions at $ \sqrt{s_{\mathrm{NN}}}=5 $ TeV}},  {\em
  JHEP} {\bf 10} (2017) 090, [\href{http://arxiv.org/abs/1707.02750}{{\tt
  arXiv:1707.02750}}].

\bibitem{Ball:2017nwa}
{\bf NNPDF} Collaboration, R.~D. Ball et~al., {\it {Parton distributions from
  high-precision collider data}},  {\em Eur. Phys. J.} {\bf C77} (2017), no.~10
  663, [\href{http://arxiv.org/abs/1706.00428}{{\tt arXiv:1706.00428}}].

\bibitem{Carrazza:2019mzf}
S.~Carrazza and J.~Cruz-Martinez, {\it {Towards a new generation of parton
  densities with deep learning models}},  {\em Eur. Phys. J. C} {\bf 79}
  (2019), no.~8 676, [\href{http://arxiv.org/abs/1907.05075}{{\tt
  arXiv:1907.05075}}].

\bibitem{Buckley:2014ana}
A.~Buckley, J.~Ferrando, S.~Lloyd, K.~Nordström, B.~Page, et~al., {\it
  {LHAPDF6: parton density access in the LHC precision era}},  {\em
  Eur.Phys.J.} {\bf C75} (2015) 132,
  [\href{http://arxiv.org/abs/1412.7420}{{\tt arXiv:1412.7420}}].

\bibitem{Eskola:2016oht}
K.~J. Eskola, P.~Paakkinen, H.~Paukkunen, and C.~A. Salgado, {\it {EPPS16:
  Nuclear parton distributions with LHC data}},  {\em Eur. Phys. J.} {\bf C77}
  (2017), no.~3 163, [\href{http://arxiv.org/abs/1612.05741}{{\tt
  arXiv:1612.05741}}].

\bibitem{Duwentaster:2021ioo}
P.~Duwent\"aster, L.~A. Husov\'a, T.~Je\v{z}o, M.~Klasen, K.~Kova\v{r}\'\i{}k,
  A.~Kusina, K.~F. Muzakka, F.~I. Olness, I.~Schienbein, and J.~Y. Yu, {\it
  {Impact of inclusive hadron production data on nuclear gluon PDFs}},  {\em
  Phys. Rev. D} {\bf 104} (2021), no.~9 094005,
  [\href{http://arxiv.org/abs/2105.09873}{{\tt arXiv:2105.09873}}].

\bibitem{NewMuon:1995cua}
{\bf New Muon} Collaboration, P.~Amaudruz et~al., {\it {A Reevaluation of the
  nuclear structure function ratios for D, He, Li-6, C and Ca}},  {\em Nucl.
  Phys. B} {\bf 441} (1995) 3--11,
  [\href{http://arxiv.org/abs/hep-ph/9503291}{{\tt hep-ph/9503291}}].

\bibitem{NewMuon:1995tgs}
{\bf New Muon} Collaboration, M.~Arneodo et~al., {\it {The Structure Function
  ratios F2(li) / F2(D) and F2(C) / F2(D) at small x}},  {\em Nucl. Phys. B}
  {\bf 441} (1995) 12--30, [\href{http://arxiv.org/abs/hep-ex/9504002}{{\tt
  hep-ex/9504002}}].

\bibitem{NewMuon:1996yuf}
{\bf New Muon} Collaboration, M.~Arneodo et~al., {\it {The A dependence of the
  nuclear structure function ratios}},  {\em Nucl. Phys. B} {\bf 481} (1996)
  3--22.

\bibitem{NewMuon:1996gam}
{\bf New Muon} Collaboration, M.~Arneodo et~al., {\it {The Q**2 dependence of
  the structure function ratio F2 Sn / F2 C and the difference R Sn - R C in
  deep inelastic muon scattering}},  {\em Nucl. Phys. B} {\bf 481} (1996)
  23--39.

\bibitem{EuropeanMuon:1988lbf}
{\bf European Muon} Collaboration, J.~Ashman et~al., {\it {Measurement of the
  Ratios of Deep Inelastic Muon - Nucleus Cross-Sections on Various Nuclei
  Compared to Deuterium}},  {\em Phys. Lett. B} {\bf 202} (1988) 603--610.

\bibitem{EuropeanMuon:1989mef}
{\bf European Muon} Collaboration, M.~Arneodo et~al., {\it {Measurements of the
  nucleon structure function in the range $0.002-{\rm GeV}^2 < x < 0.17-{\rm
  GeV}^2$ and $0.2-GeV^2 < q^2 < 8-GeV^2$ in deuterium, carbon and calcium}},
  {\em Nucl. Phys. B} {\bf 333} (1990) 1--47.

\bibitem{EuropeanMuon:1987obv}
{\bf European Muon} Collaboration, J.~J. Aubert et~al., {\it {Measurements of
  the nucleon structure functions $F2_n$ in deep inelastic muon scattering from
  deuterium and comparison with those from hydrogen and iron}},  {\em Nucl.
  Phys. B} {\bf 293} (1987) 740--786.

\bibitem{EuropeanMuon:1992pyr}
{\bf European Muon} Collaboration, J.~Ashman et~al., {\it {A Measurement of the
  ratio of the nucleon structure function in copper and deuterium}},  {\em Z.
  Phys. C} {\bf 57} (1993) 211--218.

\bibitem{Gomez:1993ri}
J.~Gomez et~al., {\it {Measurement of the A-dependence of deep inelastic
  electron scattering}},  {\em Phys. Rev.} {\bf D49} (1994) 4348--4372.

\bibitem{Alde:1990im}
D.~M. Alde et~al., {\it {Nuclear dependence of dimuon production at 800-GeV.
  FNAL-772 experiment}},  {\em Phys. Rev. Lett.} {\bf 64} (1990) 2479--2482.

\bibitem{BCDMS:1987upi}
{\bf BCDMS} Collaboration, A.~C. Benvenuti et~al., {\it {Nuclear Effects in
  Deep Inelastic Muon Scattering on Deuterium and Iron Targets}},  {\em Phys.
  Lett. B} {\bf 189} (1987) 483--487.

\bibitem{E665:1995xur}
{\bf E665} Collaboration, M.~R. Adams et~al., {\it {Shadowing in inelastic
  scattering of muons on carbon, calcium and lead at low x(Bj)}},  {\em Z.
  Phys. C} {\bf 67} (1995) 403--410,
  [\href{http://arxiv.org/abs/hep-ex/9505006}{{\tt hep-ex/9505006}}].

\bibitem{FermilabE665:1992rbv}
{\bf Fermilab E665} Collaboration, M.~R. Adams et~al., {\it {Shadowing in the
  muon xenon inelastic scattering cross-section at 490-GeV}},  {\em Phys. Lett.
  B} {\bf 287} (1992) 375--380.

\bibitem{CHORUS:2005cpn}
{\bf CHORUS} Collaboration, G.~Onengut et~al., {\it {Measurement of nucleon
  structure functions in neutrino scattering}},  {\em Phys. Lett. B} {\bf 632}
  (2006) 65--75.

\bibitem{NuTeV:2001dfo}
{\bf NuTeV} Collaboration, M.~Goncharov et~al., {\it {Precise Measurement of
  Dimuon Production Cross-Sections in $\nu_{\mu}$ Fe and $\bar{\nu}_{\mu}$ Fe
  Deep Inelastic Scattering at the Tevatron.}},  {\em Phys. Rev. D} {\bf 64}
  (2001) 112006, [\href{http://arxiv.org/abs/hep-ex/0102049}{{\tt
  hep-ex/0102049}}].

\bibitem{ATLAS:2015mwq}
{\bf ATLAS} Collaboration, G.~Aad et~al., {\it {$Z$ boson production in $p+$Pb
  collisions at $\sqrt{s_{NN}}=5.02$ TeV measured with the ATLAS detector}},
  {\em Phys. Rev. C} {\bf 92} (2015), no.~4 044915,
  [\href{http://arxiv.org/abs/1507.06232}{{\tt arXiv:1507.06232}}].

\bibitem{CMS:2015zlj}
{\bf CMS} Collaboration, V.~Khachatryan et~al., {\it {Study of Z boson
  production in pPb collisions at $\sqrt {s_{NN}} = 5.02$ TeV}},  {\em Phys.
  Lett. B} {\bf 759} (2016) 36--57,
  [\href{http://arxiv.org/abs/1512.06461}{{\tt arXiv:1512.06461}}].

\bibitem{CMS:2015ehw}
{\bf CMS} Collaboration, V.~Khachatryan et~al., {\it {Study of W boson
  production in pPb collisions at $\sqrt{s_{\mathrm{NN}}} =$ 5.02 TeV}},  {\em
  Phys. Lett. B} {\bf 750} (2015) 565--586,
  [\href{http://arxiv.org/abs/1503.05825}{{\tt arXiv:1503.05825}}].

\bibitem{CMS:2019leu}
{\bf CMS} Collaboration, A.~M. Sirunyan et~al., {\it {Observation of nuclear
  modifications in W$^\pm$ boson production in pPb collisions at
  $\sqrt{s_\mathrm{NN}} =$ 8.16 TeV}},  {\em Phys. Lett. B} {\bf 800} (2020)
  135048, [\href{http://arxiv.org/abs/1905.01486}{{\tt arXiv:1905.01486}}].

\bibitem{NewMuon:1996uwk}
{\bf New Muon} Collaboration, M.~Arneodo et~al., {\it {Accurate measurement of
  F2(d) / F2(p) and R**d - R**p}},  {\em Nucl. Phys. B} {\bf 487} (1997) 3--26,
  [\href{http://arxiv.org/abs/hep-ex/9611022}{{\tt hep-ex/9611022}}].

\bibitem{Whitlow:1991uw}
L.~W. Whitlow, E.~M. Riordan, S.~Dasu, S.~Rock, and A.~Bodek, {\it {Precise
  measurements of the proton and deuteron structure functions from a global
  analysis of the SLAC deep inelastic electron scattering cross-sections}},
  {\em Phys. Lett.} {\bf B282} (1992) 475--482.

\bibitem{BCDMS:1989qop}
{\bf BCDMS} Collaboration, A.~C. Benvenuti et~al., {\it {A High Statistics
  Measurement of the Proton Structure Functions F(2) (x, Q**2) and R from Deep
  Inelastic Muon Scattering at High Q**2}},  {\em Phys. Lett. B} {\bf 223}
  (1989) 485--489.

\bibitem{NuSea:2001idv}
{\bf NuSea} Collaboration, R.~S. Towell et~al., {\it {Improved measurement of
  the anti-d / anti-u asymmetry in the nucleon sea}},  {\em Phys. Rev. D} {\bf
  64} (2001) 052002, [\href{http://arxiv.org/abs/hep-ex/0103030}{{\tt
  hep-ex/0103030}}].

\bibitem{Nagarajan:1991zz}
M.~A. Nagarajan and J.~P. Vary, {\it {Charge-exchange effects in elastic
  scattering with radioactive beams}},  {\em Phys. Rev. C} {\bf 43} (1991)
  281--284.

\bibitem{ALICE:2016rzo}
{\bf ALICE} Collaboration, J.~Adam et~al., {\it {W and Z boson production in
  p-Pb collisions at $\sqrt{s_{\rm NN}}$ = 5.02 TeV}},  {\em JHEP} {\bf 02}
  (2017) 077, [\href{http://arxiv.org/abs/1611.03002}{{\tt arXiv:1611.03002}}].

\bibitem{LHCb:2014jgh}
{\bf LHCb} Collaboration, R.~Aaij et~al., {\it {Observation of $Z$ production
  in proton-lead collisions at LHCb}},  {\em JHEP} {\bf 09} (2014) 030,
  [\href{http://arxiv.org/abs/1406.2885}{{\tt arXiv:1406.2885}}].

\bibitem{ALICE:2020jff}
{\bf ALICE} Collaboration, S.~Acharya et~al., {\it {Z-boson production in p-Pb
  collisions at $\sqrt{s_{\mathrm{NN}}}=8.16$ TeV and Pb-Pb collisions at
  $\sqrt{s_{\mathrm{NN}}}=5.02$ TeV}},  {\em JHEP} {\bf 09} (2020) 076,
  [\href{http://arxiv.org/abs/2005.11126}{{\tt arXiv:2005.11126}}].

\bibitem{CMS:2021ynu}
{\bf CMS} Collaboration, A.~M. Sirunyan et~al., {\it {Study of Drell-Yan dimuon
  production in proton-lead collisions at $\sqrt{s_\mathrm{NN}} =$ 8.16 TeV}},
  {\em JHEP} {\bf 05} (2021) 182, [\href{http://arxiv.org/abs/2102.13648}{{\tt
  arXiv:2102.13648}}].

\bibitem{Aaboud:2019tab}
{\bf ATLAS} Collaboration, M.~Aaboud et~al., {\it {Measurement of prompt photon
  production in $\sqrt{s_\mathrm{NN}} = 8.16$ TeV $p$+Pb collisions with
  ATLAS}},  {\em Phys. Lett. B} {\bf 796} (2019) 230--252,
  [\href{http://arxiv.org/abs/1903.02209}{{\tt arXiv:1903.02209}}].

\bibitem{Ball:2010gb}
{\bf The NNPDF} Collaboration, R.~D. Ball et~al., {\it {Reweighting NNPDFs: the
  W lepton asymmetry}},  {\em Nucl. Phys.} {\bf B849} (2011) 112--143,
  [\href{http://arxiv.org/abs/1012.0836}{{\tt arXiv:1012.0836}}].

\bibitem{Ball:2011gg}
R.~D. Ball, V.~Bertone, F.~Cerutti, L.~Del~Debbio, S.~Forte, et~al., {\it
  {Reweighting and Unweighting of Parton Distributions and the LHC W lepton
  asymmetry data}},  {\em Nucl.Phys.} {\bf B855} (2012) 608--638,
  [\href{http://arxiv.org/abs/1108.1758}{{\tt arXiv:1108.1758}}].

\bibitem{Bonvini:2015ira}
M.~Bonvini, S.~Marzani, J.~Rojo, L.~Rottoli, M.~Ubiali, R.~D. Ball, V.~Bertone,
  S.~Carrazza, and N.~P. Hartland, {\it {Parton distributions with threshold
  resummation}},  {\em JHEP} {\bf 09} (2015) 191,
  [\href{http://arxiv.org/abs/1507.01006}{{\tt arXiv:1507.01006}}].

\bibitem{Forte:2010ta}
S.~Forte, E.~Laenen, P.~Nason, and J.~Rojo, {\it {Heavy quarks in
  deep-inelastic scattering}},  {\em Nucl. Phys.} {\bf B834} (2010) 116--162,
  [\href{http://arxiv.org/abs/1001.2312}{{\tt arXiv:1001.2312}}].

\bibitem{Bertone:2016lga}
V.~Bertone, S.~Carrazza, and N.~P. Hartland, {\it {APFELgrid: a high
  performance tool for parton density determinations}},  {\em Comput. Phys.
  Commun.} {\bf 212} (2017) 205--209,
  [\href{http://arxiv.org/abs/1605.02070}{{\tt arXiv:1605.02070}}].

\bibitem{Bertone:2013vaa}
V.~Bertone, S.~Carrazza, and J.~Rojo, {\it {APFEL: A PDF Evolution Library with
  QED corrections}},  {\em Comput.Phys.Commun.} {\bf 185} (2014) 1647,
  [\href{http://arxiv.org/abs/1310.1394}{{\tt arXiv:1310.1394}}].

\bibitem{NNPDF:2019ubu}
{\bf NNPDF} Collaboration, R.~Abdul~Khalek et~al., {\it {Parton Distributions
  with Theory Uncertainties: General Formalism and First Phenomenological
  Studies}},  {\em Eur. Phys. J. C} {\bf 79} (2019), no.~11 931,
  [\href{http://arxiv.org/abs/1906.10698}{{\tt arXiv:1906.10698}}].

\bibitem{NNPDF:2019vjt}
{\bf NNPDF} Collaboration, R.~Abdul~Khalek et~al., {\it {A first determination
  of parton distributions with theoretical uncertainties}},  {\em Eur. Phys.
  J.} {\bf C} (2019) 79:838, [\href{http://arxiv.org/abs/1905.04311}{{\tt
  arXiv:1905.04311}}].

\bibitem{Campbell:1999ah}
J.~M. Campbell and R.~K. Ellis, {\it {An Update on vector boson pair production
  at hadron colliders}},  {\em Phys. Rev. D} {\bf 60} (1999) 113006,
  [\href{http://arxiv.org/abs/hep-ph/9905386}{{\tt hep-ph/9905386}}].

\bibitem{Campbell:2011bn}
J.~M. Campbell, R.~K. Ellis, and C.~Williams, {\it {Vector boson pair
  production at the LHC}},  {\em JHEP} {\bf 1107} (2011) 018,
  [\href{http://arxiv.org/abs/1105.0020}{{\tt arXiv:1105.0020}}].

\bibitem{Campbell:2015qma}
J.~M. Campbell, R.~K. Ellis, and W.~T. Giele, {\it {A Multi-Threaded Version of
  MCFM}},  {\em Eur. Phys. J. C} {\bf 75} (2015), no.~6 246,
  [\href{http://arxiv.org/abs/1503.06182}{{\tt arXiv:1503.06182}}].

\bibitem{Campbell:2018wfu}
J.~M. Campbell, J.~Rojo, E.~Slade, and C.~Williams, {\it {Direct photon
  production and PDF fits reloaded}},  {\em Eur. Phys. J.} {\bf C78} (2018),
  no.~6 470, [\href{http://arxiv.org/abs/1802.03021}{{\tt arXiv:1802.03021}}].

\bibitem{Nagy:2001fj}
Z.~Nagy, {\it {Three jet cross-sections in hadron hadron collisions at
  next-to-leading order}},  {\em Phys.Rev.Lett.} {\bf 88} (2002) 122003,
  [\href{http://arxiv.org/abs/hep-ph/0110315}{{\tt hep-ph/0110315}}].

\bibitem{Gehrmann-DeRidder:2019ibf}
A.~Gehrmann-De~Ridder, T.~Gehrmann, E.~W.~N. Glover, A.~Huss, and J.~Pires,
  {\it {Triple Differential Dijet Cross Section at the LHC}},  {\em Phys. Rev.
  Lett.} {\bf 123} (2019), no.~10 102001,
  [\href{http://arxiv.org/abs/1905.09047}{{\tt arXiv:1905.09047}}].

\bibitem{AbdulKhalek:2020jut}
R.~Abdul~Khalek et~al., {\it {Phenomenology of NNLO jet production at the LHC
  and its impact on parton distributions}},  {\em Eur. Phys. J. C} {\bf 80}
  (2020), no.~8 797, [\href{http://arxiv.org/abs/2005.11327}{{\tt
  arXiv:2005.11327}}].

\bibitem{LHCb:2016ikn}
{\bf LHCb} Collaboration, R.~Aaij et~al., {\it {Measurements of prompt charm
  production cross-sections in pp collisions at $ \sqrt{s}=5 $ TeV}},  {\em
  JHEP} {\bf 06} (2017) 147, [\href{http://arxiv.org/abs/1610.02230}{{\tt
  arXiv:1610.02230}}].

\bibitem{Gauld:2015lxa}
R.~Gauld, {\it {Forward $D$ predictions for $p\rm Pb$ collisions, and
  sensitivity to cold nuclear matter effects}},  {\em Phys. Rev. D} {\bf 93}
  (2016), no.~1 014001, [\href{http://arxiv.org/abs/1508.07629}{{\tt
  arXiv:1508.07629}}].

\bibitem{Nason:2004rx}
P.~Nason, {\it {A New method for combining NLO QCD with shower Monte Carlo
  algorithms}},  {\em JHEP} {\bf 11} (2004) 040,
  [\href{http://arxiv.org/abs/hep-ph/0409146}{{\tt hep-ph/0409146}}].

\bibitem{Frixione:2007vw}
S.~Frixione, P.~Nason, and C.~Oleari, {\it {Matching NLO QCD computations with
  Parton Shower simulations: the POWHEG method}},  {\em JHEP} {\bf 11} (2007)
  070, [\href{http://arxiv.org/abs/0709.2092}{{\tt arXiv:0709.2092}}].

\bibitem{Alioli:2010xd}
S.~Alioli, P.~Nason, C.~Oleari, and E.~Re, {\it {A general framework for
  implementing NLO calculations in shower Monte Carlo programs: the POWHEG
  BOX}},  {\em JHEP} {\bf 1006} (2010) 043,
  [\href{http://arxiv.org/abs/1002.2581}{{\tt arXiv:1002.2581}}].

\bibitem{Sjostrand:2014zea}
T.~Sj\"ostrand, S.~Ask, J.~R. Christiansen, R.~Corke, N.~Desai, P.~Ilten,
  S.~Mrenna, S.~Prestel, C.~O. Rasmussen, and P.~Z. Skands, {\it {An
  introduction to PYTHIA 8.2}},  {\em Comput. Phys. Commun.} {\bf 191} (2015)
  159--177, [\href{http://arxiv.org/abs/1410.3012}{{\tt arXiv:1410.3012}}].

\bibitem{Skands:2014pea}
P.~Skands, S.~Carrazza, and J.~Rojo, {\it {Tuning PYTHIA 8.1: the Monash 2013
  Tune}},  {\em European Physical Journal} {\bf 74} (2014) 3024,
  [\href{http://arxiv.org/abs/1404.5630}{{\tt arXiv:1404.5630}}].

\bibitem{Gauld:2016kpd}
R.~Gauld and J.~Rojo, {\it {Precision determination of the small-$x$ gluon from
  charm production at LHCb}},  {\em Phys. Rev. Lett.} {\bf 118} (2017), no.~7
  072001, [\href{http://arxiv.org/abs/1610.09373}{{\tt arXiv:1610.09373}}].

\bibitem{Eskola:2019bgf}
K.~J. Eskola, I.~Helenius, P.~Paakkinen, and H.~Paukkunen, {\it {A QCD analysis
  of LHCb D-meson data in p+Pb collisions}},  {\em JHEP} {\bf 05} (2020) 037,
  [\href{http://arxiv.org/abs/1906.02512}{{\tt arXiv:1906.02512}}].

\bibitem{Helenius:2018uul}
I.~Helenius and H.~Paukkunen, {\it {Revisiting the D-meson hadroproduction in
  general-mass variable flavour number scheme}},  {\em JHEP} {\bf 05} (2018)
  196, [\href{http://arxiv.org/abs/1804.03557}{{\tt arXiv:1804.03557}}].

\bibitem{LHCb:2013xam}
{\bf LHCb} Collaboration, R.~Aaij et~al., {\it {Prompt charm production in pp
  collisions at sqrt(s)=7 TeV}},  {\em Nucl. Phys. B} {\bf 871} (2013) 1--20,
  [\href{http://arxiv.org/abs/1302.2864}{{\tt arXiv:1302.2864}}].

\bibitem{LHCb:2015swx}
{\bf LHCb} Collaboration, R.~Aaij et~al., {\it {Measurements of prompt charm
  production cross-sections in $pp$ collisions at $ \sqrt{s}=13 $ TeV}},  {\em
  JHEP} {\bf 03} (2016) 159, [\href{http://arxiv.org/abs/1510.01707}{{\tt
  arXiv:1510.01707}}]. [Erratum: JHEP 09, 013 (2016), Erratum: JHEP 05, 074
  (2017)].

\bibitem{Zenaiev:2015rfa}
{\bf PROSA} Collaboration, O.~Zenaiev et~al., {\it {Impact of heavy-flavour
  production cross sections measured by the LHCb experiment on parton
  distribution functions at low x}},  {\em Eur. Phys. J.} {\bf C75} (2015),
  no.~8 396, [\href{http://arxiv.org/abs/1503.04581}{{\tt arXiv:1503.04581}}].

\bibitem{Carrazza:2021yrg}
S.~Carrazza, J.~M. Cruz-Martinez, and R.~Stegeman, {\it {A data-based
  parametrization of parton distribution functions}},
  \href{http://arxiv.org/abs/2111.02954}{{\tt arXiv:2111.02954}}.

\bibitem{Ball:2009qv}
{\bf The NNPDF} Collaboration, R.~D. Ball et~al., {\it {Fitting Parton
  Distribution Data with Multiplicative Normalization Uncertainties}},  {\em
  JHEP} {\bf 05} (2010) 075, [\href{http://arxiv.org/abs/0912.2276}{{\tt
  arXiv:0912.2276}}].

\bibitem{Ball:2012wy}
R.~D. Ball, S.~Carrazza, L.~Del~Debbio, S.~Forte, J.~Gao, et~al., {\it {Parton
  Distribution Benchmarking with LHC Data}},  {\em JHEP} {\bf 1304} (2013) 125,
  [\href{http://arxiv.org/abs/1211.5142}{{\tt arXiv:1211.5142}}].

\bibitem{Faura:2020oom}
F.~Faura, S.~Iranipour, E.~R. Nocera, J.~Rojo, and M.~Ubiali, {\it The
  strangest proton?},  {\em Eur.Phys.J.} {\bf C80} (2020) 1168,
  [\href{http://arxiv.org/abs/2009.00014}{{\tt arXiv:2009.00014}}].

\bibitem{10.1214/aoms/1177728190}
M.~Rosenblatt, {\it {Remarks on Some Nonparametric Estimates of a Density
  Function}},  {\em The Annals of Mathematical Statistics} {\bf 27} (1956),
  no.~3 832 -- 837.

\bibitem{10.1214/aoms/1177704472}
E.~Parzen, {\it {On Estimation of a Probability Density Function and Mode}},
  {\em The Annals of Mathematical Statistics} {\bf 33} (1962), no.~3 1065 --
  1076.

\bibitem{Eskola:2021nhw}
K.~J. Eskola, P.~Paakkinen, H.~Paukkunen, and C.~A. Salgado, {\it {EPPS21: A
  global QCD analysis of nuclear PDFs}},
  \href{http://arxiv.org/abs/2112.12462}{{\tt arXiv:2112.12462}}.

\bibitem{Kusina:2017gkz}
A.~Kusina, J.-P. Lansberg, I.~Schienbein, and H.-S. Shao, {\it {Gluon Shadowing
  in Heavy-Flavor Production at the LHC}},  {\em Phys. Rev. Lett.} {\bf 121}
  (2018), no.~5 052004, [\href{http://arxiv.org/abs/1712.07024}{{\tt
  arXiv:1712.07024}}].

\bibitem{Arneodo:1996kd}
{\bf New Muon} Collaboration, M.~Arneodo et~al., {\it {Accurate measurement of
  $F_2^d/F_2^p$ and $R_d-R_p$}},  {\em Nucl. Phys.} {\bf B487} (1997) 3--26,
  [\href{http://arxiv.org/abs/hep-ex/9611022}{{\tt hep-ex/9611022}}].

\bibitem{Arneodo:1996qe}
{\bf New Muon} Collaboration, M.~Arneodo et~al., {\it {Measurement of the
  proton and deuteron structure functions, $F_2^p$ and $F_2^d$, and of the
  ratio $\sigma_L/\sigma_T$}},  {\em Nucl. Phys.} {\bf B483} (1997) 3--43,
  [\href{http://arxiv.org/abs/hep-ph/9610231}{{\tt hep-ph/9610231}}].

\bibitem{Benvenuti:1989rh}
{\bf BCDMS} Collaboration, A.~C. Benvenuti et~al., {\it {A High Statistics
  Measurement of the Proton Structure Functions $F_2(x, Q^2)$ and $R$ from Deep
  Inelastic Muon Scattering at High $Q^2$}},  {\em Phys. Lett.} {\bf B223}
  (1989) 485.

\bibitem{Amaudruz:1995tq}
{\bf New Muon} Collaboration, P.~Amaudruz et~al., {\it {A Reevaluation of the
  nuclear structure function ratios for D, He, Li-6, C and Ca}},  {\em
  Nucl.Phys.} {\bf B441} (1995) 3--11,
  [\href{http://arxiv.org/abs/hep-ph/9503291}{{\tt hep-ph/9503291}}].

\bibitem{Arneodo:1995cs}
{\bf New Muon} Collaboration, M.~Arneodo et~al., {\it {The Structure Function
  ratios F2(li) / F2(D) and F2(C) / F2(D) at small x}},  {\em Nucl.Phys.} {\bf
  B441} (1995) 12--30, [\href{http://arxiv.org/abs/hep-ex/9504002}{{\tt
  hep-ex/9504002}}].

\bibitem{Arneodo:1996rv}
{\bf New Muon} Collaboration, M.~Arneodo et~al., {\it {The A dependence of the
  nuclear structure function ratios}},  {\em Nucl.Phys.} {\bf B481} (1996)
  3--22.

\bibitem{Ashman:1988bf}
{\bf European Muon} Collaboration, J.~Ashman et~al., {\it {Measurement of the
  Ratios of Deep Inelastic Muon - Nucleus Cross-Sections on Various Nuclei
  Compared to Deuterium}},  {\em Phys. Lett.} {\bf B202} (1988) 603--610.

\bibitem{Arneodo:1989sy}
{\bf European Muon} Collaboration, M.~Arneodo et~al., {\it {Measurements of the
  nucleon structure function in the range $0.002-{\rm GeV}^2 < x < 0.17-{\rm
  GeV}^2$ and $0.2-GeV^2 < q^2 < 8-GeV^2$ in deuterium, carbon and calcium}},
  {\em Nucl. Phys.} {\bf B333} (1990) 1--47.

\bibitem{Adams:1995is}
{\bf E665} Collaboration, M.~R. Adams et~al., {\it {Shadowing in inelastic
  scattering of muons on carbon, calcium and lead at low x(Bj)}},  {\em Z.
  Phys.} {\bf C67} (1995) 403--410,
  [\href{http://arxiv.org/abs/hep-ex/9505006}{{\tt hep-ex/9505006}}].

\bibitem{Aubert:1987da}
{\bf European Muon} Collaboration, J.~J. Aubert et~al., {\it {Measurements of
  the nucleon structure functions $F2_n$ in deep inelastic muon scattering from
  deuterium and comparison with those from hydrogen and iron}},  {\em Nucl.
  Phys.} {\bf B293} (1987) 740--786.

\bibitem{Benvenuti:1987az}
{\bf BCDMS} Collaboration, A.~C. Benvenuti et~al., {\it {Nuclear Effects in
  Deep Inelastic Muon Scattering on Deuterium and Iron Targets}},  {\em Phys.
  Lett.} {\bf B189} (1987) 483--487.

\bibitem{Ashman:1992kv}
{\bf European Muon} Collaboration, J.~Ashman et~al., {\it {A Measurement of the
  ratio of the nucleon structure function in copper and deuterium}},  {\em
  Z.Phys.} {\bf C57} (1993) 211--218.

\bibitem{Arneodo:1996ru}
{\bf New Muon} Collaboration, M.~Arneodo et~al., {\it {The $Q^2$ dependence of
  the structure function ratio $F_2^{\rm Sn}/F_2^{C}$ in deep inelastic muon
  scattering}},  {\em Nucl.Phys.} {\bf B481} (1996) 23--39.

\bibitem{Adams:1992vm}
{\bf Fermilab E665} Collaboration, M.~R. Adams et~al., {\it {Shadowing in the
  muon xenon inelastic scattering cross-section at 490-GeV}},  {\em Phys.
  Lett.} {\bf B287} (1992) 375--380.

\bibitem{Goncharov:2001qe}
{\bf NuTeV} Collaboration, M.~Goncharov et~al., {\it {Precise measurement of
  dimuon production cross-sections in $\nu_{\mu}$Fe and $\bar{\nu}_{\mu}$Fe
  deep inelastic scattering at the Tevatron}},  {\em Phys. Rev.} {\bf D64}
  (2001) 112006, [\href{http://arxiv.org/abs/hep-ex/0102049}{{\tt
  hep-ex/0102049}}].

\bibitem{Onengut:2005kv}
{\bf CHORUS} Collaboration, G.~Onengut et~al., {\it {Measurement of nucleon
  structure functions in neutrino scattering}},  {\em Phys. Lett.} {\bf B632}
  (2006) 65--75.

\bibitem{Moreno:1990sf}
G.~Moreno et~al., {\it {Dimuon production in proton - copper collisions at
  $\sqrt{s}$ = 38.8-GeV}},  {\em Phys. Rev.} {\bf D43} (1991) 2815--2836.

\bibitem{Webb:2003ps}
{\bf NuSea} Collaboration, J.~C. Webb et~al., {\it {Absolute Drell-Yan dimuon
  cross sections in 800-GeV/c p p and p d collisions}},
  \href{http://arxiv.org/abs/hep-ex/0302019}{{\tt hep-ex/0302019}}.

\bibitem{Webb:2003bj}
J.~C. Webb, {\it {Measurement of continuum dimuon production in 800-GeV/c
  proton nucleon collisions}},  \href{http://arxiv.org/abs/hep-ex/0301031}{{\tt
  hep-ex/0301031}}.

\bibitem{Towell:2001nh}
{\bf FNAL E866/NuSea} Collaboration, R.~S. Towell et~al., {\it {Improved
  measurement of the anti-d/anti-u asymmetry in the nucleon sea}},  {\em Phys.
  Rev.} {\bf D64} (2001) 052002,
  [\href{http://arxiv.org/abs/hep-ex/0103030}{{\tt hep-ex/0103030}}].

\bibitem{Aad:2015gta}
{\bf ATLAS} Collaboration, G.~Aad et~al., {\it {$Z$ boson production in $p+$Pb
  collisions at $\sqrt{s_{NN}}=5.02$ TeV measured with the ATLAS detector}},
  {\em Phys. Rev.} {\bf C92} (2015), no.~4 044915,
  [\href{http://arxiv.org/abs/1507.06232}{{\tt arXiv:1507.06232}}].

\bibitem{Khachatryan:2015pzs}
{\bf CMS} Collaboration, V.~Khachatryan et~al., {\it {Study of Z boson
  production in pPb collisions at $\sqrt {s NN }$=5.02TeV}},  {\em Phys. Lett.}
  {\bf B759} (2016) 36--57, [\href{http://arxiv.org/abs/1512.06461}{{\tt
  arXiv:1512.06461}}].

\bibitem{Khachatryan:2015hha}
{\bf CMS} Collaboration, V.~Khachatryan et~al., {\it {Study of W boson
  production in pPb collisions at $\sqrt{s_{\mathrm{NN}}} =$ 5.02 TeV}},  {\em
  Phys. Lett.} {\bf B750} (2015) 565--586,
  [\href{http://arxiv.org/abs/1503.05825}{{\tt arXiv:1503.05825}}].

\bibitem{Sirunyan:2019dox}
{\bf CMS} Collaboration, A.~M. Sirunyan et~al., {\it {Observation of nuclear
  modifications in W$^\pm$ boson production in pPb collisions at
  $\sqrt{s_\mathrm{NN}} =$ 8.16 TeV}},  {\em Phys. Lett.} {\bf B800} (2020)
  135048, [\href{http://arxiv.org/abs/1905.01486}{{\tt arXiv:1905.01486}}].

\bibitem{Anastasiou:2003yy}
C.~Anastasiou, L.~J. Dixon, K.~Melnikov, and F.~Petriello, {\it {Dilepton
  rapidity distribution in the Drell-Yan process at NNLO in QCD}},  {\em Phys.
  Rev. Lett.} {\bf 91} (2003) 182002,
  [\href{http://arxiv.org/abs/hep-ph/0306192}{{\tt hep-ph/0306192}}].

\bibitem{Kusina:2016fxy}
A.~Kusina, F.~Lyonnet, D.~B. Clark, E.~Godat, T.~Jezo, K.~Kovarik, F.~I.
  Olness, I.~Schienbein, and J.~Y. Yu, {\it {Vector boson production in pPb and
  PbPb collisions at the LHC and its impact on nCTEQ15 PDFs}},  {\em Eur. Phys.
  J.} {\bf C77} (2017), no.~7 488, [\href{http://arxiv.org/abs/1610.02925}{{\tt
  arXiv:1610.02925}}].

\bibitem{Kusina:2020lyz}
A.~Kusina et~al., {\it {Impact of LHC vector boson production in heavy ion
  collisions on strange PDFs}},  {\em Eur. Phys. J. C} {\bf 80} (2020), no.~10
  968, [\href{http://arxiv.org/abs/2007.09100}{{\tt arXiv:2007.09100}}].

\bibitem{Kovarik:2015cma}
K.~Kovarik et~al., {\it {nCTEQ15 - Global analysis of nuclear parton
  distributions with uncertainties in the CTEQ framework}},  {\em Phys. Rev.}
  {\bf D93} (2016), no.~8 085037, [\href{http://arxiv.org/abs/1509.00792}{{\tt
  arXiv:1509.00792}}].

\bibitem{deFlorian:2003qf}
D.~de~Florian and R.~Sassot, {\it {Nuclear parton distributions at next to
  leading order}},  {\em Phys. Rev.} {\bf D69} (2004) 074028,
  [\href{http://arxiv.org/abs/hep-ph/0311227}{{\tt hep-ph/0311227}}].

\bibitem{deFlorian:2011fp}
D.~de~Florian, R.~Sassot, P.~Zurita, and M.~Stratmann, {\it {Global Analysis of
  Nuclear Parton Distributions}},  {\em Phys. Rev.} {\bf D85} (2012) 074028,
  [\href{http://arxiv.org/abs/1112.6324}{{\tt arXiv:1112.6324}}].

\bibitem{Eskola:2008ca}
K.~J. Eskola, H.~Paukkunen, and C.~A. Salgado, {\it {An Improved global
  analysis of nuclear parton distribution functions including RHIC data}},
  {\em JHEP} {\bf 07} (2008) 102, [\href{http://arxiv.org/abs/0802.0139}{{\tt
  arXiv:0802.0139}}].

\bibitem{Eskola:2009uj}
K.~J. Eskola, H.~Paukkunen, and C.~A. Salgado, {\it {EPS09: A New Generation of
  NLO and LO Nuclear Parton Distribution Functions}},  {\em JHEP} {\bf 04}
  (2009) 065, [\href{http://arxiv.org/abs/0902.4154}{{\tt arXiv:0902.4154}}].

\bibitem{Walt:2019slu}
M.~Walt, I.~Helenius, and W.~Vogelsang, {\it {Open-source QCD analysis of
  nuclear parton distribution functions at NLO and NNLO}},  {\em Phys. Rev.}
  {\bf D100} (2019), no.~9 096015, [\href{http://arxiv.org/abs/1908.03355}{{\tt
  arXiv:1908.03355}}].

\bibitem{Helenius:2021tof}
I.~Helenius, M.~Walt, and W.~Vogelsang, {\it {TUJU21: NNLO nuclear parton
  distribution functions with electroweak-boson production data from the LHC}},
   \href{http://arxiv.org/abs/2112.11904}{{\tt arXiv:2112.11904}}.

\bibitem{Eskola:2019dui}
K.~J. Eskola, P.~Paakkinen, and H.~Paukkunen, {\it {Non-quadratic improved
  Hessian PDF reweighting and application to CMS dijet measurements at 5.02
  TeV}},  {\em Eur. Phys. J.} {\bf C79} (2019), no.~6 511,
  [\href{http://arxiv.org/abs/1903.09832}{{\tt arXiv:1903.09832}}].

\bibitem{Paukkunen:2020rnb}
H.~Paukkunen and P.~Zurita, {\it {Can we fit nuclear PDFs with the high-x CLAS
  data?}},  {\em Eur. Phys. J. C} {\bf 80} (2020), no.~5 381,
  [\href{http://arxiv.org/abs/2003.02195}{{\tt arXiv:2003.02195}}].

\bibitem{Segarra:2020gtj}
E.~P. Segarra et~al., {\it {Extending nuclear PDF analyses into the high-$x$ ,
  low-$Q^2$ region}},  {\em Phys. Rev. D} {\bf 103} (2021), no.~11 114015,
  [\href{http://arxiv.org/abs/2012.11566}{{\tt arXiv:2012.11566}}].

\bibitem{Rojo:2014kta}
J.~Rojo, {\it {Constraints on parton distributions and the strong coupling from
  LHC jet data}},  {\em Int. J. Mod. Phys.} {\bf A30} (2015) 1546005,
  [\href{http://arxiv.org/abs/1410.7728}{{\tt arXiv:1410.7728}}].

\bibitem{GRAND:2018iaj}
{\bf GRAND} Collaboration, J.~\'Alvarez-Mu\~niz et~al., {\it {The Giant Radio
  Array for Neutrino Detection (GRAND): Science and Design}},  {\em Sci. China
  Phys. Mech. Astron.} {\bf 63} (2020), no.~1 219501,
  [\href{http://arxiv.org/abs/1810.09994}{{\tt arXiv:1810.09994}}].

\bibitem{Olinto:2017xbi}
A.~V. Olinto et~al., {\it {POEMMA: Probe Of Extreme Multi-Messenger
  Astrophysics}},  {\em PoS} {\bf ICRC2017} (2018) 542,
  [\href{http://arxiv.org/abs/1708.07599}{{\tt arXiv:1708.07599}}].

\bibitem{Denton:2020jft}
P.~B. Denton and Y.~Kini, {\it {Ultra-High-Energy Tau Neutrino Cross Sections
  with GRAND and POEMMA}},  {\em Phys. Rev. D} {\bf 102} (2020) 123019,
  [\href{http://arxiv.org/abs/2007.10334}{{\tt arXiv:2007.10334}}].

\bibitem{Gauld:2019pgt}
R.~Gauld, {\it {Precise predictions for multi-TeV and PeV energy neutrino
  scattering rates}},  {\em Phys. Rev. D} {\bf 100} (2019), no.~9 091301,
  [\href{http://arxiv.org/abs/1905.03792}{{\tt arXiv:1905.03792}}].

\bibitem{Ball:2011mu}
{\bf {The NNPDF }} Collaboration, R.~D. Ball et~al., {\it {Impact of Heavy
  Quark Masses on Parton Distributions and LHC Phenomenology}},  {\em Nucl.
  Phys.} {\bf B849} (2011) 296, [\href{http://arxiv.org/abs/1101.1300}{{\tt
  arXiv:1101.1300}}].

\bibitem{Gauld:2021zmq}
R.~Gauld, {\it {A massive variable flavour number scheme for the Drell-Yan
  process}},  {\em SciPost Phys.} {\bf 12} (2022), no.~1 024,
  [\href{http://arxiv.org/abs/2107.01226}{{\tt arXiv:2107.01226}}].

\bibitem{Seckel:1997kk}
D.~Seckel, {\it {Neutrino photon reactions in astrophysics and cosmology}},
  {\em Phys. Rev. Lett.} {\bf 80} (1998) 900--903,
  [\href{http://arxiv.org/abs/hep-ph/9709290}{{\tt hep-ph/9709290}}].

\bibitem{Alikhanov:2015kla}
I.~Alikhanov, {\it {Hidden Glashow resonance in neutrino\textendash{}nucleus
  collisions}},  {\em Phys. Lett. B} {\bf 756} (2016) 247--253,
  [\href{http://arxiv.org/abs/1503.08817}{{\tt arXiv:1503.08817}}].

\bibitem{Zhou:2019vxt}
B.~Zhou and J.~F. Beacom, {\it {Neutrino-nucleus cross sections for W-boson and
  trident production}},  {\em Phys. Rev. D} {\bf 101} (2020), no.~3 036011,
  [\href{http://arxiv.org/abs/1910.08090}{{\tt arXiv:1910.08090}}].

\bibitem{Zhou:2019frk}
B.~Zhou and J.~F. Beacom, {\it {W-boson and trident production in
  TeV\textendash{}PeV neutrino observatories}},  {\em Phys. Rev. D} {\bf 101}
  (2020), no.~3 036010, [\href{http://arxiv.org/abs/1910.10720}{{\tt
  arXiv:1910.10720}}].

\bibitem{Paakkinen:2021jjp}
P.~Paakkinen, {\it {Light-nuclei gluons from dijet production in proton-oxygen
  collisions}},  \href{http://arxiv.org/abs/2111.05368}{{\tt
  arXiv:2111.05368}}.

\bibitem{Ethier:2017zbq}
J.~Ethier, N.~Sato, and W.~Melnitchouk, {\it {First simultaneous extraction of
  spin-dependent parton distributions and fragmentation functions from a global
  QCD analysis}},  {\em Phys. Rev. Lett.} {\bf 119} (2017), no.~13 132001,
  [\href{http://arxiv.org/abs/1705.05889}{{\tt arXiv:1705.05889}}].

\bibitem{Sato:2019yez}
{\bf JAM} Collaboration, N.~Sato, C.~Andres, J.~Ethier, and W.~Melnitchouk,
  {\it {Strange quark suppression from a simultaneous Monte Carlo analysis of
  parton distributions and fragmentation functions}},  {\em Phys. Rev. D} {\bf
  101} (2020), no.~7 074020, [\href{http://arxiv.org/abs/1905.03788}{{\tt
  arXiv:1905.03788}}].

\bibitem{Moffat:2021dji}
{\bf Jefferson Lab Angular Momentum (JAM)} Collaboration, E.~Moffat,
  W.~Melnitchouk, T.~C. Rogers, and N.~Sato, {\it {Simultaneous Monte~Carlo
  analysis of parton densities and fragmentation functions}},  {\em Phys. Rev.
  D} {\bf 104} (2021), no.~1 016015,
  [\href{http://arxiv.org/abs/2101.04664}{{\tt arXiv:2101.04664}}].

\bibitem{Albacete:2013ei}
J.~L. Albacete et~al., {\it {Predictions for $p+$Pb Collisions at sqrt s\_NN =
  5 TeV}},  {\em Int. J. Mod. Phys. E} {\bf 22} (2013) 1330007,
  [\href{http://arxiv.org/abs/1301.3395}{{\tt arXiv:1301.3395}}].

\bibitem{HADJIDAKIS20211}
C.~H. et~al., {\it A fixed-target programme at the lhc: Physics case and
  projected performances for heavy-ion, hadron, spin and astroparticle
  studies},  {\em Physics Reports} {\bf 911} (2021) 1--83. A Fixed-Target
  Programme at the LHC: Physics Case and Projected Performances for Heavy-Ion,
  Hadron, Spin and Astroparticle Studies.

\end{thebibliography}\endgroup

\end{document}